%% file: main.tex
\documentclass[12pt]{iopart}
\usepackage{comment,mfirstuc}
\usepackage{graphicx}% Include figure files
\usepackage{grffile}
\usepackage{algorithm}% http://ctan.org/pkg/algorithm
\usepackage{algpseudocode}% http://ctan.org/pkg/algorithmicx
\usepackage{hyperref}% add hypertext capabilities
\usepackage{soul}
\usepackage{listings}
\usepackage{multirow}
\usepackage{dcolumn}% Align table columns on decimal point
\usepackage{bm}% bold math
\usepackage{placeins}
\expandafter\let\csname equation*\endcsname\relax
\expandafter\let\csname endequation*\endcsname\relax
\usepackage{amsmath}
\usepackage{amssymb}
\usepackage{mathtools}
\usepackage[bottom=1.0in]{geometry}
\usepackage{cite}
\usepackage{lineno}
\RequirePackage{xcolor}
\usepackage[misc]{ifsym}
\usepackage{lipsum} 
\usepackage{subcaption}
\usepackage{tcolorbox} % Create coloured boxes (e.g. the one for the key-words)
\usepackage{stfloats} % Correct position of the tables

\usepackage{siunitx}
\DeclareSIUnit\eVperc{\eV\per\clight}
\DeclareSIUnit\clight{\text{\ensuremath{c}}}

\usepackage{xspace}
\usepackage{colortbl}
\usepackage{array}
\newcolumntype{P}[1]{>{\centering\arraybackslash}p{#1}}
\newcolumntype{M}[1]{>{\centering\arraybackslash}m{#1}}
\hypersetup{ 
    pdfnewwindow=true,      % links in new window
    colorlinks=true,       % false: boxed links; true: colored links
    linkcolor=blue,         % color of internal links
    citecolor=blue,        % color of links to bibliography
    filecolor=blue,      % color of file links
    urlcolor=blue        % color of external links
}  

\algblock{Input}{EndInput}
\algnotext{EndInput}
\algblock{Output}{EndOutput}
\algnotext{EndOutput}

\def\be{\begin{eqnarray} &&} 
 
\def\ee{\end{eqnarray}}

\makeatletter
\newcommand{\mainmatter}{%
  \setcounter{footnote}{0}%
  \patchcmd{\@makefntext}{\fnsymbol}{\arabic}{}{}%
  \patchcmd{\@thefnmark}{\fnsymbol}{\arabic}{}{}%
  \def\@makefnmark{\textsuperscript{\arabic{footnote}}}
  \long\def\@makefntext##1{\parindent 1em\noindent
        \hb@xt@1.8em{%
            \hss\@textsuperscript{\normalfont\@thefnmark}}##1}%
}
\makeatother

\newcommand{\addComment}[2]{
  \expandafter\newcommand\csname #1\endcsname[1]{{\bf \color{#2} \capitalisewords{#1}:\,##1}}
  \expandafter\newcommand\csname #1cor\endcsname[2]{{\color{#2} \capitalisewords{#1}:\,\st{##1}{\bf ##2}}}
  \expandafter\newcommand\csname #1color\endcsname{#2}
}
\addComment{cris}{blue} 
\addComment{james}{red}
\addComment{mike}{magenta}

\newcommand{\gluex}{\textsc{GlueX}\xspace} 
\newcommand{\geant}{\textsc{Geant4}\xspace} 
\newcommand{\D}{\mathrm{d}}

\begin{document}

% Keywords command
\providecommand{\keywords}[1]
{
  \small
  \textbf{Keywords:}  {\color{blue}#1
  }
}

%\title[\scriptsize{Fast Simulation for the hpDIRC at the Electron Ion Collider}]{Fast Simulation for the hpDIRC at the Electron Ion Collider} 

\title[\scriptsize{Generative Models for Fast Simulation of Cherenkov Detectors at the Electron-Ion Collider}]{Generative Models for Fast Simulation of Cherenkov Detectors at the Electron-Ion Collider}

\author{J. Giroux$^{1,\star}$, M. Martinez$^{1}$, C. Fanelli$^{1,\star}$} 

\address{
$^{1}$ William \& Mary, Department of Data Science, Williamsburg, VA 23185, USA\\
$^{\star}$ Author to whom any correspondence should be addressed.
}

\ead{{\color{blue}
jgiroux@wm.edu,
cfanelli@wm.edu
}}

\vspace{10pt}
\begin{indented}
\item[]\today
\end{indented}

%\linenumbers
\begin{abstract}
\begin{comment}
The integration of Deep Learning (DL) in experimental nuclear and particle physics has led to significant advancements in simulation and reconstruction pipelines. However, the computational complexity of traditional simulation tools, such as \geant, presents challenges in generating large datasets, particularly for Cherenkov detectors.  These detectors represent a major bottleneck in simulation pipelines, as they require the tracking of a large number of optical photons through complex geometries and reflective surfaces—making the process highly resource-intensive.
To address this, we present a fully open fast simulation tool for the Detection of Internally Reflected Cherenkov Light (DIRC) detectors, specifically the High Performance DIRC (hpDIRC) at the Electron Ion Collider (EIC). We provide a suite of generative models designed to accelerate Particle Identification (PID) tasks by offering a more efficient alternative to \geant simulations, leveraging the high parallelization capabilities of GPUs. Our approach provides a user-friendly simulation package that allows DL experts and physicists to generate data on demand, overcoming barriers imposed by complex physics software. 
This standalone simulation allows to generate data that can be used to develop new DL-inspired PID methods and benchmark their performance. 
In addition, our fast simulation pipeline is the first step towards enabling PID methods developed by physicists at EIC that rely on potentially unlimited simulated samples over the entire acceptance range of the detector.
\end{comment}
%
The integration of Deep Learning (DL) into experimental nuclear and particle physics has driven significant progress in simulation and reconstruction workflows. However, traditional simulation frameworks such as \geant remain computationally intensive, especially for Cherenkov detectors, where simulating optical photon transport through complex geometries and reflective surfaces introduces a major bottleneck.
To address this, we present an open, standalone fast simulation tool for Detection of Internally Reflected Cherenkov Light (DIRC) detectors, with a focus on the High-Performance DIRC (hpDIRC) at the future Electron-Ion Collider (EIC). Our framework incorporates a suite of generative models tailored to accelerate particle identification (PID) tasks by offering a scalable, GPU-accelerated alternative to full \geant-based simulations.
Designed with accessibility in mind, our simulation package enables both DL researchers and physicists to efficiently generate high-fidelity large-scale datasets on demand, without relying on complex traditional simulation stacks. This flexibility supports the development and benchmarking of novel DL-driven PID methods.
Moreover, this fast simulation pipeline represents a critical step toward enabling EIC-wide PID strategies that depend on virtually unlimited simulated samples, spanning the full acceptance of the hpDIRC.
\end{abstract}

% \begin{abstract}
% We introduce a suite of generative models developed for the High Performance DIRC (hpDIRC) simulation framework at the Electron Ion Collider (EIC). The primary goal of this work is to provide an open-source, fast simulation tool for Pion and Kaon ($\pi^{\pm}/\mathcal{K}^{\pm}$)  detector response in Detection of Internally Reflected Cherenkov Light (DIRC) detectors. Our approach includes the development of user-friendly simulation software that facilitates the integration of Deep Learning (DL) techniques, enabling both physicists and DL experts to leverage the power of generative models in particle identification (PID). Additionally, we address the challenge of data scalability by allowing the generation of datasets on demand, which is crucial for the diverse needs of DL research. This framework is equipped with a comprehensive suite of state-of-the-art generative models and provides benchmarking tools, making it a valuable resource for the physics community. Ongoing updates will incorporate the latest advancements in both simulation and DL methodologies, ensuring continuous improvement in PID performance.
% \end{abstract}


%\maketitle

\mainmatter

\input{1_introduction}

\input{2_data}

\input{3_architecture}

\input{4_analysis}

\input{5_impacts}

%\clearpage

\input{6_summary}

\section*{Code Availability}
The code is publicly available at \href{https://github.com/wmdataphys/Deep\_hpDIRC}{https://github.com/wmdataphys/Deep\_hpDIRC}.

%\clearpage

\section*{Acknowledgments}

% this is alphabetical reflecting author list
%
We thank William \& Mary for supporting the work of CF and JG through CF's start-up funding. 
We gratefully acknowledge Roman Dzhygadlo for his valuable guidance on implementing and utilizing the standalone Geant4 simulation of the hpDIRC. 
We also thank Justin Stevens for insightful discussions and contributions regarding DIRC detector concepts.
%
%\cris{Thank Roman Dzyghadlo}
%
The authors acknowledge  William  \&  Mary  Research  Computing for providing computational resources and technical support that have contributed to the results reported within this article.

\clearpage
\section*{References}
\bibliographystyle{iopart-num}
\bibliography{biblio}

\clearpage
\input{Appendix}

\end{document}

%% file: 1_introduction.tex
\section{Introduction}\label{sec:intro}

The fields of experimental nuclear and particle physics are increasingly embracing the integration of Deep Learning (DL) into existing reconstruction pipelines. A key component of these pipelines is the reliance on simulated data generated by \geant. \geant has inherent computational complexity that cannot be reduced, and as a result a high computational budget must be expected to produce large volumes of data. 
Moreover, full simulation of sub-detector systems and their underlying physical processes, can involve excessive stepping action within \geant causing non-linear time complexity to arise over different sub-detector systems. 
A prime example of such a case is the simulation of Cherenkov detector systems. In particular, the simulation of photon propagation within a Detection of Internally Reflected Cherenkov Light (DIRC) detectors, presents a significant challenge due to the intricate reflection patterns that occur as photons travel through individual bars. Each Cherenkov photon undergoes multiple total internal reflections along the length of the bar, bouncing off the internal surfaces at various angles before eventually reaching the readout system. These reflections depend on the photon emission angle, the refractive index of the medium, and any imperfections or surface treatments applied to the bar. As a result, even a single photon’s trajectory becomes a complex sequence of reflections, refractions, and potential absorption or scattering events. Given that hundreds of photons can be produced per charged particle, the computational burden scales combinatorially with the number of emitted photons and their unique propagation paths, making full \geant simulations of these detectors prohibitively expensive for large-scale applications.

To combat this, the usage of Generative Artificial Intelligence has become increasingly popular. These algorithms are flexible and can be deployed in a range of simulations tasks such as jet physics \cite{Mikuni_2023,araz2024point,mikuni2024omnilearn,birk2024omnijet,birk2025flow,Buhmann_2023}, calorimeter simulations \cite{birk2025omnijet,Krause_2023,Krause_2023_accel,favaro2025calodream,Paganini_2018,Mikuni_2022,Diefenbacher_2023,fanelli2022flux+,devlin2024diffusion} and Cherenkov detectors simulations \cite{fanelli2024deep,maevskiy2020fast}. These methods of fast simulation have provided invaluable speed ups over \geant given their capability to utilize the high parallelization potential of Graphical Processing Units (GPUs), in contrast to the Central Processing Unit (CPU) bound nature of \geant. 
In their current form, these models may not serve as direct replacements for \geant, as they exhibit limitations in achieving uniform reliability across the full phase space.\footnote{Generative models often struggle in low-density regions of the training distribution, potentially introducing biases in applications that demand high-precision datasets.} Moreover, at least one full \geant-based simulation is always required to generate the initial training data necessary for the models to learn the underlying physics and detector response.

However, these models directly serve the purpose of expediting research of both classical and DL based reconstruction methods, \textit{e.g.}, Particle Identification (PID) tasks. Modern DL approaches require large corpus of data, which can be more efficiently simulated through the usage of generative models on GPUs. This also allows users with limited computational resources, \textit{i.e.,} those without access to hundreds of CPU cores, to produce their own data in reasonable amounts of time on a singular (or multiple if available) GPU, a reasonable philosophy given the state of modern personal computing and workstations available to those in the DL community. Such a philosophy also allows those not familiar with common software frameworks used in the physics community such as ROOT or \geant, often serving as a ``barrier to entry''. 

In this work, we present a fully open fast simulation framework for pions and kaons in DIRC detectors—specifically, the High-Performance DIRC (hpDIRC) at the Electron-Ion Collider (EIC)~\cite{kalicy2024high}, which is designed primarily for $\pi^{\pm}/K^{\pm}$ separation.
Our work encompasses four main goals:

\begin{itemize}
    \item Develop an \textbf{open} and \textbf{user-friendly simulation package} designed to be accessible to both physicists and DL experts.
    \item \textbf{Accelerate DL research} in PID for Cherenkov detectors by eliminating the barrier posed by complex physics software, making the problem more accessible to DL experts from adjacent fields.
    \item Move \textbf{beyond fixed datasets} — different DL approaches require varying amounts of data. Our fast simulation package enables users to generate data on demand, ensuring scalability for diverse research needs.
    \item Establish a comprehensive suite of state-of-the-art (SOTA) generative models and provide \textbf{benchmarking} for the physics community, with ongoing updates to incorporate the latest advancements.
\end{itemize}

As a byproduct of our work, we also enable the deployment of more classical reconstruction approaches such as time imaging \cite{Dzhygadlo_2020} and FastDIRC \cite{hardin2016fastdirc}, previously limited by storage and time constraints needed to generate reference distributions with \geant. We also construct a pythonic version of the FastDIRC's PID method, capable of leveraging GPUs in traditional DL environments, and coupled to fast simulation algorithms at the photon level.

 In the future, we will provide baseline methods and benchmark datasets for PID performance using data from our most optimal fast simulation methods produced herein, thereby promoting a competitive and continually evolving space of DL-based PID methods in the space of DIRC detectors, which are currently only represented by the two seminal works \textit{DeepRICH} \cite{fanelli2020deeprich} and it's successor \textit{Deep(er)RICH} \cite{fanelli2024deep}, in which the latter lays the foundation for our contribution. These methods can then be trained in more optimal settings by those who wish to deploy them, \textit{i.e.,} by physicists using simulation from \geant or high purity samples from real data.

%% file: 2_data.tex
\section{Data Preparation}\label{sec:data}

In this section, we provide a description of the data and the preprocessing steps used to train the models included in our suite.

\begin{figure}[!]
    \centering 
    \includegraphics[width=0.6\textwidth]{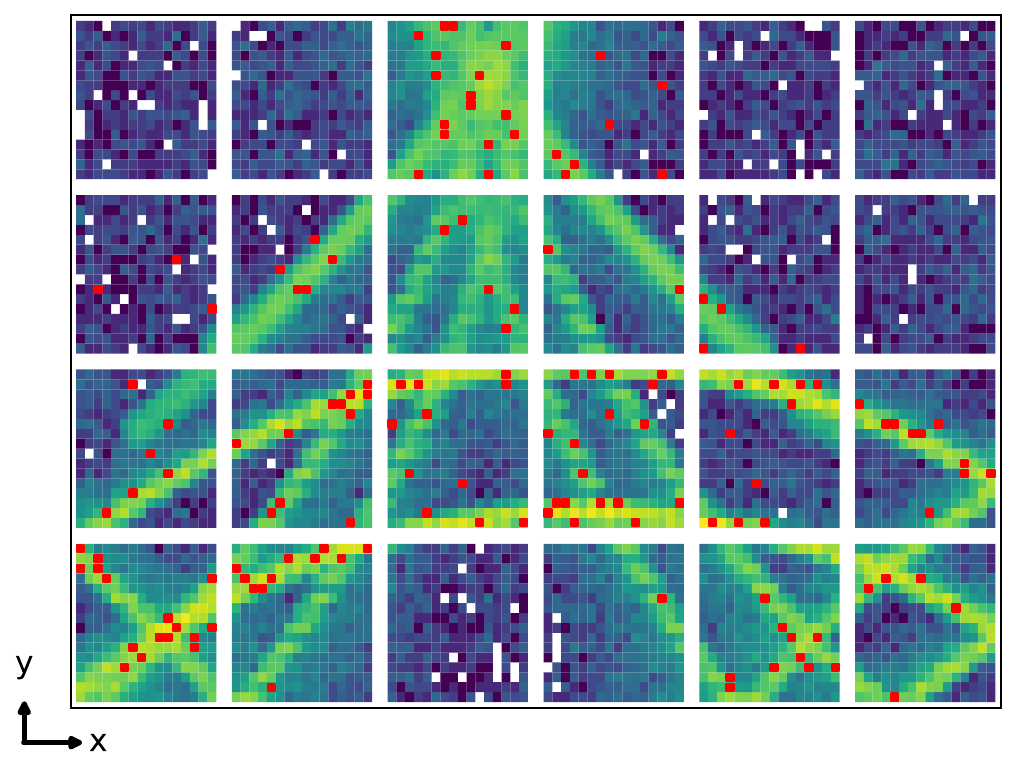} \\
    \caption{\textbf{Optical box output:} Individual tracks leave sparse hit patterns (red points) integrated over time on the hpDIRC readout plane. 
    The denser hit pattern is obtained by accumulating multiple tracks with the same kinematics .}
    \label{fig:sparse_hits}
\end{figure}

Following the approach outlined in Fanelli et al. \cite{fanelli2024deep}, we analyze Cherenkov photon hit patterns recorded by an electronic readout composed of PMT arrays \cite{kalicy2024high}. Each hit pattern corresponds to a single charged track with specific kinematics. As shown in Figure \ref{fig:sparse_hits}, an individual track produces a sparse and approximately stochastic subset of hits (highlighted in red). However, when aggregating hits over multiple tracks with identical kinematics—momentum ($|\vec{p}|$) and polar angle ($\theta$)—the true probability density function (PDF) emerges. The number of hits per track is not fixed, leading to a vast combinatorial space of possible hit distributions, a challenge that we address in later sections within the context of our model design.\footnote{Further discussion on the combinatorial nature of hit distributions and its implications for our modeling approach is provided in a later section.}  

\paragraph{General Processing}

We utilize the standalone \geant simulation \cite{eicdirc} developed for studies in \cite{kalicy2024high,Kalicy_2020} with simulation provided through \textit{Particle Gun} generators. 
\textit{Particle Gun} samples (\textit{i.e.}, simulations of individual tracks with specified kinematics), provide full control over the generated phase-space, in which we generate charged pions and kaons approximately uniformly, and continuously, over the acceptance of the hpDIRC, corresponding to the ranges $1<|\vec{p}|<10$~GeV/c, $25<\theta<160^\circ$, resulting in $\sim$ 5 million tracks for each PID, which is split into the general $70\% / 15\% / 15\%$ train, validation and test subsets. Such a scheme is optimal when the objective of the generative models is to learn detector response, irrespective of the physics. Given the cylindrical geometry of the hpDIRC, we invoke symmetry arguments under the azimuthal ($\phi$) coordinate and generate at the central region of an individual bar as our working point. As such, the azimuthal coordinate is fixed and the conditioning scheme deployed during the learning phase is only dependent on $|\vec{p}|$ and $\theta$, differing from the fast simulation posed in \cite{fanelli2024deep}. This conditional scheme can be altered to encapsulate the azimuthal coordinate in future studies and expansions of our fast simulation package. Moreover, we utilize no magnetic field as our initial working point and will introduce this in later updates, providing control (\textit{e.g.}, on or off) to the user. We also include no dark rate from the PMTs themselves, but provide the ability to include it in the fast simulations produced by our models.

For the datasets used to validate the fast simulation methods developed within, we deploy two methods of data production (i) fixed kinematic generations, \textit{e.g.}, fixing the momentum at 6 GeV/c and scanning $\theta$ in bins of \SI{5}{\degree}, and (ii) continuous generation over the full phase-space. With regard to the former, the reasoning behind this validation scheme is two-fold:

\begin{itemize}
    \item The Cherenkov ring structures exhibit a direct dependence on the kinematics, making it essential to perform comparisons within fixed bins to assess the agreement between the underlying PDFs.
    \item Comparing across fixed kinematic bins validates that, after training over a continuous phase space, there is minimal to no degradation in the quality of the fast simulation in specific regions. This ensures that the underlying distribution across varying kinematics is accurately learned, \textit{i.e.}, diverse and representative models are achieved.
\end{itemize}

Utilizing validation sets generated continuously over the full phase-space allows a comprehensive measure of agreement between our fast simulation methods.

 \paragraph{Translation to sensor coordinate system}

In what follows, we leverage the methods developed in \cite{fanelli2024deep} and describe the translation from the \gluex DIRC, to the hpDIRC at EIC.
Fast simulation of imaging Cherenkov detectors faces two primary challenges: non-uniform input sizes and pixelized readout systems. To address the former (non-uniform input sizes), the authors leverage the approximate independence of Cherenkov photons at the track level, enabling the decomposition of photon data into a tabular format where each photon is conditioned on the kinematics of its parent track. This effectively constructs posterior distributions over the photon space as a function of track kinematics.

The second challenge arises from the discretized nature of pixelized readout systems. Complicating the application of generative AI methods, particularly those designed to work in continuous spaces. This is resolved by introducing two key mappings. The first, Eq. \ref{eq:rowcol_transformation}, converts pixelized integer inputs into a structured index representation $(i,j)$ corresponding to an image of the readout system, where $M_{PMT}$ and $N_{pixel}$ denote the photomultiplier tube (PMT) and pixel indices, respectively.\footnote{The $\lfloor \ \rfloor$ notation represents floor division.}
\begin{equation}\label{eq:rowcol_transformation}
    D_{i,j} =
    \begin{cases}
        \begin{aligned}
        & \lfloor M_{PMT.}/6 \rfloor \cdot 16 + \lfloor N_{pixel.} / 16 \rfloor \\
        & (M_{PMT.} \; \% \; 6) \cdot 16 + (N_{pixel.} \; \% \; 16) 
        \end{aligned}
    \end{cases}
\end{equation}
We then deploy Eq. \ref{eq:xy_transformation}, in which we transform the discretized image into a x,y coordinate system taking into account physical pixel sizes ($p_{width.,height.}$), spacings between PMTs ($\text{gap}_{x,y}$) and offsets due to the PMT housing. Note that all values in Eq. \ref{eq:xy_transformation} are in millimeters ($mm$).

\begin{equation}\label{eq:xy_transformation}
        \begin{aligned}
        & x = 2mm + D_j \cdot p_{width.} \, + (M_{PMT.} \; \% \; 6) \cdot \text{gap}_x \, + \frac{1}{2}p_{width.} \\
        & y = 2mm + D_i \cdot p_{height.} + \lfloor M_{PMT.} \; / \; 6 \rfloor \cdot  \text{gap}_y + \frac{1}{2}p_{height.}
        \end{aligned}
\end{equation}

We address the challenge of discrete pixel values through a controlled smearing operation. Since the probability of a photon striking any point on a pixel face in the $x,y$ coordinate system is approximately uniform, we introduce random uniform noise within the physical sensor resolution:  

\[
\delta (x) \sim \text{Uniform}\left(-\frac{p_{width}}{2},\frac{p_{width}}{2}\right), \quad 
\delta (y) \sim \text{Uniform}\left(-\frac{p_{height}}{2},\frac{p_{height}}{2}\right)
\]

This transformation preserves the underlying data structure while mitigating the instability observed when training directly on discrete values, which often leads to large fluctuations. Importantly, smearing is applied only during training, while evaluation is performed using the central representations.

%% file: 3_architecture.tex
\section{Methods}\label{sec:methods} 

\subsection{General Philosophy}

As in \cite{fanelli2024deep}, we have deliberately chosen to move away from the most "natural" representation of Cherenkov photons at the track level—point clouds. While modern techniques exist for reliable point cloud generation, we argue that our chosen approach is optimal for this specific problem. In what follows expand on the rational behind such choices.

The distribution of photon hits at the track level, and consequently the generative process, is entirely dependent on the kinematics of the parent track. Furthermore, these kinematic variables are continuous, making the learning problem inherently sparse. To illustrate this, consider discretizing the kinematic space into bins of $\Delta (|\vec{p}|) \sim \SI[per-mode=symbol]{100}{\mega\eVperc}$, $\Delta (\theta) \sim \SI{1}{\degree}$. Based on the training statistics discussed earlier, each bin contains approximately 400 training samples.
Now, consider the combinatorial complexity at the track level. The hpDIRC detector consists of 6144 individual pixels. If we make a simplifying assumption regarding the underlying PDF — that a Cherenkov photon can arrive at roughly half of these pixels, we obtain a combinatorial space on the order of $3072!/(N_{\gamma}! (3072-N_{\gamma})!)$.
%
% $\binom{3072}{N_{\gamma}} = \frac{3072!}{N_{\gamma}!(3072-N_{\gamma})!}$. 
%
%\[
%\binom{3072}{N_{\gamma}} = \frac{3072!}{N_{\gamma}!(3072-N_{\gamma})!}
%\]
%
The result is an unfeasible number of tracks needed to sufficiently cover the combinatorics at a given kinematics. Therefore, it is more optimal to learn approximate posterior distributions (decomposition of tracks) and resort to a probabilistic sampling procedure through the generative model. 

\subsection{Deep Learning based Generation}

%%%%%%%%%%%%%%%%%%%%%%%%%%%%%%%%%%%%%%%%%%%%%%%%%%%%%%%%%%

In this section we detail the inner workings of the generative models used within this work. Note that we drop vector notation for simplicity.

%%%%%%%%%%%%%%%%%%%%%%%%%%%%%%%%%%%%%%%%%%%%%%%%%%%%%%%%%%
\subsection*{Discrete Normalizing Flows}

Discrete Normalizing Flows (DNF) employ a series of bijective transformations, \( f_{\theta} \), to map an unknown PDF \( p(x) \), to a known and easily sampled distribution \( q(z) \). The bijective nature of these transformations enables exact likelihood computation via a change of variables. However, this constraint becomes increasingly restrictive in higher-dimensional spaces. Furthermore, DNFs can be conditioned on external parameters, allowing for a controlled generative process.
Let \( x \in \boldsymbol{X} \) be an element from a set of vectors following an unknown probability distribution \( p(x | k) \), where \( k \in \boldsymbol{K} \) denotes the conditional vector representing the kinematics of \( x \), and let \( z \in \boldsymbol{Z} \) correspond to a Gaussian representation of \( x \) via a learnable transformation function \( f_{\theta} \) \cite{fanelli2022flux+,fanelli2024deep}.
A conditional flow with N layers can be described by:
\begin{equation}\label{eq:trans}
x_{k} = f_{\theta}(z,k) = f_{\theta_N} \circ f_{\theta_{N-1}} \circ ... f_{\theta_1}(z_{0},k),
\end{equation}
where the function $f(z,k)$ is commonly represented by Affine transformations \cite{Affine}.

The logarithm of the transformed probability is then given by Eq. \ref{eq:trans_prob}, where $q(\ast)$ denotes the probability under a known distribution: 
\begin{equation}\label{eq:trans_prob}
    \log p(x|k) = \log q (f_{\theta}^{-1}(x) | k) + \sum_{i=1}^{N} \log \left|det \left(\frac{\partial f_{\theta_i}^{-1}(x)}{\partial x} \right)\right|
\end{equation}
The loss function is then given by the negative log-likelihood:

\begin{equation}\label{eq:nf_loss}
    %\mathcal{L}_{LKD.} = -\frac{1}{|\boldsymbol{X}|}\sum_{\boldsymbol{x},\boldsymbol{k} \in \boldsymbol{X}} \log p(\boldsymbol{x}|\boldsymbol{k})
    %min_{\theta} \; \mathbb{E}_{\boldsymbol{x} \sim p_{data}(\boldsymbol{x})}[\log p(\boldsymbol{x} | \boldsymbol{k})]
    min_{\theta} \; \mathbb{E}_{x \sim p_{data}(x)}[-\log p_{\theta}(x | k)]
\end{equation}

In our study, we utilize twenty Affine Coupling bijections parameterized by Residual Networks \cite{he2016deep}, consisting of two residual blocks, each with 128 hidden nodes per layer. For our base distribution $q(z|k)$, we use a Mixture of Gaussian (MoG) with $k=40$, parameterized through 2-6 residual blocks. We found performance in this range to be similar, with our main criteria being computational complexity, \textit{i.e.}, generation and training time. In the case of DNF, the overhead is low enough that such an addition of complexity is negligible.\footnote{In contrast to CNF, where given numerical integration at both training and inference time, we minimize complexity in every aspect.} While a simpler base distribution, such as a fully factorized Gaussian, does produce quality generations, we found the scheme utilizing a MoG is capable of better capturing the kinematic dependency of the photons.

\subsection*{Continuous Normalizing Flows}

Continuous Normalizing Flows were first introduced in \cite{NEURIPS2018_69386f6b}, and convert the set of discrete transformations of traditional NF to a series of intermediate states following an ordinary differential equation (ODE). The time evolution of a data point through the flow $z(x,k,t)$ can be represented through Eq. \ref{eq:diff_eq}, where the function $v_t$ is represented by a neural network (commonly residual neural networks), where $k$ denotes conditional kinematic parameters.

\begin{equation}\label{eq:diff_eq}
\begin{aligned}
    \frac{\D z}{\D t}  & = v_t(z_t,k ; \theta) \\
    z_{t_0} &  = x
\end{aligned}
\end{equation}

The trajectory of the data point from and unknown PDF to a base distribution is then given as the solution of the ODE. At each time step, the change in probability is governed by the instantaneous change in the determinant of the Jacobian:

\begin{equation}
    \frac{\D \log p (z_t|k)}{\D t} = - \nabla_z \cdot v_t(z_t,k;\theta)
\end{equation}

It is then possible to define the probability flow of the data point through the vector field and derive the change of variables through a continuous transformation in Eq. \ref{eq:cont_change}, and train the model under traditional Maximum Likelihood Estimation (MLE).

\begin{equation}\label{eq:cont_change}
        \log p(z_{t_0}|k) = \log q (z_{t_1} | k) - \int_{t_0}^{t_1} \nabla_{z_t} \cdot v_t(z_t,k;\theta) \D t
\end{equation}

In general, the computation of the above integral formulation is expensive and computationally infeasible within high-dimensional spaces. As such, approximation techniques such as the Skilling-Hutchinson trace estimator are used \cite{Skilling1989,Hutchinson01011990}. We directly compute this integral given the low dimensionality space were working in.

\paragraph{Optimal Transport Constraints}

In general, the training of continuous normalizing flows under MLE requires careful tuning of hyperparameters in order to prevent stiff dynamics in the neural ODE. Moreover, the trajectory of individual elements through the vector field is ill-defined, meaning there are infinitely many flows that achieve the same likelihood for a given input. In practice, this amounts to excessive computation as the model explores many trajectories making training on large and complex datasets infeasible. To remedy this, Onken et. al \cite{onken2021otflowfastaccuratecontinuous} propose imposing constraints on the underlying vector field through the lens of optimal transport. First, a traditional optimal transport cost, Eq. \ref{eq:opt_trans}, is added to penalize the arc length of trajectories through the field, essentially regularizing the model to select the flow yielding the shortest path lengths.

\begin{equation}\label{eq:opt_trans}
    \mathcal{L}_{OT} = \int_{t_0}^{t_1}\frac{1}{2}||v_t(z_t,k,t)||^2 \D t
\end{equation}

Where the integral is computed as a part of the divergence accumulation in Eq. \ref{eq:cont_change}. \cite{onken2021otflowfastaccuratecontinuous}. Moreover, Onken et al. further regularize through a potential function, Eq. \ref{eq:potential}, such that their transformation satisfies the Pontryagin Maximum Principle \cite{evans_2013}. Stating that there exists a potential function $\Phi: \mathcal{R}^d \times [0,T] \rightarrow \mathcal{R}$ such that 

\begin{equation}\label{eq:potential}
    v(z,t ; \theta) = -\nabla \Phi (z,t ;\theta)
\end{equation}

The derivatives of the potential function govern the underlying dynamics of the transformation, implying that samples within the vector field move in such a way to minimize their potential, satisfying the Hamilton-Jacobi-Bellman (HJB) equations \cite{onken2021otflowfastaccuratecontinuous}. The additional regularization term can then be computed with limited additional computation given the previous computation of the divergence. The regularization term is then given by Eq. \ref{eq:HJB}.

\begin{equation}\label{eq:HJB}
    \mathcal{L}_{HJB} = \int_{t_0}^{t_1}|\partial_t \Phi(z_t,k,t) - \frac{1}{2}||\nabla \Phi (z_t,k,t;\theta)||^2 | \D t
\end{equation}

The full loss contribution is then given by the linear combination of the three previous terms in Eq. \ref{eq:cnf_loss}.

\begin{equation}\label{eq:cnf_loss}
    min_{\theta} \; \mathbb{E}_{x \sim p_{data}(x)} \left[ \lambda_1 \mathcal{L}_{LKD.} + \lambda_2\mathcal{L}_{OT.} + \lambda_3 \mathcal{L}_{HJB.} \right]
\end{equation}

Such a formulation drastically increases training efficiency and allows the utilization of CNFs on large and complex datasets. In the case of traditional CNFs such as FJJORD \cite{grathwohl2018ffjordfreeformcontinuousdynamics}, the associated computation time is so large that we are not able to obtain any reasonable results.

In our study, we parameterize the velocity field through a Residual Network \cite{onken2021otflowfastaccuratecontinuous}. The network consists of 6 blocks, with 128 hidden units per layer. For the base distribution $q(z|k)$, we again use a Mixture of Gaussian (MoG) with $k=40$, parameterized through 2 residual blocks to limit computational overhead. We also use a trainable integration window (a trainable upper bound for time), in which we utilize 12 integration steps and a 4 point Runge-Kutta (RK4) integration method during training. During generation, we increase the number of integration steps to 20 to provide higher fidelity generations. We found that increasing the number of integration steps during training had limited effect on the models capability, but drastically increased computational time (as expected). In general, we argue that likelihood based learning for CNF models is simply infeasible on large datasets (even with the improvements discussed prior) and it is included more for completeness rather than its effectiveness.

\subsection*{Conditional Flow Matching}
Flow Matching (FM) \cite{lipman2023flowmatchinggenerativemodeling,tong2024improvinggeneralizingflowbasedgenerative} provides a simulation free method of training CNFs by directly learning the probability path $q_t(x)$, through a vector field $u_t(x)$ such that $q_{t_0} \sim N(0,I)$, $q_{t_1} \sim p(x)$ and $t \sim \mathcal{U}[0,1]$. In essence, the FM objective defines a regression problem over the vector field $u_t$ through a neural network $v_t$, Eq. \ref{eq:flowmatch}, where we include traditional transformations on our kinematic parameters $k$.

\begin{equation}\label{eq:flowmatch}
    \mathcal{L}_{FM}(\theta) = \mathbb{E}_{t,q_t(x|k)}\left[||v_t(x,k) - u_t(x,k)||^2\right]
\end{equation}

In practice, the choice of an appropriate vector field $u_t$ and probability path $q_t$ in Eq. \ref{eq:flowmatch} is unknown apriori, resulting in an intractable loss formulation given that many choices of probability path satisfy the conditions. Instead, Lipman et al. \cite{lipman2023flowmatchinggenerativemodeling} show that one may construct the probability path $q_t$, through a mixture of conditional probability paths such that $q_{t_0}(x|x_{t_0}) = q_{t_0}(x)$ and $q_{t_1}(x|x_{t_1}) = N(x|x_{t_1},\sigma^2 I)$, where sufficiently small $\sigma > 0$ provides $x_{t_1} \sim x$. Such a formulation can be cast into a simpler, tractable objective known as Conditional Flow Matching (CFM), Eq. \ref{eq:cfm}. 

\begin{equation}\label{eq:cfm}
\mathcal{L}_{CFM}(\theta) = \mathbb{E}_{t,p(x_{t_1}|k),q_t(x|x_{t_1},k)}\left[||v_t(x,k) - u_t(x|x_{t_1},k)||^2 \right]
\end{equation}

The formulation in Eq. \ref{eq:cfm} provides identical gradients of that in Eq. \ref{eq:flowmatch}, resulting in equivalent optimization under expectation \cite{lipman2023flowmatchinggenerativemodeling}. We use an Affine probability path such that $x_t = \sigma_t x_{t_0} + \alpha_t x_{t_1}$, where $\sigma_t = 1 - t$ and $\alpha_t = t$. This formulation defines the transition from the intial state $x_{t_0}$ (Gaussian) to the final state $x_{t_1}$ over the time interval $t \in [0,1]$.

In our study, we utilize the same Residual Network as the CNF method, although now we are able to greatly increase its size given the likelihood free learning scheme of flow matching (\textit{i.e.}, no numerical integration during training). We increase the number of blocks to 20, with the same 128 hidden nodes per layer as mentioned for the CNF network. At generation, we deploy a midpoint solver with 20 integration steps (consistent with CNF). We found that increasing the order of accuracy of the integration method in the ODE solver through methods such as Dormand-Prince 5(4) (Dopri5) produced marginal improvements in the generation fidelity, but increased generation time by a factor of $2-3\times$. 

% \begin{equation}
%     x_t = \sigma_t x_0 + \alpha_t x_1.
% \end{equation}

% $x_t = x_0 - tx_0 + tx_1 = x_0 + t(x_1 - x_0)$
% $\alpha_t = t$
% $\sigma_t = 1 - t$

% \begin{equation}
%     u_t = \frac{d x_t}{d t} = \dot{\sigma}_t x_0 + \dot{\alpha}_t x_1.
% \end{equation}

\subsection*{Denoising Diffusion Probabilistic Models}

Denoising Diffusion Probabilistic Models (DDPM) \cite{sohldickstein2015deepunsupervisedlearningusing,ho2020denoisingdiffusionprobabilisticmodels} are a class of generative models that transform intractable data distributions into a simple latent distribution by slowly perturbing the data with Gaussian noise.
%(a flowchart is described in Fig. \ref{fig:diff_flowchart}). 
%
%
%\james{This is nice, but I would remove it as we do not have images for the other methods. So breaks up the continuity of the section IMO.} \cris{We should make a flow chart for NF+DIRC. It shouldn't be too difficult.} \james{I will work on an improved version then.}
%\begin{figure}[!h]
%    \centering
%    \includegraphics[width=\textwidth, trim= 0cm 1cm 0cm 0cm,
%    clip]{Figures/diffusion_schematic.pdf}
%    \caption{\textbf{Diffusion Process Flowchart:} Input data is progressively diffused over $T$ timesteps by adding Gaussian noise until the input is fully white noise. In the reverse process, a latent input $x_T$ is sampled from a Gaussian distribution, and each transformation learned in the forward process is reversed, allowing us to retrieve a generated sample $x_0$. Note that Score Based Generative Models can be represented similarly by using SDEs instead to parametrize the forward and backward processes.}
%    \label{fig:diff_flowchart}
%\end{figure}
%
At the same time, a neural network is trained to learn the noise added at each step. This is known as the forward diffusion process, which can be parametrized as a Markov chain with noise added according to a predetermined variance scheduler $\beta(t)$. Letting $x\in \boldsymbol{X}$ be an element of the dataset, $x_1,...,x_T$ be latent variables transformed by the Markov process, and $k\in \boldsymbol{K}$ be the conditional vector for the kinematics of x, the Markov chain can be described through Eq. \ref{eq:ddpmforward}.

\begin{equation}\label{eq:ddpmforward}
    q(x_{1:T}|x_0, k):=\prod_{t=1}^{T}q(x_t|x_{t-1},k), \quad q(x_t|x_{t-1},k):=\mathcal{N}(x_t;\sqrt{1-\beta_t}x_{t-1},\beta_t \mathbf{I})
\end{equation}

To sample from the final latent distribution, the reverse diffusion process, Eq. \ref{eq:ddpmbackward}, can also be parametrized as a Markov chain where each transformation is learned, Gaussian, and conditioned on the kinematics $k$.

\begin{equation}\label{eq:ddpmbackward}
    p_\theta(x_{0:T}|k) := p(x_T)\prod_{t=1}^Tp_\theta(x_{t-1}|x_t,k), \quad p_\theta(x_{t-1}|x_t,k):=\mathcal{N}(x_{t-1};\mu_\theta(x_t,t,k),\Sigma_\theta(x_t,t,k))
\end{equation}

 Setting $\alpha_t := 1-\beta_t$ and $\bar{\alpha}_t := \prod_{s=1}^t\alpha_s$ allows us to sample any $x_t$ at any arbitrary timestep $t$ in closed-form. Ho et al. \cite{ho2020denoisingdiffusionprobabilisticmodels} show that using this parametrization, the loss function can be simplified from a complex variational lower bound term to the mean squared error between the true noise and the predicted noise added during the forward pass, Eq. \ref{eq:ddpm_loss}.

\begin{equation}\label{eq:ddpm_loss}
    \mathbb{E}_{t,x_0,\epsilon}[||\epsilon -  \epsilon_\theta(\sqrt{\bar{\alpha_t}}x_0 + \sqrt{1-\bar{\alpha_t}}\epsilon, t, k)||^2]
\end{equation}

Where $\epsilon \sim \mathcal{N}(0, \mathbf{I})$ and $e_\theta(x_t,t, k)$ is estimated using a neural network. 
In our study, we parametrize the Markov chain through a Residual Network \cite{he2016deep} to learn the noise added at each step. The network consists of 4 blocks, with 128 hidden nodes per layer. In the network, our conditional kinematics are injected using a network with 2 layers with 16 hidden nodes. For the discrete process, we use 100 time steps with a cosine noise schedule for $\beta(t)$. 

\subsection*{Denoising Score-Based Generative Models}

Song et al. \cite{song2021scorebasedgenerativemodelingstochastic} construct a diffusion process similar to DDPM, but in continuous time. As the step size $\Delta t \rightarrow0$ and the number of time steps $T\rightarrow\infty$, the forward Markov chain can be approximated as a stochastic differential equation (SDE), Eq. \ref{eq:sdeforward}.

\begin{equation}\label{eq:sdeforward}
    \mathrm{d}x = f(x,t)\mathrm{d}t + g(t)\mathrm{d}w
\end{equation}

Where $w$ is a standard Wiener process, $f(x,t)$ represents the drift coefficient, deterministically guiding the data to the latent distribution, and $g(t)$ controls the level of noise added from the Wiener process at each step. The reverse process is then given by Eq. \ref{eq:sdebackward}.
%To create the reverse process, Anderson \cite{Anderson_1982} derives the general form of any general SDE of the form given in Eq. \ref{eq:sdeforward}: 

\begin{equation}\label{eq:sdebackward}
    \mathrm{d}x = [f(x,t)-g(t)^2\nabla_x\log p_t(x)]\mathrm{d}t + g(t)\mathrm{d}\bar{w}
\end{equation}

Where time flows backward from $T\rightarrow0$ and $\nabla_\mathbf{x}\log p_t(\mathbf{x})$ is the score function, which represents the direction of the steepest increase of $p_t(x)$ to transform the noisy data back to the original data manifold.  Using the variance scheduler of $\beta(t)$ outlined in DDPM and the transformation in Eq. \ref{eq:ddpmforward}, it can be shown that $f(x,t)=-\frac{1}{2}\beta(t)x{}$ and $g(t)=\sqrt{\beta(t)}$. Importantly, maintaining the drift coefficient as Affine ensures that each transformation is Gaussian and can be computed tractably.
Let $\tilde{x}$ denote a data point $x \in \boldsymbol{X}$ perturbed by random Gaussian noise $\epsilon$. Given the conditional kinematics $k$, let $s_{\theta}(\tilde{x}, t, k)$ represent the model's predicted score. The loss function is defined in Eq. \ref{eq:sde_loss}, where the noise is given by $\epsilon = \frac{1}{\sigma_t} (\tilde{x} - x)$, in which a similar assumption to that of DDPM (\textit{i.e.}, Gaussian transformations) provides a relaxed loss formulation.

\begin{equation}\label{eq:sde_loss}
    \frac{1}{2} \mathbb{E}_{t, x, \epsilon} [ \| s_\theta(\tilde{x}, t, k) - \frac{\epsilon}{\sigma_{t}} \|^2 ]
\end{equation}

In our study, we parameterize the SDE through a Residual Network \cite{he2016deep} consisting of 6 blocks, with 128 hidden nodes per layer. The kinematics are injected in the same manner as that of the discrete time model (DDPM). For the continuous process, we use 100 time steps and a cosine noise schedule.

\subsection{Generation Procedure}

Transforming input distributions from a discrete pixel space in 
$x$, $y$ to a continuous space necessitates addressing the constraints imposed by the DIRC’s pixelized readout. Additionally, ensuring the physical validity of generated samples requires incorporating the sensor's phsysical limitations. This can be interpreted as generating under a prior defined by the physical dimensions of the DIRC. To achieve this, we apply the resampling technique introduced in \cite{fanelli2024deep}, which enforces consistency with the detector’s resolution and constraints. In what follows, we quantify this effect for each model.
Note we take this a step further and provide two ``priors'' - a simple prior over the physical bounds of the DIRC (the outer limits in x and y), and a more fine grained prior that takes into account individual PMT spacings. The two differ in the fact that under the former, these generations are simply set to the closest pixel, where as in the latter, they are treated as unphysical and resampled. This fine grained prior provides higher fidelity generations, specifically at the boundaries of the PMTs, although it does increase the generation time per track by a factor of $\sim 2$.

\subsection{Photon Yield Sampling}

Up to this point, we have not addressed the final steps needed for a holistic simulation package. In the current form, our models are capable of generating individual photons conditional to the parameters of the parent track. However, our generators have no information of the photon yield, \textit{i.e.}, how many photons should the model generate and aggregate to form a single track? What we are looking for is a function $f: \mathbb{R}^2 \rightarrow \mathbb{Z}_{+}$, mapping from the space of real valued kinematics, to the space of positive integers. While there are many ways to find such a function (\textit{e.g.}, a Neural Network), we impose computational constraints such that the function should have minimal computational overhead, and be CPU bound. This can be realized through a simple approximation using a Look-Up-Table (LUT) in a discretized space. Specifically, we form a LUT for the photon yield using our training datasets in bins of $\Delta (|\vec{p}|) \sim \SI[per-mode=symbol]{100}{\mega\eVperc}$, $\Delta (\theta) \sim \SI{1}{\degree}$. For each bin, we obtain the corresponding frequencies (probability) of each photon yield with respect to all possible values (contained in our reference dataset). We then apply a simple Gaussian kernel to smooth the probability space. The result is an efficient method of photon yield estimation, and allows us to create holistic simulation framework.  Note that this LUT can continually be improved and have its resolutions increased through the inclusion of more data, whether from \geant simulations or experimental data (if applicable). In later sections we demonstrate the validity of our approach.

%% file: 4_analysis.tex
\section{Analysis and Results}\label{sec:results}

In what follows, we provide full evaluations of our generative models, and photon yield sampling process. Note that we have produced fast simulations across the entire phase space and have chosen to display a subset of regions with complex structures in both spatial and time coordinates. Given the large phase space we are working with, showing all regions is infeasible and therefore we must select working points to be contained within the main body of the document. Specifically, we validate our models at the central region in momentum ($\SI[per-mode=symbol]{6}{\giga\eVperc}$) through a series of histograms and ratios. Given also the large selection of models we have included, we have opted to show only one in this case, in which additional figures for the other models can be found in \ref{app:6GeV}. Plots for additional kinematics (\textit{e.g.}, 3 GeV/c and 9 GeV/c) can be found in \ref{app:3GeV} and \ref{app:9GeV}. This will be then followed by an intrinsic model comparison, in which we utilize proxies from the prior section, along with generation timing studies to select our most optimal model for further evaluation.

After selecting the optimal model for fast simulation, we investigate the classification potential at fixed kinematics by deploying our GPU-accelerated version of FastDIRC for PID~\cite{hardin2016fastdirc}. Its adaptation and deployment to the hpDIRC is a natural byproduct of this work.
FastDIRC allows a non-parametric classification approach through Kernel Density Estimation (KDE), in which we form reference PDF's from both fast simulated and \geant samples. Note that the combination of fast simulation of Cherenkov detectors allows the realization of FastDIRC and time imaging methods \cite{Dzhygadlo_2020} for DIRC detectors, previously limited by the computation time, or vast amounts of simulation (and therefore storage) required for adequate coverage of the full momentum and angular acceptance ranges.
This evaluation is further enhanced through classifier agreement techniques, where we employ the Delta-LogLikelihood (DLL) approach introduced in~\cite{fanelli2024deep} to quantify discrepancies between fast-simulated tracks and those generated with \geant, in terms of pion–kaon separation. This allows us to estimate average classifier behavior across the full phase space.

\subsection*{Histogram Level Generative Model Evaluation}
\begin{figure}[h]
    \centering
    \includegraphics[width=0.49\textwidth]{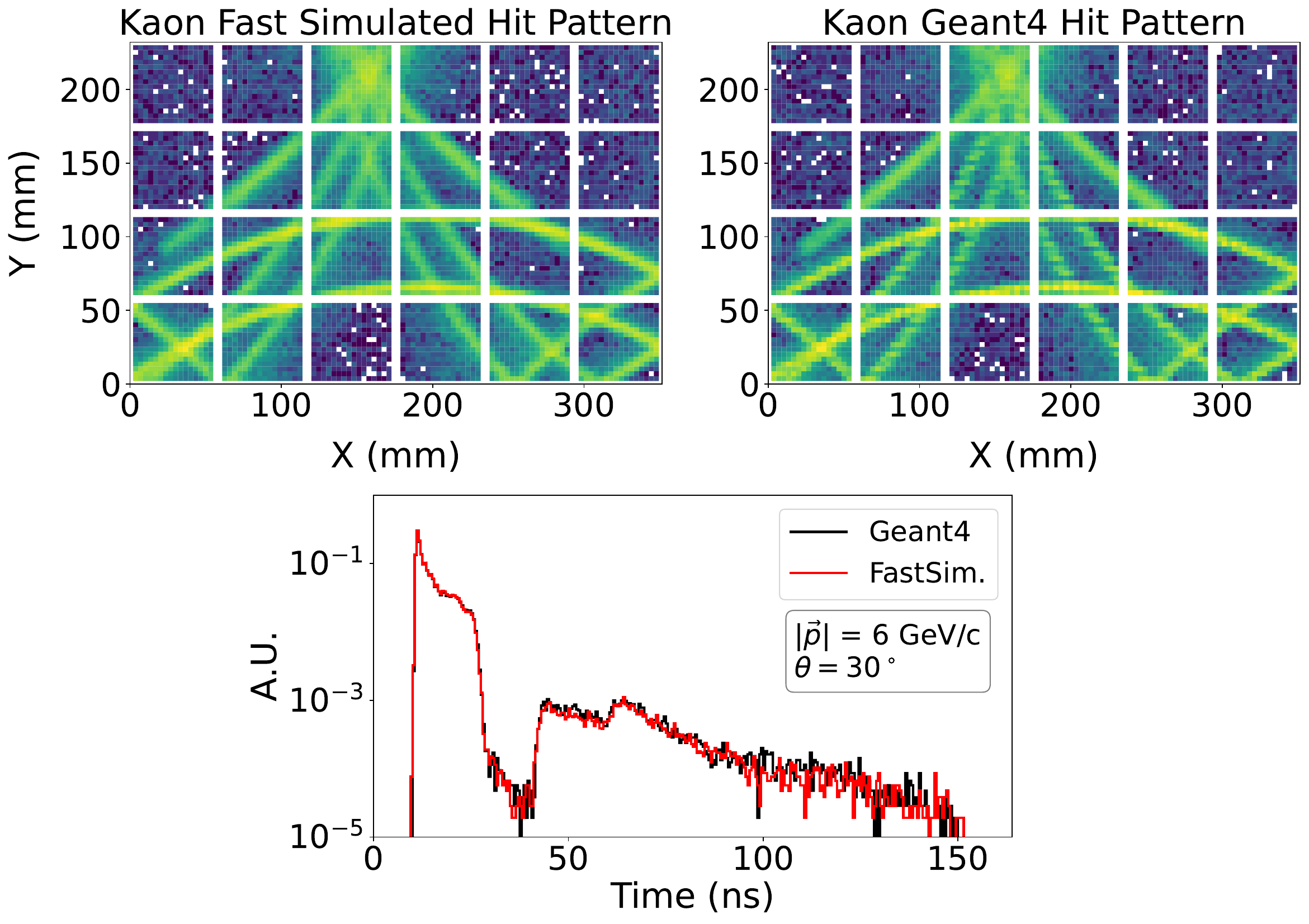}% 
   \includegraphics[width=0.49\textwidth]{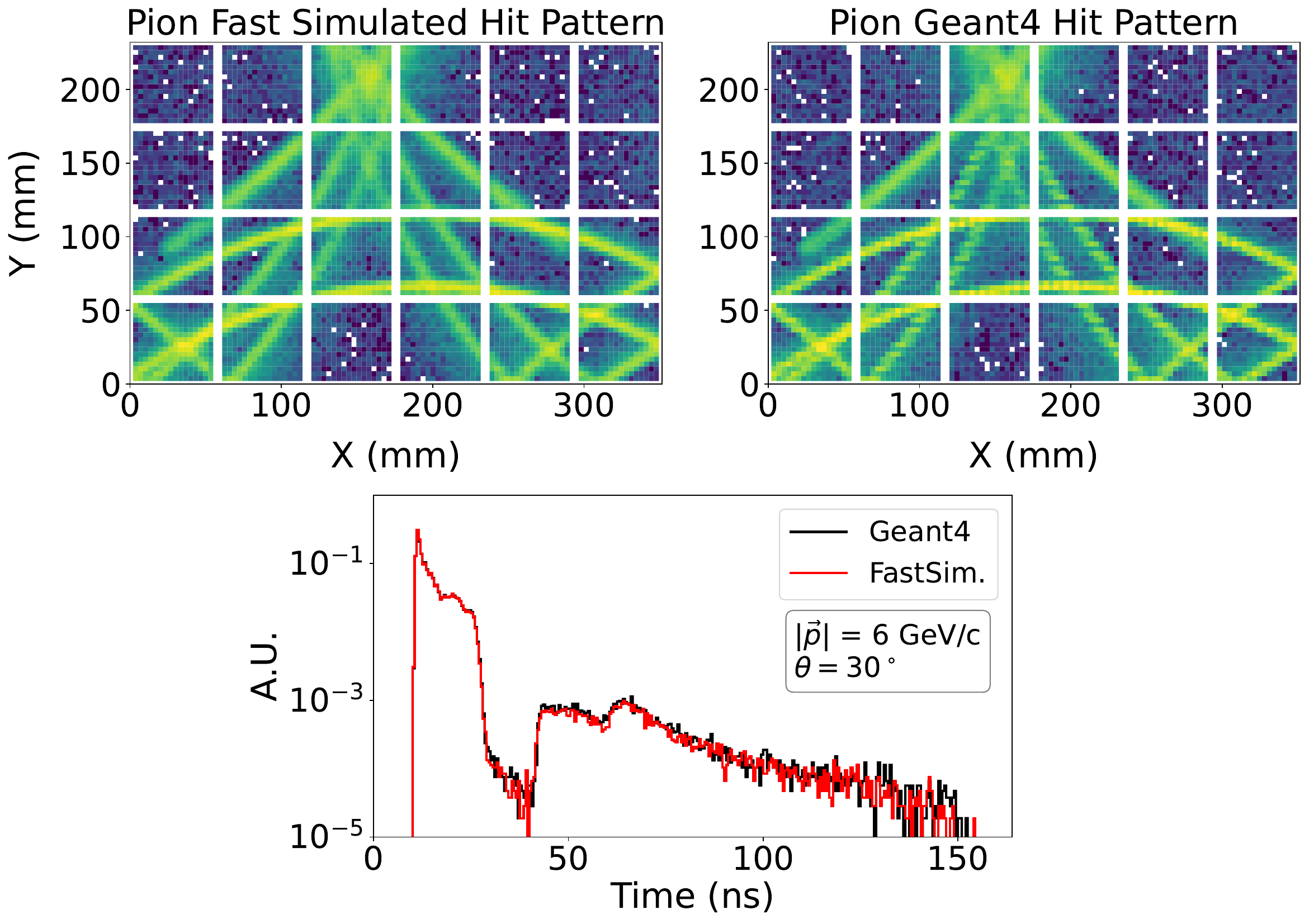} \\
    \includegraphics[width=0.49\textwidth]{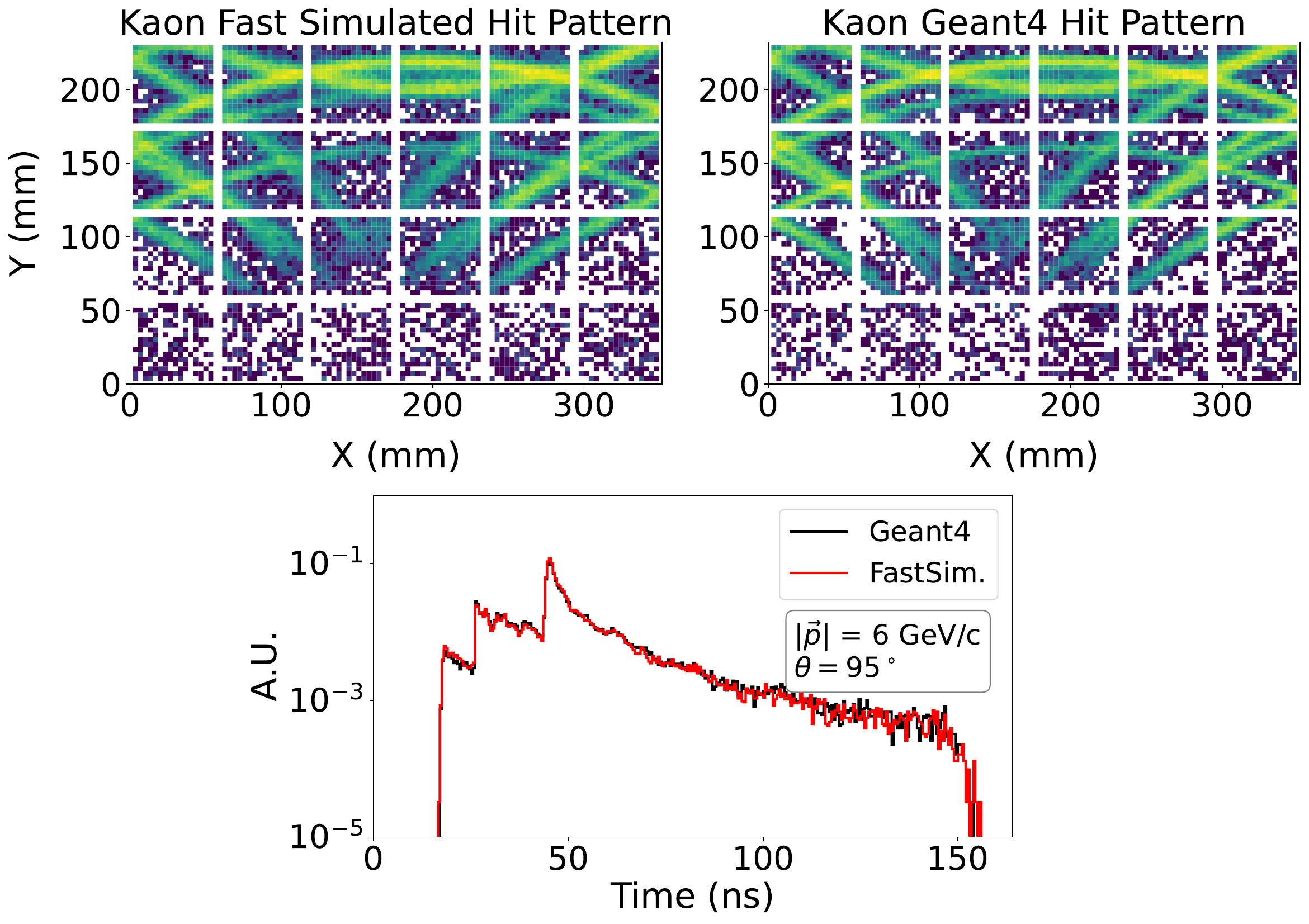} %
    \includegraphics[width=0.49\textwidth]{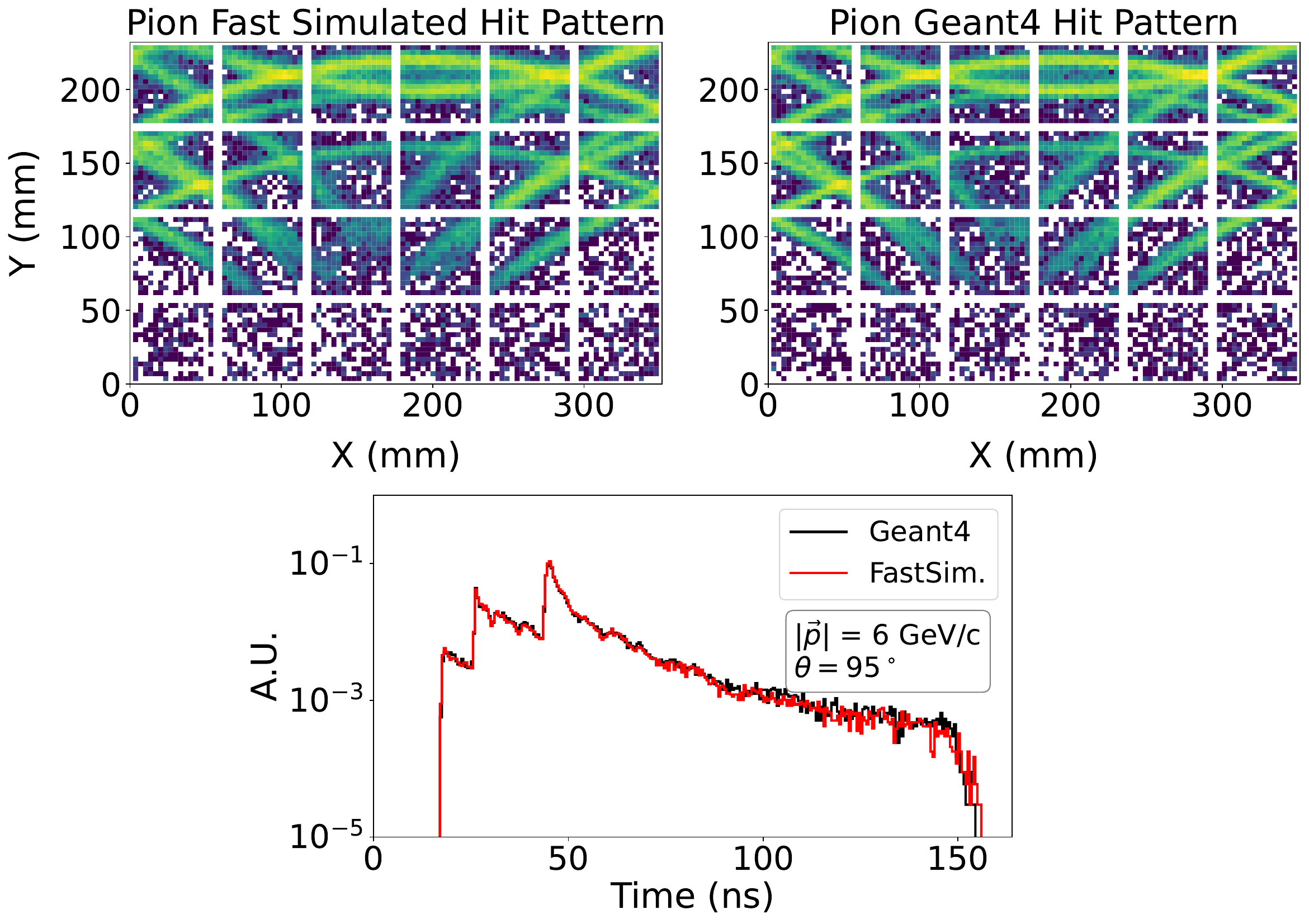} \\
    \includegraphics[width=0.49\textwidth]{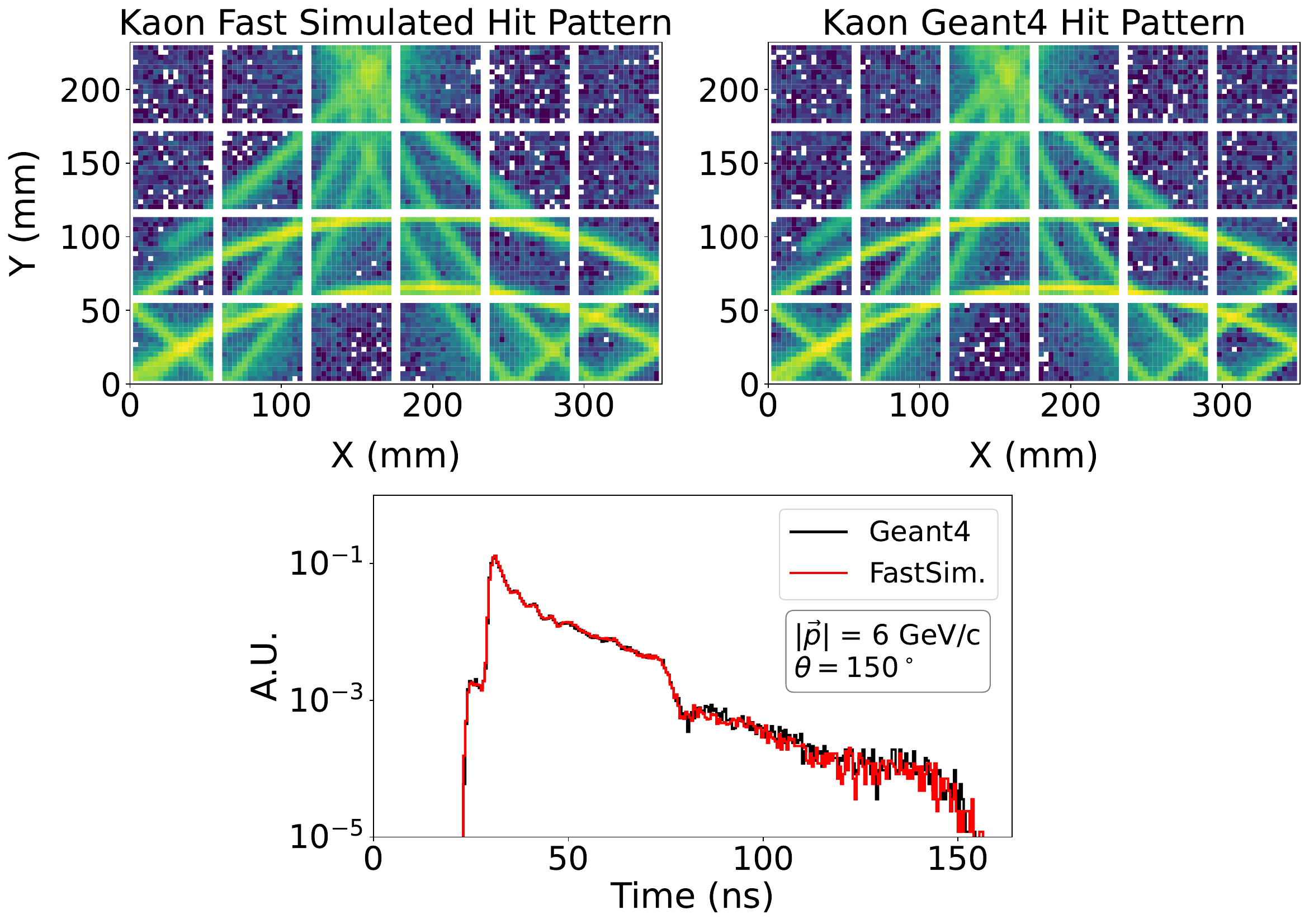} %
    \includegraphics[width=0.49\textwidth]{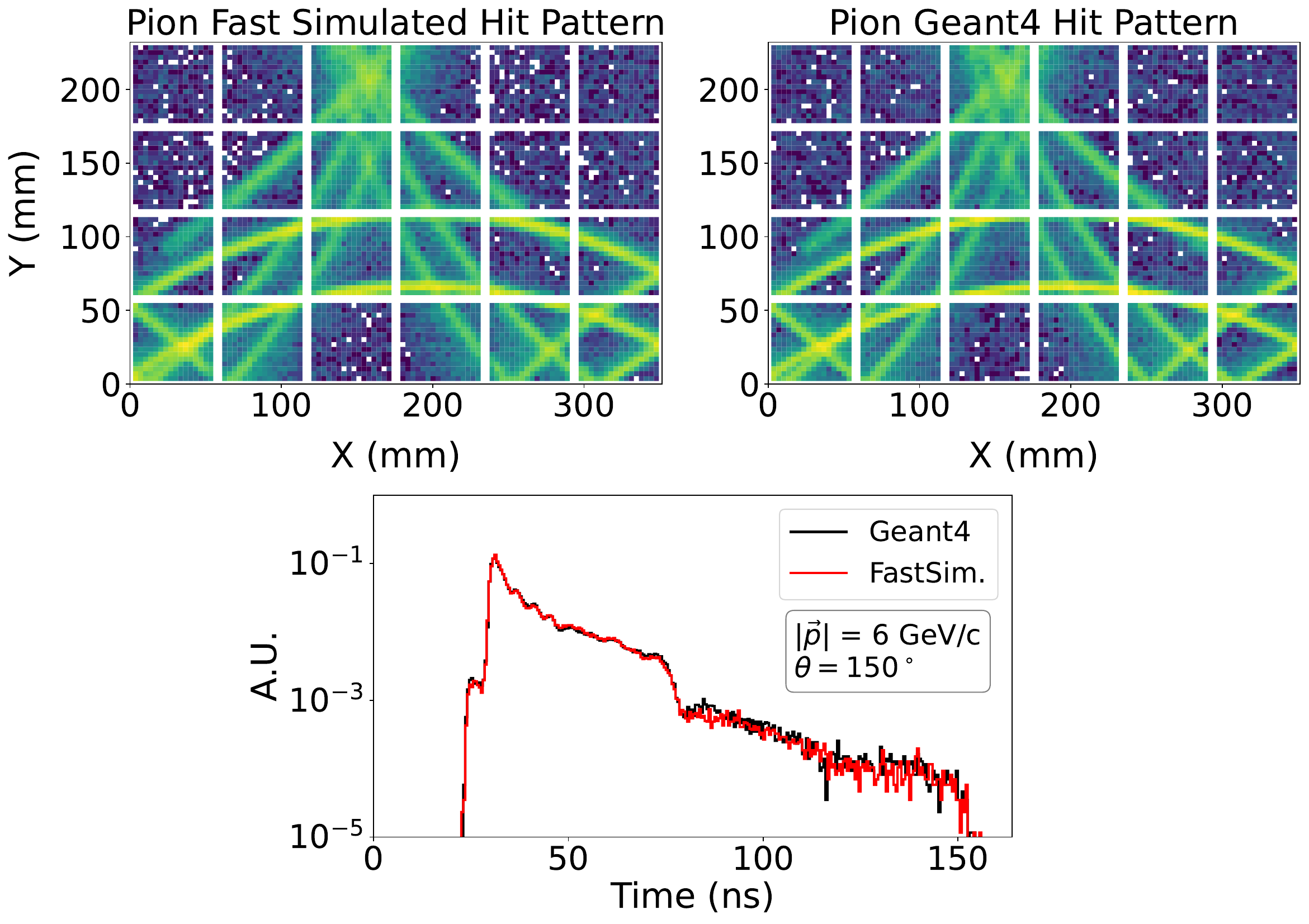} %
    \caption{
    \textbf{Fast Simulation with Discrete Normalizing Flows:} Fast Simulation of kaons (left column of plots), and pions (right column of plots) at 6 GeV/c and various polar angles using DNF.}
    \label{fig:DNF_Generations_6GeV}
\end{figure}

We begin with a visual comparison of the models and their ability to replicate the Cherenkov ring structures at various values of polar angle, when the momentum is fixed. We again remind the reader that the model is trained over a continuous phase space. While not quantitative in nature, a visual inspection of the generation quality at fixed kinematics does yield insight into potential (or lack thereof) mode collapse, further testing our generation philosophy (refer Sec. \ref{sec:methods}). Specifically, we show generations from our model near the two ends of the bar (large, or small polar angle), along with the central region of the bar. The former demonstrates the ability of the models to remain $\sim$ symmetric in their ring structures, while learning the correct shifts in arrival time, and the latter demonstrates the ability of the models to capture highly multi-modal distributions in arrival time due to reflection paths differing by a bar length. Some photons are directed directly towards the optical box ($\sim 1/2 \times$ bar length), while others must reflect from the opposite end and travel the entire distance of the bar to the optical box ($\sim 1.5 \times$ bar length). Note that these generations are produced using the ``ground truth'' photon yield, \textit{i.e.}, that corresponding to the parent track from \geant. Fig. \ref{fig:DNF_Generations_6GeV} depicts generations of kaons (left column of images) and pions (right column of images) for various polar angles ($\SI{30}{\degree}$ top row, $\SI{95}{\degree}$ middle row and $\SI{150}{\degree}$ bottom row) for the DNF model. We remind the reader that additional figures for the other models can be found in \ref{app:6GeV}. 
%Figs. \ref{fig:CNF_Generations_6GeV}, \ref{fig:FlowMatching_Generations_6GeV}, \ref{fig:DDPM_Generations_6GeV}, and \ref{fig:score_Generations_6GeV} show the same generation scheme for the CNF, Flow Matching, DDPM and Scored-Based approaches respectively. 

%%%%%%%%%%%%%%%%%%% DNF %%%%%%%%%%%%%%%%%%%%%%%%%%%%%%%%%%%%%

% %%%%%%%%%%%%%%%%%%% Score Based %%%%%%%%%%%%%%%%%%%%%%%%%%%%%%%%%%%%%

In our initial qualitative assessment we draw two major conclusions:

\begin{itemize}
    \item All methods are able to capture the complex kinematic dependencies in both time and spatial components.
    \item Likelihood free methods generate ``noisier'' ring structures upon aggregation of multiple photons.
\end{itemize}

\begin{figure}[h]
    \centering
    \begin{subfigure}[b]{0.49\textwidth}
        \centering
        \includegraphics[width=\textwidth]{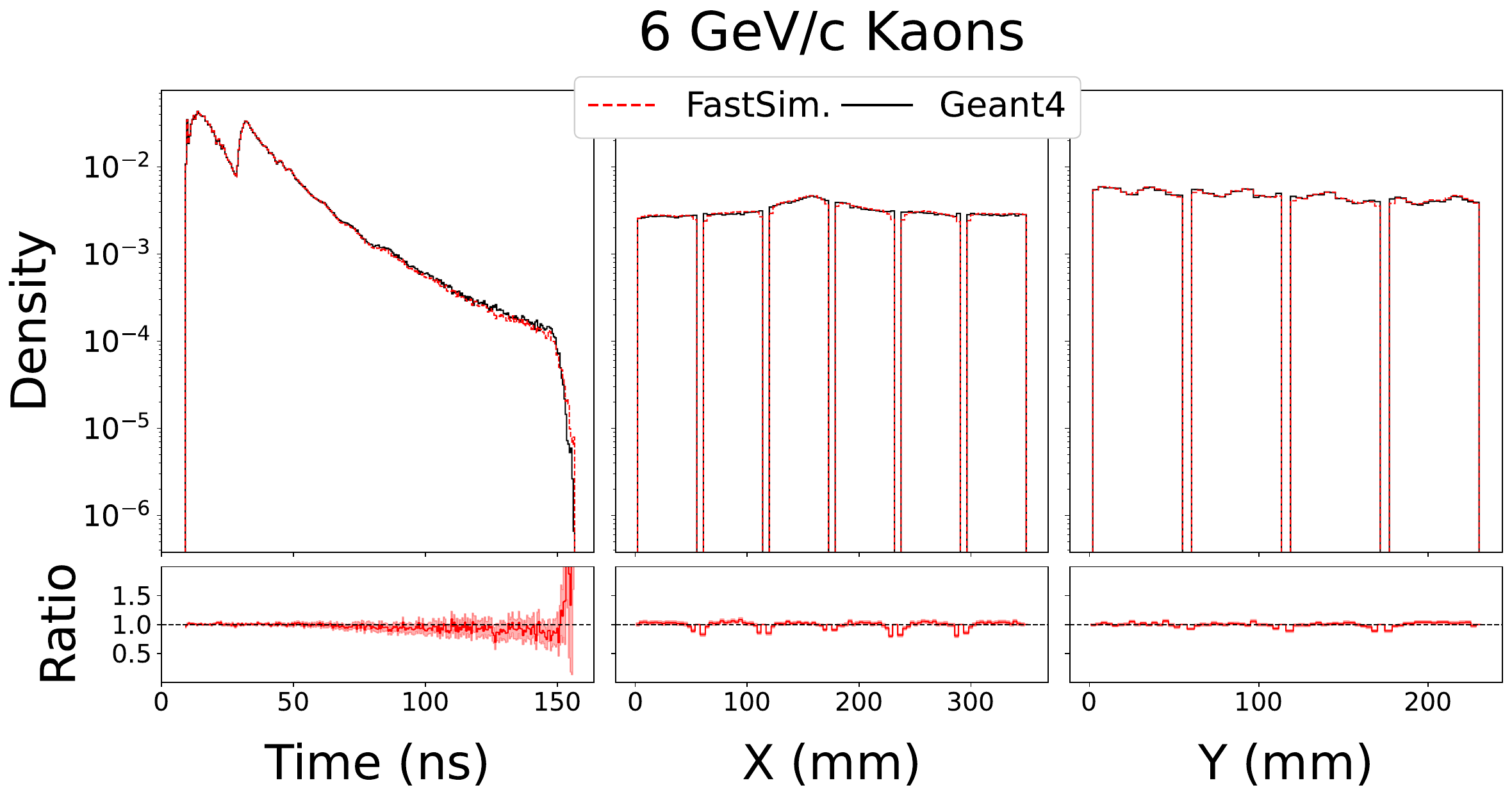}
        \caption{DNF}
    \end{subfigure}
    \begin{subfigure}[b]{0.49\textwidth}
        \centering
        \includegraphics[width=\textwidth]{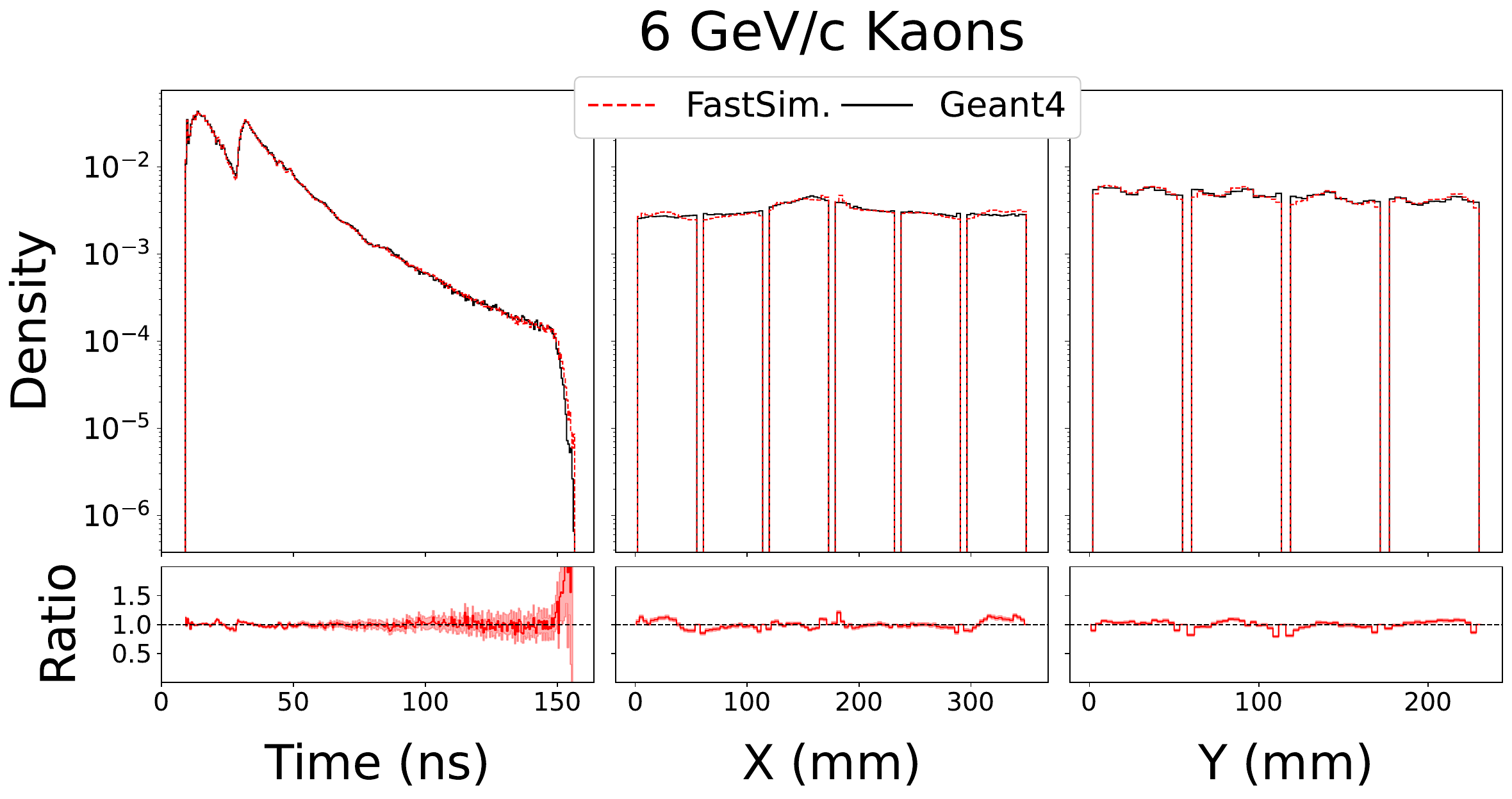}
        \caption{CNF}
    \end{subfigure} \\   
    
    \begin{subfigure}[b]{0.49\textwidth}
        \centering
        \includegraphics[width=\textwidth]{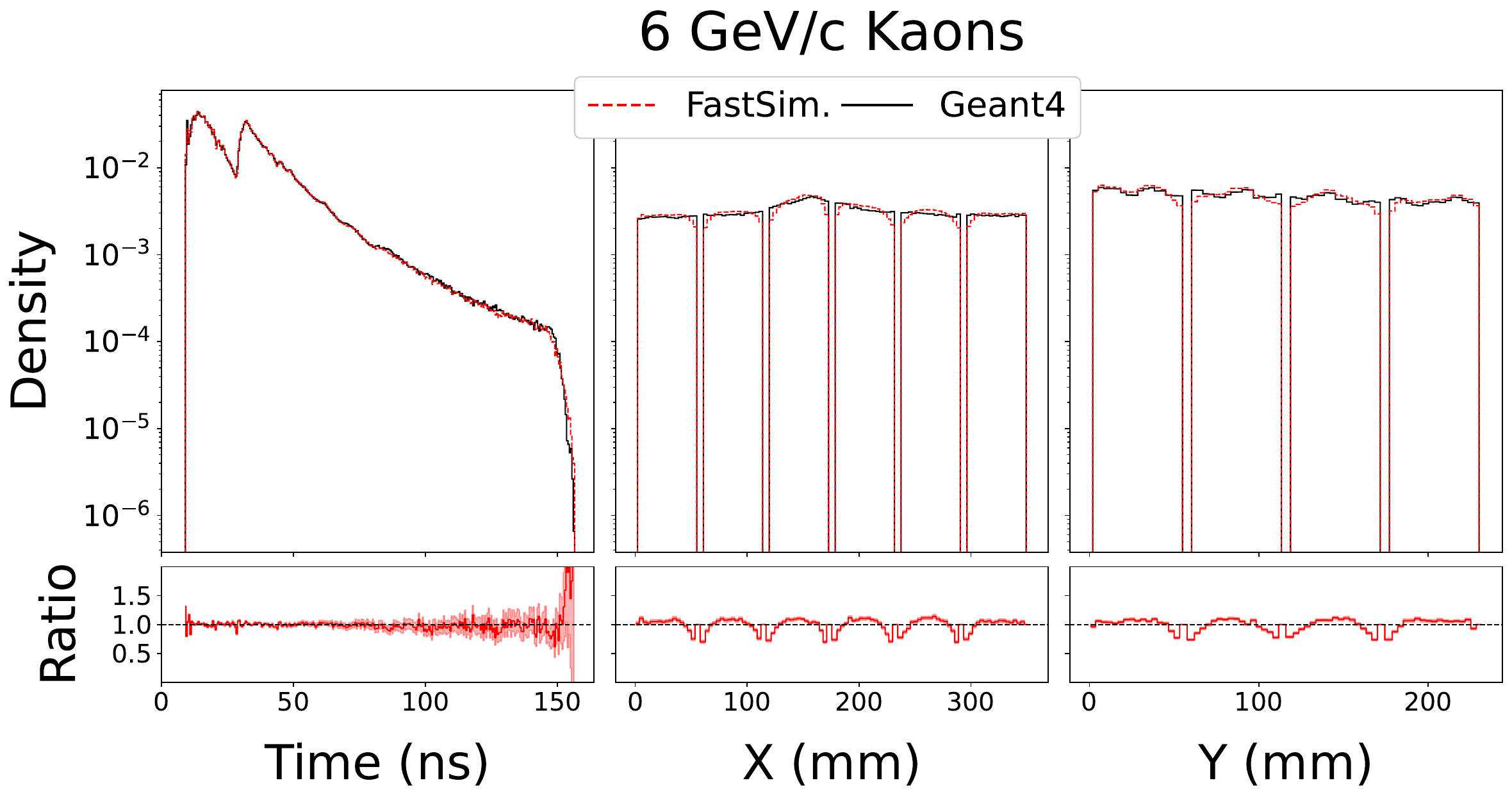}
        \caption{Flow Matching}
    \end{subfigure}
    \begin{subfigure}[b]{0.49\textwidth}
        \centering
        \includegraphics[width=\textwidth]{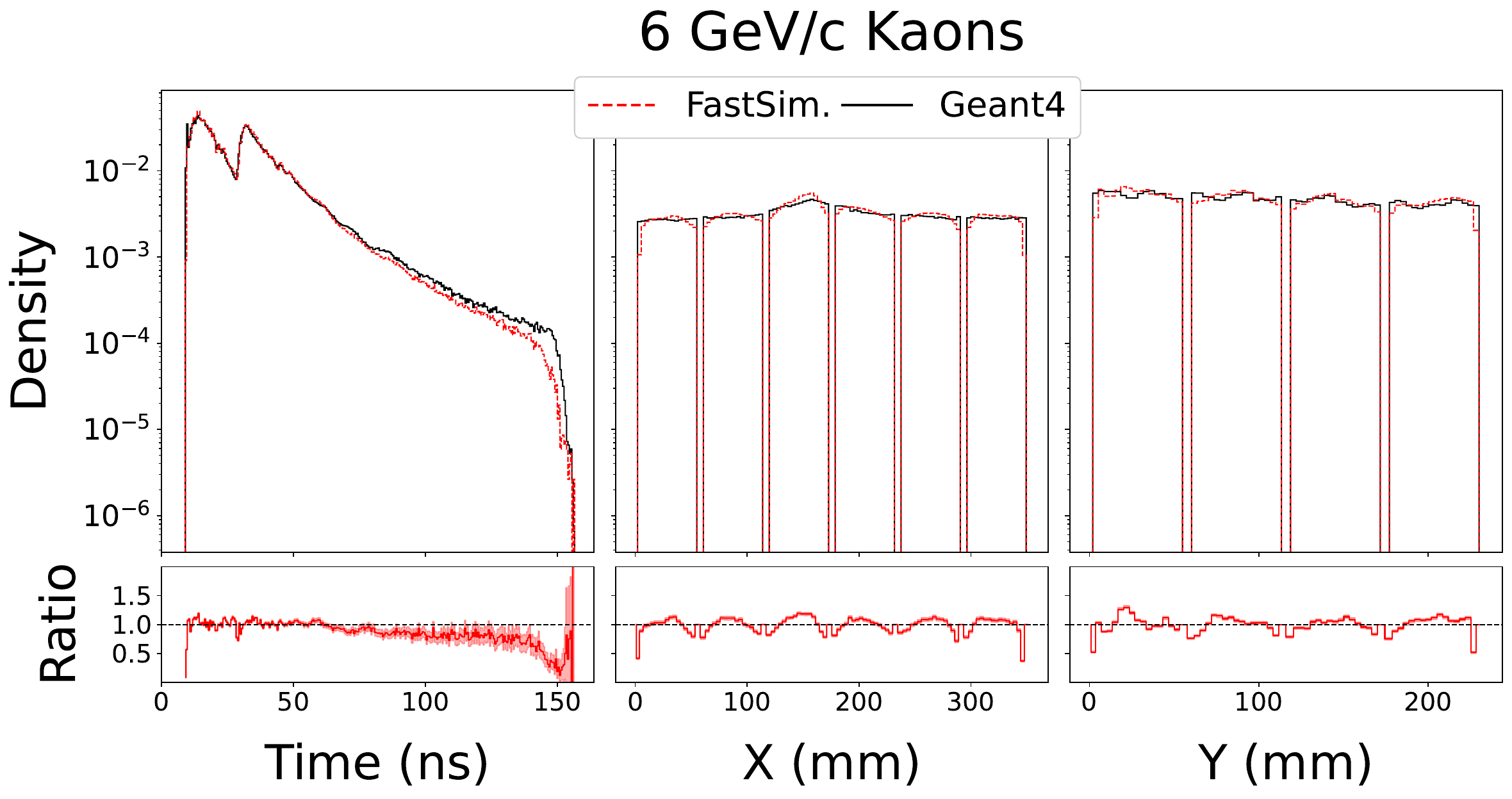}
        \caption{DDPM}
    \end{subfigure} \\  
    
    \begin{subfigure}[b]{0.49\textwidth}
        \centering
        \includegraphics[width=\textwidth]{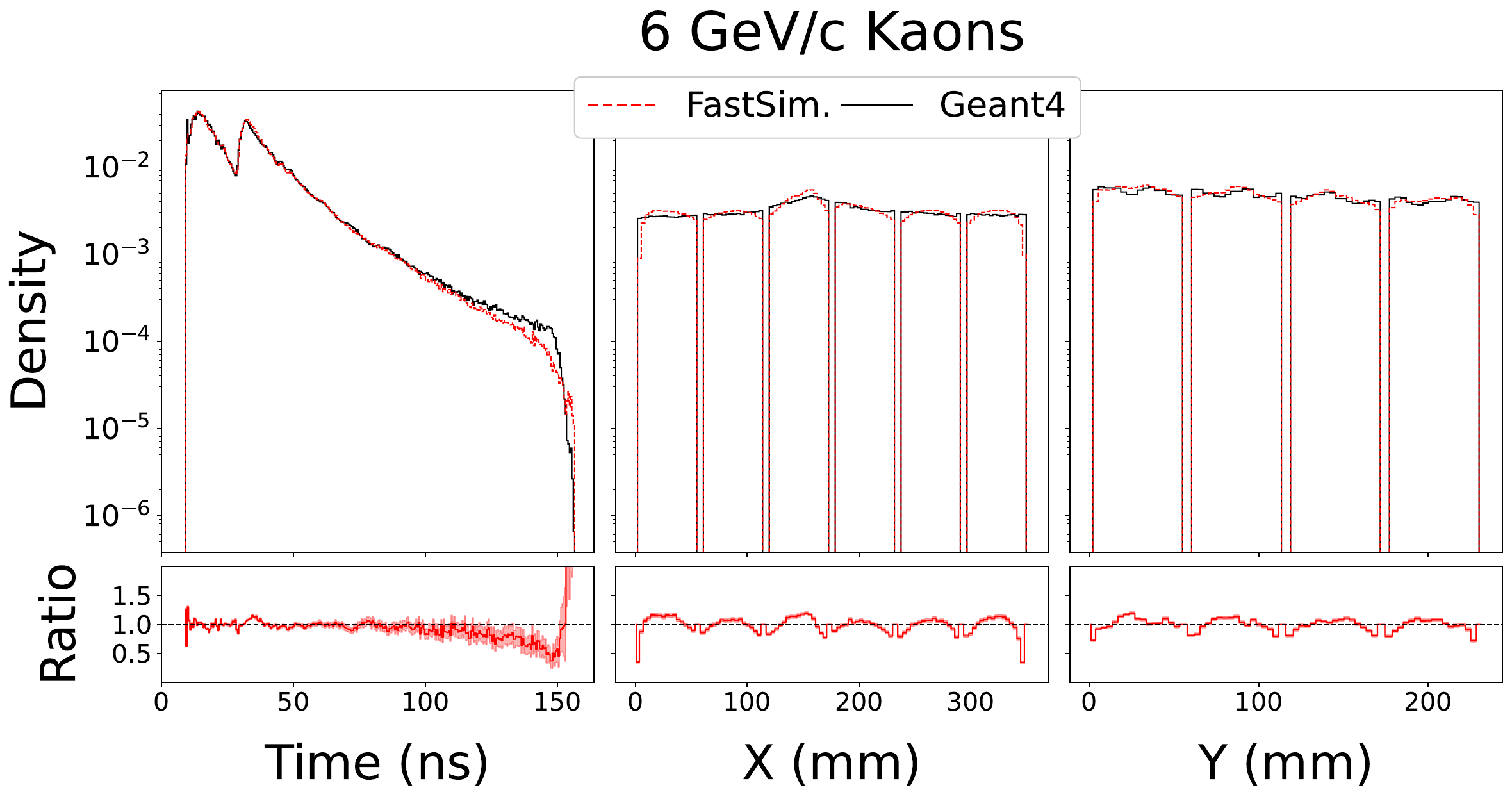}
        \caption{Score-Based}
    \end{subfigure}
    % \begin{subfigure}[b]{0.49\textwidth}
    %     \centering
    %     \includegraphics[width=\textwidth]{Figures/Generations/GSGM/6GeV/Ratios_Kaon.pdf}
    %     \caption{GSGM}
    % \end{subfigure}
    \caption{\textbf{Ratio Plots at 6 GeV/c for kaons:} Ratio plots for kaons using the various different models (a) Discrete Normalizing Flows (DNF), (b) Continuous Normalizing Flows (b), (c) Flow Matching, (d) Denoising Diffusion Probabilistic Models (DDPM), and (e) Score-Based Generative Models at 6 GeV/c, integrated over the polar angle. }
    \label{fig:ratio_plots_kaon_6GeV}
\end{figure}

\begin{figure}
    \centering
    \begin{subfigure}[b]{0.49\textwidth}
        \centering
        \includegraphics[width=\textwidth]{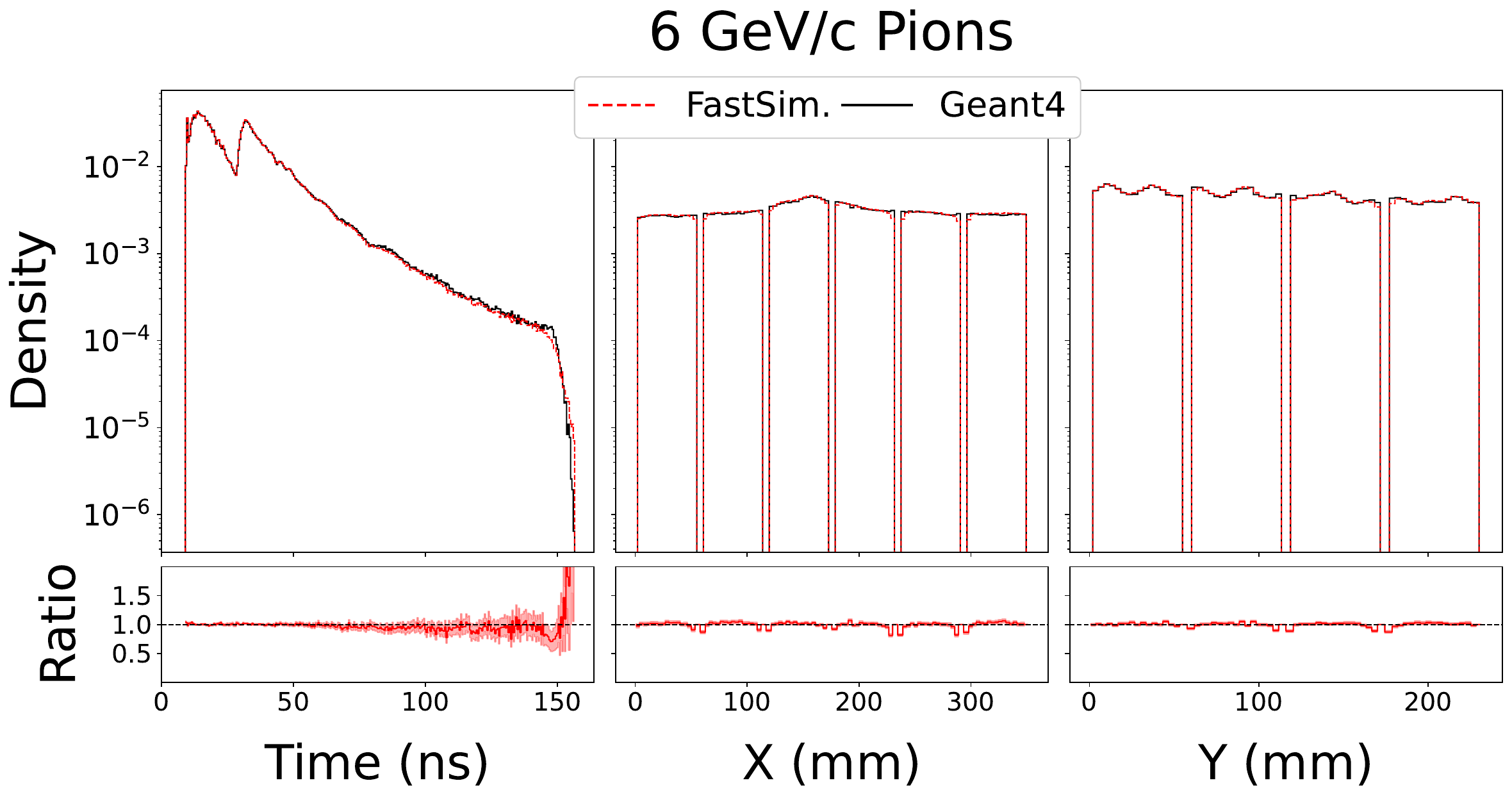}
        \caption{DNF}
    \end{subfigure}
    \begin{subfigure}[b]{0.49\textwidth}
        \centering
        \includegraphics[width=\textwidth]{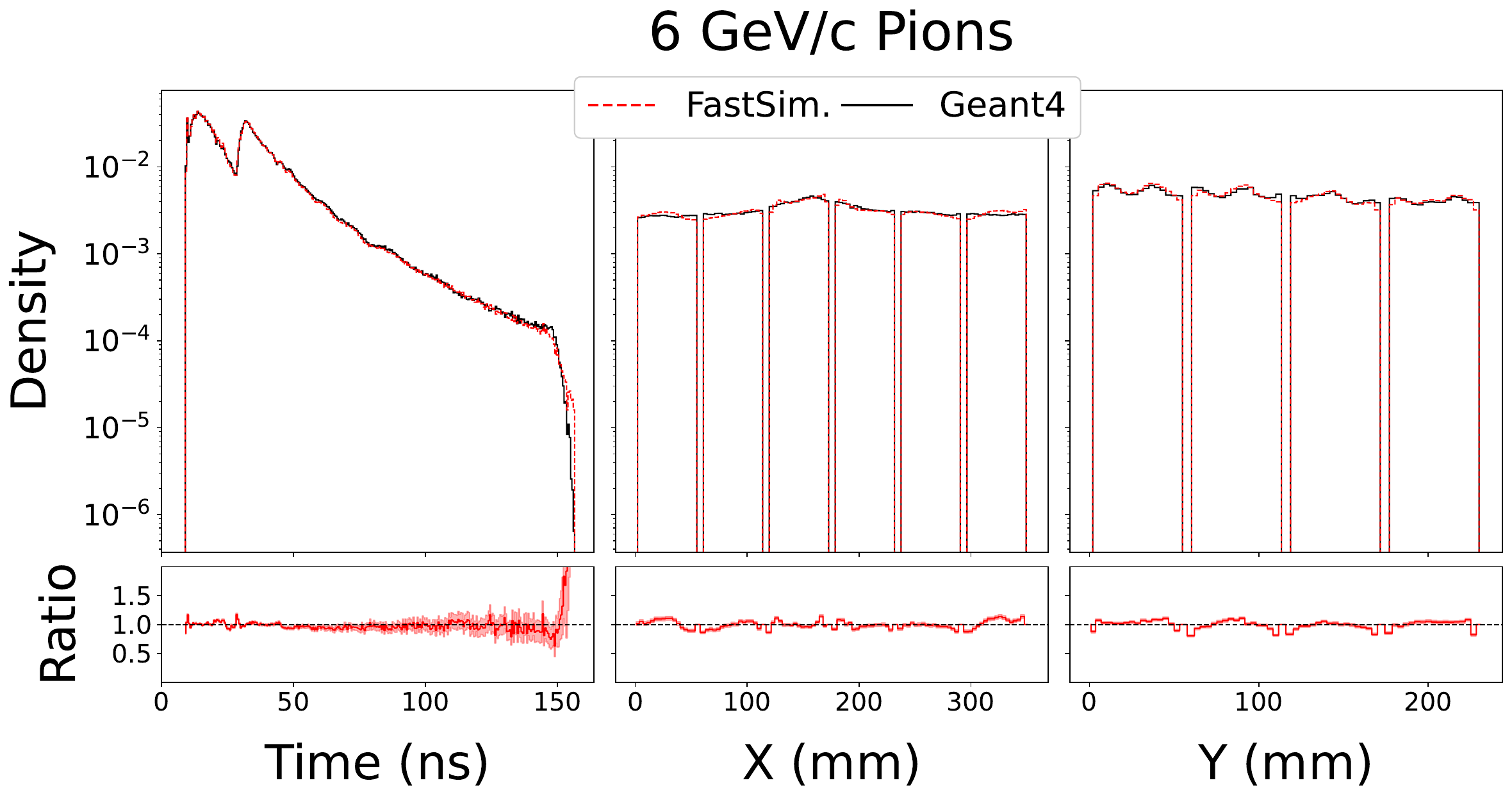}
        \caption{CNF}
    \end{subfigure} \\   
    
    \begin{subfigure}[b]{0.49\textwidth}
        \centering
        \includegraphics[width=\textwidth]{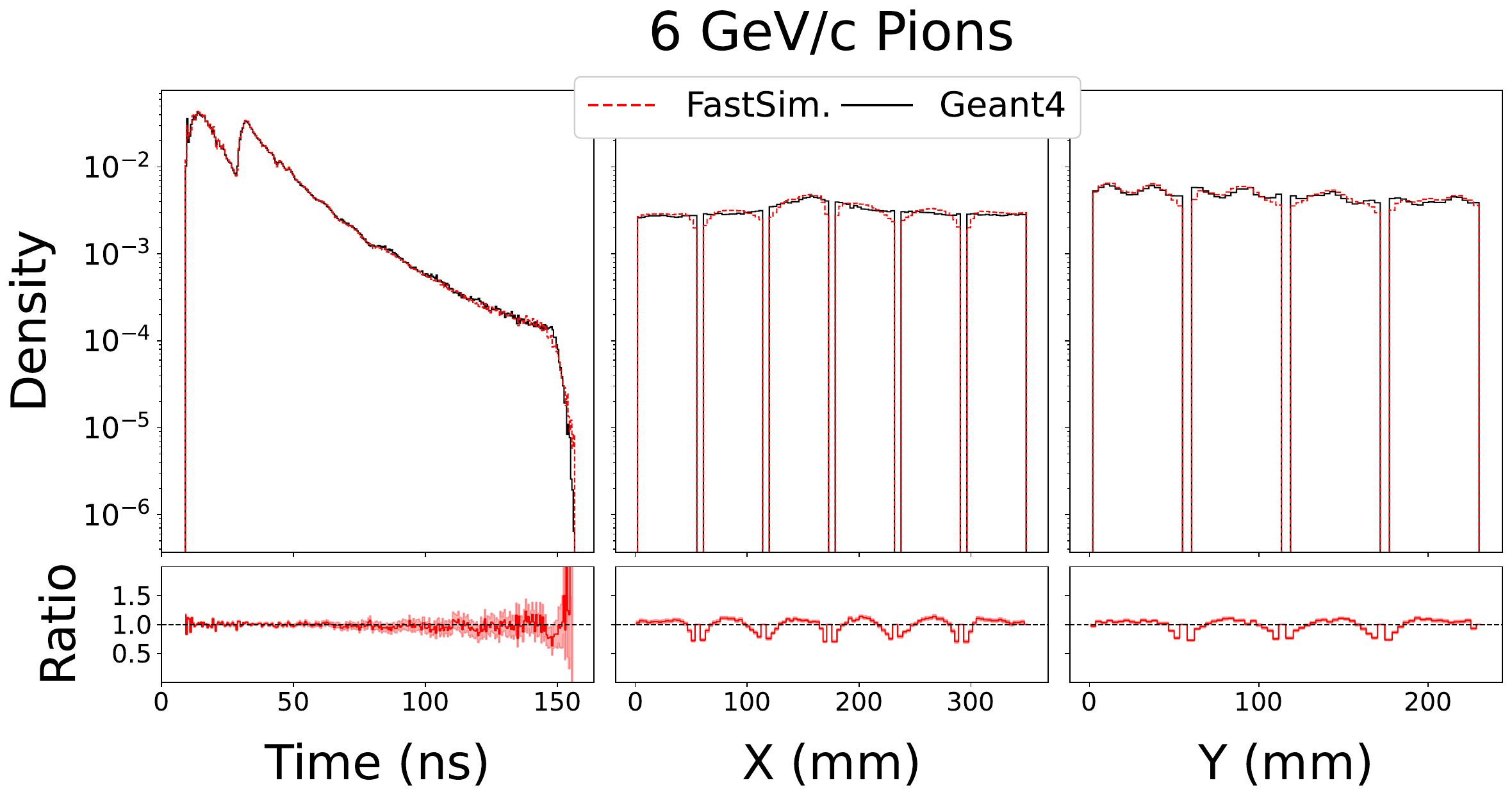}
        \caption{Flow Matching}
    \end{subfigure}
    \begin{subfigure}[b]{0.49\textwidth}
        \centering
        \includegraphics[width=\textwidth]{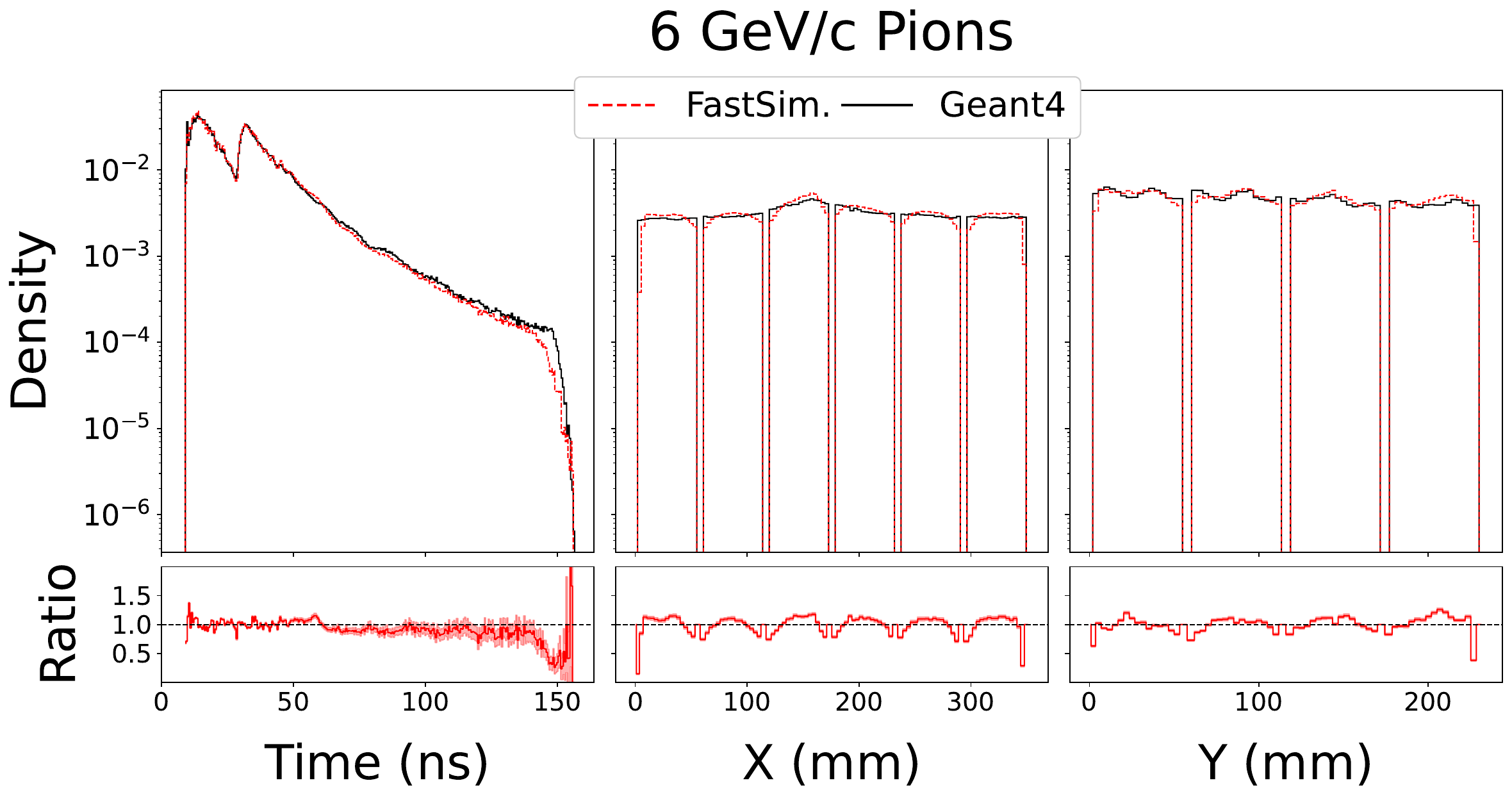}
        \caption{DDPM}
    \end{subfigure} \\  
    
    \begin{subfigure}[b]{0.49\textwidth}
        \centering
        \includegraphics[width=\textwidth]{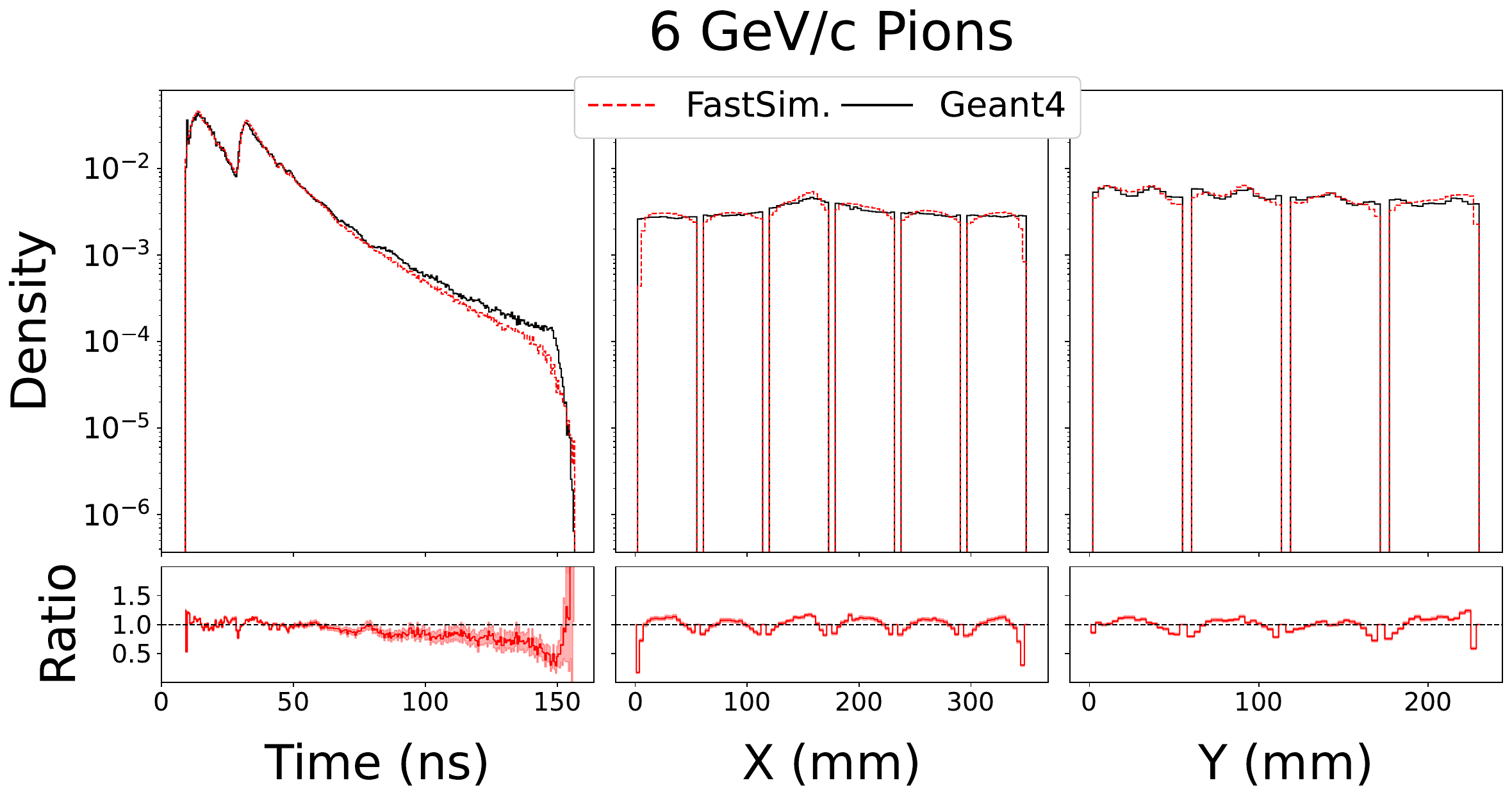}
        \caption{Score-Based}
    \end{subfigure}
    % \begin{subfigure}[b]{0.49\textwidth}
    %     \centering
    %     \includegraphics[width=\textwidth]{Figures/Generations/GSGM/6GeV/Ratios_Pion.pdf}
    %     \caption{GSGM}
    % \end{subfigure}
    \caption{\textbf{Ratio Plots at 6 GeV/c for pions:} Ratio plots pions using the various different models (a) Discrete Normalizing Flows (DNF), (b) Continuous Normalizing Flows (b), (c) Flow Matching, (d) Denoising Diffusion Probabilistic Models (DDPM), and (e) Score-Based Generative Models at 6 GeV/c, integrated over the polar angle. }
    \label{fig:ratio_plots_pion_6GeV}
\end{figure}

With respect to the last point, although we cannot prove it formally, we can provide a series of intuitive explanations for why likelihood based methods are perhaps more optimal for our task.\footnote{Note that for the CNF, even though it is likelihood based, the quality of the model we are able to fit is of lower quality than DNF simply due to their complications with training discussed prior. We expect that in theory, their performance should exhibit more convergent behavior.} This effect can be seen further through Fig. \ref{fig:ratio_plots_kaon_6GeV} and \ref{fig:ratio_plots_pion_6GeV} in which the marginal distributions for the likelihood free methods are generally worse than that of DNF.

%%%%%%%%%%%%%%%%%%%%%%%%%%%%%%%% Ratios %%%%%%%%%%%%%%%%%%%%%%%%%%%%%%%


%%%%%%%%%%%%%%%%%%%%%%%%%%%%%%%%%%%%%%%%%%%%%%%%%%%%%%%%%%%%%%%%%%%%%%

In the case of our Cherenkov data (spatial x,y and time), the data manifold is well-defined and low-dimensional, making likelihood-based methods particularly suitable. Approaches such as Diffusion Models and Flow Matching are often preferred when the underlying data manifold is ill-defined (\textit{e.g.}, images), as they implicitly learn data structure through iterative refinement and remove the strict implications of suitable DL algorithms that can be used (in reference to the bijectivity constraint of DNF). Moreover, likelihood-based learning provides a more explicit framework for learning the marginal distributions under the joint distribution. Since likelihood-based models, such as DNF, explicitly parameterize the data distribution and enforce bijectivity, they inherently preserve marginal consistency. This is in contrast to implicit generative models like diffusion models or flow matching, where marginals emerge from the learned joint distribution but are not directly optimized, which in turn could result in ``blurrier'' generations upon the aggregation of multiple photons. 

 We note that in general the tails of the distributions are the least performant, which is an expected result under the context of modern generative models. It is also important to consider that in the spatial components there exists multiple, and periodic ``tails'' due to PMT spacing, which increases the complexity of the both the learning and generation procedures. %While our smearing operation removes the discretization effects due to the pixelized readout, we are only allowed to smear within the physical limitations of individual sensors.
We also note the effects at the boundaries of PMT's is not uniform across the kinematics, indicating some regions of the phase space are easier to learn than others. This can be seen through further comparison of ratios in the \ref{app:3GeV} and \ref{app:9GeV}.

\subsection*{Model Comparisons}

\begin{figure}[h]
    \centering
    \begin{subfigure}[b]{0.49\textwidth}
        \centering
        \includegraphics[width=\textwidth]{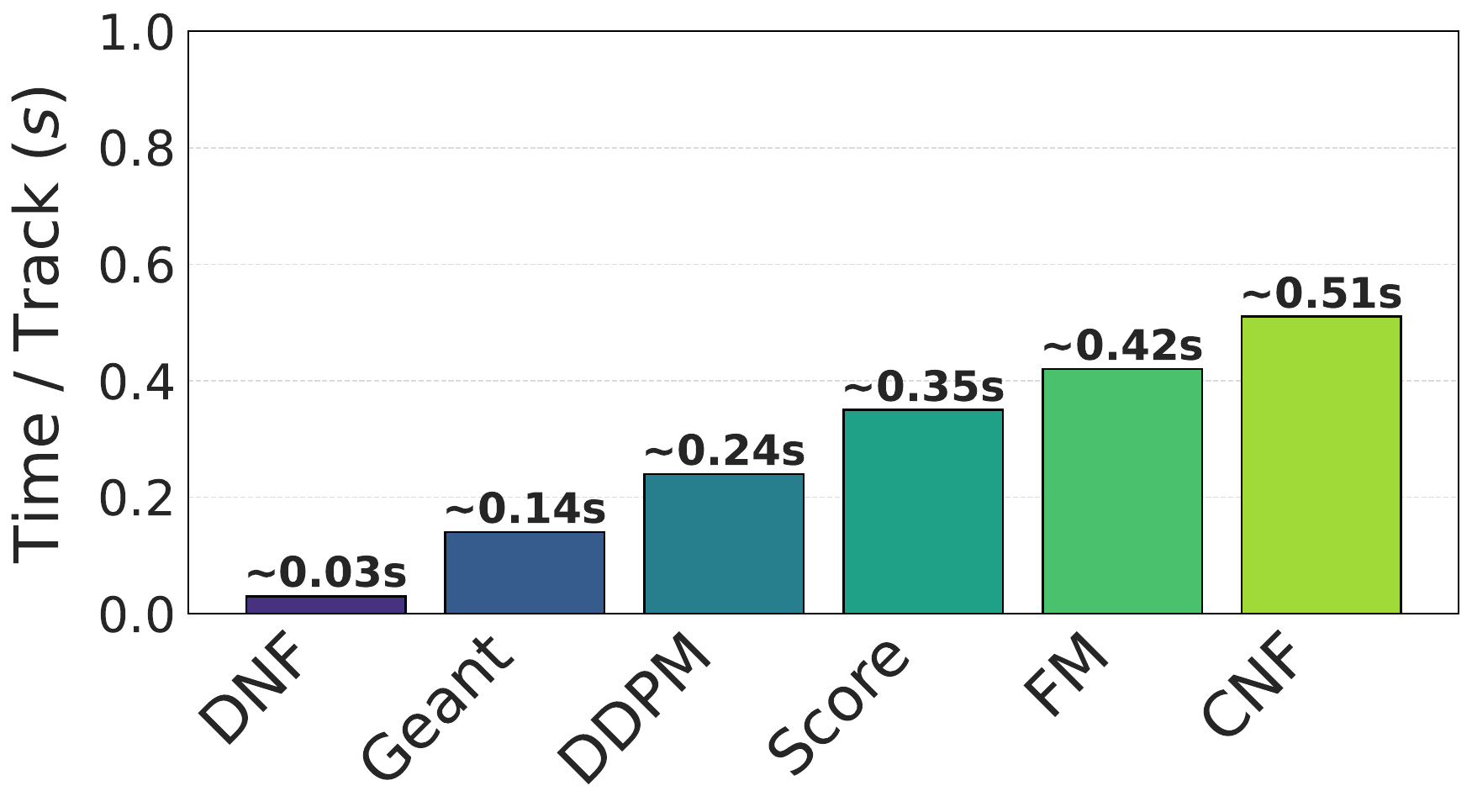}
        \caption{Track Level}
    \end{subfigure}
    \begin{subfigure}[b]{0.49\textwidth}
        \centering
        \includegraphics[width=\textwidth]{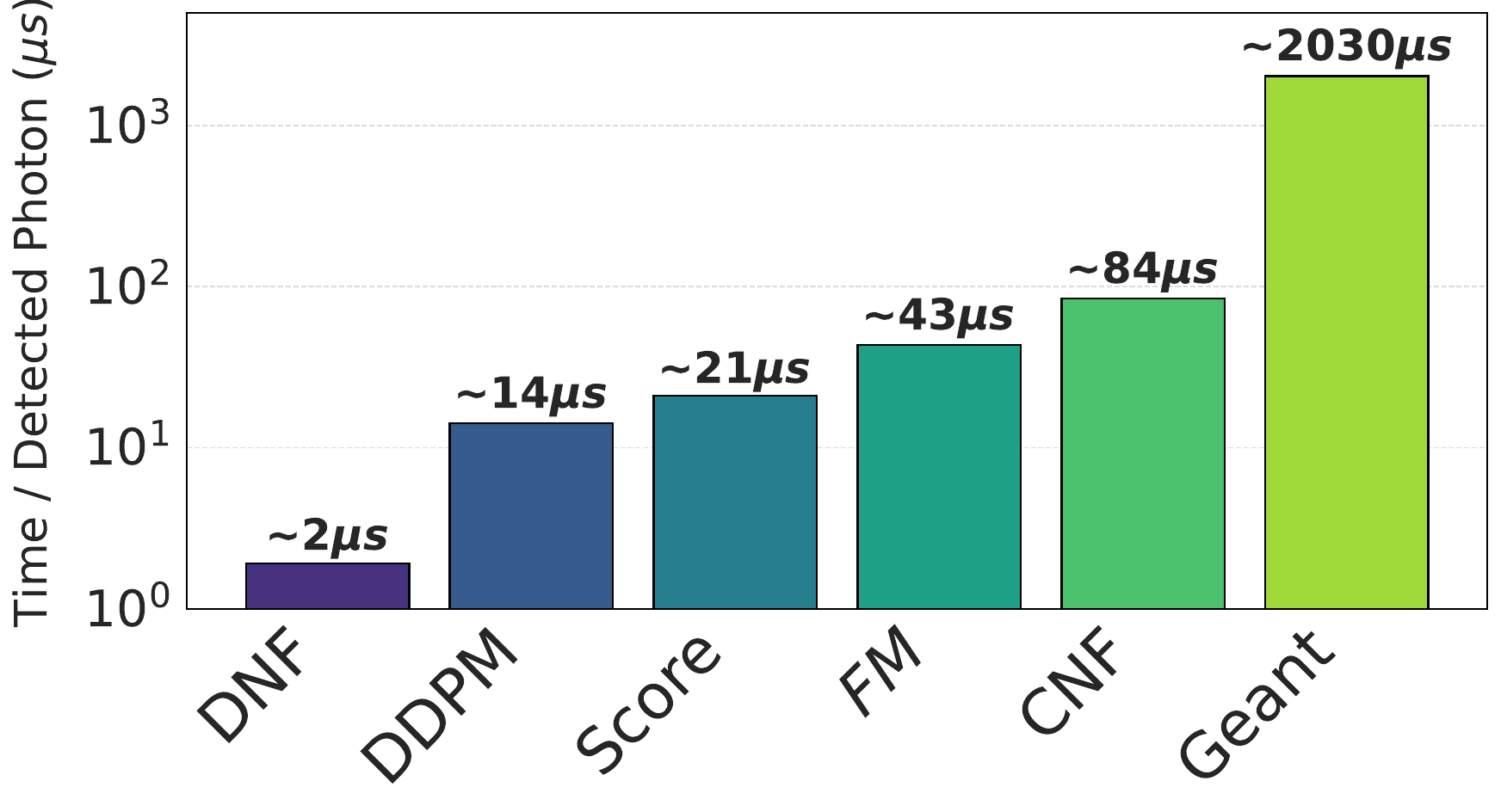}
        \caption{Photon Level}
    \end{subfigure} 
    \begin{subfigure}[b]{0.49\textwidth}
        \centering
        \includegraphics[width=\textwidth]{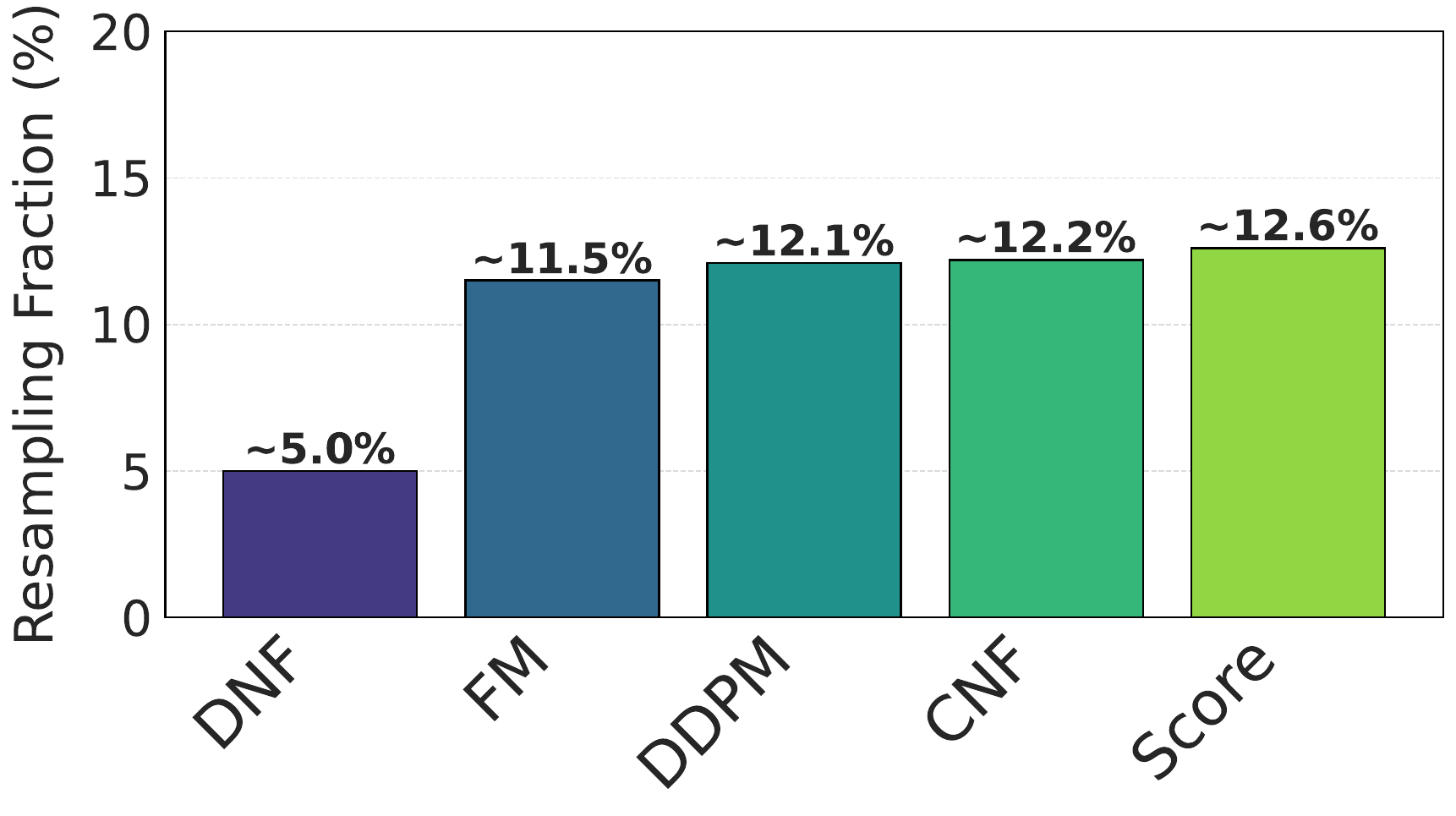}
        \caption{Resampling Fraction}
    \end{subfigure}
    \caption{\textbf{Average generation metrics for various models:} Average generation metrics for the various models, (a) average generation time per track, (b) effective generation time per detected photon, and (c) average resampling fraction. Values are estimates from generations over the entire phase-space. Models on the x-axis are sorted in ascending order for ease of viewing. Model timings are obtained with a single Nvidia A40 GPU, \geant timing are obtained with an AMD EPYC 7313P processor (single core).}
    \label{fig:generation_metrics}
\end{figure}

%Apart from the obvious question of individual model fidelity, 
There also exists other crucial aspects one must consider, such as generation time and the resampling fraction incurred by our models under our prior, the former being a limiting factor in cases of both fast simulation for dataset creation and potential PID applications, whereas the latter provides clear indication of the model's ability to correctly learn the underlying distribution in the presence of discontinuities caused by the physical sensor layout. In the ideal case, no prior would need to be imposed at generation, although this is not realistic given the smooth function the architectures will inherently learn. Fig. \ref{fig:generation_metrics} (a) shows the generation time per track, (b) effective generation time per detected photon and (c) the resampling fraction for each model, averaged across the phase space.

It is clear that DNF has a drastic advantage over other methods in terms of average computational time per track, effective detected photon time, and average resampling\footnote{Note that a resampling fraction does not pertain to \geant.}, noting order(s) of magnitude reduction in computational time in comparison to all methods, including \geant. 
 Given that we are able to generate batches of arbitrary numbers of photons, crucial for tasks such as reference PDF creation (\textit{e.g.}, those in \ref{app:fastDIRC_PDFs}), we note a generation time per detected photon for DNF of $\sim \SI{2}{\micro\second}$, in comparison to \geant which is on the order of $\sim \SI{2000}{\micro\second}$ - a reduction of $\mathcal{O}(100-1000 \times)$.
 This improvement is partly due to the fact that our architectures inherently bypass detector inefficiencies. In contrast, \geant simulates a significantly larger number of Cherenkov photons than are ultimately detected, due to losses from transport inefficiency, quantum efficiency, and other instrumental effects. As a result, the time required to simulate a \textit{detected} photon in \geant does not scale linearly as one may intuitively expect.
 While this is not an explicitly fair comparison given we are exploiting parallel computation of our models on a GPU, an equivalent timing per photon can be achieved using $\sim $ hundreds CPU cores with \geant. 

%In all cases, the simplest of the three models exhibits an order of magnitude difference.
In light of this, along with the results shown prior, we conclude the the simple DNF is the most optimal choice given our task and will serve as the baseline for evaluations to follow. Note that it may be possible for others to create higher performance models of various types, in which we remind the reader given the open and community oriented philosophy we follow, such advancements are welcomed and encouraged.

\subsection*{Evaluation with FastDIRC}

One of the primary focuses of the seminal FastDIRC \cite{hardin2016fastdirc} work is increased PID performance in comparison to traditional geometrical LUT methods for pions and kaons. Attempting to provide real time simulations, therfore removing the necessity for storage of large reference datasets, and allowing the building of per-track PDFs (as opposed to per-bar).
For a given track, $\mathcal{O}(10000) \times N_{\gamma}$ photons are simulated under each mass hypothesis, \textit{e.g.}, pion or kaon, where $N_{\gamma}$ corresponds to the number of photons produced by the inference track. These reference populations can then be used to form a DLL and obtain PID estimates for a given track through Kernel Density Estimation (KDE). In their formulation, the authors propose a diagonal Gaussian Kernel for the KDE estimates, in which the covariance matrix corresponds to one, or two times the individual dimensions of pixels (along x and y), and some scaling of the timing resolution. For a given Cherenkov photon $\vec{x}_i$ within a track, the log-likelihood is proportional to that in Eq. \ref{eq:fastDIRC_LL}, where $\vec{\mu}_{K \pi}$ represents the support distribution of either pions or kaons. 

\begin{equation}\label{eq:fastDIRC_LL}
   log ~ p(\vec{x}_i|\vec{k})_{K \pi} \propto   (\vec{x}_i - \vec{\mu}_{K \pi })^T \boldsymbol{\Sigma}^{-1} (\vec{x}_i - \vec{\mu}_{K \pi})\, , \; \boldsymbol{\Sigma} =
    \begin{bmatrix}
    p_{width}^2 & 0 & 0 \\
    0 & p_{height}^2 & 0 \\
    0 & 0 & \frac{\sigma_t^2}{2} \\
    \end{bmatrix}
\end{equation}

The resulting likelihoods for each mass hypothesis can then be used to compute a DLL and perform PID. To quantify the separation power, we fit Gaussian distributions to the respective PIDs and calculate the distance between their means through Eq. \ref{eq:sigma_sep}.\footnote{The definition of the separation power described by Eq. \ref{eq:sigma_sep} is consistent with what is used in the Yellow Report \cite{khalek2022science}.}  

\begin{equation}\label{eq:sigma_sep}
    \sigma_{sep.} = 2\frac{|\mu_{K} - \mu_{\pi}|}{\sigma_{K} + \sigma{\pi}}
\end{equation}

In our case, we do not simulate $\mathcal{O}(10000) \times N_{\gamma}$ but rather compute various fixed sized reference distributions for each particle at inference in order to provide performance comparisons between the fast simulated PDFs and those from \geant. 
We only utilize the largest fast simulated PDF for comparison to \geant, whereas for \geant we consider multiple reference population sizes (indicated in the legend). This choice is motivated by the potential for intrinsic differences in the simulation methods (which will be discussed in detail). Since the fast simulation is designed to generate large numbers of photons efficiently, we assume that using the largest available reference set provides the best performance without computational constraints. 
However, for \geant, we wish to investigate whether there exists a point of ``convergence'', where a sufficiently large but finite number of fast simulation photons yields equivalent performance to that of \geant. In both cases, we deploy a time cut of $\SI{100}{\nano\second}$.

\begin{figure}[!]
    \centering
    \includegraphics[width=0.7\textwidth]{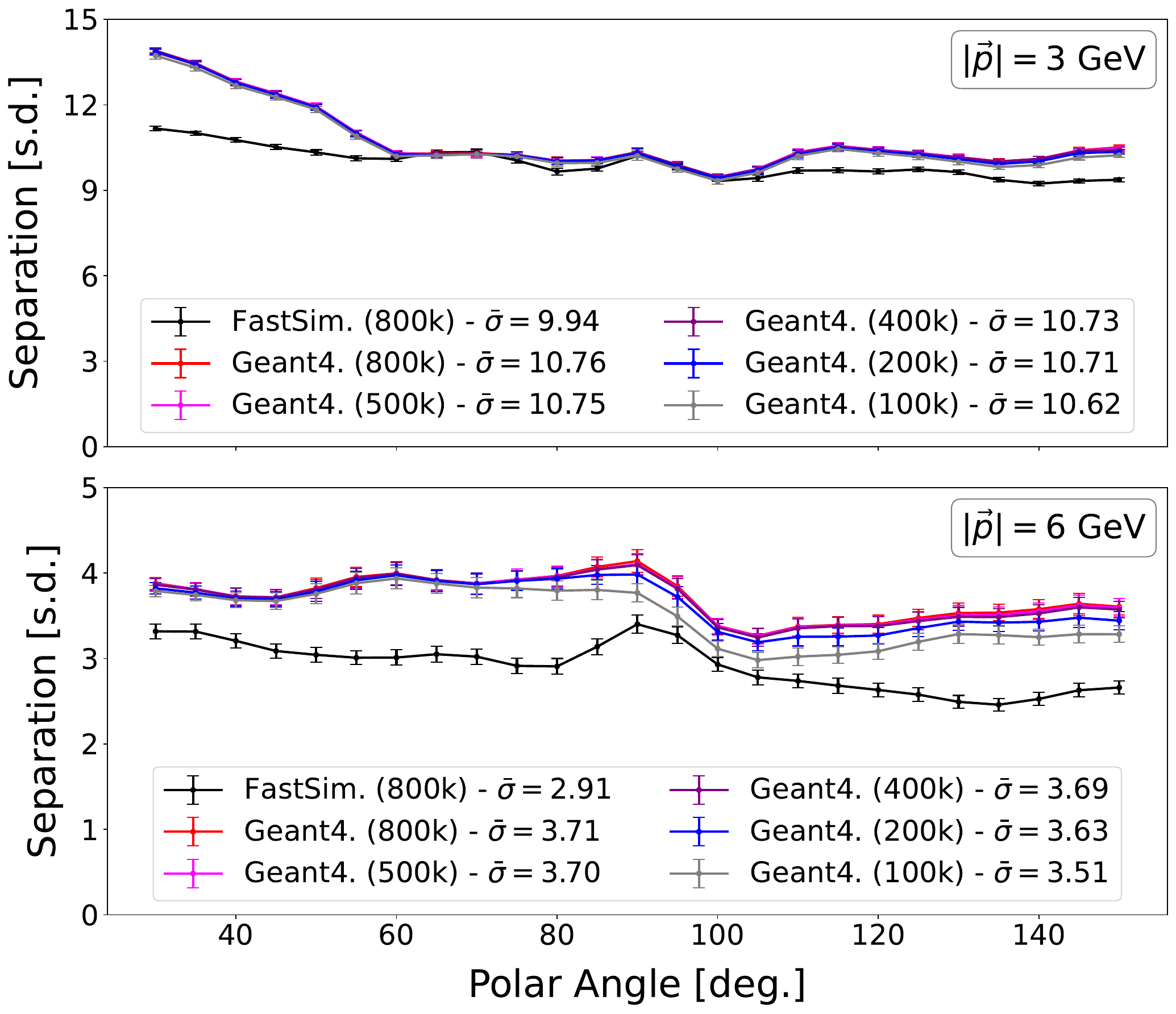}
    \caption{\textbf{FastDIRC Performance Comparison:} Particle Identification performance of pions and kaons using FastDIRC with both fast simulated, and various sized \geant reference populations (indicated in the legend) at 3 GeV/c (top) and 6 GeV/c (bottom) for various values of the polar angle.}
    \label{fig:fastDIRC_performance}
\end{figure}

In reality, we wish to utilize as many photons as possible, which is only limited by the available VRAM of the individual GPU. However, as seen in Fig. \ref{fig:fastDIRC_performance}, the performance quickly saturates, suggesting that beyond a certain reference size, additional photons contribute minimally to the accuracy of the reconstruction. Our version of FastDIRC provides an inference time on the order of $\mathcal{O}(\SI{0.03}{\second})$ per track with a reference PDF of 800k photons. Fig. \ref{fig:fastDIRC_performance} depicts the performance comparison for $\SI[per-mode=symbol]{3}{\giga\eVperc}$ (top) and $\SI[per-mode=symbol]{6}{\giga\eVperc}$ (bottom), for various values of the polar angle (in bins of $\SI{5}{\degree}$). 

From inspection of Fig. \ref{fig:fastDIRC_performance}, it is clear that the reference PDFs provided by the fast simulation are less performant than their \geant counterparts, in which the relative separation power decreases in an inversely proportional fashion to momentum again promoting the idea that some regions of the phase space are more difficult to learn than others.
While not initially obvious, we argue that our fast simulation does in fact achieve high fidelity, accurately reproducing the expected distributions with minimal deviation. However, a key consequence of our learned PDFs through generative models is their increased smoothness, which, while reducing statistical noise, also makes the generated pion and kaon distributions more similar on average. This smoothing effect can diminish the separation between particle species, therefore impacting classification performance. Given that at regions of the phase space (\textit{i.e.}, as momentum increases), the difference in ring structure between pions and kaons only differs by a few pixels spatially, any amount of smoothing will show reduced performance under perfectly clean inference tracks from \geant. 
Nevertheless, we are able to approach the required separation power of $3\sigma$ at $\sim \SI[per-mode=symbol]{6}{\giga\eVperc}$, as outlined in the EIC Yellow Report \cite{khalek2022science}.
Additionally, while the fast simulation captures the overall structure well, it is unable to fully reproduce certain complex, discrete geometrical effects inherent to the detailed detector modeling in full \geant-based simulations \cite{va2008simulation,va2014optical}. An example of such an effect is the Kaleidoscopic effect, which translates to appeared ``pixel preferences'' at the readout due to the lack of total internal reflection during photon propagation, with the effect being more pronounced the closer a track hits to the expansion volume. This effect is visualized in \ref{app:fastDIRC_PDFs} (Fig. \ref{fig:example_PDFs}, top left) for pions at $\SI{40}{\degree}$, in which other reference polar angles for pions and kaons can be found in therein.

% \begin{figure}
%     \centering
%     \includegraphics[width=\textwidth]{Figures/FastDIRC/ExamplePDFs/Example_PDFs_theta_40.0_p_6.0_PID_pion.pdf} %
%     \caption{\textbf{Example Probability Density Functions for FastDIRC:}}
%     \label{fig:example_PDFs_40deg}
% \end{figure}

In essence, we have ``noisier'' simulations of the two particle types given the smoothing effect induced by the generative models. In light of this, the correct structure of the Cherenkov rings in both spatial and time components is preserved by our generative models, indicating high fidelity with limitations on ability to capture low level detector effects and increased smoothness.

%%%%%%%%%%%%%%%%%%%%%%%%%%% Photon Yield Studies %%%%%%%%%%%%%%%%%%%%%%%%%%%%%%%
\subsection*{Photon Yield Sampling}

\begin{figure}
    \centering
    \begin{subfigure}[b]{0.9\textwidth}
        \includegraphics[width=0.45\textwidth]{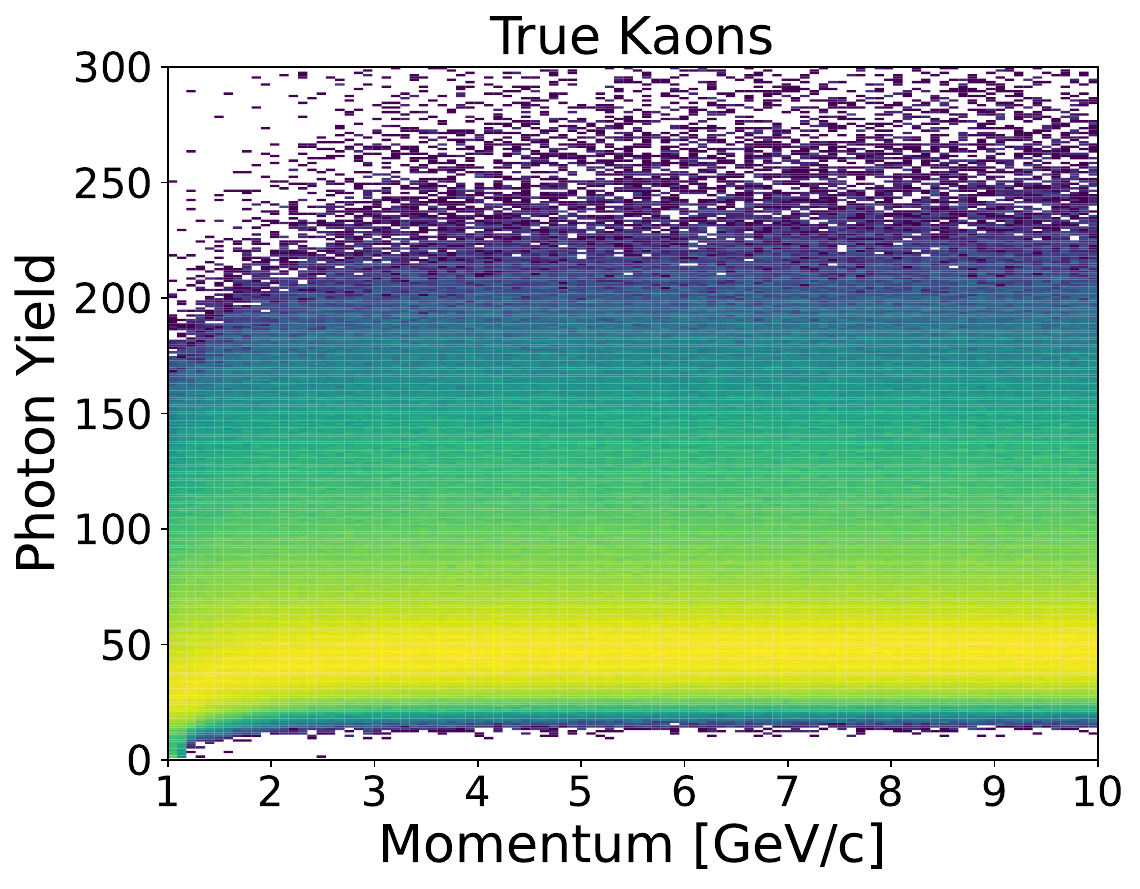} %
        \includegraphics[width=0.44\textwidth]{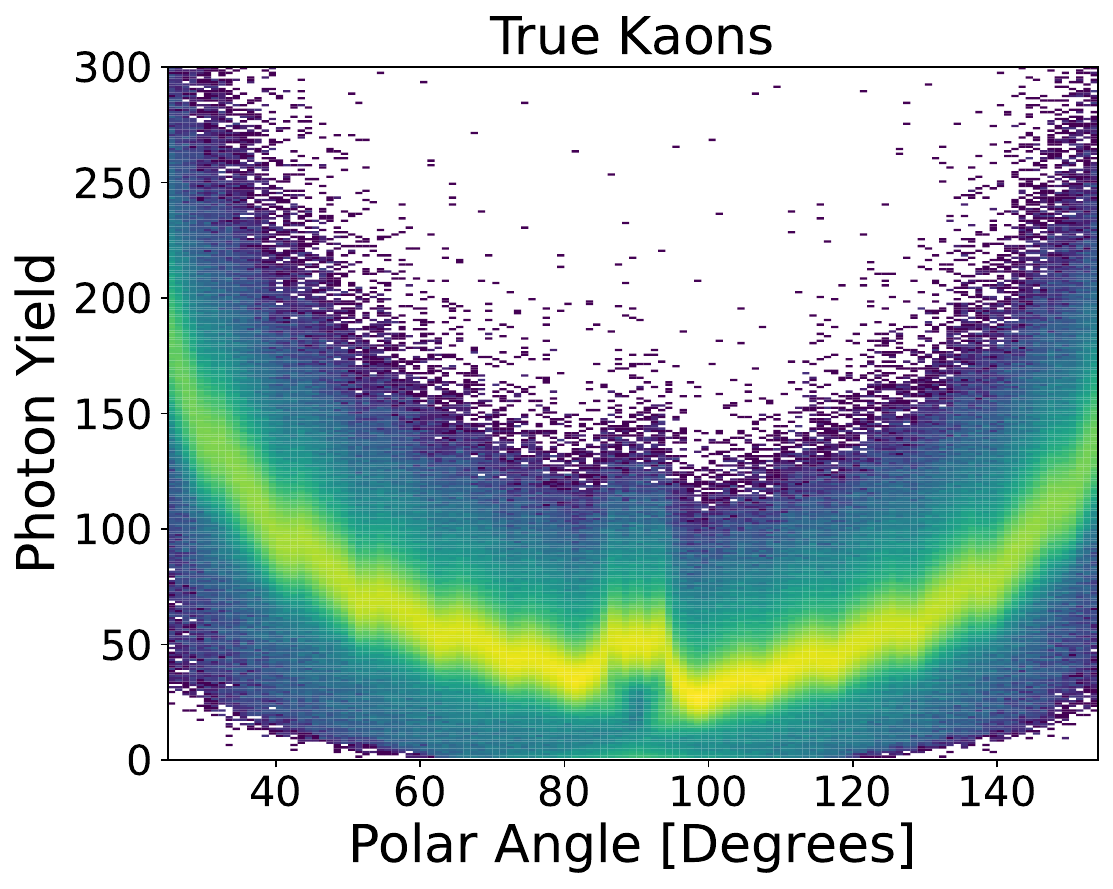} \\
        \includegraphics[width=0.45\textwidth]{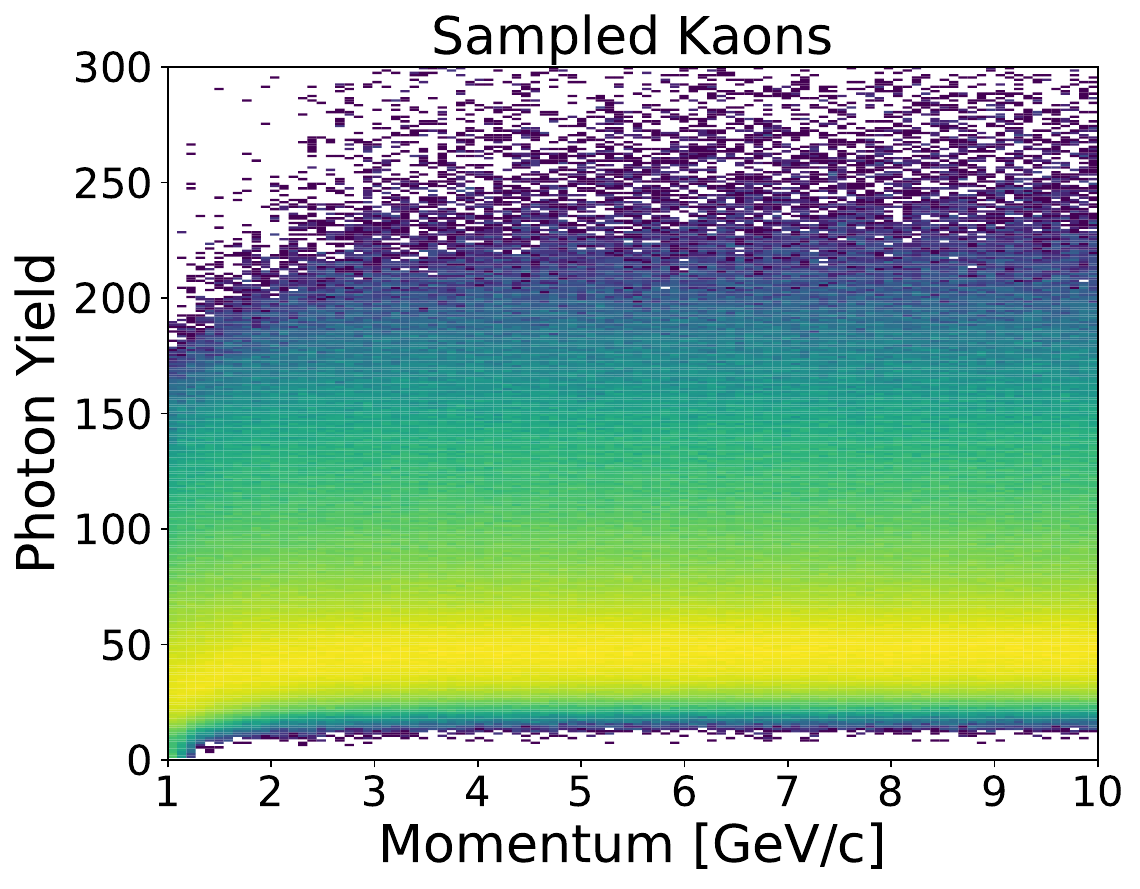} %
        \includegraphics[width=0.44\textwidth]{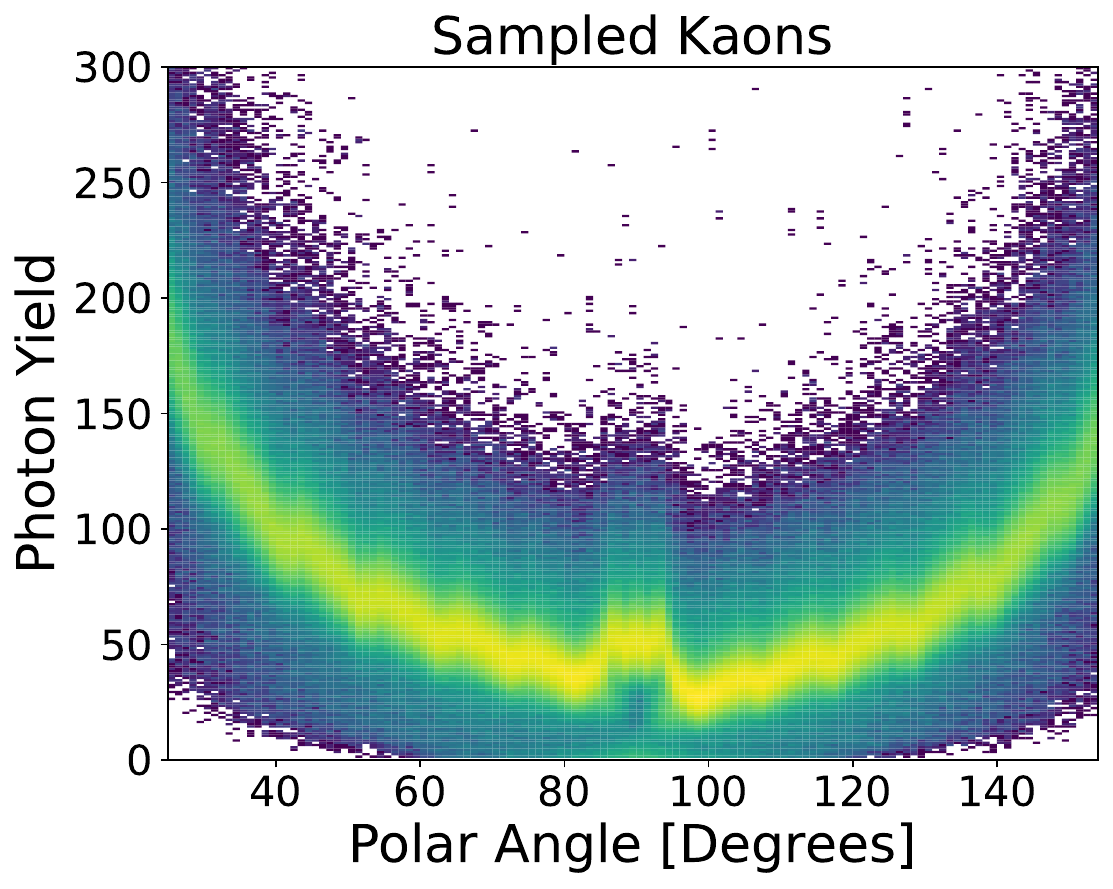}
        \caption{Kaons}
    \end{subfigure}
    \begin{subfigure}[b]{0.9\textwidth}
        \includegraphics[width=0.45\textwidth]{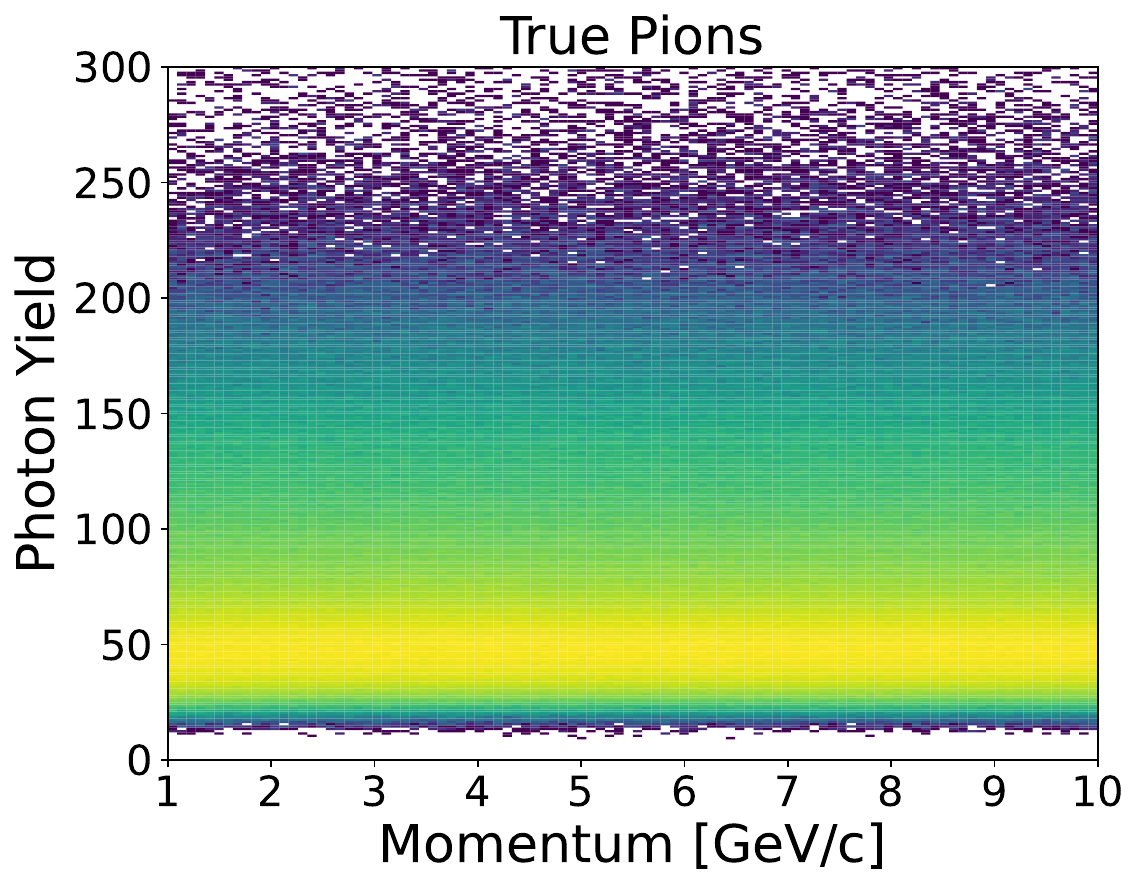} %
        \includegraphics[width=0.44\textwidth]{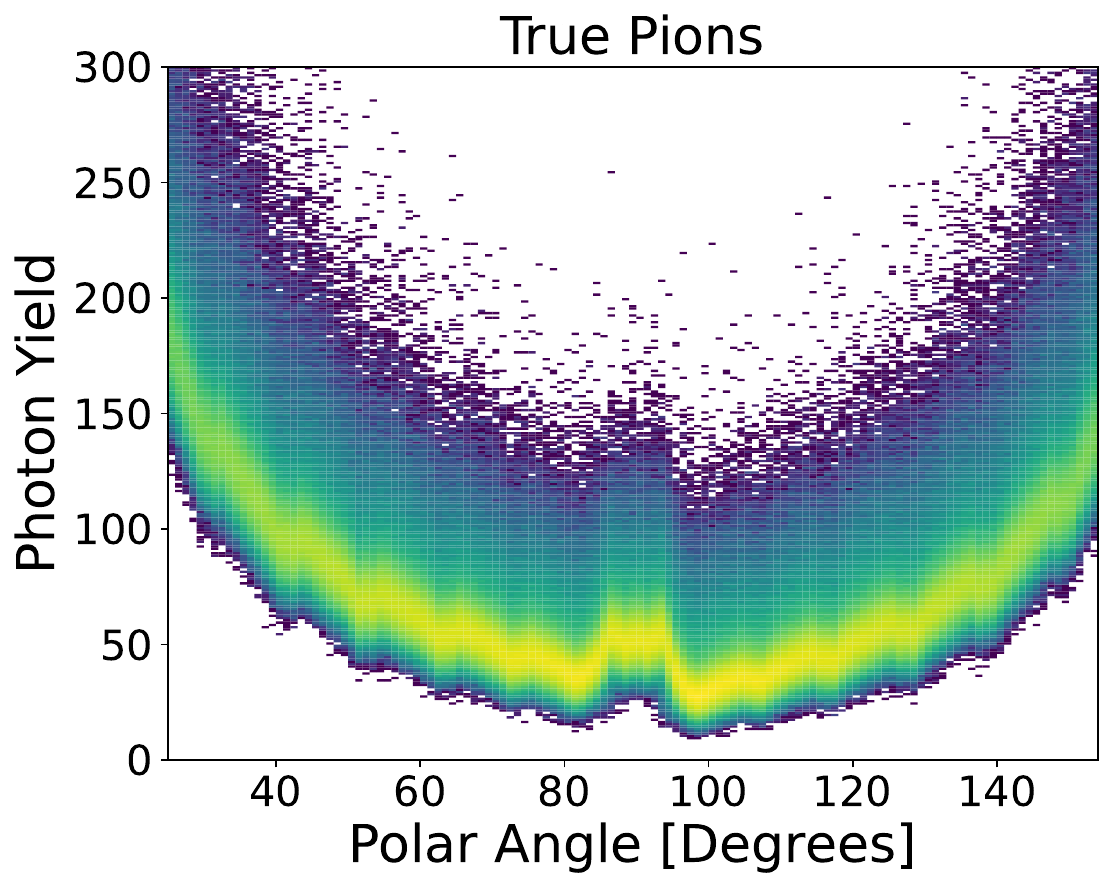} \\
        \includegraphics[width=0.45\textwidth]{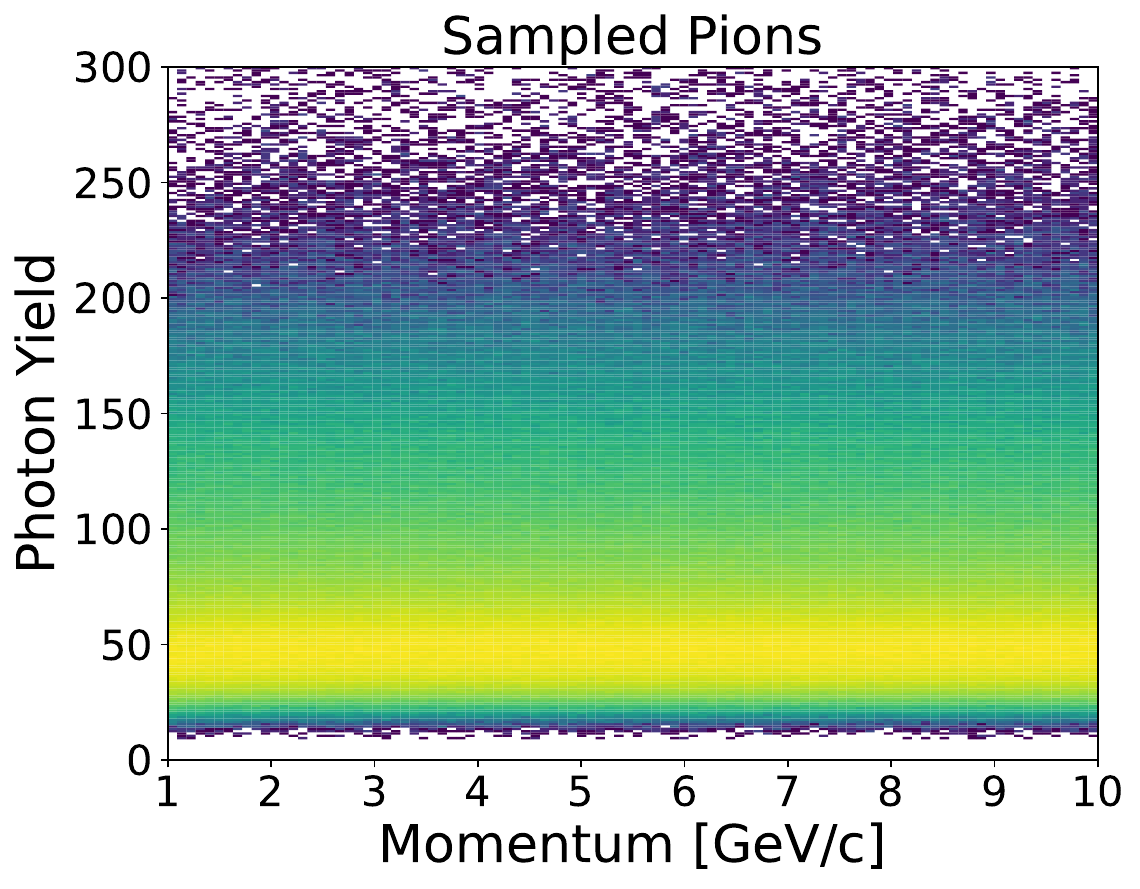} %
        \includegraphics[width=0.44\textwidth]{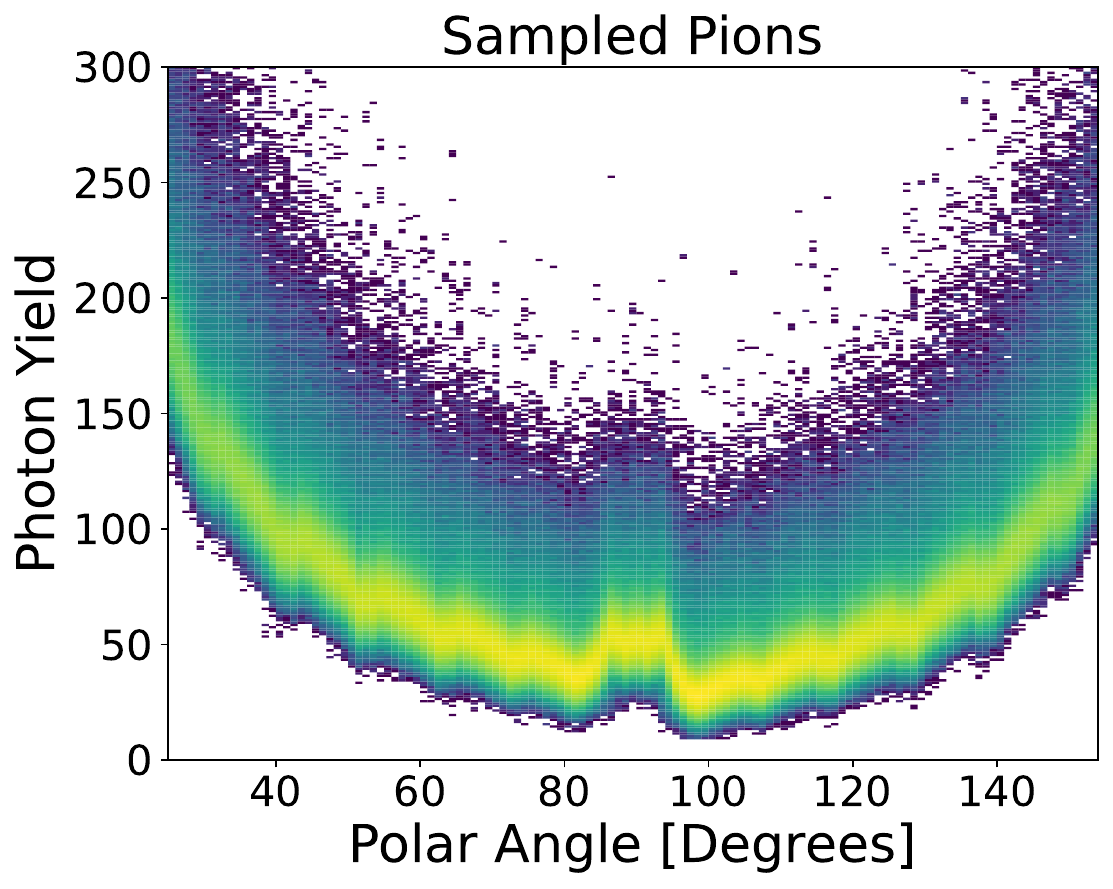} 
        \caption{Pions}
    \end{subfigure}
    \caption{\textbf{Comparison of Photon Yield Distributions as a function of the phase space:} Histogram comparison of our photon yield sampling procedure to the ground truth from \geant as a function of momentum (left column), and polar angle (right column) for both kaons (a) and pions(b).}
    \label{fig:LUT_hists}
\end{figure}

In what follows, we validate our photon yield sampling procedure through a series of closure tests. First, we compare at the histogram level the generated photon yield to the ground truth from \geant. Given the explicit dependence of the photon yield, we perform this as a function of the momentum and polar angle, for both pions and kaons in Fig. \ref{fig:LUT_hists}

\begin{figure}
    \centering
    \includegraphics[width=0.45\textwidth]{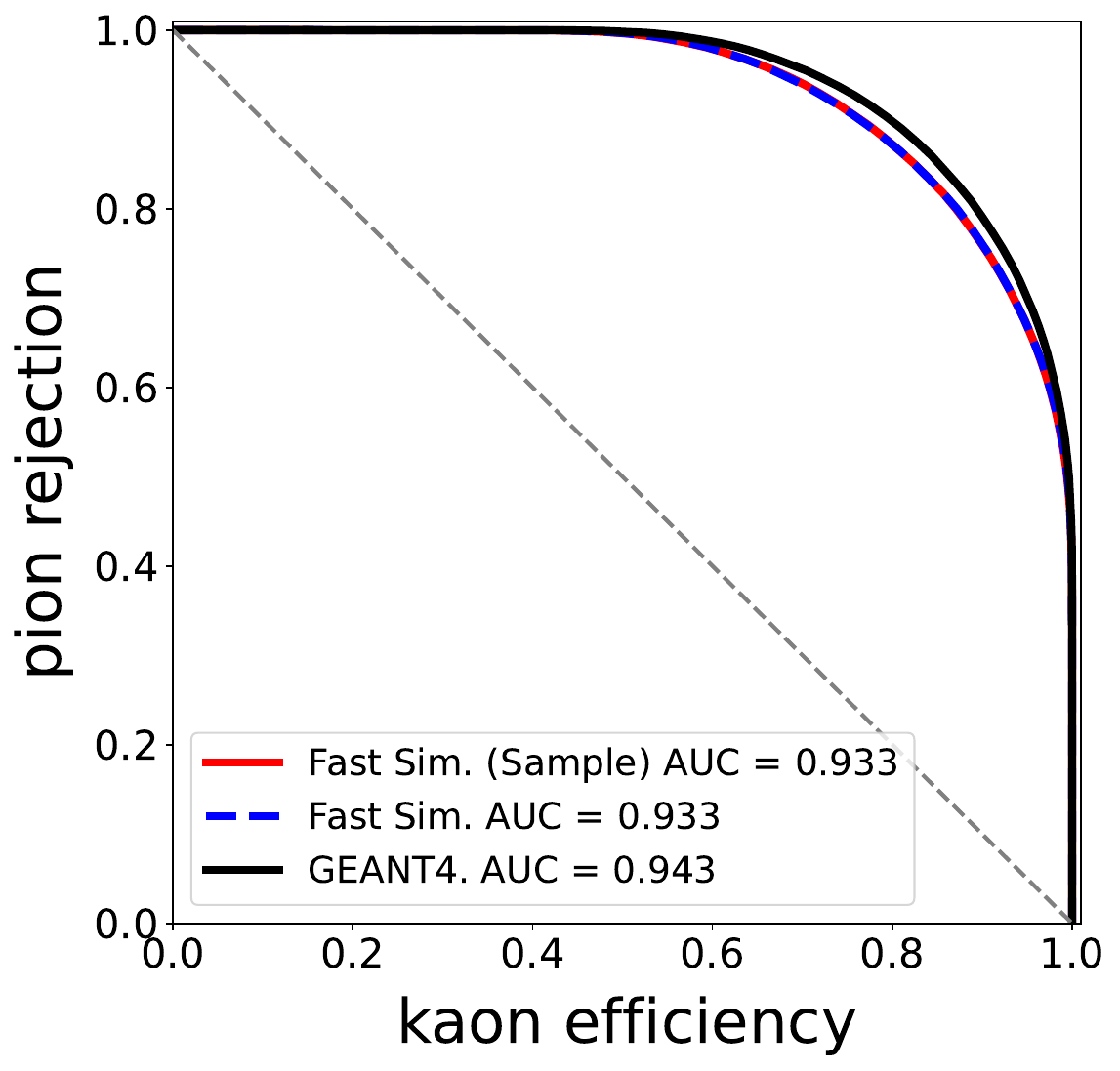} %
    \includegraphics[width=0.53\textwidth]{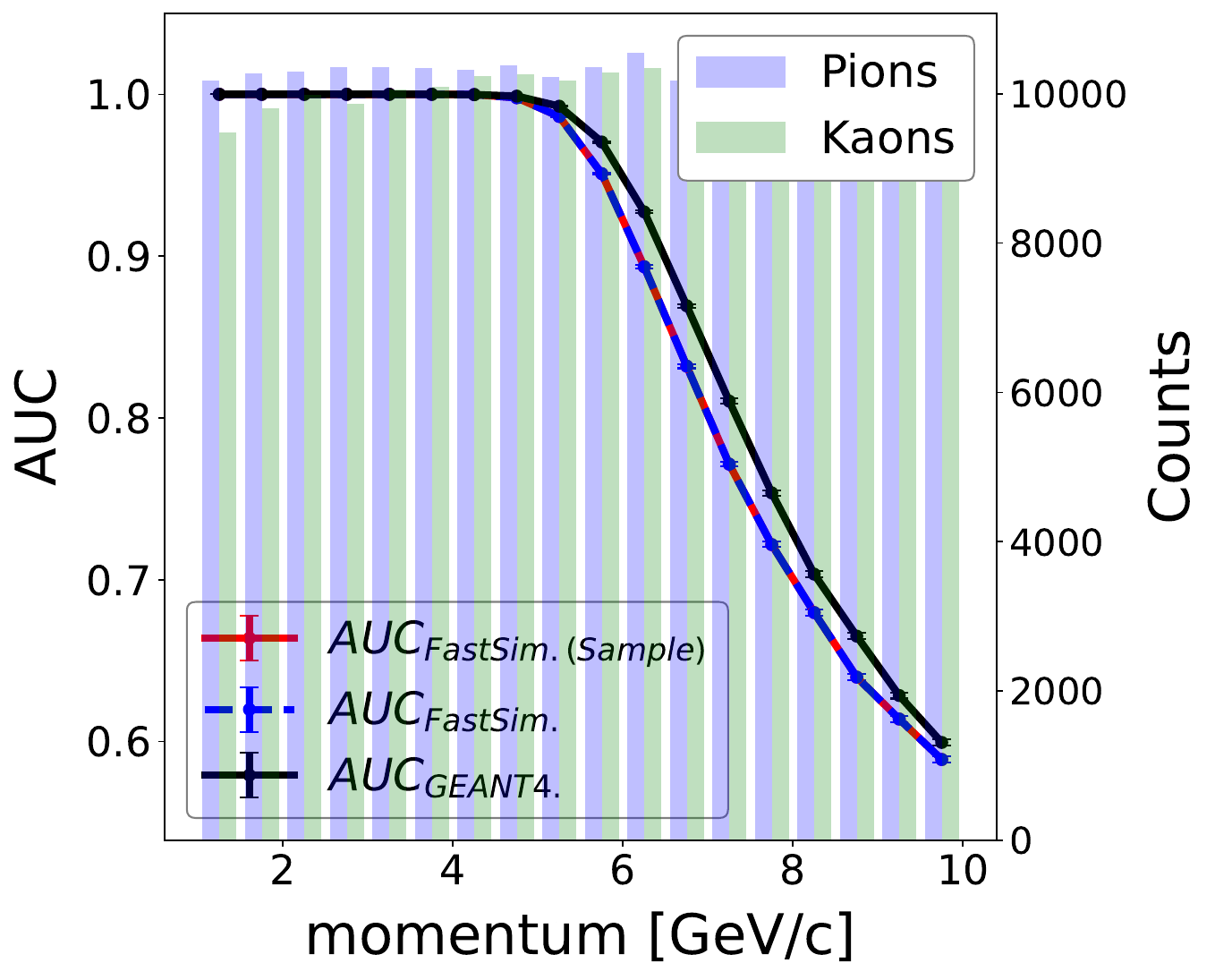} %    
    \caption{\textbf{Performance Comparison over the entire Phase Space:} Pion rejection as a function of kaon efficiency, integrated over the phase space (left), comparing the CNF performance on (i) fast simulated samples with photon yields matching those from the corresponding \geant tracks, (ii) fast simulated samples with photon yields derived from our LUT approximation (indicated as "Sample" in the legend), and (iii) an independent \geant sample. (Right) Area Under the Curve (AUC) of the left plot, shown as a function of momentum, highlighting the explicit momentum dependence of the Cherenkov angle for pions and kaons.}
    \label{fig:CNF_eval}
\end{figure}

From visual inspection, we can see that our histograms match to a high degree, providing initial validation of our procedure. Specifically, we note the agreement as a function of the polar angle, in which their exists a high degree of kinematic dependence. In the case of momentum, the photon yield tends to saturate giving rise to a lesser dependence, although it still remains. To further validate our method, under the same working point of continuously across the phase space, we employ the DLL method with generative models developed in \cite{fanelli2024deep} using the CNF. While this model is not optimal in terms of separation power, it provides a reliable proxy given that the method treats the likelihood of each individual hit independently through a summation. This allows us to identify possible over or under sampling of the photon yield causing distortions in the total likelihood of a track, and therefore the separation power of the two particles ($\pi / K$). Moreover, this method further validates the fidelity of our fast simulation in terms of more average behaviors over the phase space. Fig. \ref{fig:CNF_eval} (left) shows the performance of our model in terms of pion rejection as a function of kaon efficiency, integrated over the phase space, where we have included the performance of the CNF on (i) fast simulated samples whose photon yield corresponds to that of the associated track in \geant (those from (iii)), (ii) fast simulated samples whose photon yield is derived from our LUT approximation (indicated by ``Sample'' in the legend), and (iii) an independent \geant sample. Fig. \ref{fig:CNF_eval} (right) shows the Area Under the Curve (AUC) (referring to the plot on the left), as a function of momentum, given the explicit momentum dependence of the Cherenkov angle for pions and kaons.

First, we note there exists zero difference in our fast simulated data with, or without photon yield sampling, indicated through both the curves and their associated AUC, a clear indication that our LUT method suffices as a reliable method. Second, we note that performance of our fast simulated data differs on the order of $\mathcal{O}(1\%)$ from that of \geant, another indication that our model is able to produce high fidelity generations. Note that in prior classification based evaluation methods, we utilized fixed point kinematics which essentially tests the model under extreme cases.\footnote{We again remind the reader that the model is trained over a continuous phase space.} In downstream tasks where researchers use our fast simulated data to train their models over continuous phase spaces (\textit{e.g.}, complete, or restricted kinematic coverage), the resulting proxy remains even more valid.

%% file: 5_impacts.tex
\section{Broader Impact}\label{sec:impacts}

The impact of our contribution is multidisciplinary, reaching audiences in both the physics and DL communities alike. We have introduced the first holistic simulation pipeline for DIRC detectors at EIC (and DIRC detectors in general) using a deep generative model approach, capable of reducing the computational burden and time associated with current \geant based simulation packages, providing potential directions for increased research productivity in both classical and DL-inspired PID methods through our surrogate model proxies. Moreover, our fast simulation pipeline is the first step towards enabling high performance methods developed by physicists such as FastDIRC \cite{hardin2016fastdirc} and time imaging \cite{Dzhygadlo_2020} at EIC, removing the requirement for large file storage systems to house reference datasets. Instead, these datasets are capable of being generated in real time at inference, noting a effective generation time of $\mathcal{O}(\SI{2}{\micro\second})$ per photon in construction of the FastDIRC PDFs with DNF using a single GPU. These methods, specifically our version of  the FastDIRC PID method using a GPU, are highly parallelizable across modern compute facilities with hundreds of GPUs, providing an inference time on the order of $\mathcal{O}(\SI{0.03}{\second})$ per track with a reference PDF of 800k photons. These results being inherent byproducts of our work and the foundation laid in \cite{fanelli2024deep}.

Our holistic simulation pipeline inherently removes the barrier to entry imposed by software packages such as \geant and $\textsc{ROOT}$ to those in the broader DL community, a point that should not be underestimated or overlooked by those within the physics community. Our package puts both readily available and user controlled datasets directly into the hands of the individual, providing potential for the development of DL algorithms with increased reconstruction performance from those outside the physics community, ultimately increasing the rate and precision at which physicists can perform invaluable science.

%% file: 6_summary.tex
\section{Summary and Conclusions}\label{sec:summary}

We have introduced the first fully open fast simulation framework for pions and kaons in DIRC detectors, with a focus on the hpDIRC at EIC. This work presents a comprehensive suite of state-of-the-art generative models designed to accelerate PID research by enabling efficient, on-demand data generation. By abstracting away the computational and software complexities associated with traditional simulation packages such as \geant, our framework significantly lowers the barrier to entry for researchers, particularly those in the DL community.

Our self-contained simulation package incorporates a robust and flexible method for modeling photon yields across the relevant phase space, representing a critical improvement over prior works. This addition enables a complete and holistic fast simulation pipeline suitable for both classical and ML-based reconstruction workflows.

We demonstrate that our approach can simulate Cherenkov detector responses at a fraction of the computational cost of \geant, achieving an order-of-magnitude speedup at the track level and multiple orders of magnitude at the photon level, in which our framework supports on-the-fly data generation, eliminating the need for large-scale storage of reference datasets used in more traditional PID methods. These improvements are primarily enabled by the high degree of parallelism afforded by modern GPUs. Furthermore, through rigorous closure tests, we show that our generative models produce high-fidelity simulations, with the expected smoothing behavior typical of deep generative models.

Crucially, the modular and extensible nature of our framework allows it to be adapted to varying detector geometries and resolutions. This positions our method as a promising pathway toward fast simulation of the true detector response, especially when trained on high-purity real data from experiment. We believe this work lays the groundwork for a new era of scalable, GPU-native simulation and reconstruction in Cherenkov detector systems and beyond.

%% file: Appendix.tex
%\newpage
\appendix

%%%%%%%%%%% 3 GeV %%%%%%%%%%%%%%%%%%%%%
\section{Evaluation at $\SI[per-mode=symbol]{3}{\giga\eVperc}$} \label{app:3GeV}

%%%%%%%%%%%%%%%%%%%%%%% Cherenkov Ring Plots %%%%%%%%%%%%%%%%%%%%%%%%%%%%%%%

\begin{figure}[h]
    \centering
    \includegraphics[width=0.49\textwidth]{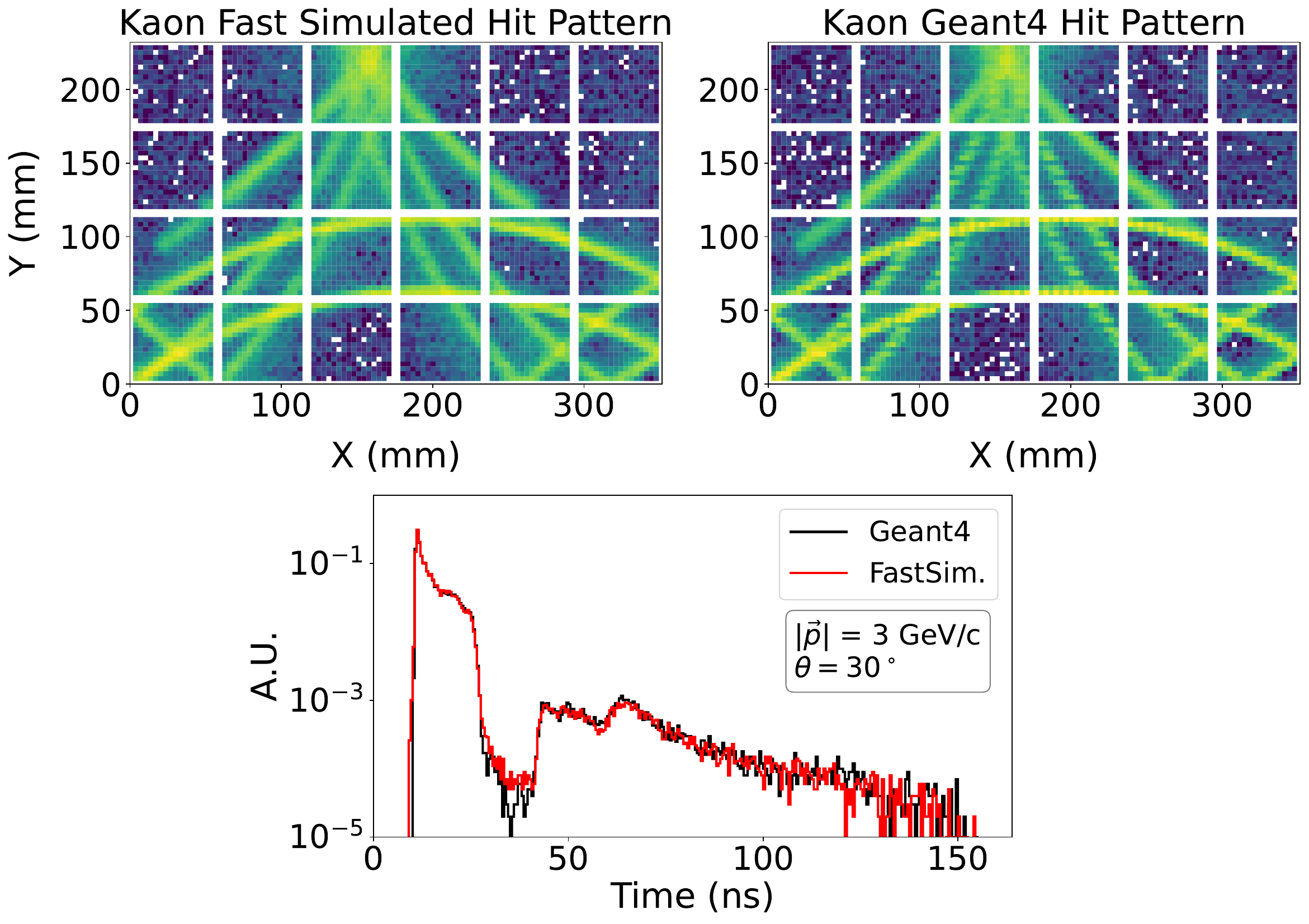}% 
   \includegraphics[width=0.49\textwidth]{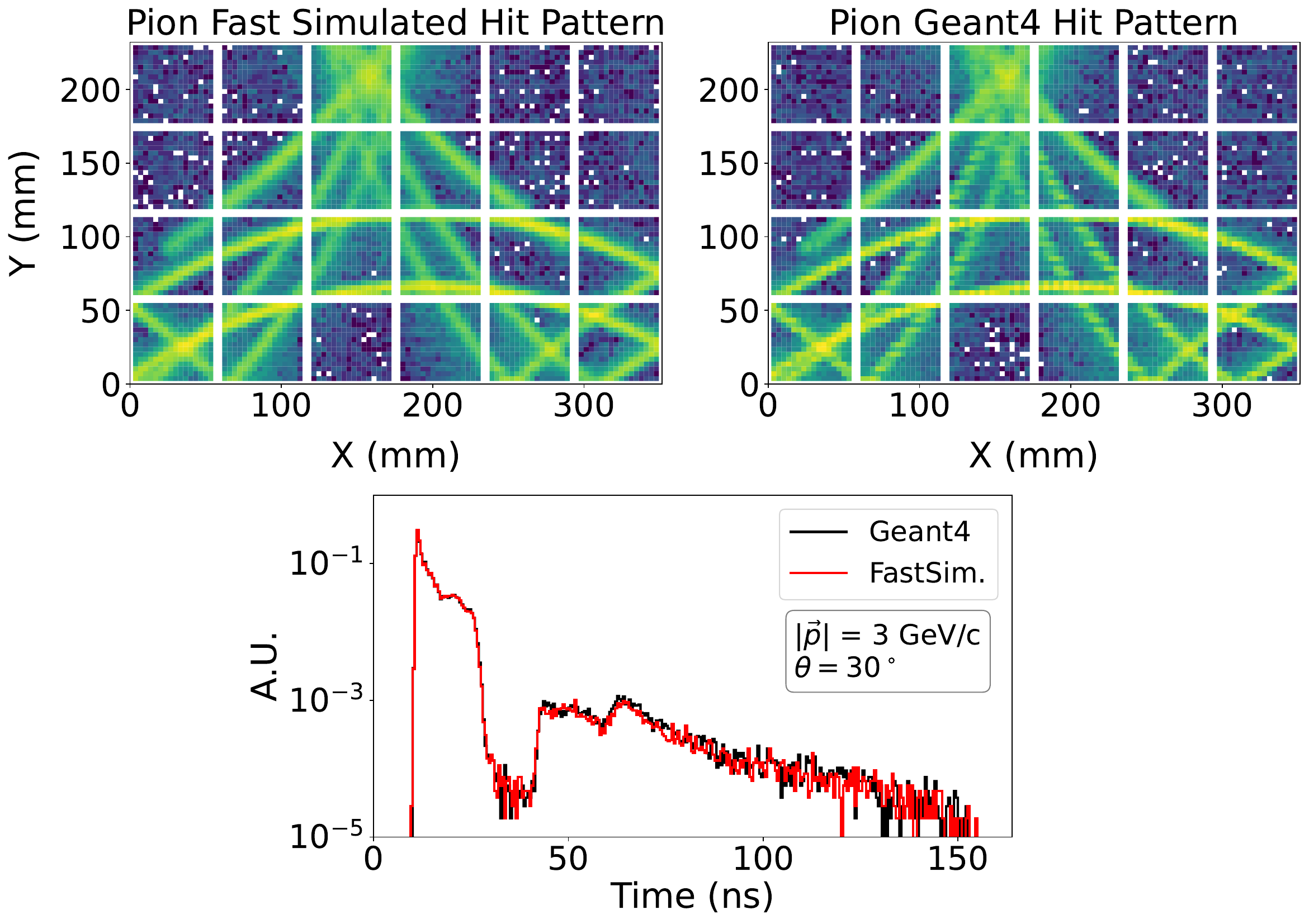} \\
    \includegraphics[width=0.49\textwidth]{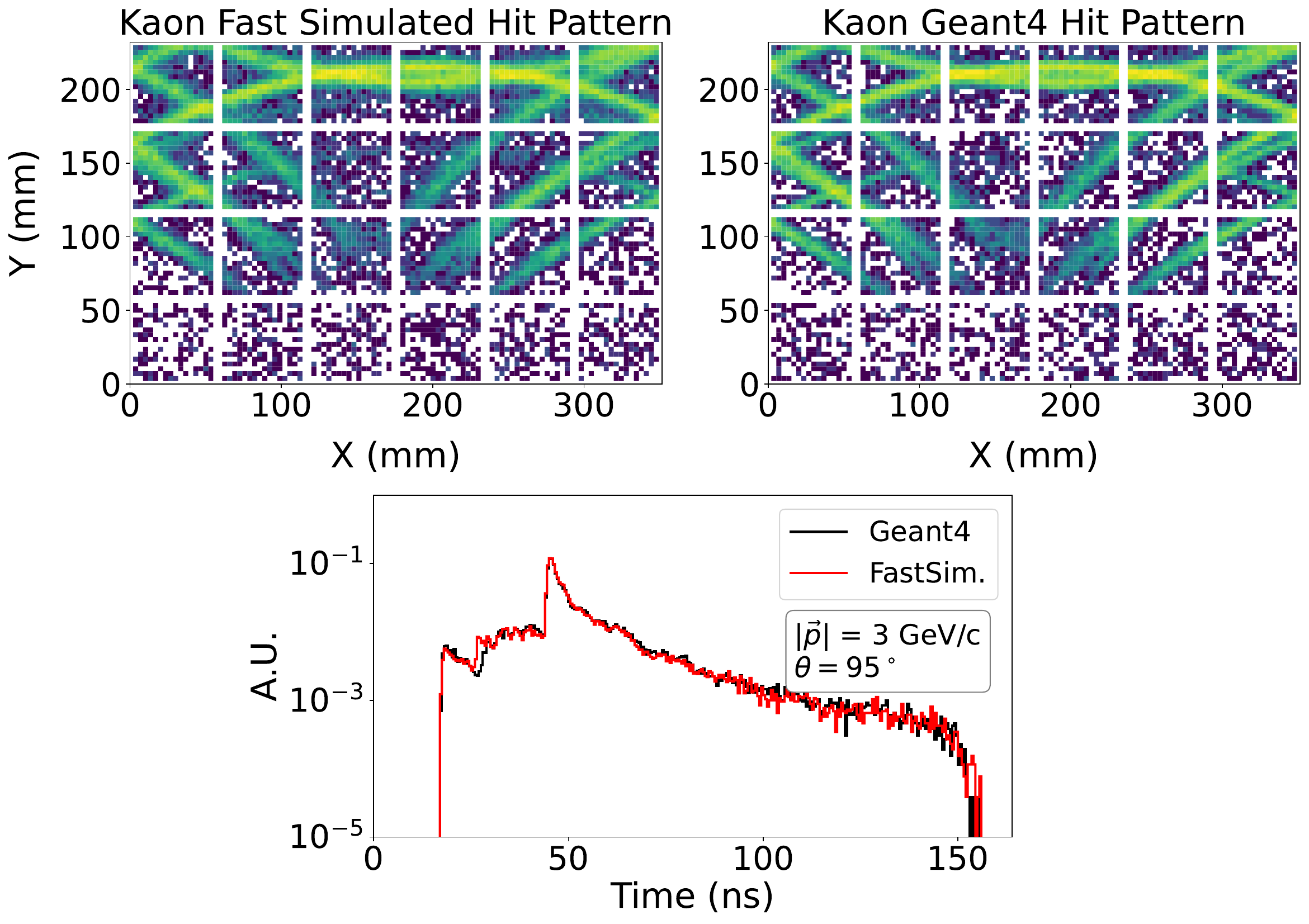} %
    \includegraphics[width=0.49\textwidth]{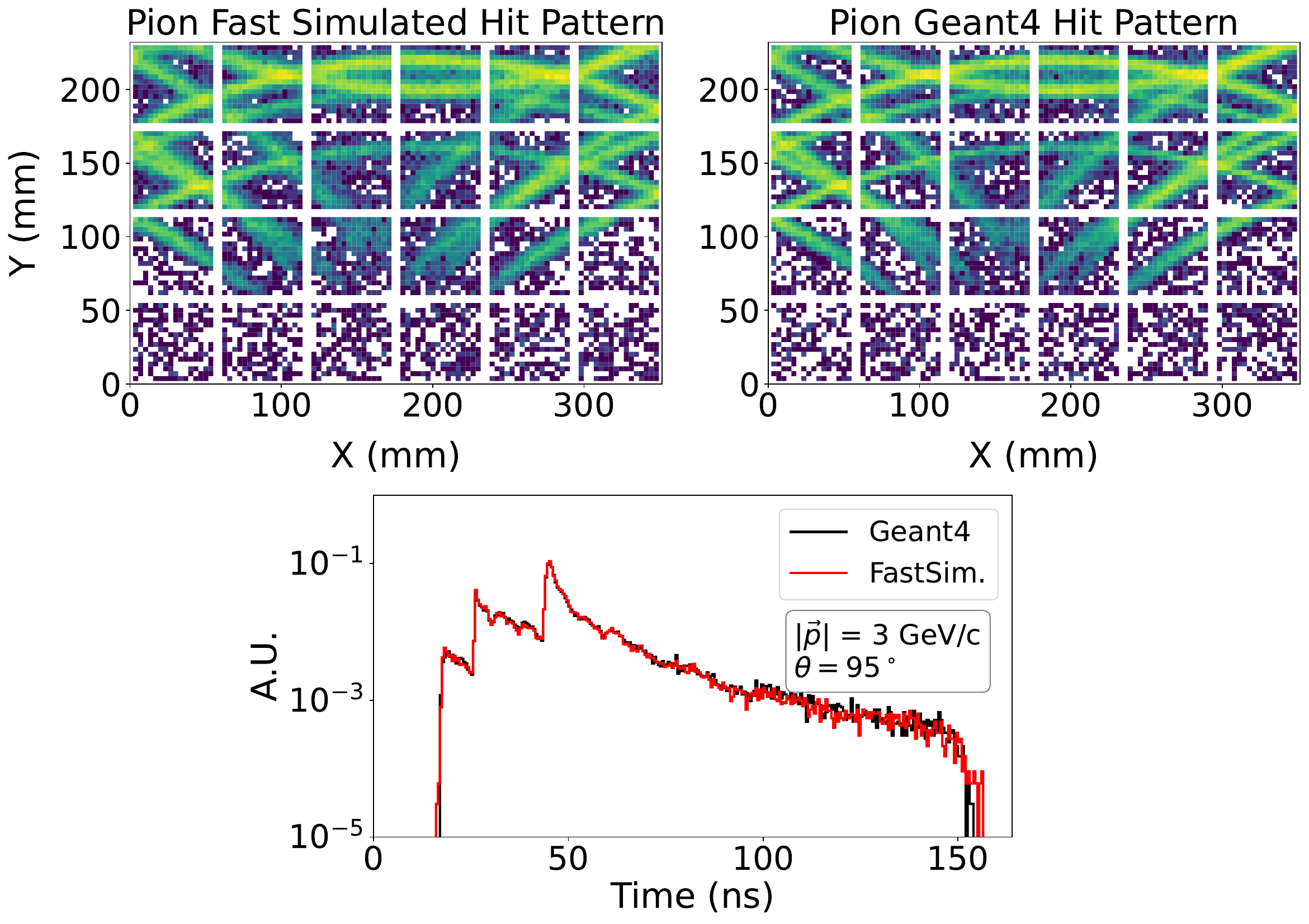} \\
    \includegraphics[width=0.49\textwidth]{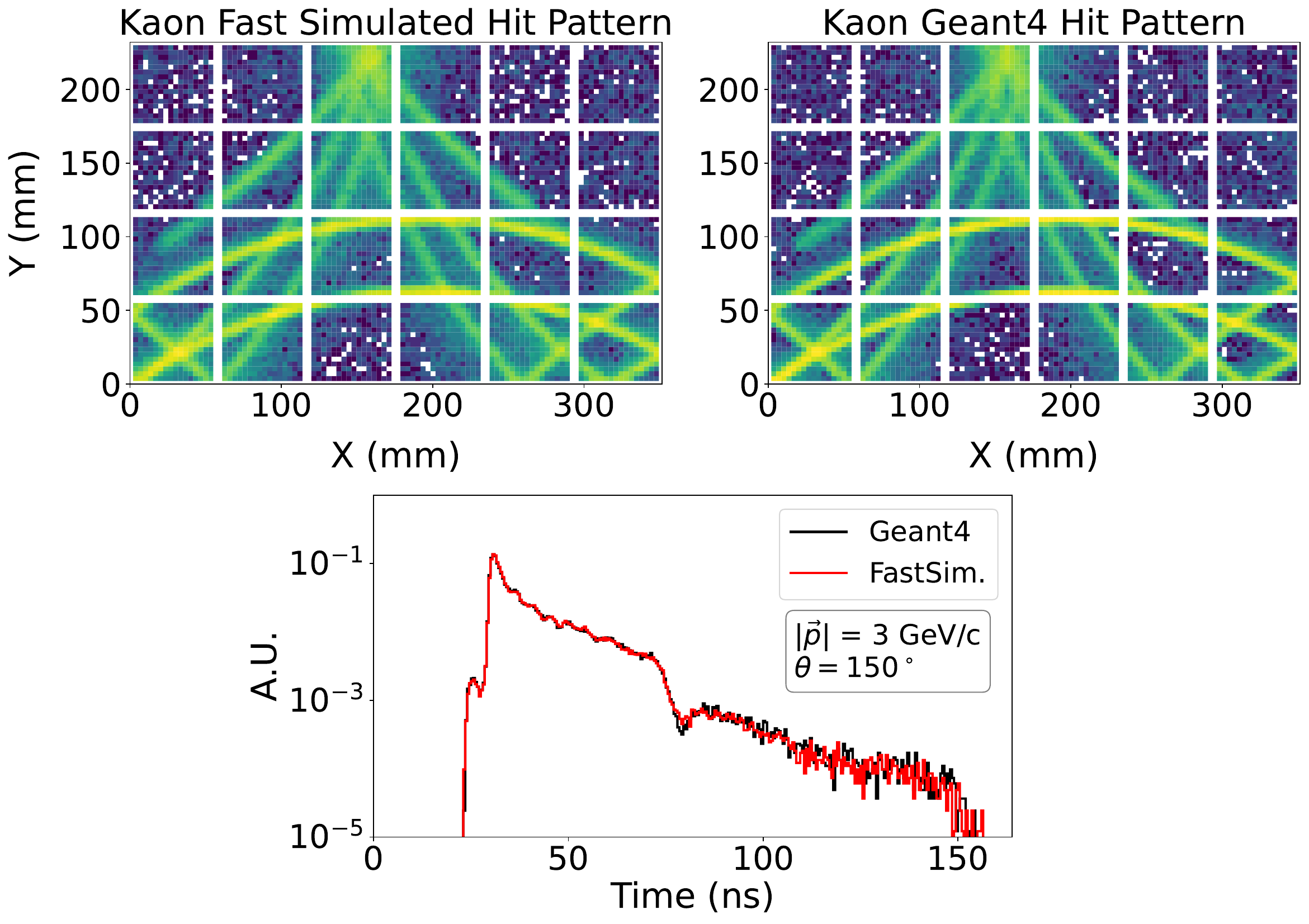} %
    \includegraphics[width=0.49\textwidth]{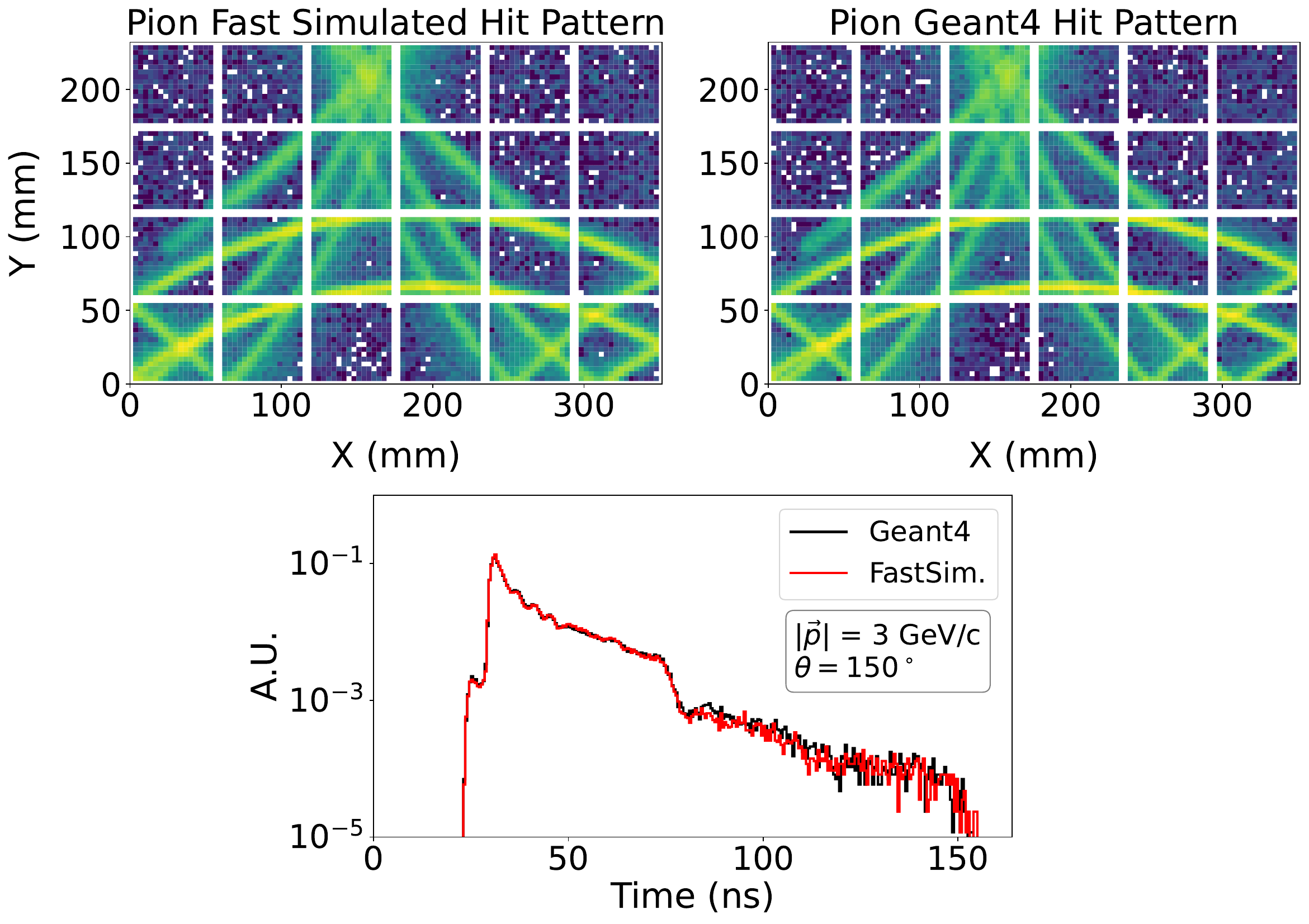} %
    \caption{
    \textbf{Fast Simulation with Discrete Normalizing Flows:} Fast Simulation of Kaons (left column of plots), and Pions (right column of plots) at 3 GeV/c and various polar angles.}
    \label{fig:DNF_Generations_3GeV}
\end{figure}

\begin{figure}[h]
    \centering
    \includegraphics[width=0.49\textwidth]{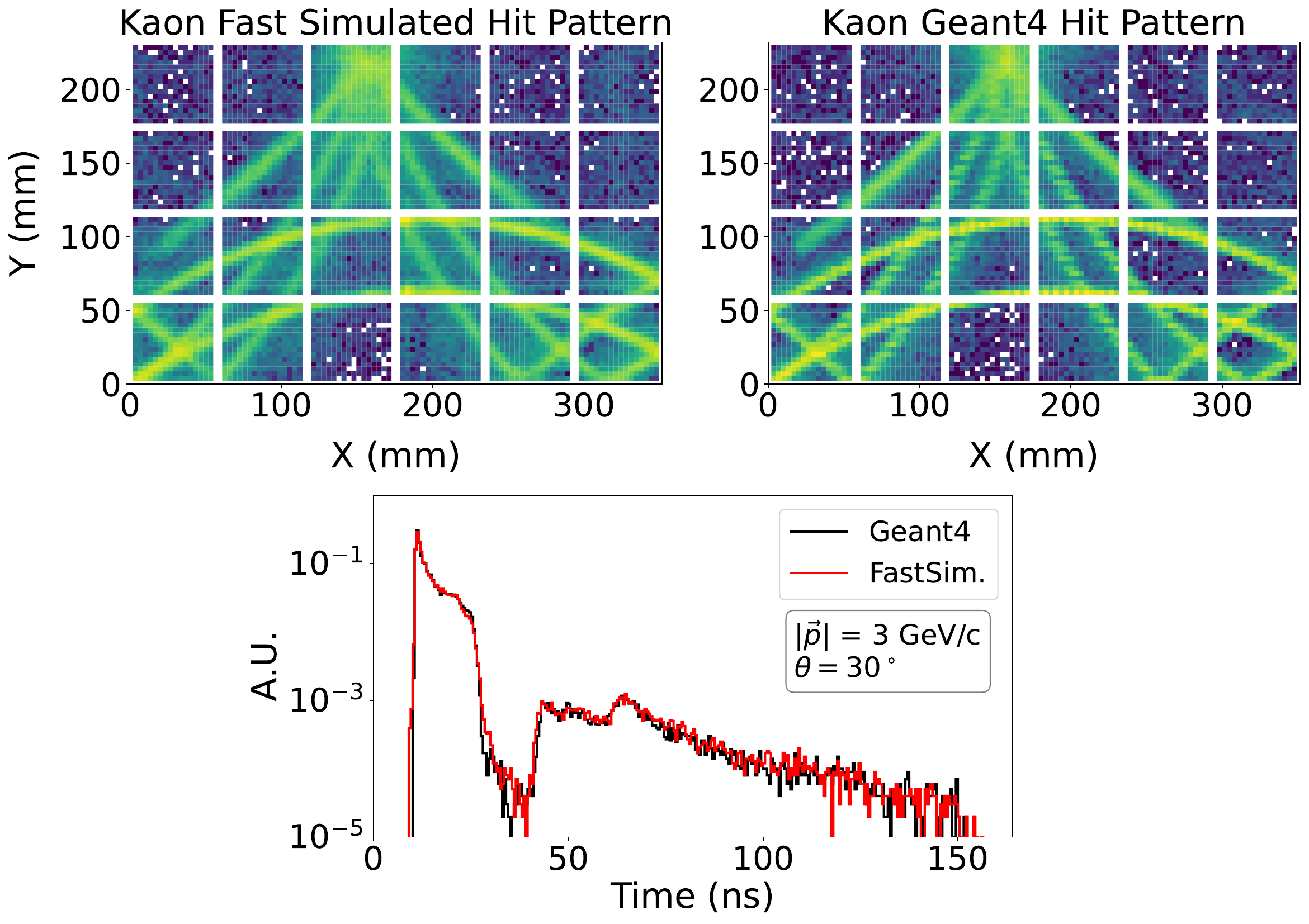}% 
   \includegraphics[width=0.49\textwidth]{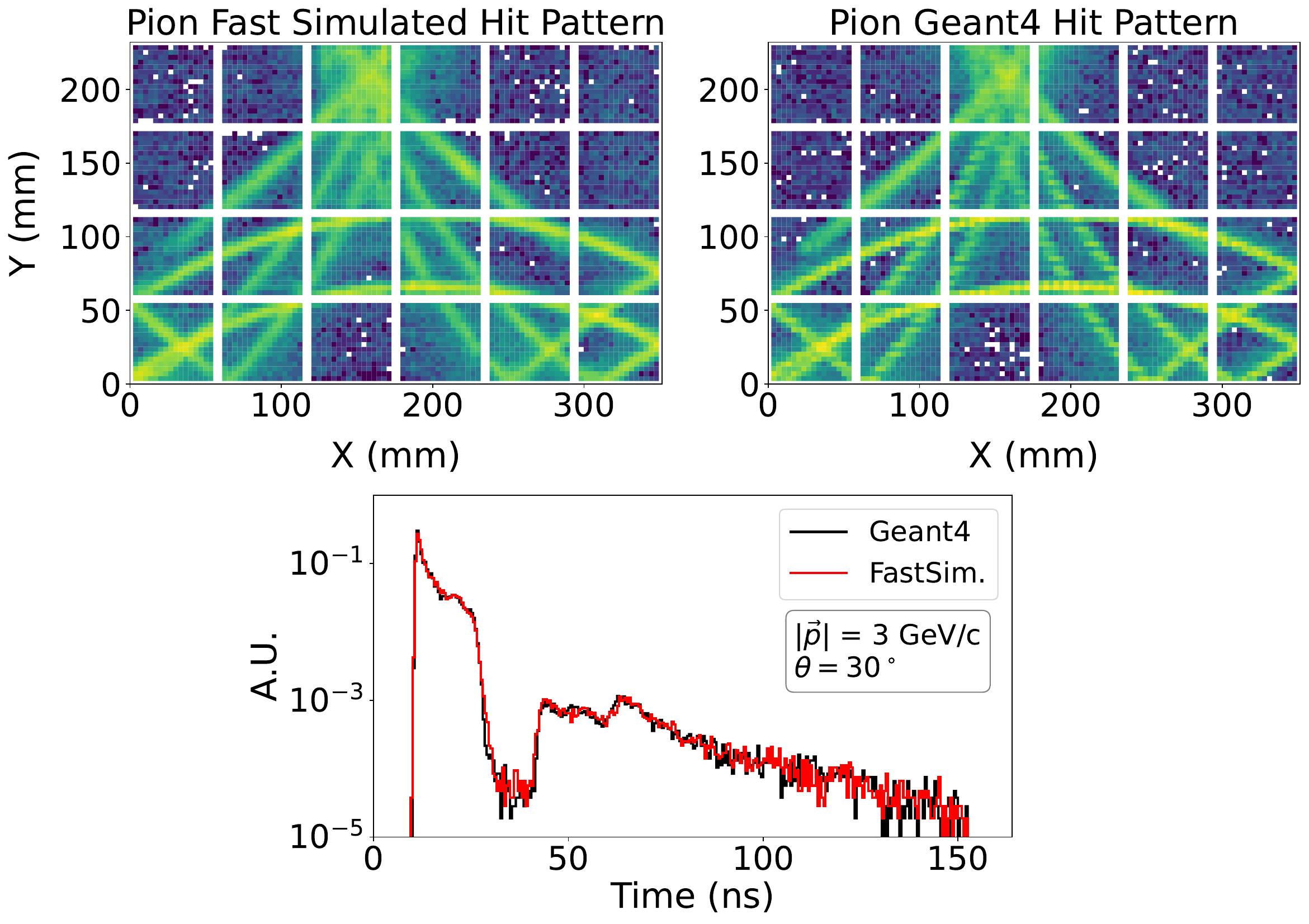} \\
    \includegraphics[width=0.49\textwidth]{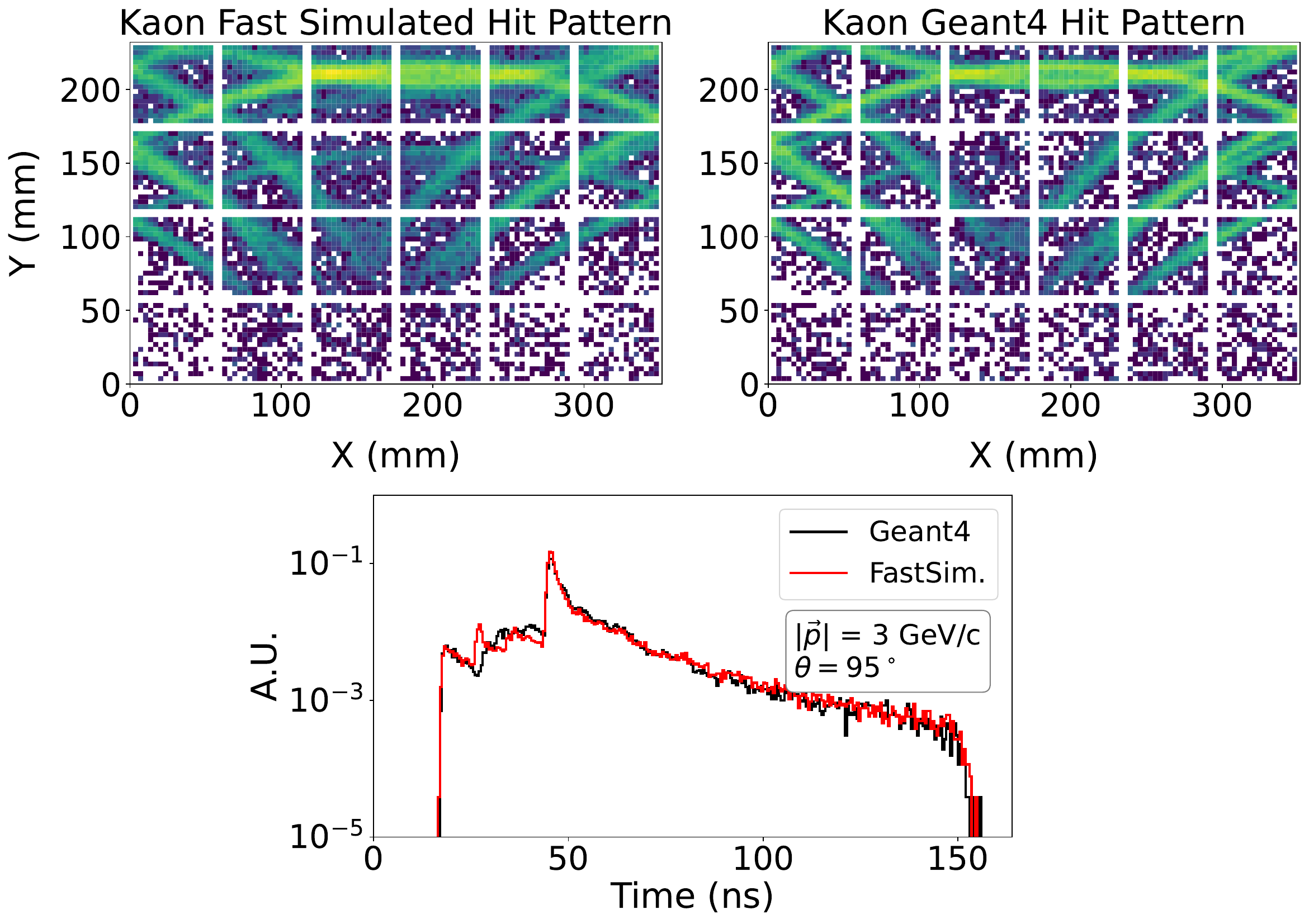} %
    \includegraphics[width=0.49\textwidth]{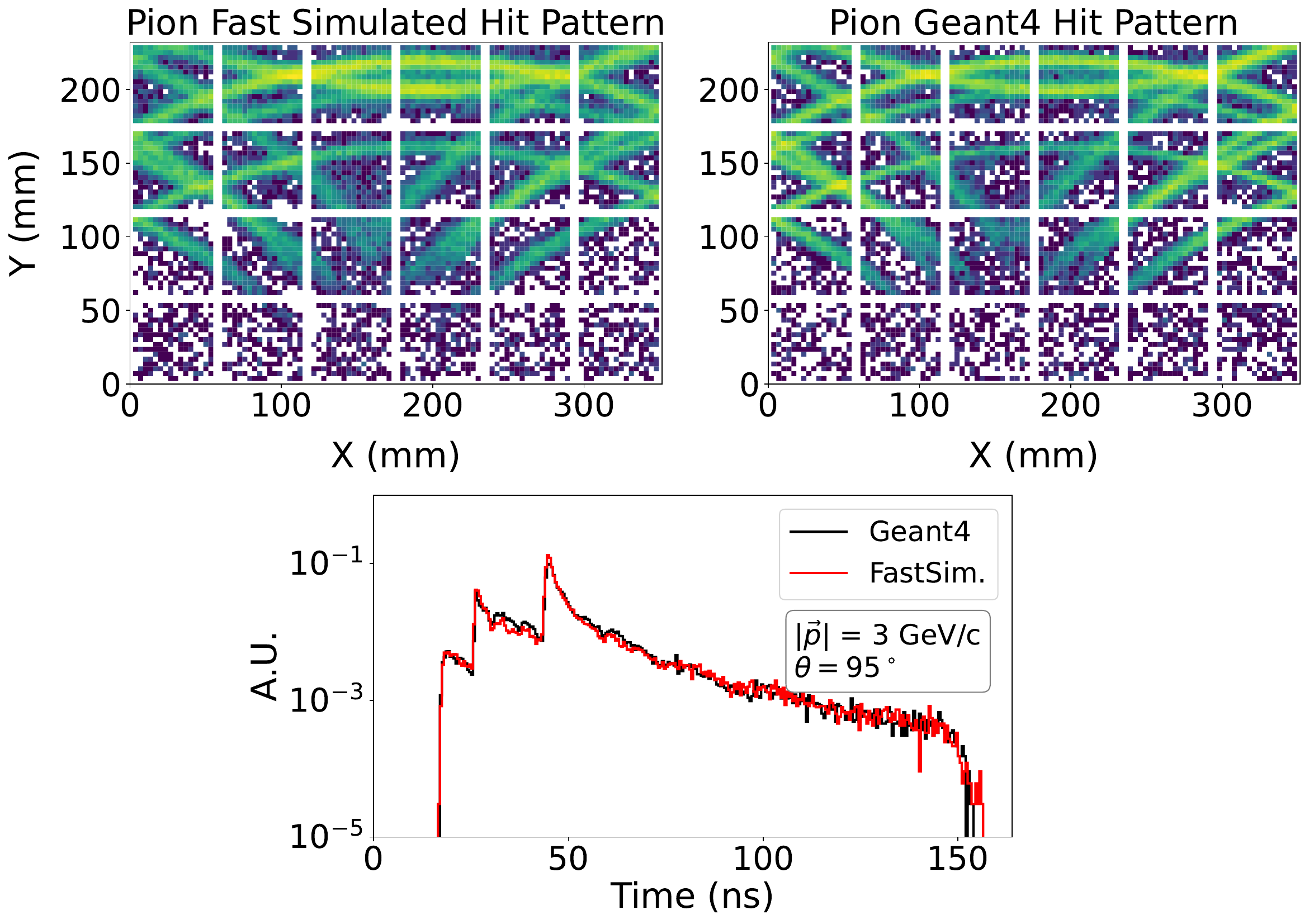} \\
    \includegraphics[width=0.49\textwidth]{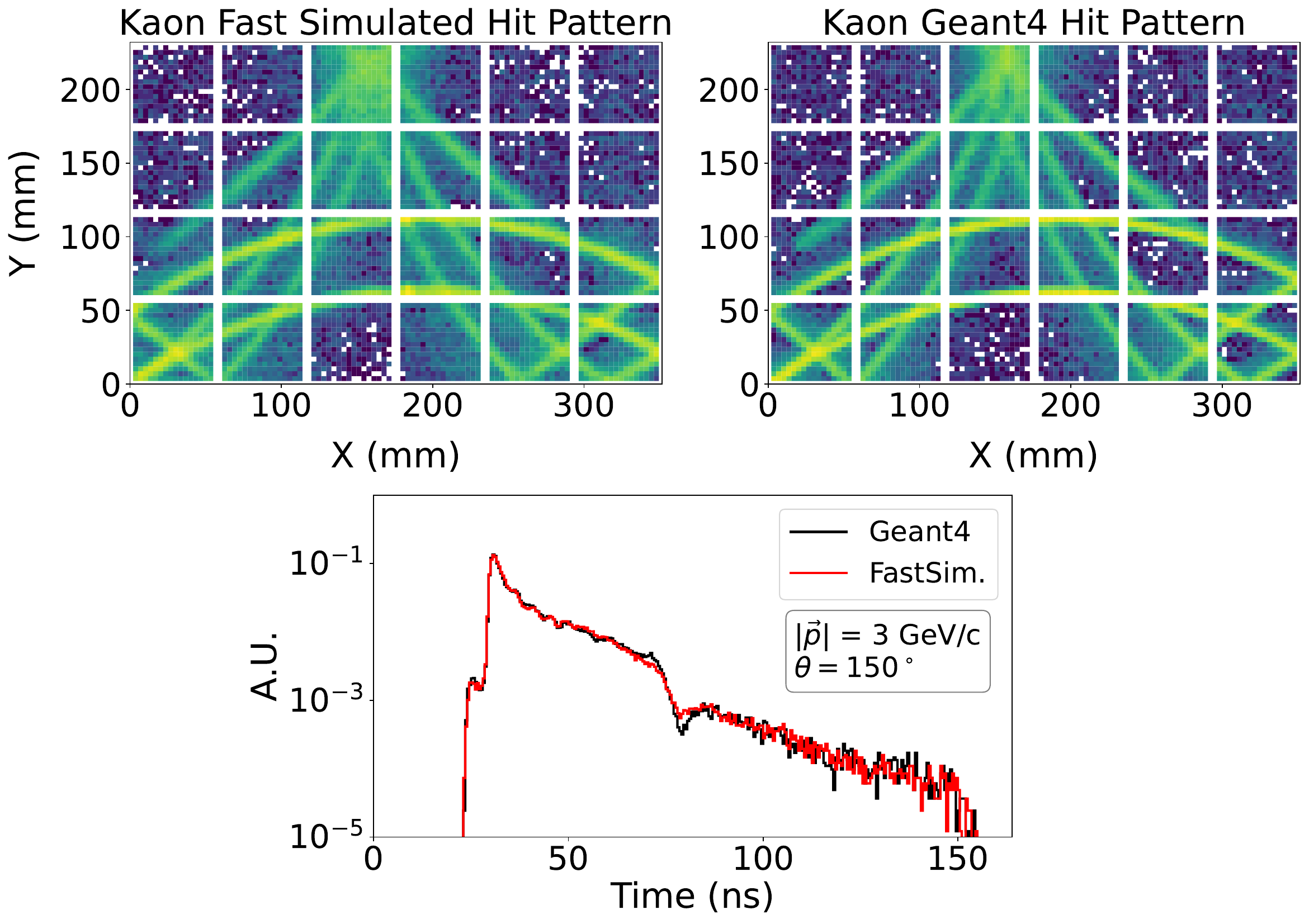} %
    \includegraphics[width=0.49\textwidth]{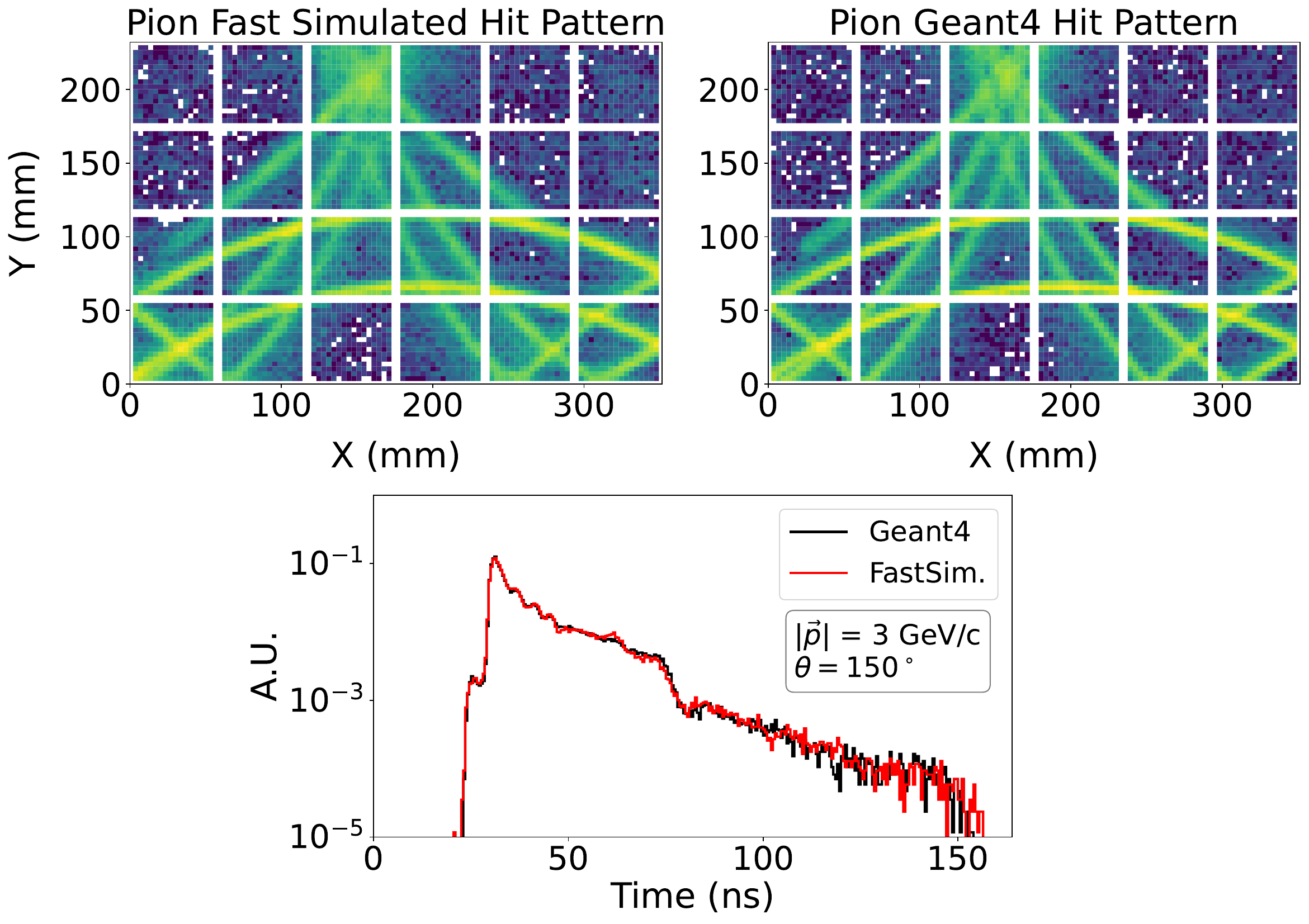} %
    \caption{
    \textbf{Fast Simulation with Continuous Normalizing Flows:} Fast Simulation of Kaons (left column of plots), and Pions (right column of plots) at 3 GeV/c and various polar angles. 20 integration steps with midpoint solver.}
    \label{fig:CNF_Generations_3GeV}
\end{figure}

\begin{figure}[h]
    \centering
    \includegraphics[width=0.49\textwidth]{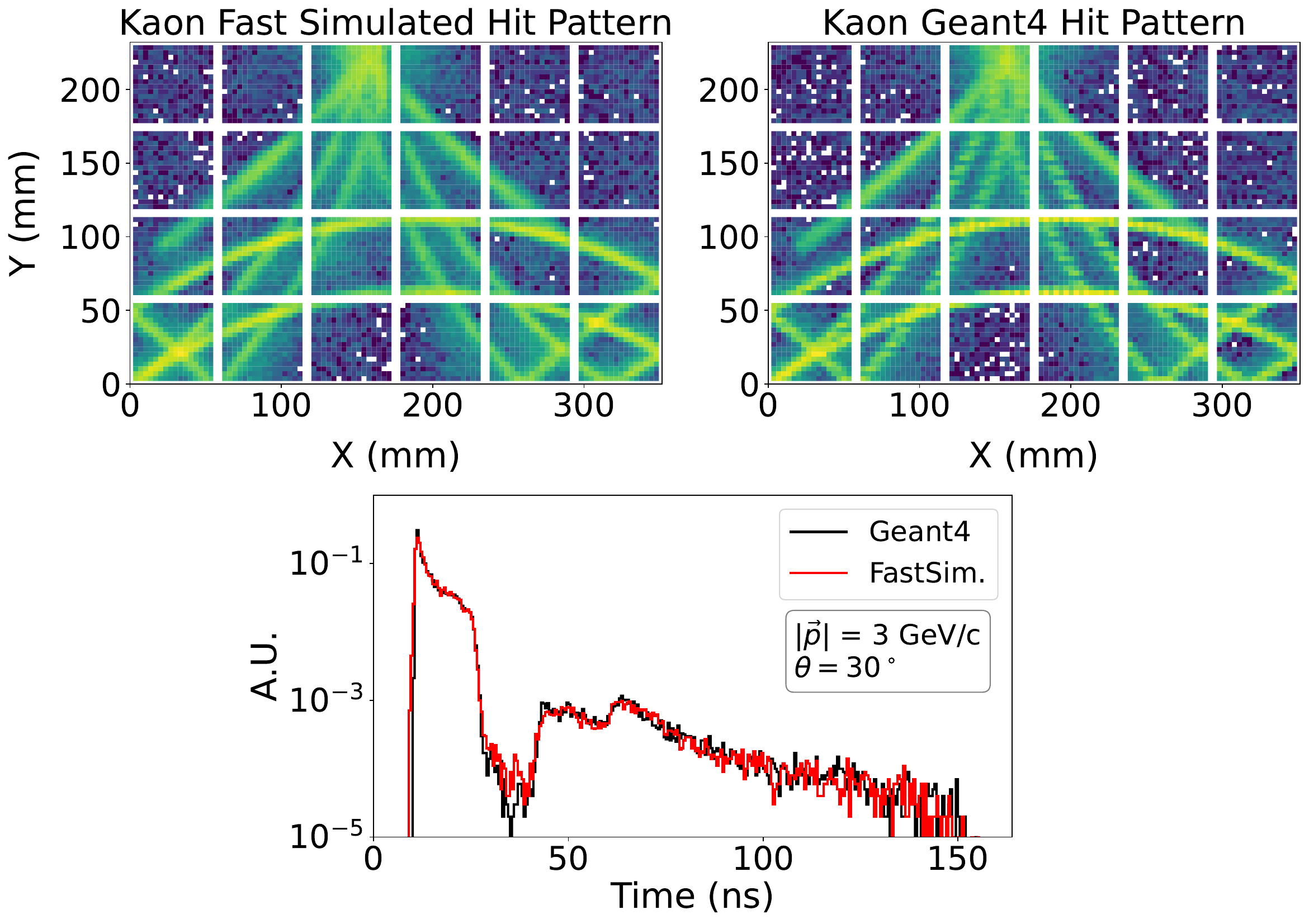}% 
   \includegraphics[width=0.49\textwidth]{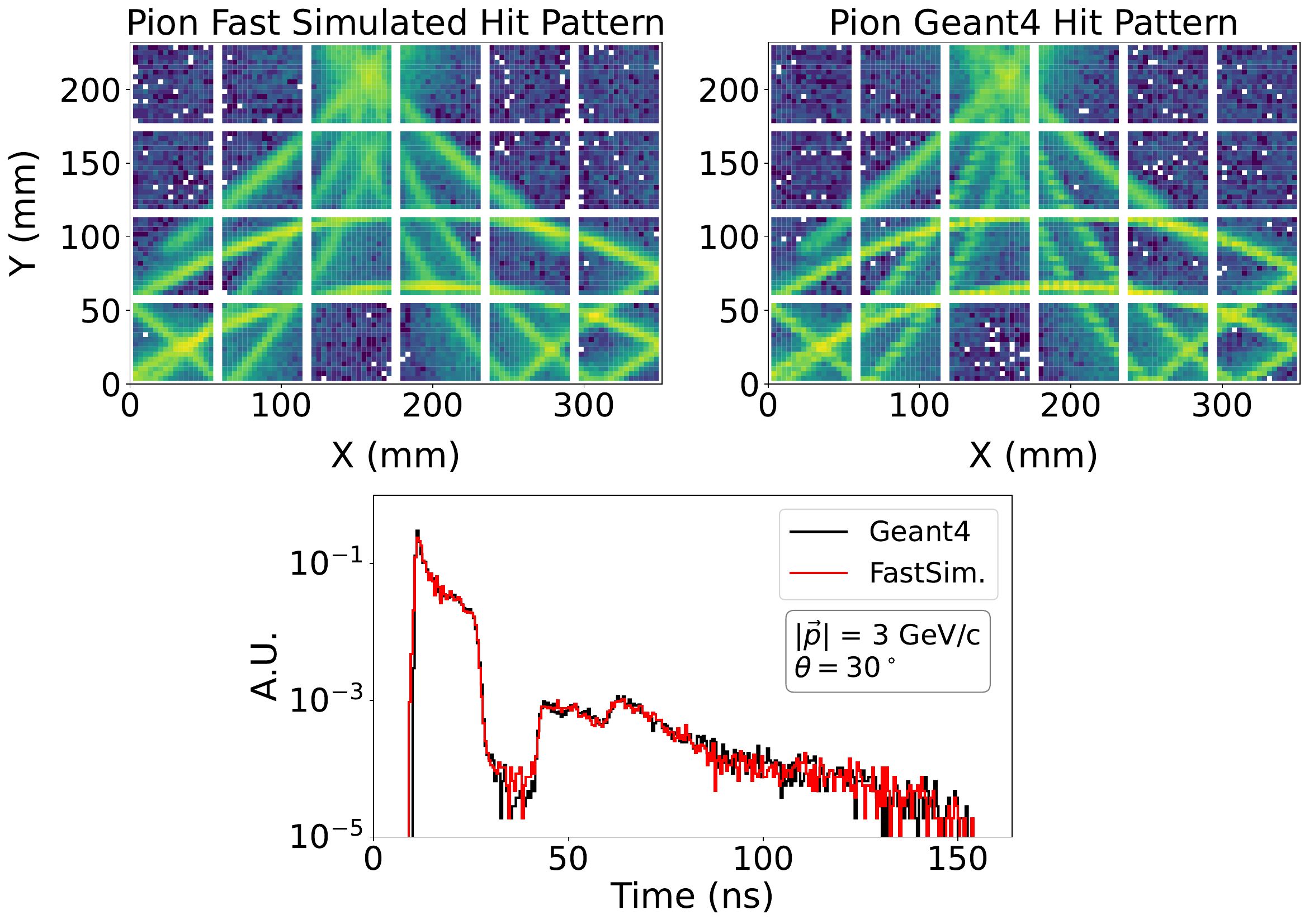} \\
    \includegraphics[width=0.49\textwidth]{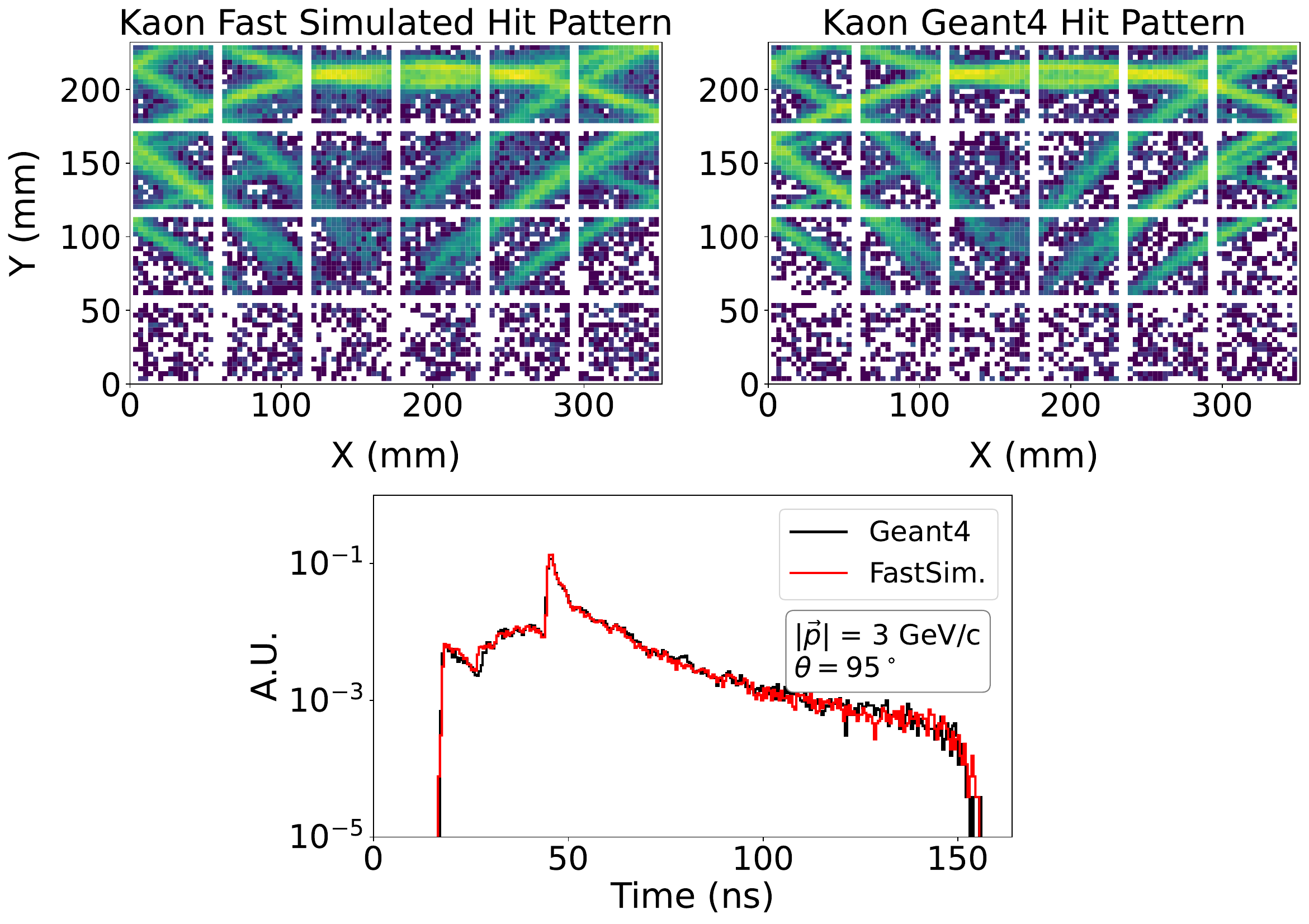} %
    \includegraphics[width=0.49\textwidth]{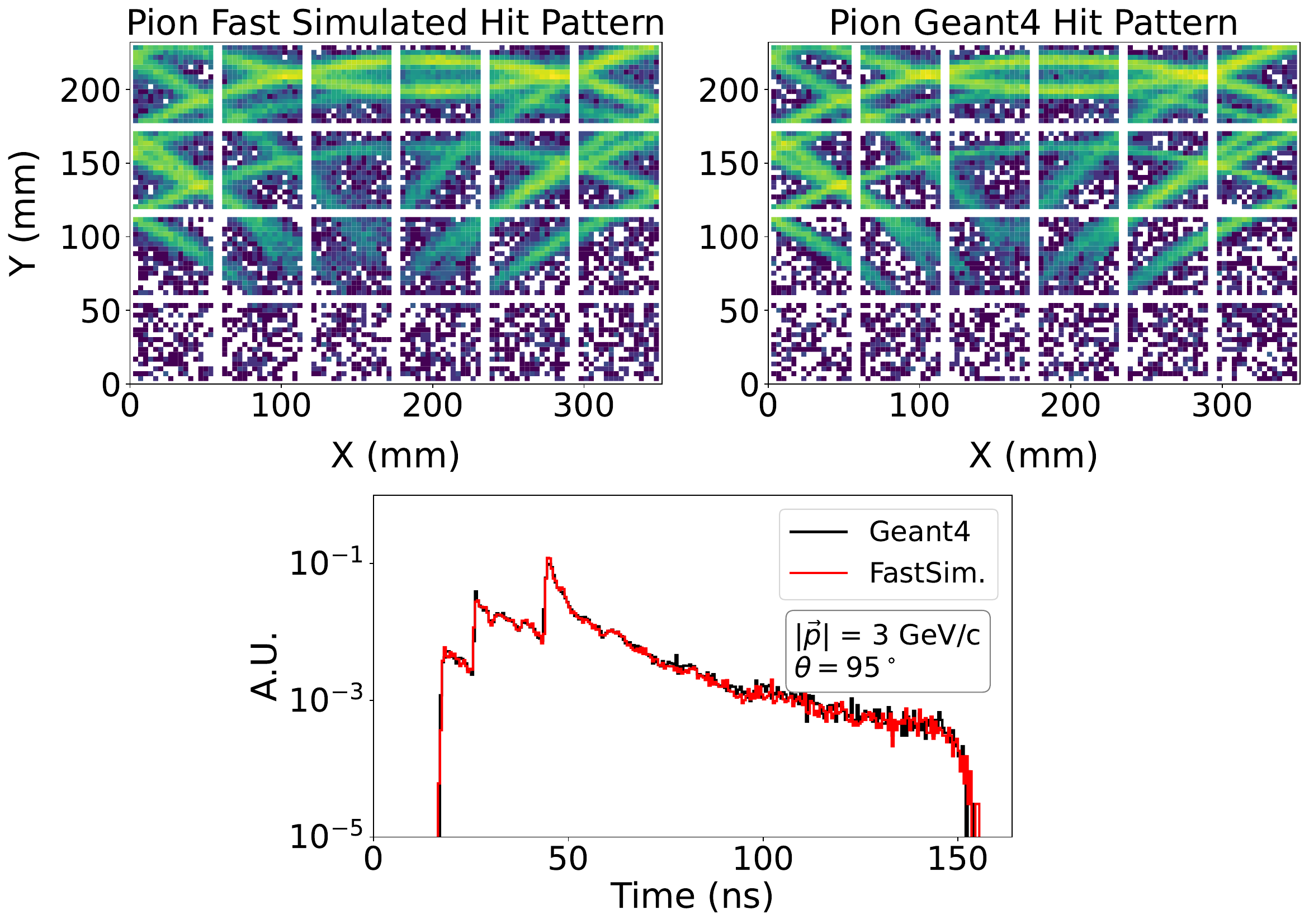} \\
    \includegraphics[width=0.49\textwidth]{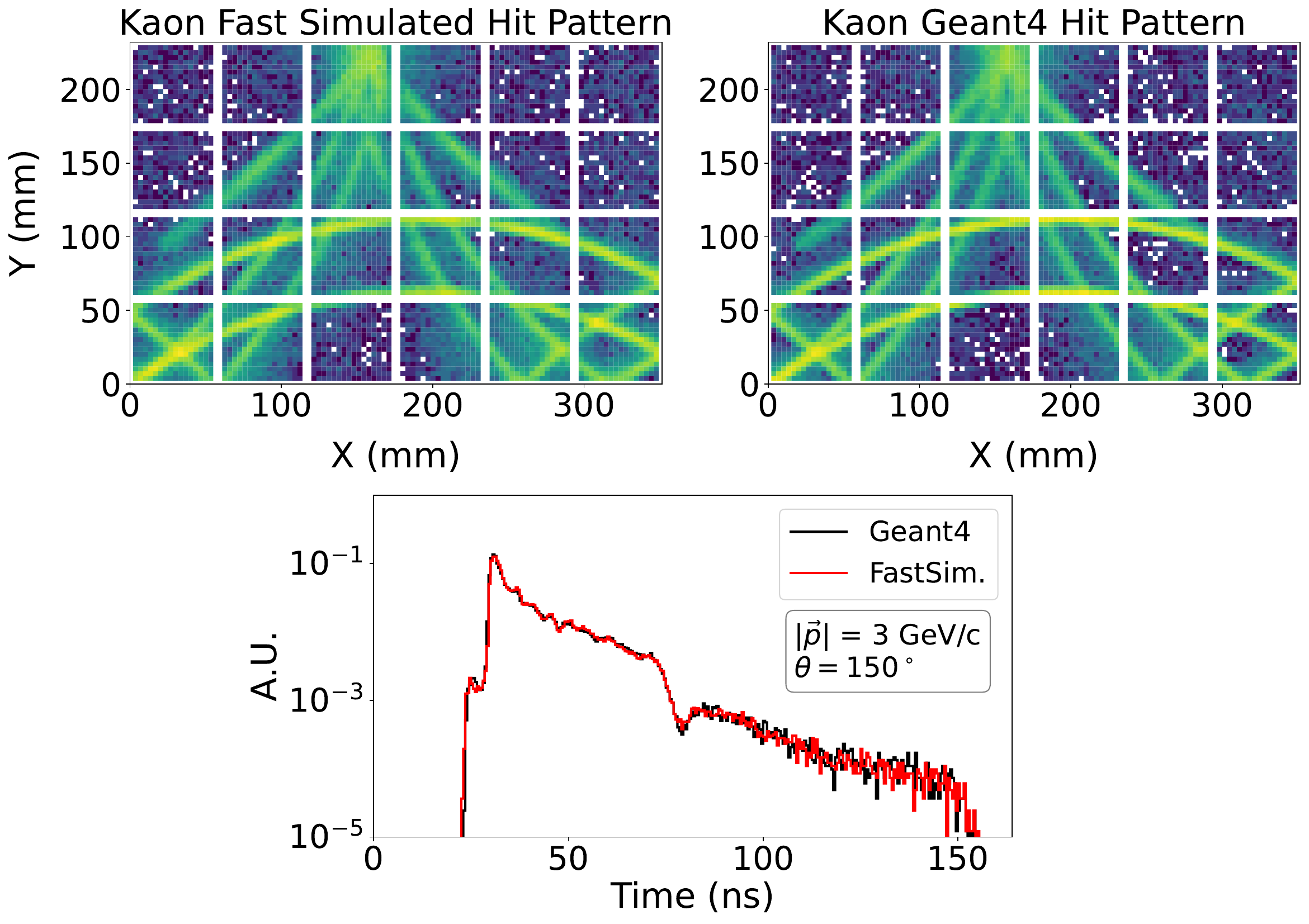} %
    \includegraphics[width=0.49\textwidth]{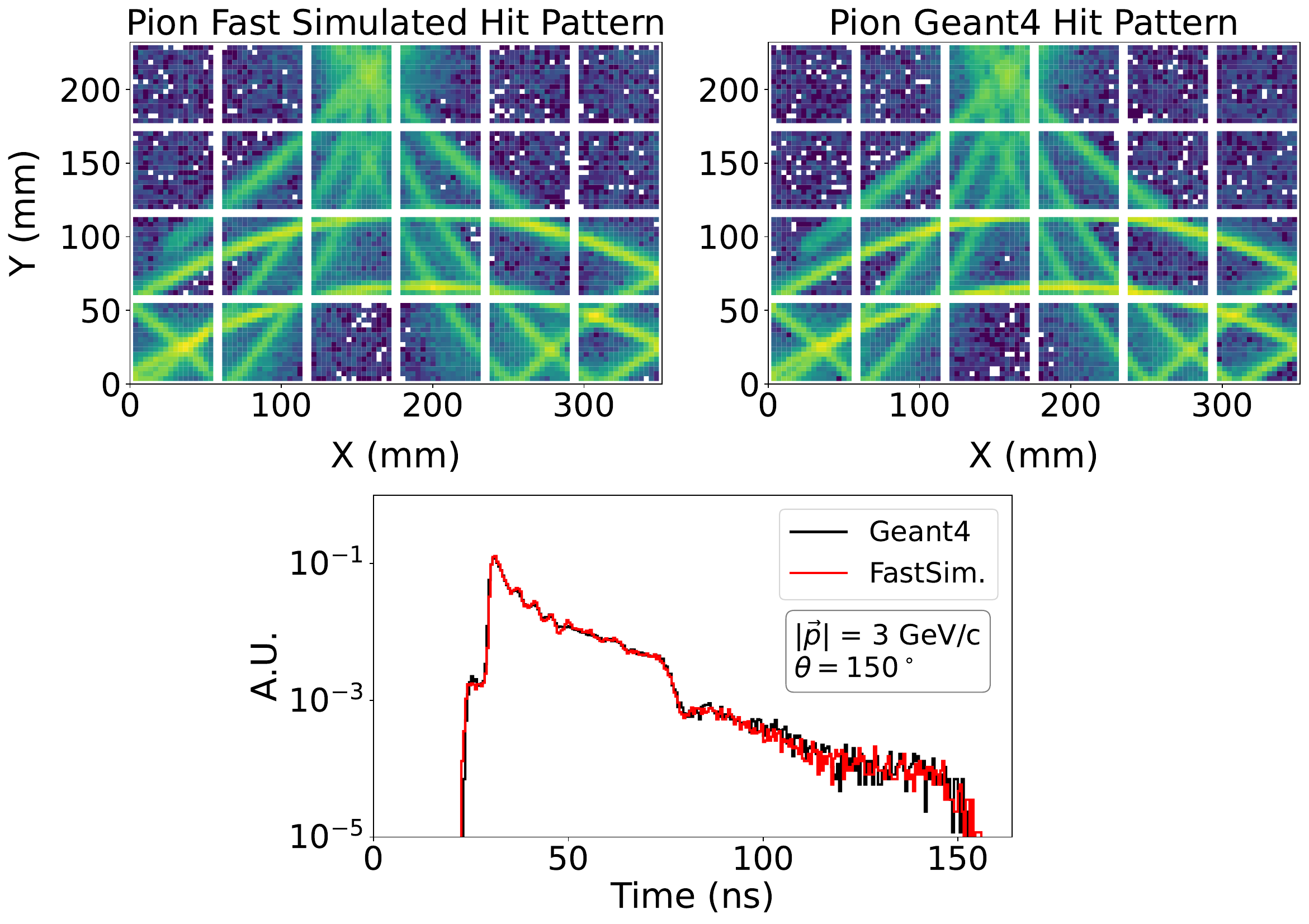} %
    \caption{
    \textbf{Fast Simulation with Flow Matching:} Fast Simulation of Kaons (left column of plots), and Pions (right column of plots) at 3 GeV/c and various polar angles. 20 integration steps with midpoint solver.}
    \label{fig:FlowMatching_Generations_3GeV}
\end{figure}

%%%%%%%%%%%%%%%%%%% DDPM %%%%%%%%%%%%%%%%%%%%%%%%%%%%%%%%%%%%%
\begin{figure}[h]
    \centering
    \includegraphics[width=0.49\textwidth]{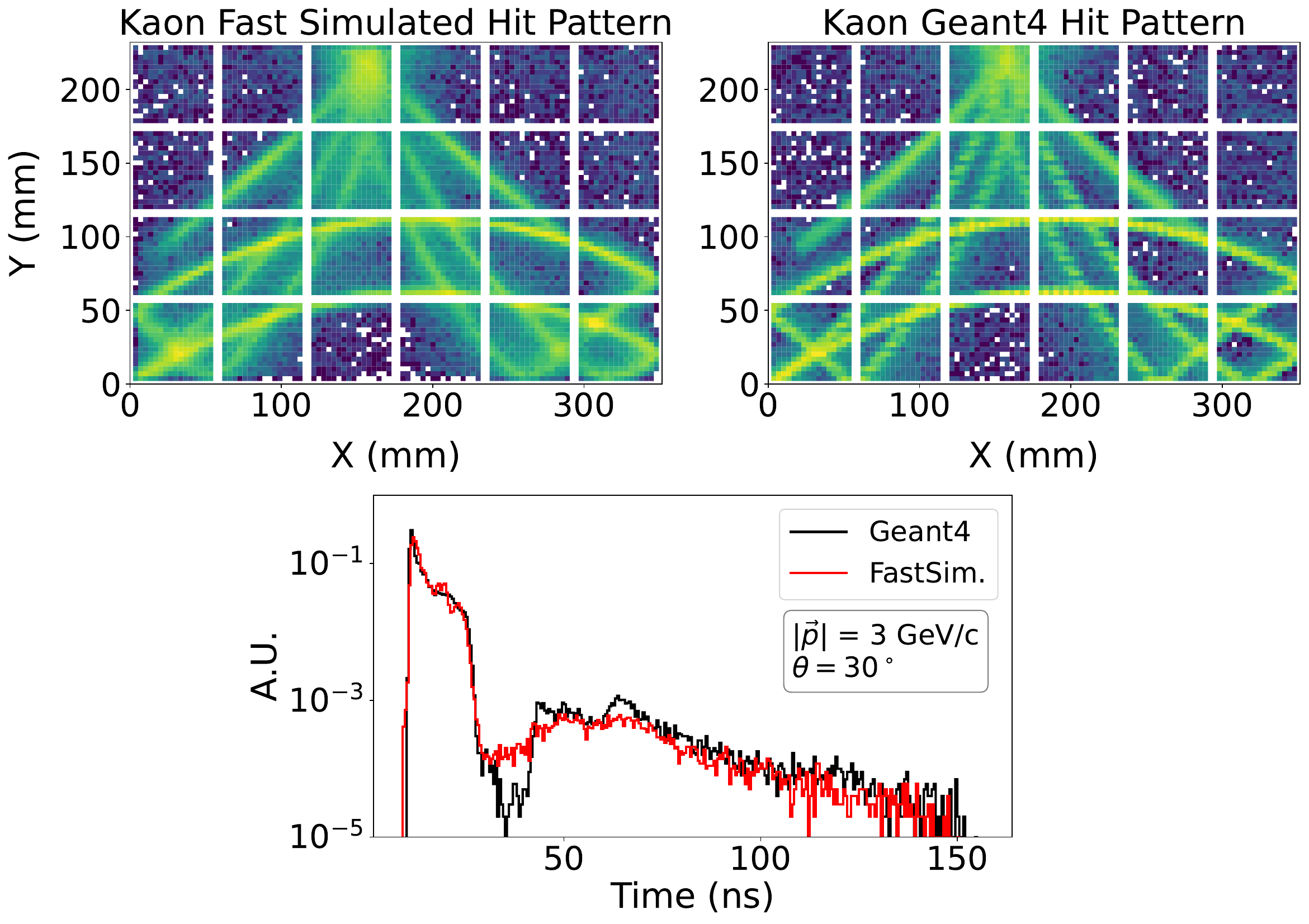}% 
   \includegraphics[width=0.49\textwidth]{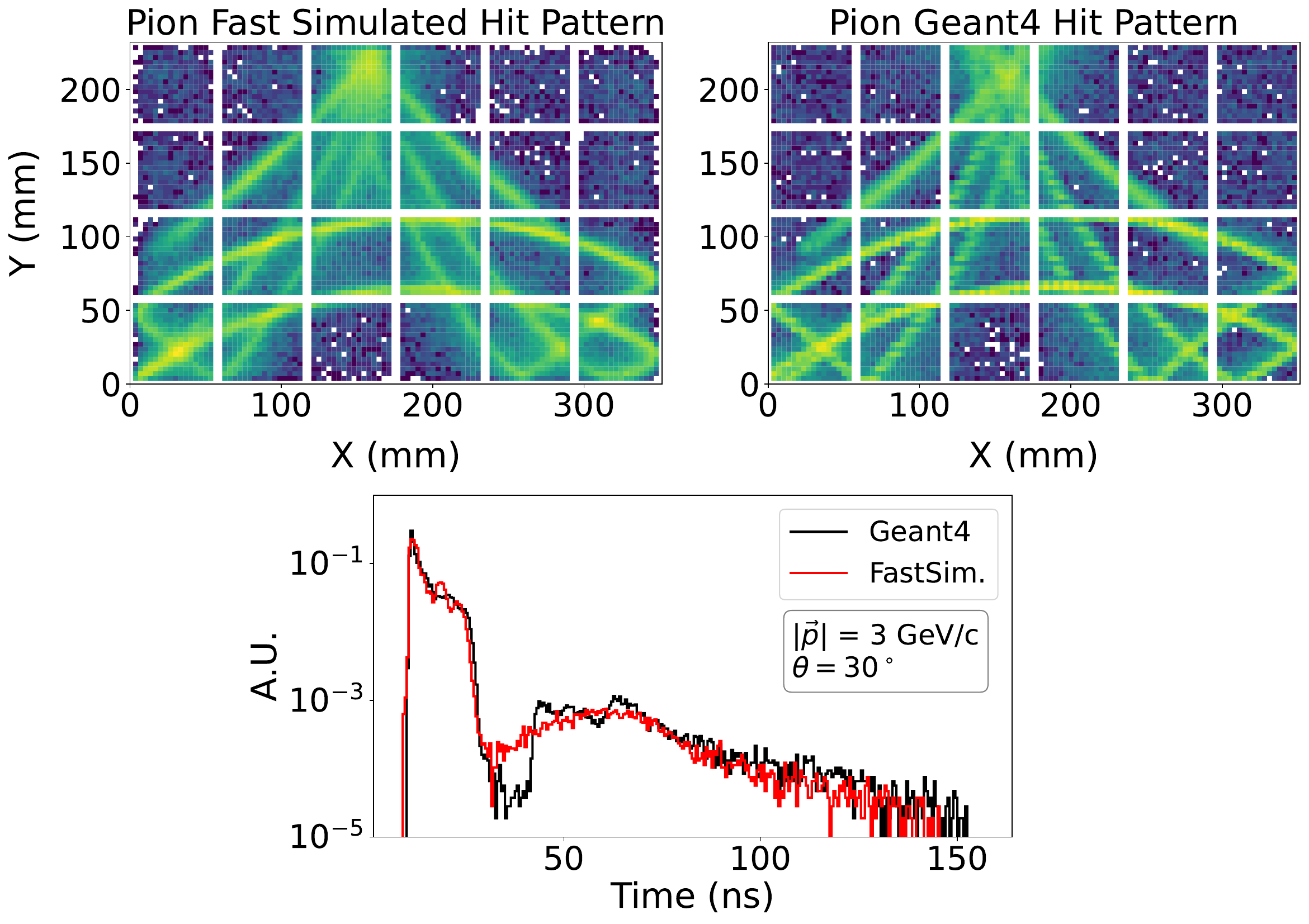} \\
    \includegraphics[width=0.49\textwidth]{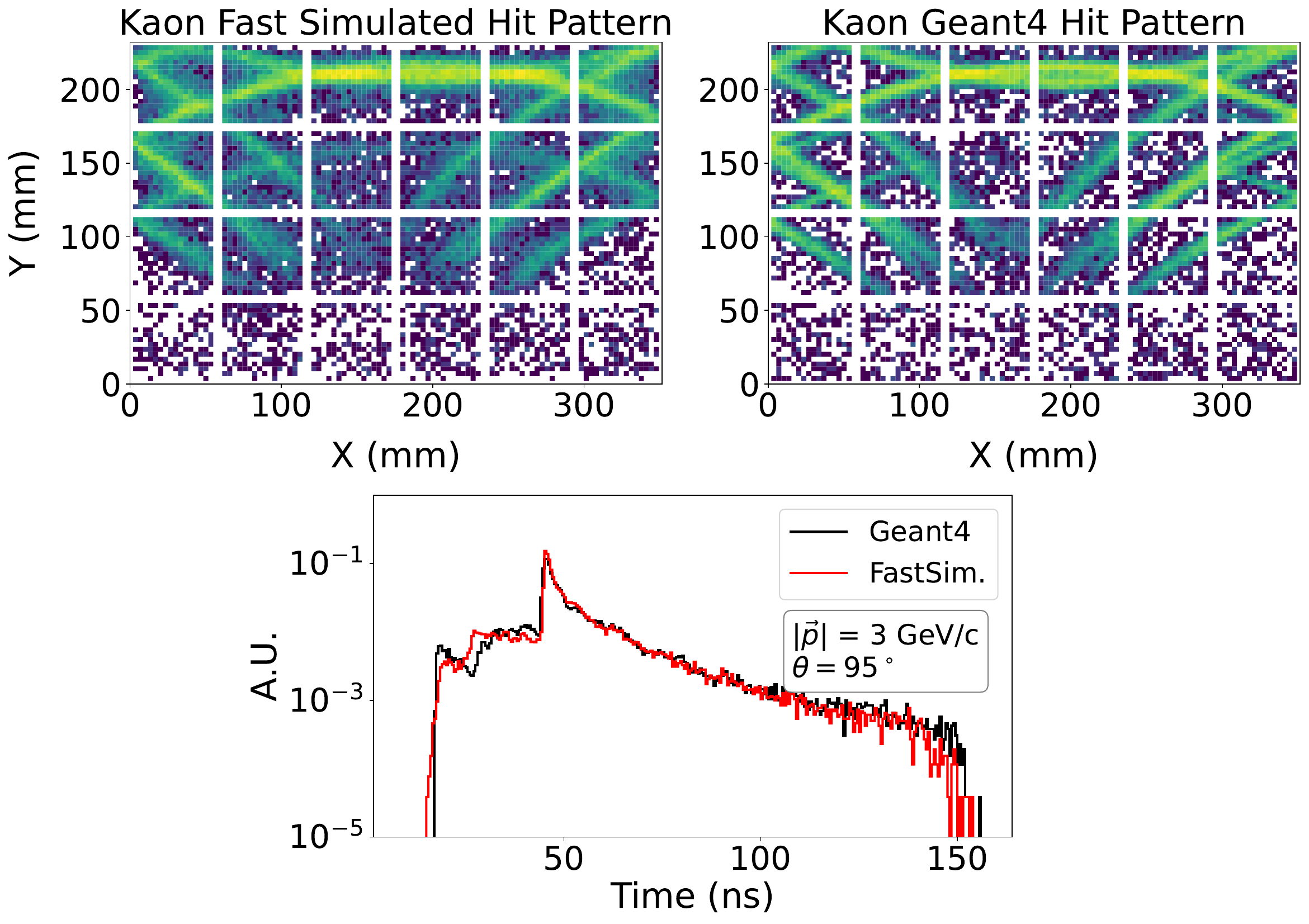} %
    \includegraphics[width=0.49\textwidth]{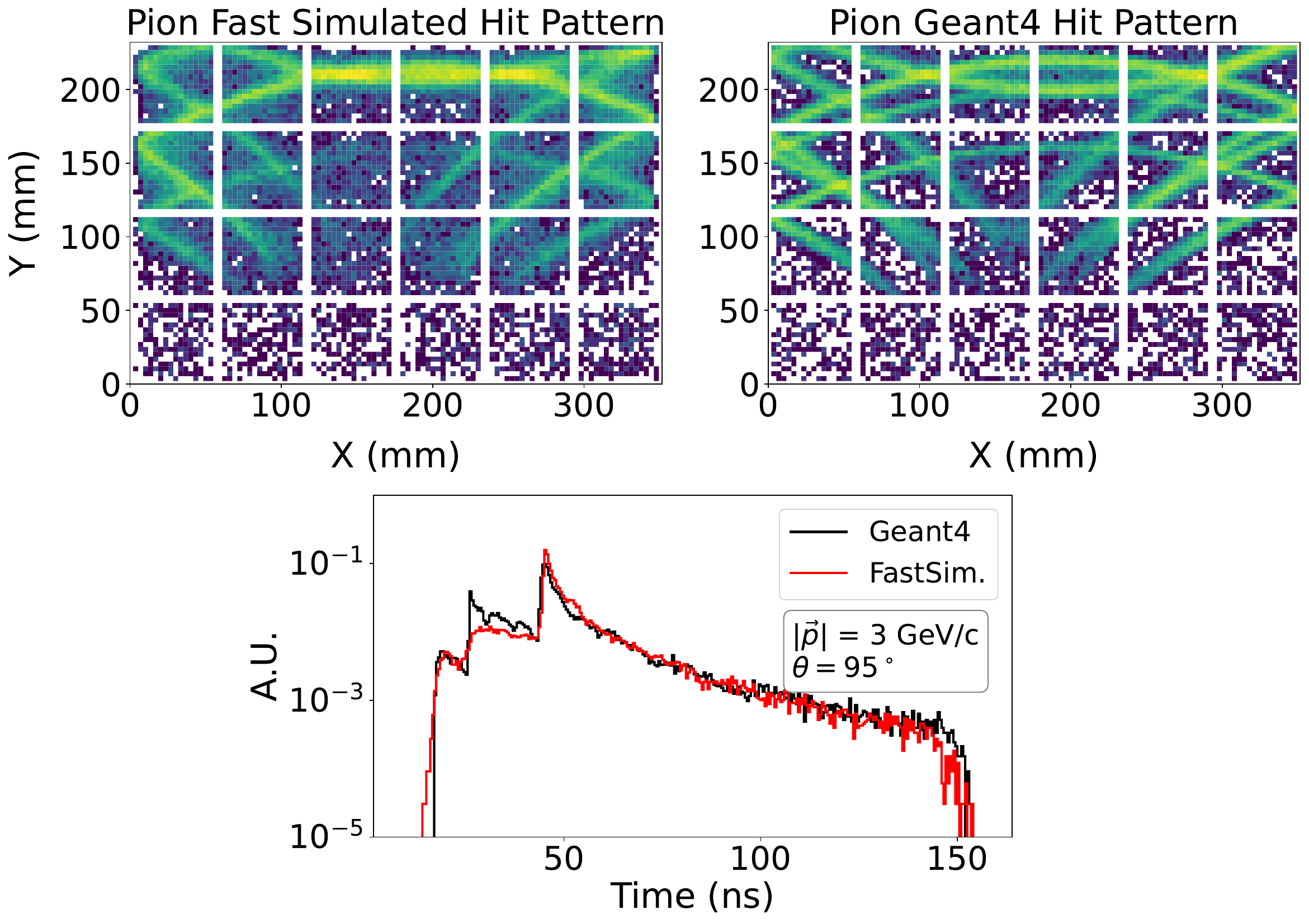} \\
    \includegraphics[width=0.49\textwidth]{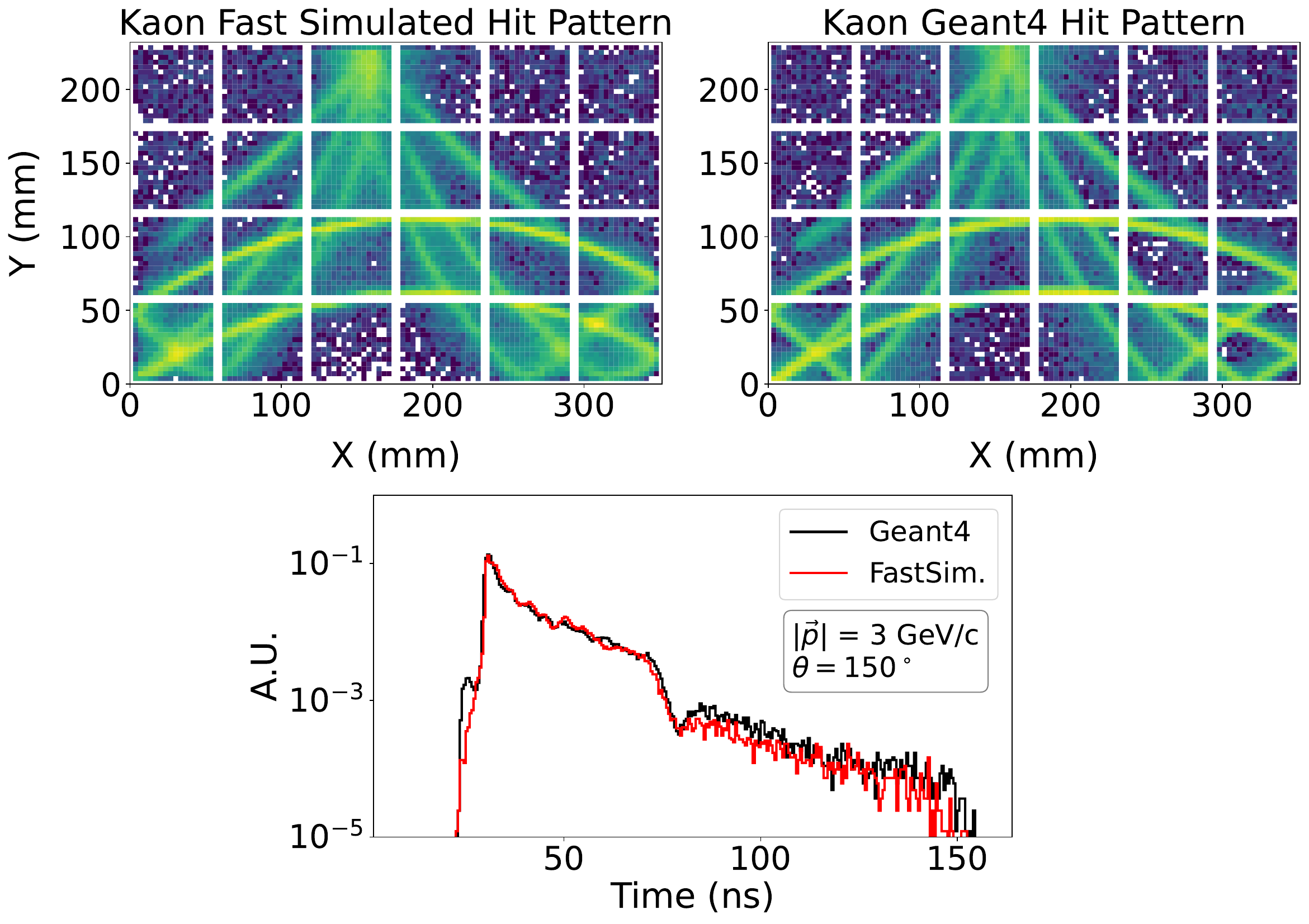} %
    \includegraphics[width=0.49\textwidth]{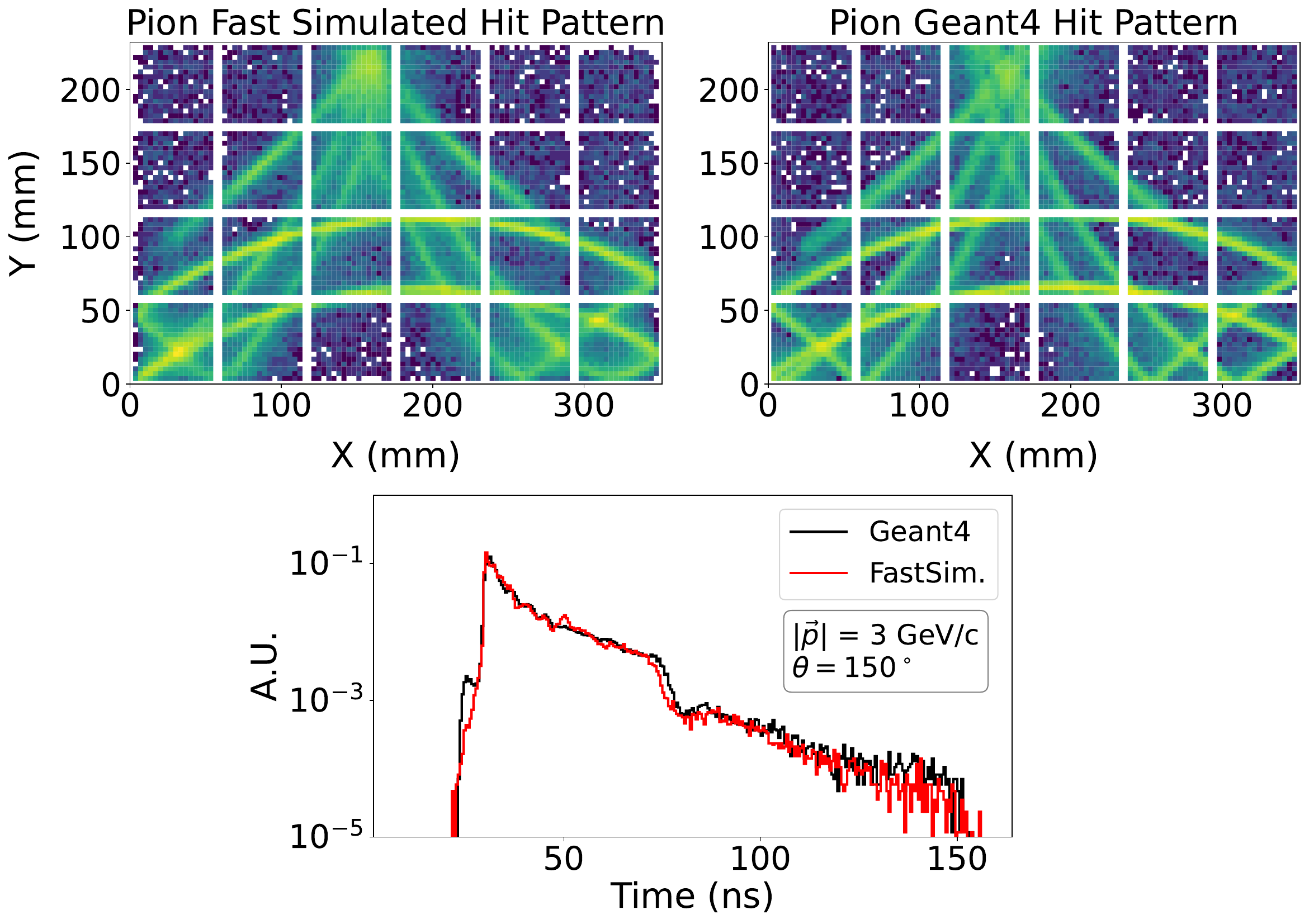} %
    \caption{
    \textbf{Fast Simulation with DDPM:} Fast Simulation of Kaons (left column of plots), and Pions (right column of plots) at 3 GeV/c and various polar angles using DDPM.}
    \label{fig:DDPM_Generations_3GeV}
\end{figure}

%%%%%%%%%%%%%%%%%%% Score Based %%%%%%%%%%%%%%%%%%%%%%%%%%%%%%%%%%%%%
\begin{figure}[h]
    \centering
    \includegraphics[width=0.49\textwidth]{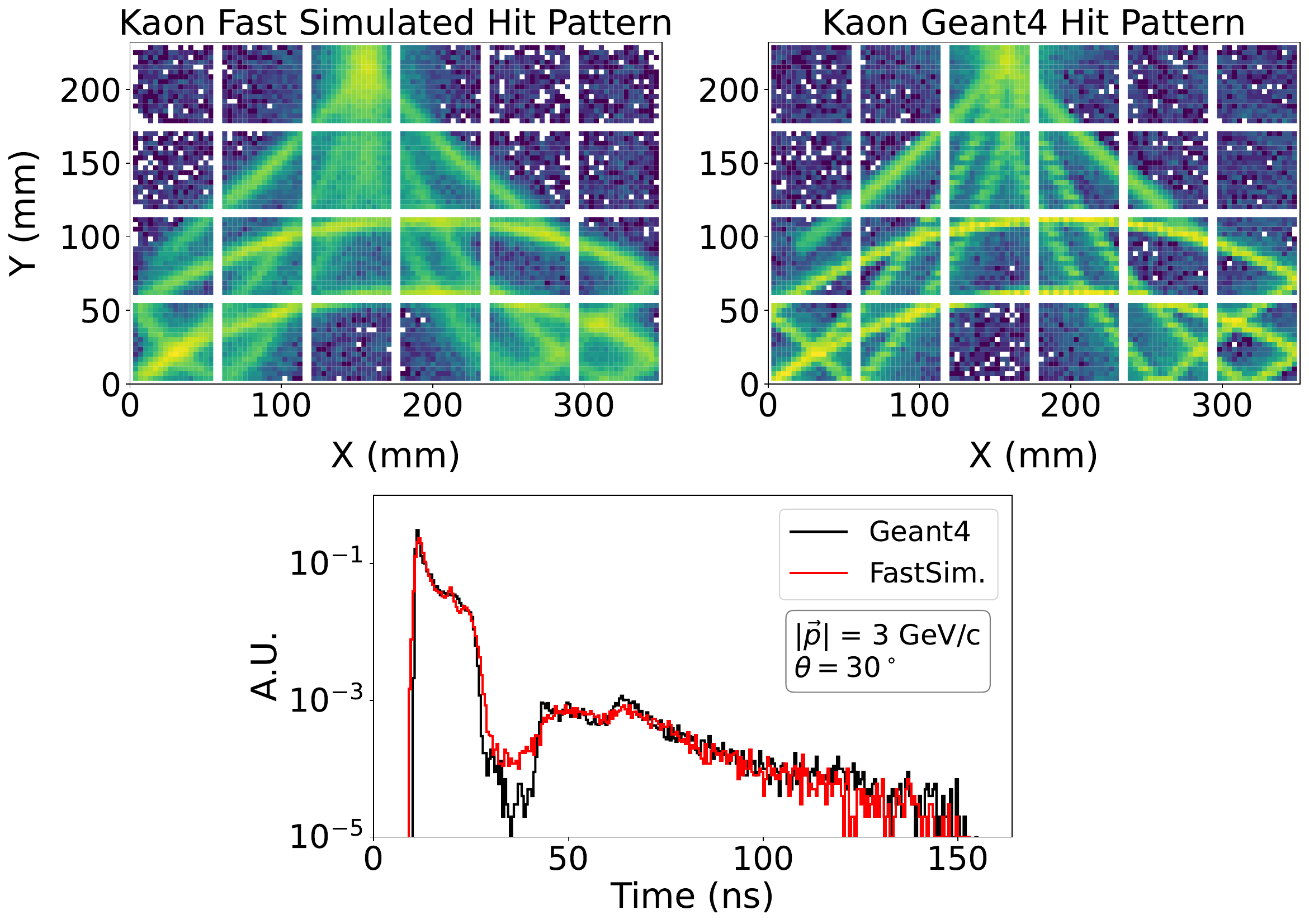}% 
   \includegraphics[width=0.49\textwidth]{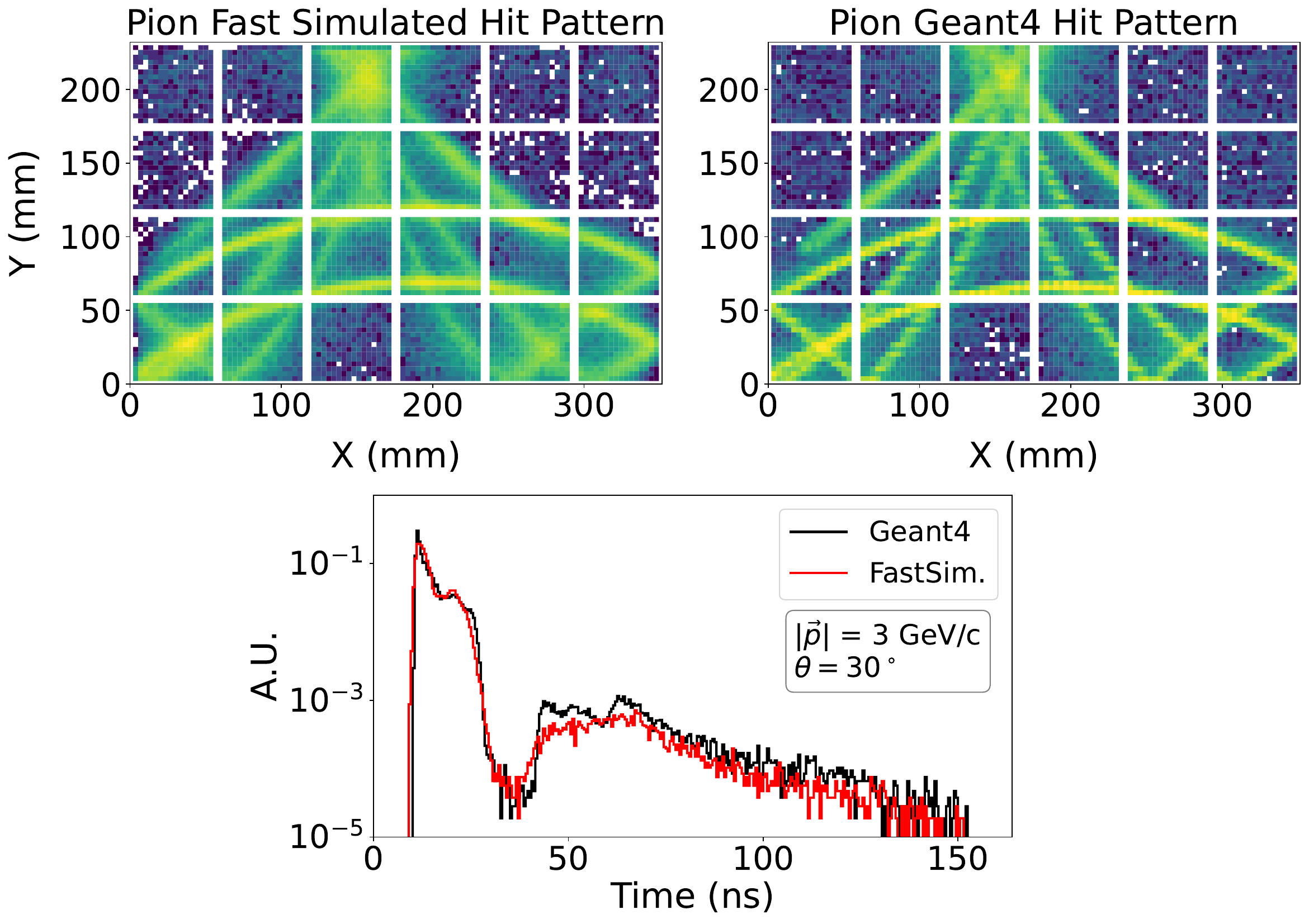} \\
    \includegraphics[width=0.49\textwidth]{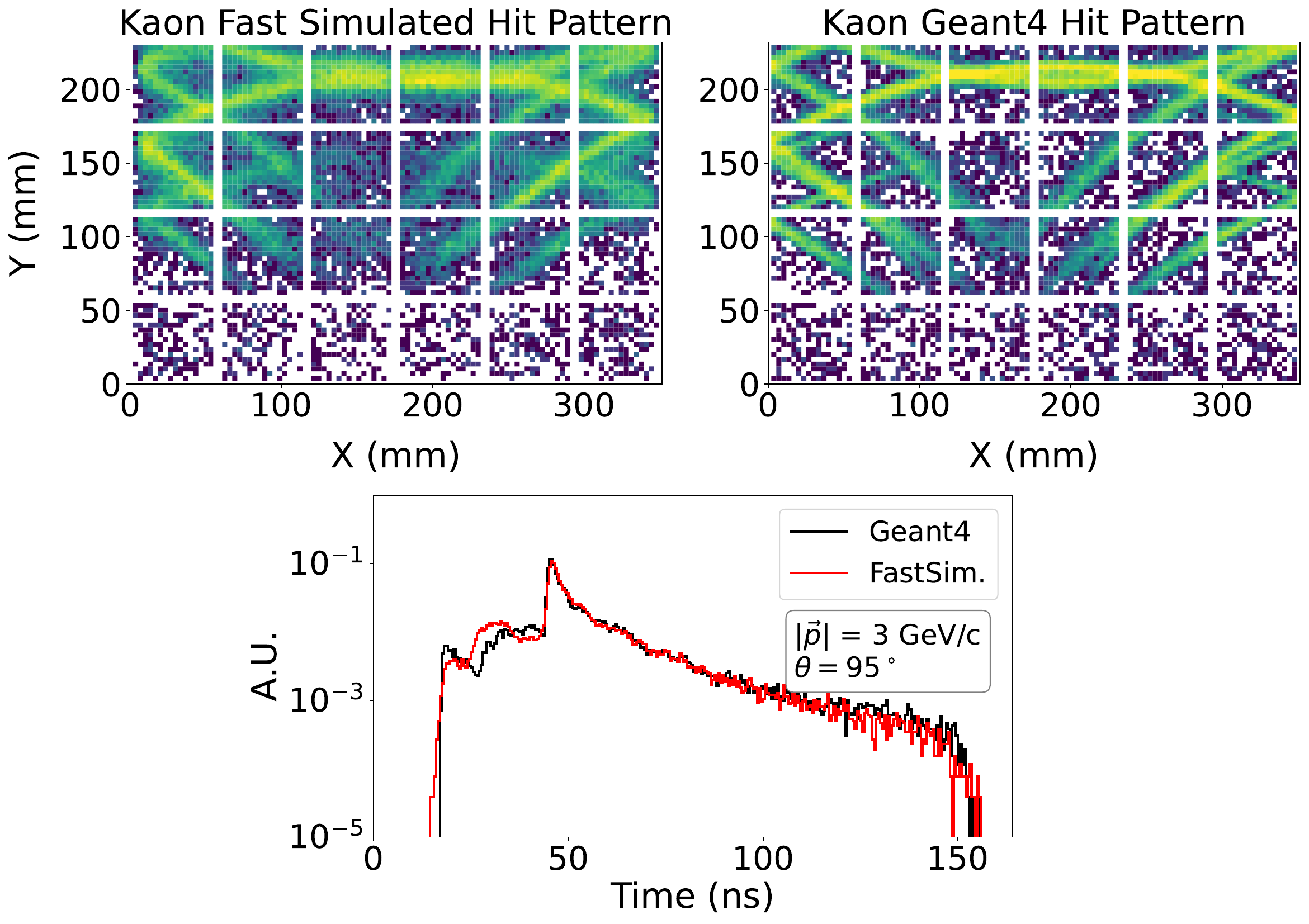} %
    \includegraphics[width=0.49\textwidth]{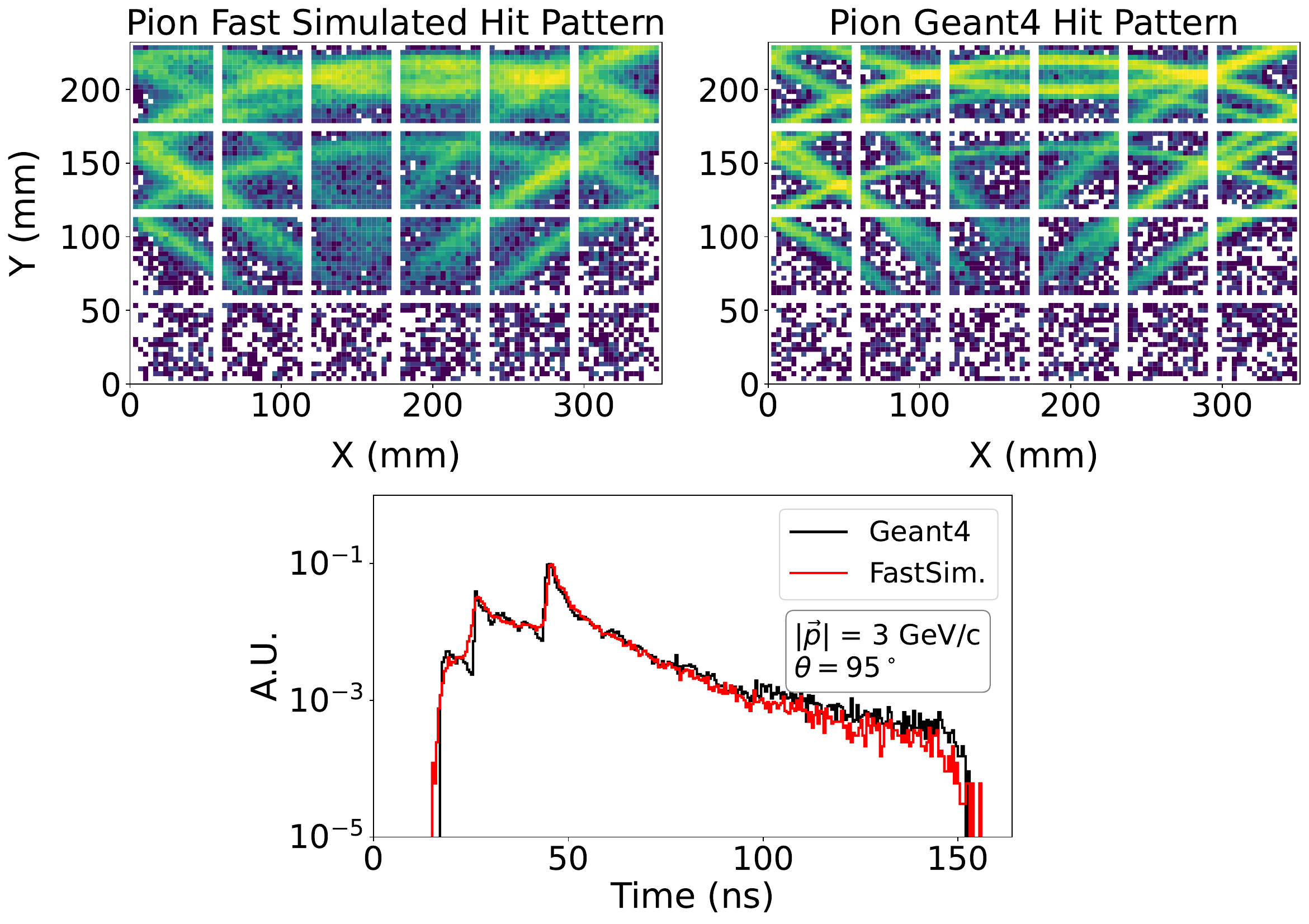} \\
    \includegraphics[width=0.49\textwidth]{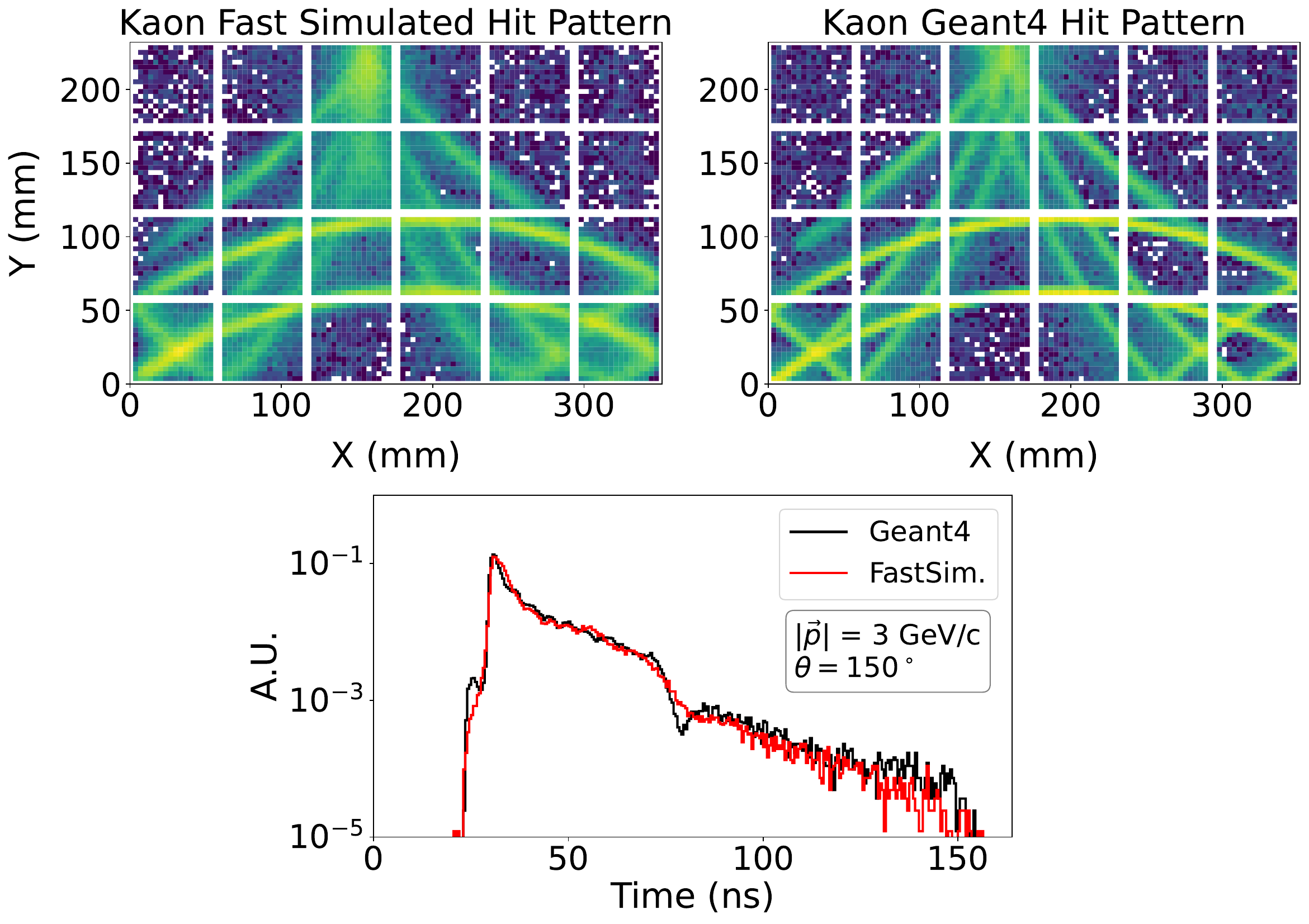} %
    \includegraphics[width=0.49\textwidth]{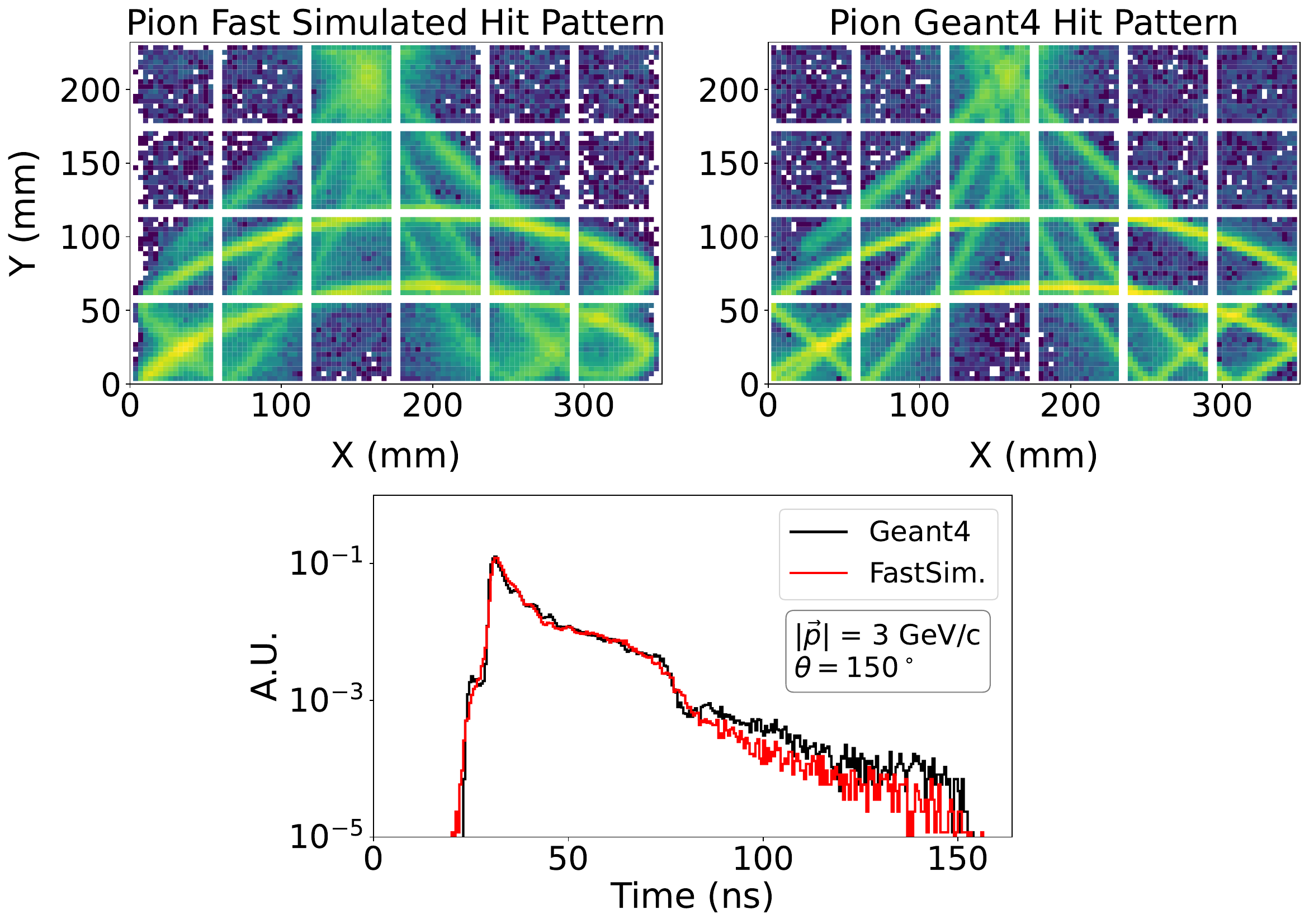} %
    \caption{
    \textbf{Fast Simulation with Score-Based Models:} Fast Simulation of Kaons (left column of plots), and Pions (right column of plots) at 3 GeV/c and various polar angles using Score-Based models.}
    \label{fig:score_Generations_3GeV}
\end{figure}

%%%%%%%%%%%%%%%%%%%%%%% Histogram and Ratios %%%%%%%%%%%%%%%%%%%%%%%%%%%%%%%%%%%

\begin{figure}[h]
    \centering
    \begin{subfigure}[b]{0.49\textwidth}
        \centering
        \includegraphics[width=\textwidth]{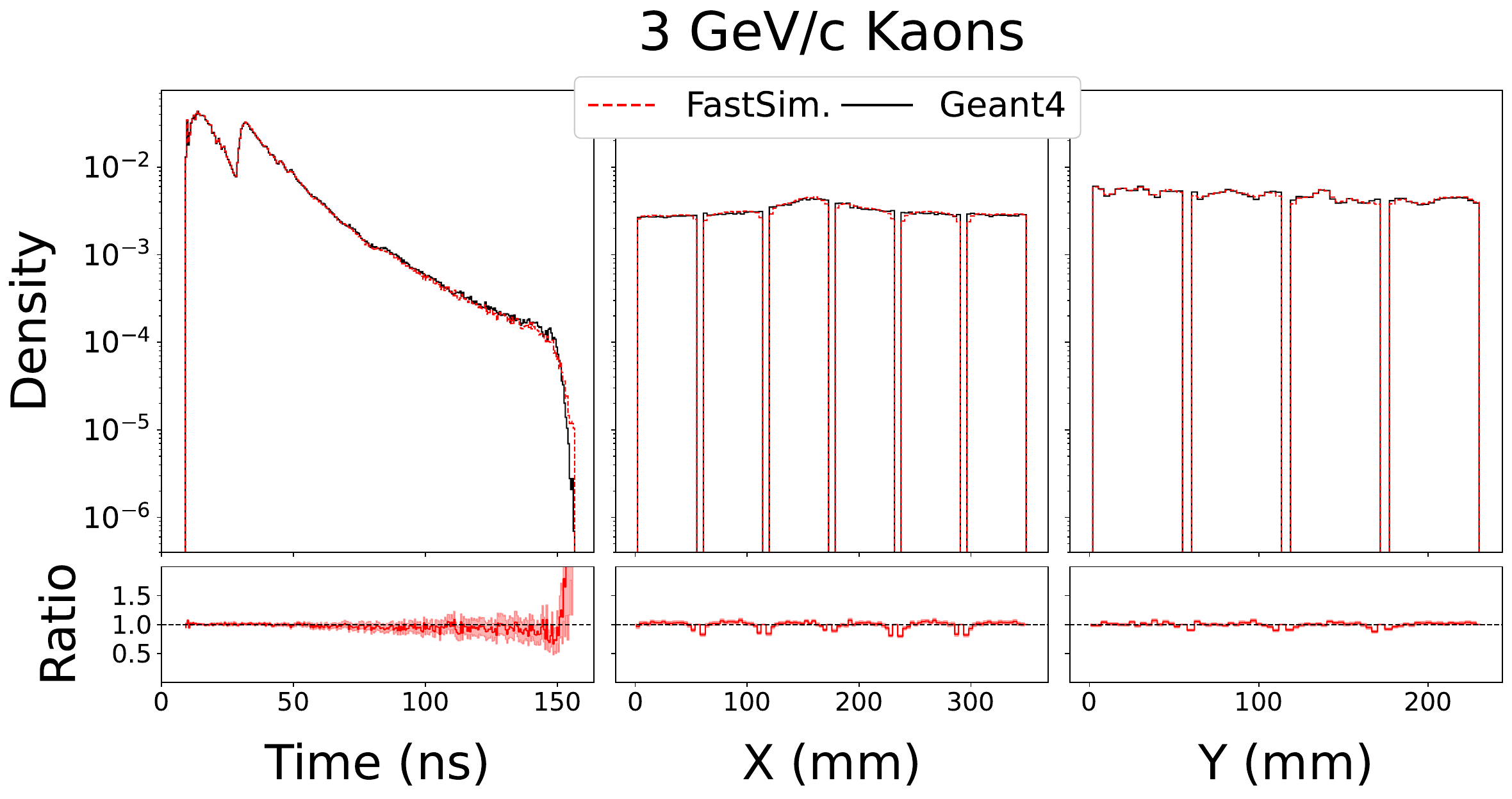}
        \caption{DNF}
    \end{subfigure}
    \begin{subfigure}[b]{0.49\textwidth}
        \centering
        \includegraphics[width=\textwidth]{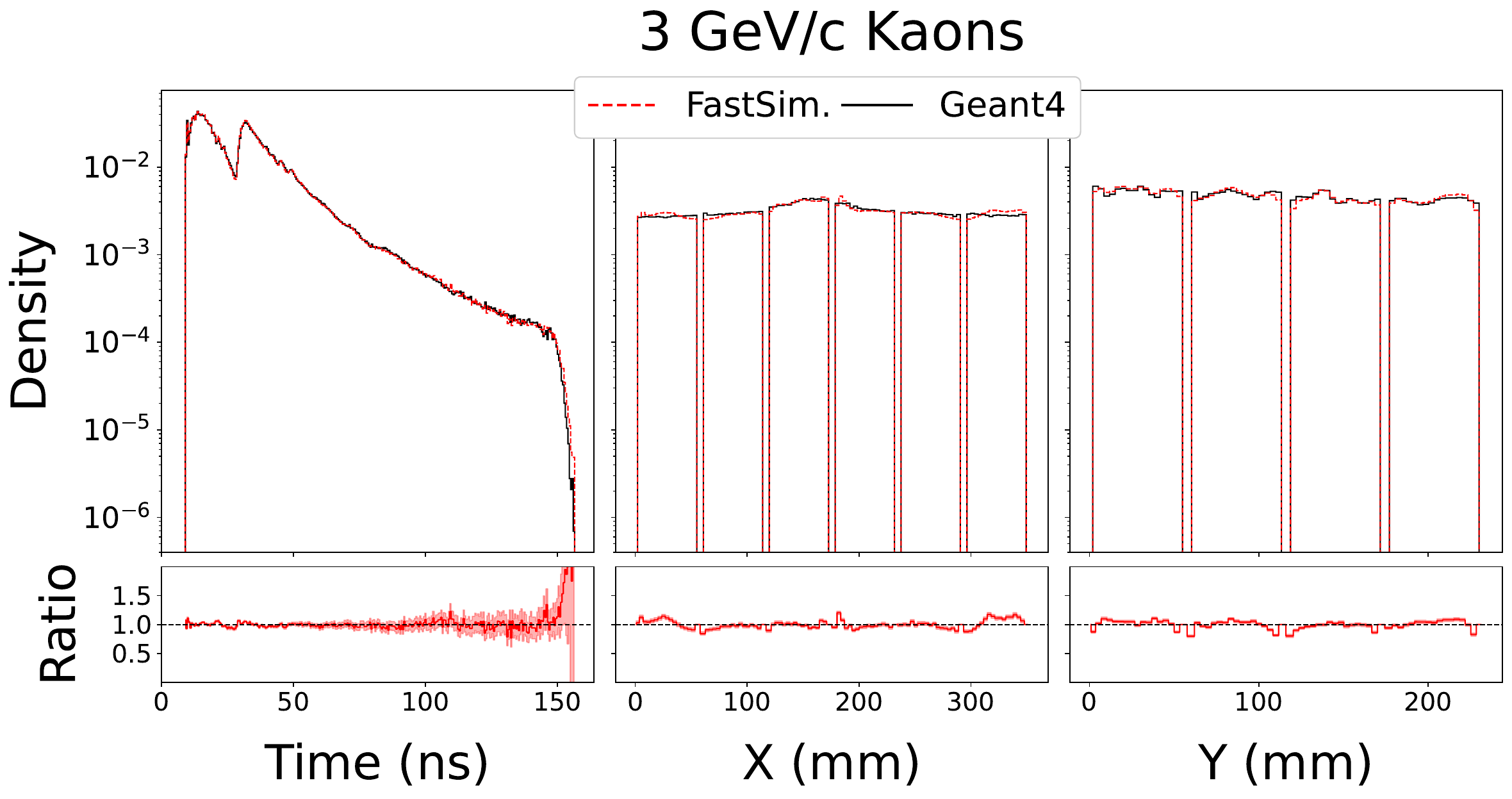}
        \caption{CNF}
    \end{subfigure} \\   
    
    \begin{subfigure}[b]{0.49\textwidth}
        \centering
        \includegraphics[width=\textwidth]{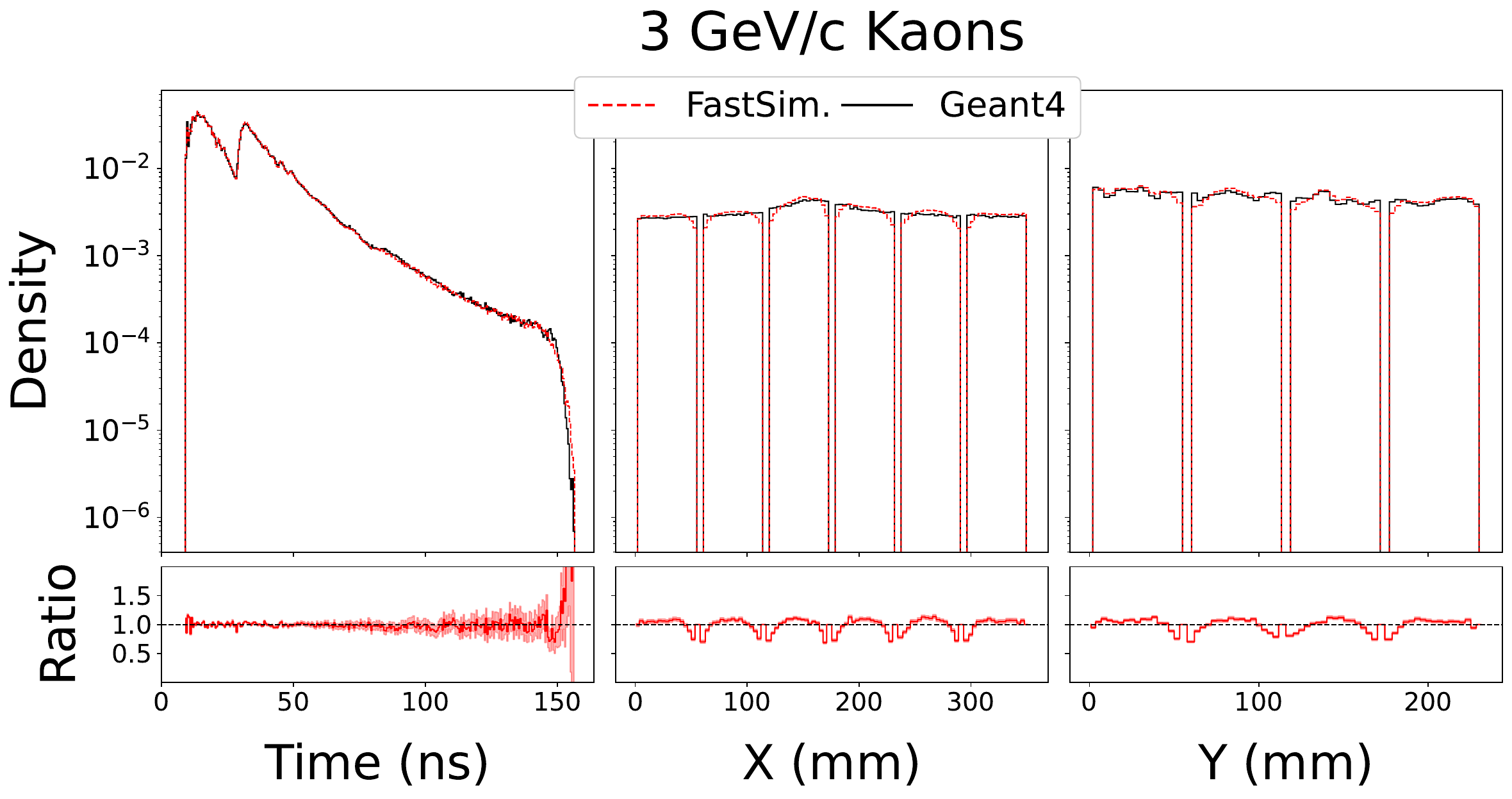}
        \caption{Flow Matching}
    \end{subfigure}
    \begin{subfigure}[b]{0.49\textwidth}
        \centering
        \includegraphics[width=\textwidth]{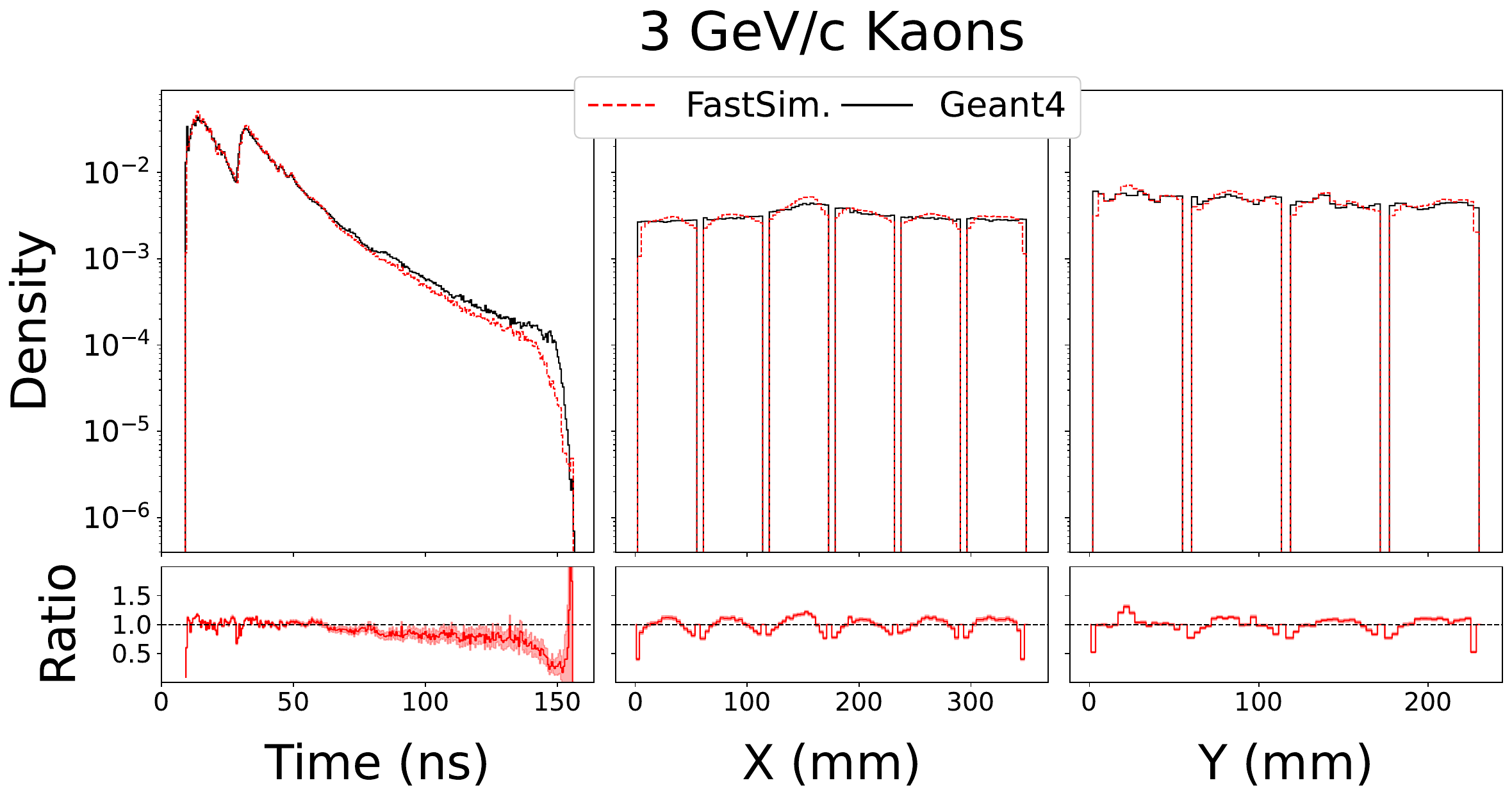}
        \caption{DDPM}
    \end{subfigure} \\  
    
    \begin{subfigure}[b]{0.49\textwidth}
        \centering
        \includegraphics[width=\textwidth]{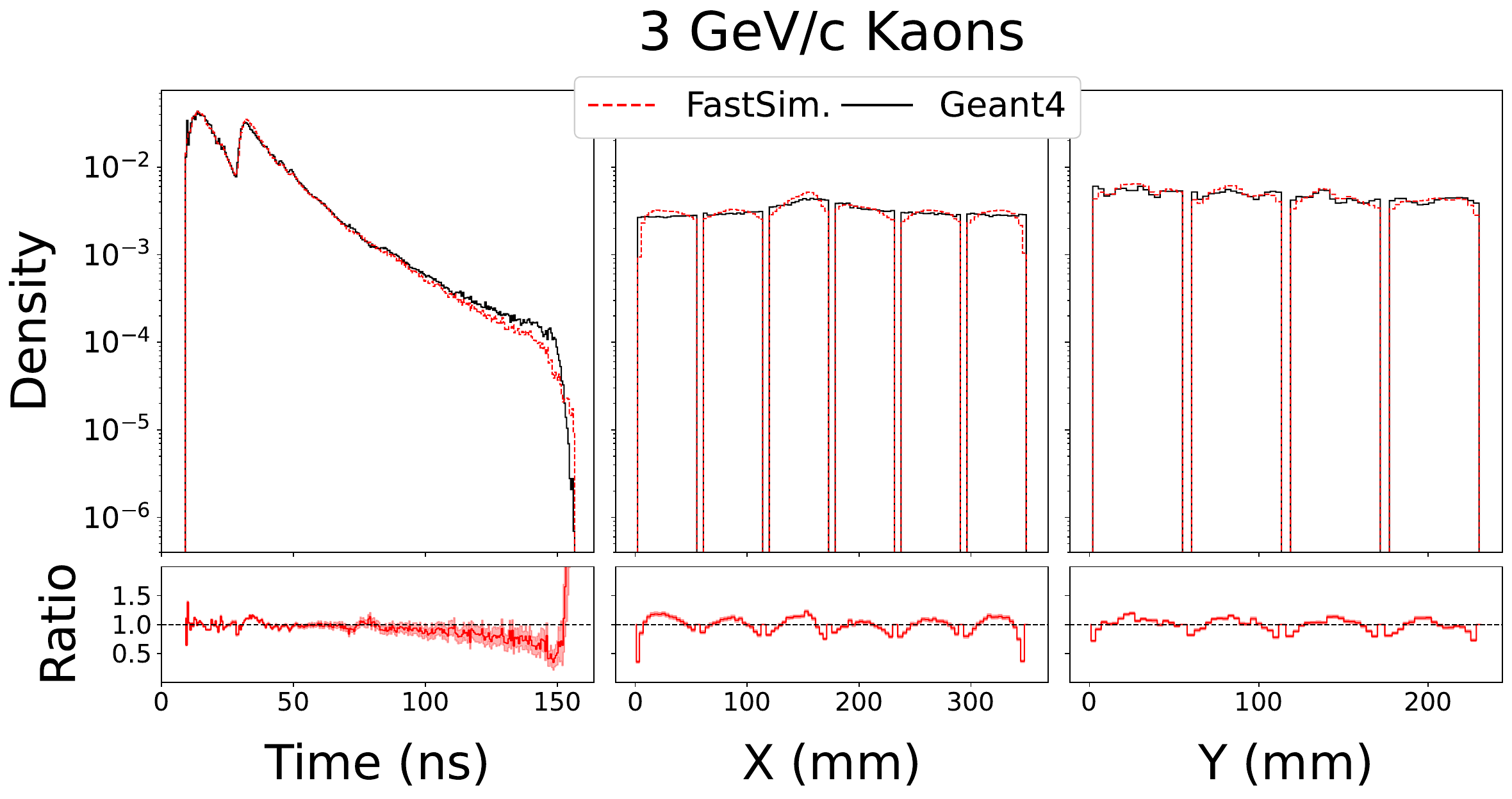}
        \caption{Score-Based}
    \end{subfigure}
    % \begin{subfigure}[b]{0.49\textwidth}
    %     \centering
    %     \includegraphics[width=\textwidth]{Figures/Generations/GSGM/3GeV/Ratios_Kaon.pdf}
    %     \caption{GSGM}
    % \end{subfigure}
    \caption{\textbf{Ratio Plots at 3 GeV/c for Kaons:} Ratio plots for Kaons using the various different models (a) Discrete Normalizing Flows (DNF), (b) Continuous Normalizing Flows (b), (c) Flow Matching, (d) Denoising Diffusion Probabilistic Models (DDPM), and (e) Score-Based Generative Models at 3 GeV/c, integrated over the polar angle.}
    \label{fig:ratio_plots_kaon_3GeV}
\end{figure}

\begin{figure}
    \centering
    \begin{subfigure}[b]{0.49\textwidth}
        \centering
        \includegraphics[width=\textwidth]{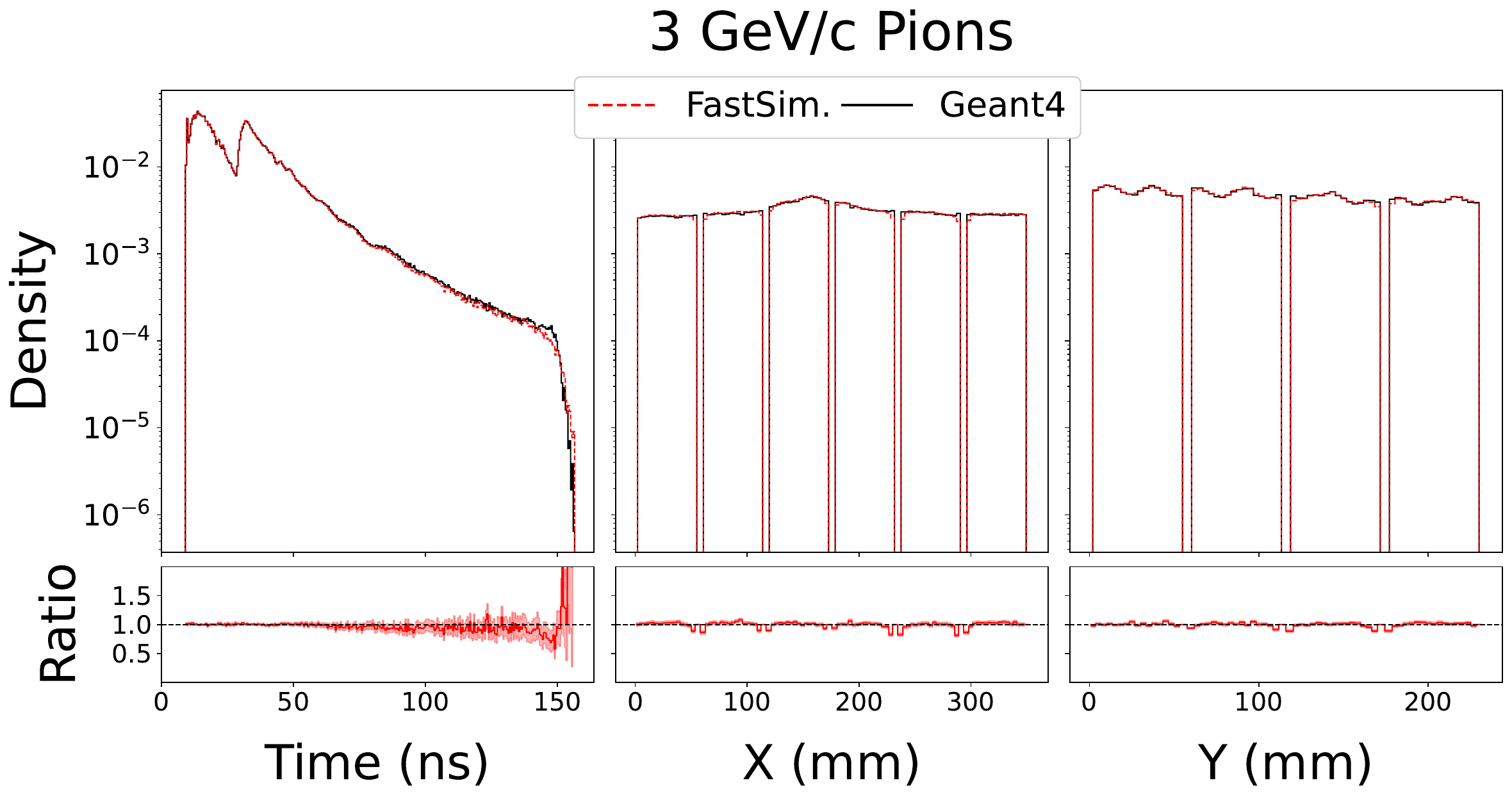}
        \caption{DNF}
    \end{subfigure}
    \begin{subfigure}[b]{0.49\textwidth}
        \centering
        \includegraphics[width=\textwidth]{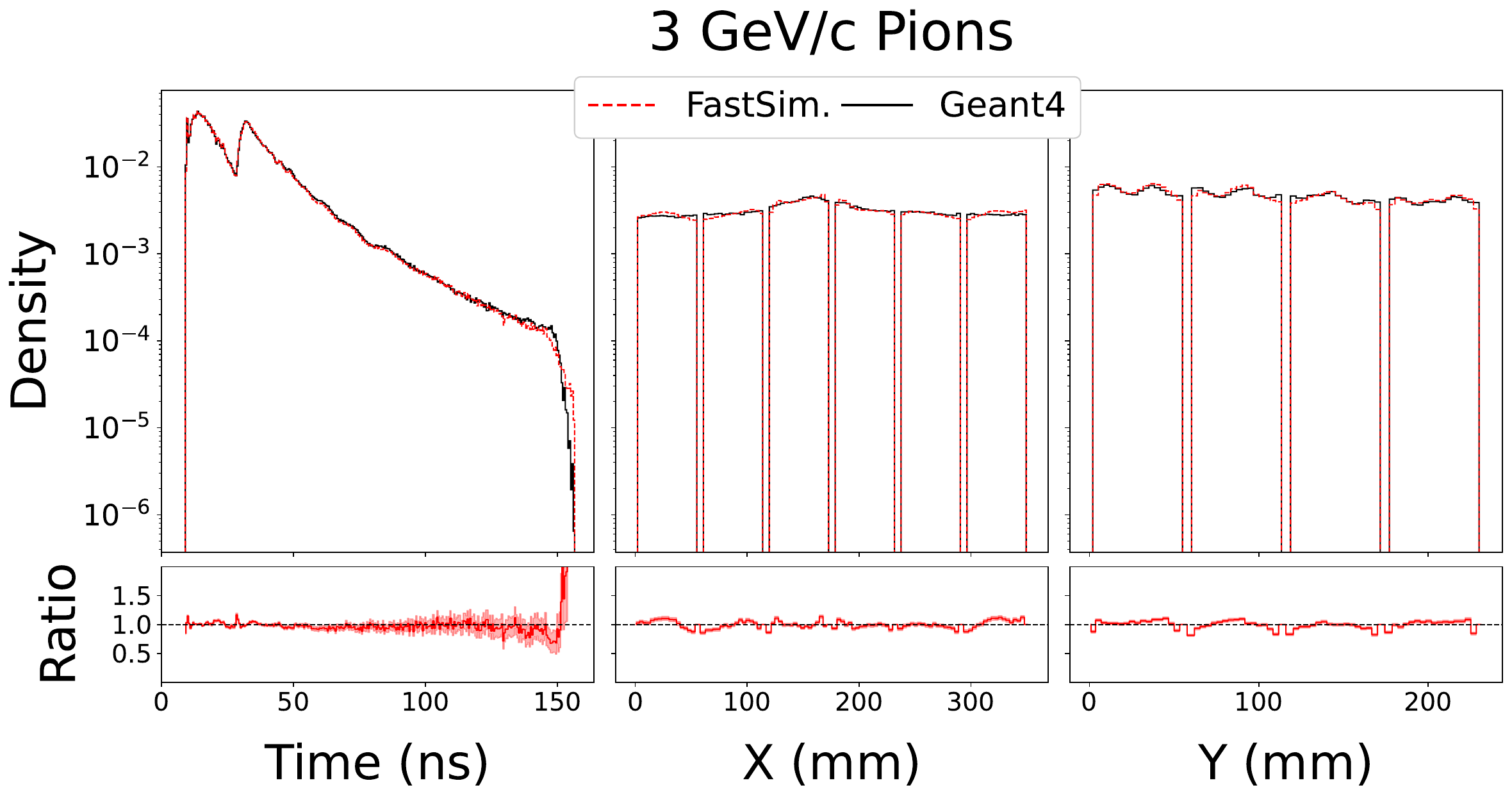}
        \caption{CNF}
    \end{subfigure} \\   
    
    \begin{subfigure}[b]{0.49\textwidth}
        \centering
        \includegraphics[width=\textwidth]{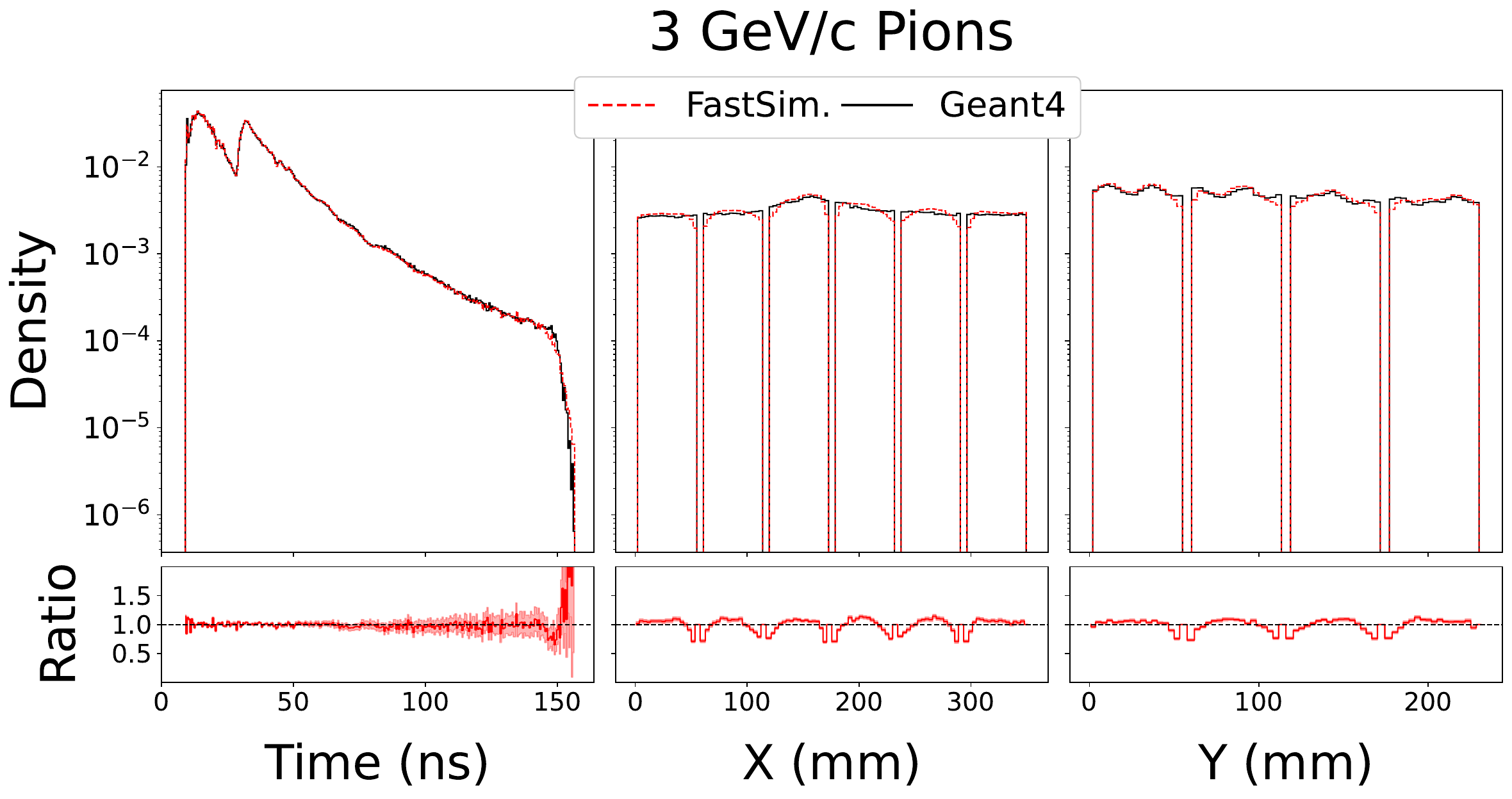}
        \caption{Flow Matching}
    \end{subfigure}
    \begin{subfigure}[b]{0.49\textwidth}
        \centering
        \includegraphics[width=\textwidth]{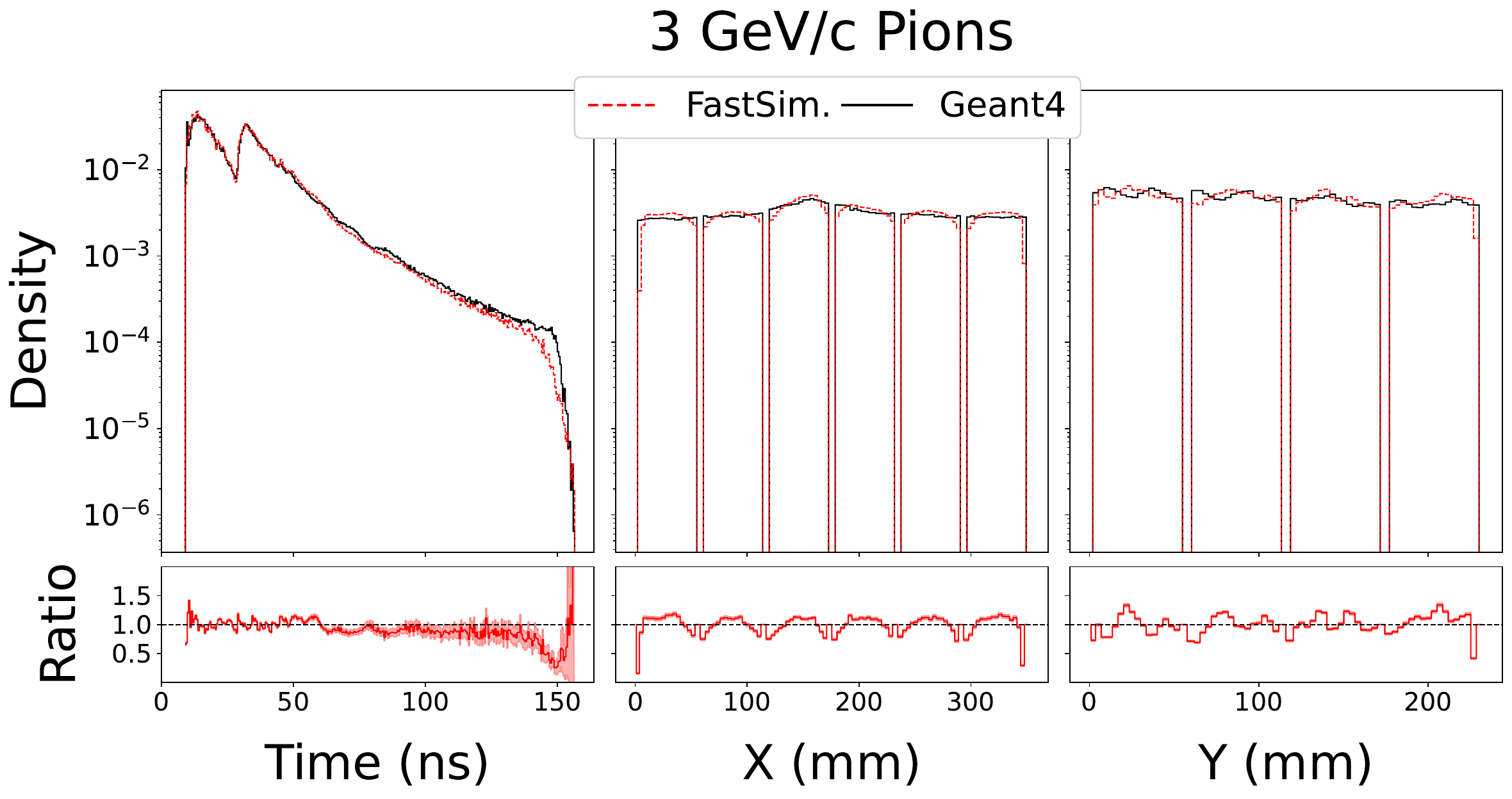}
        \caption{DDPM}
    \end{subfigure} \\  
    
    \begin{subfigure}[b]{0.49\textwidth}
        \centering
        \includegraphics[width=\textwidth]{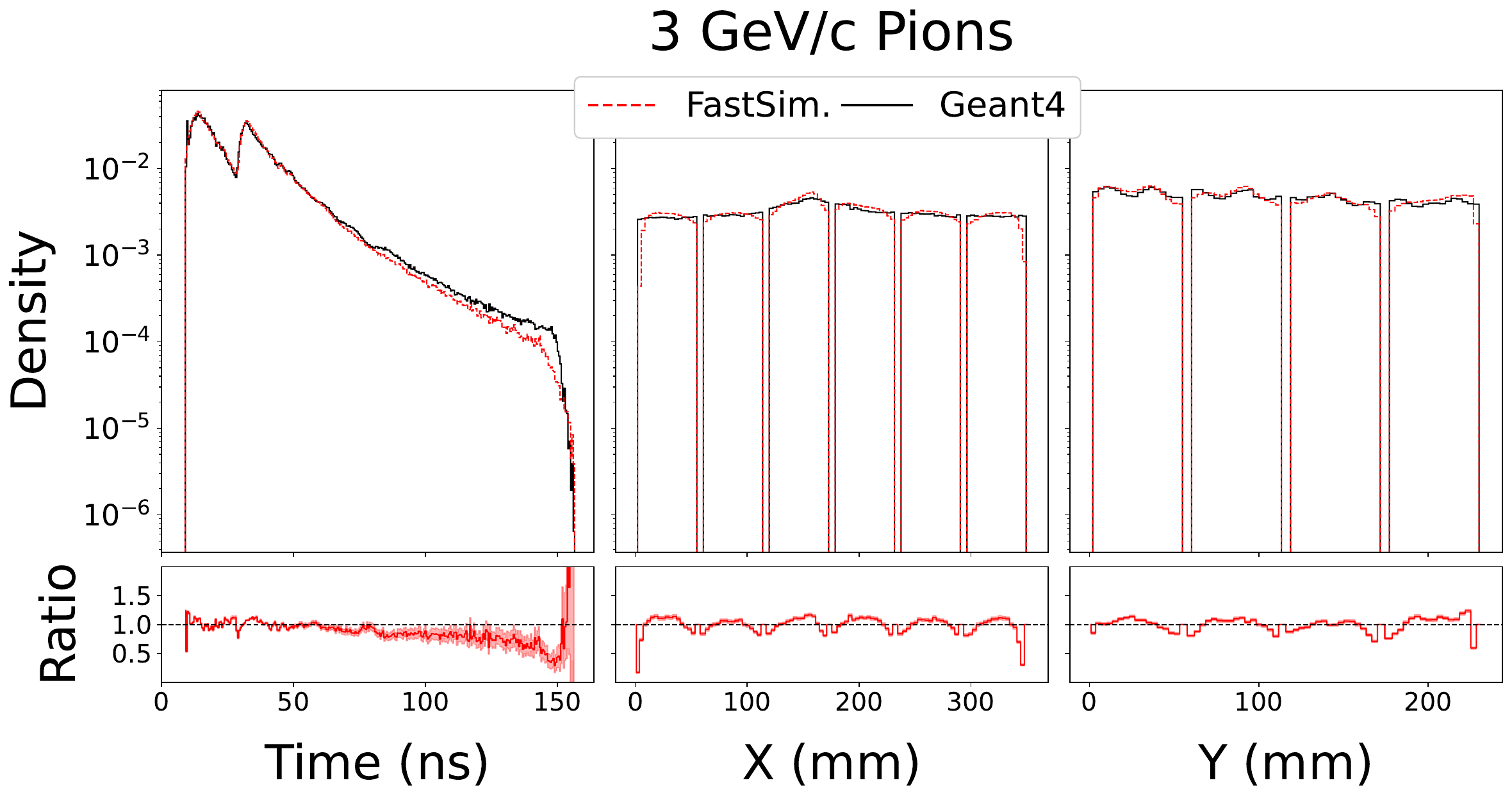}
        \caption{Score-Based}
    \end{subfigure}
    % \begin{subfigure}[b]{0.49\textwidth}
    %     \centering
    %     \includegraphics[width=\textwidth]{Figures/Generations/GSGM/3GeV/Ratios_Pion.pdf}
    %     \caption{GSGM}
    % \end{subfigure}
    \caption{\textbf{Ratio Plots at 3 GeV/c for Pions:} Ratio plots Pions using the various different models (a) Discrete Normalizing Flows (DNF), (b) Continuous Normalizing Flows (b), (c) Flow Matching, (d) Denoising Diffusion Probabilistic Models (DDPM), and (e) Score-Based Generative Models at 3 GeV/c, integrated over the polar angle..}
    \label{fig:ratio_plots_pion_3GeV}
\end{figure}

\clearpage
%%%%%%%%%%% 6 GeV %%%%%%%%%%%%%%%%%%%%%
\section{Additional Model Generations at $\SI[per-mode=symbol]{6}{\giga\eVperc}$}\label{app:6GeV}

%%%%%%%%%%%%%%%%%%% CNF %%%%%%%%%%%%%%%%%%%%%%%%%%%%%%%%%%%%%
\begin{figure}[h]
    \centering
    \includegraphics[width=0.49\textwidth]{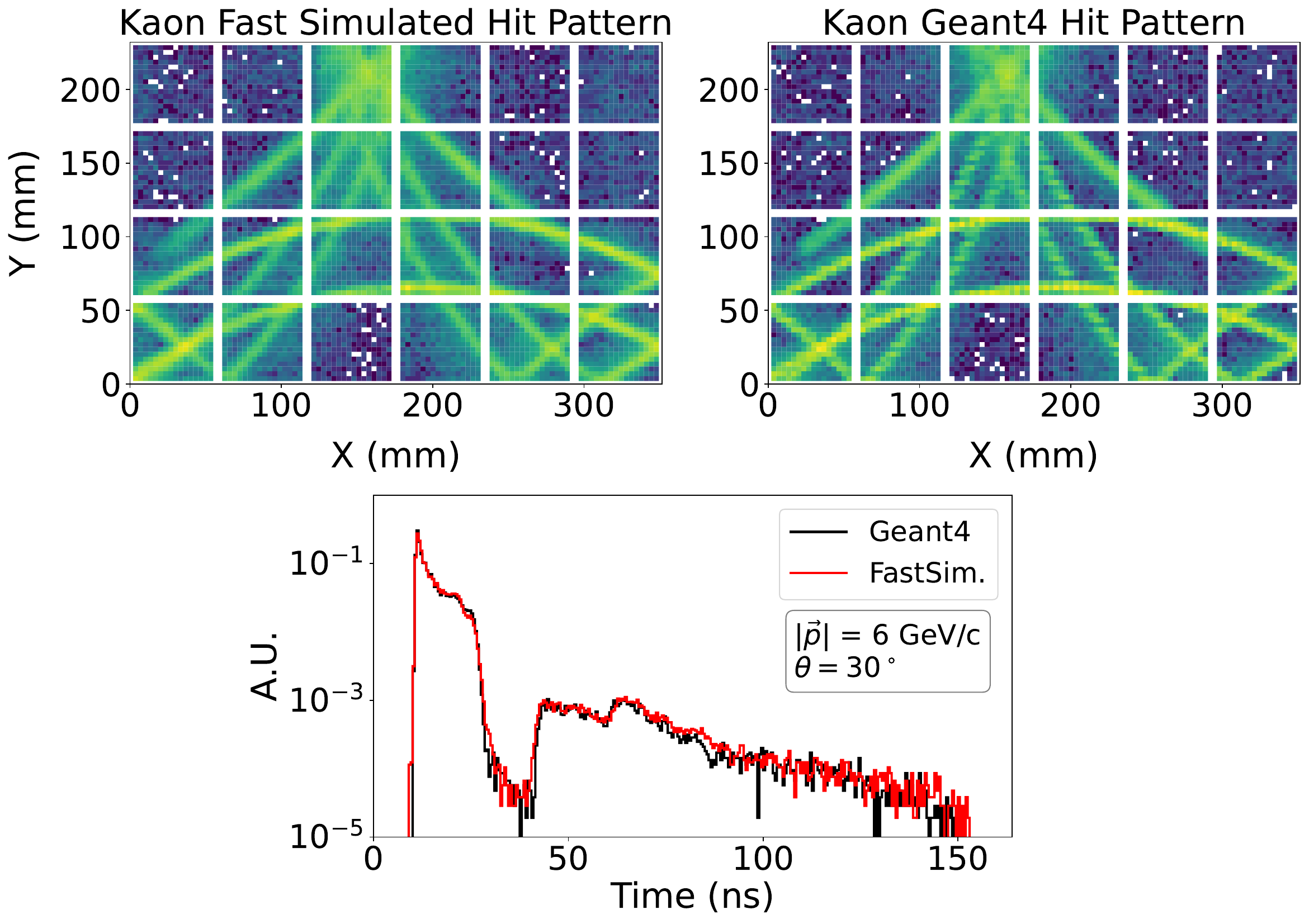}% 
   \includegraphics[width=0.49\textwidth]{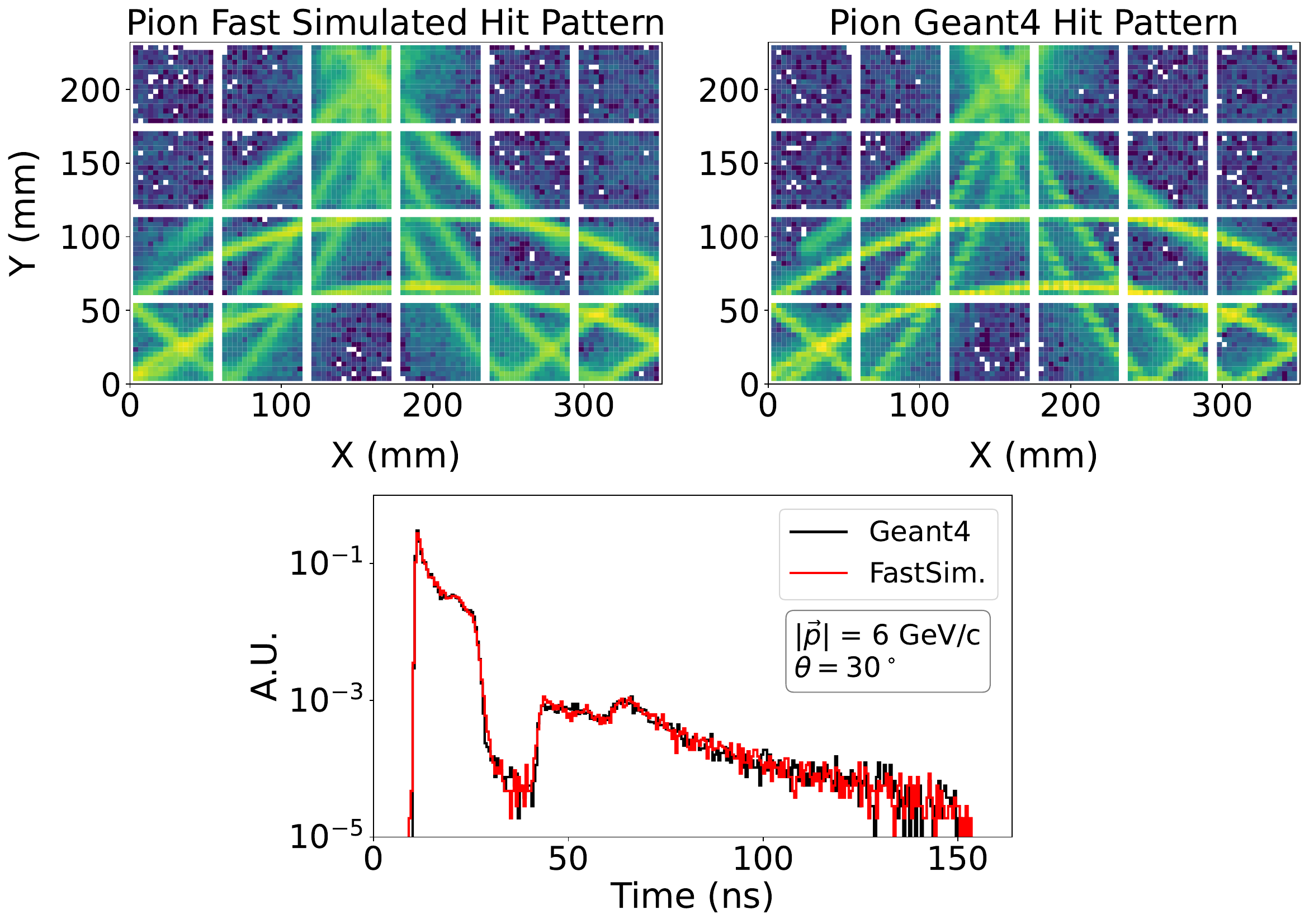} \\
    \includegraphics[width=0.49\textwidth]{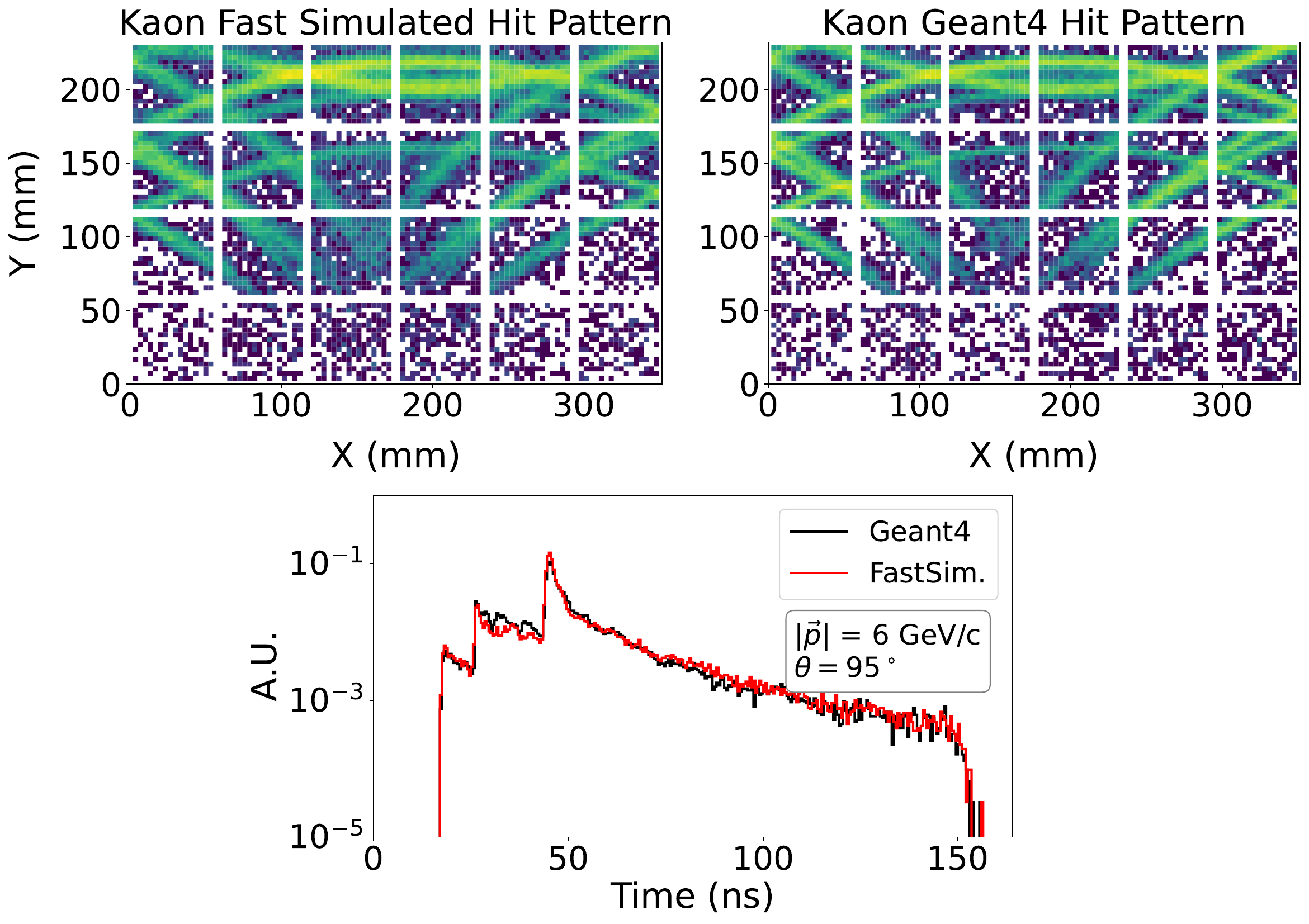} %
    \includegraphics[width=0.49\textwidth]{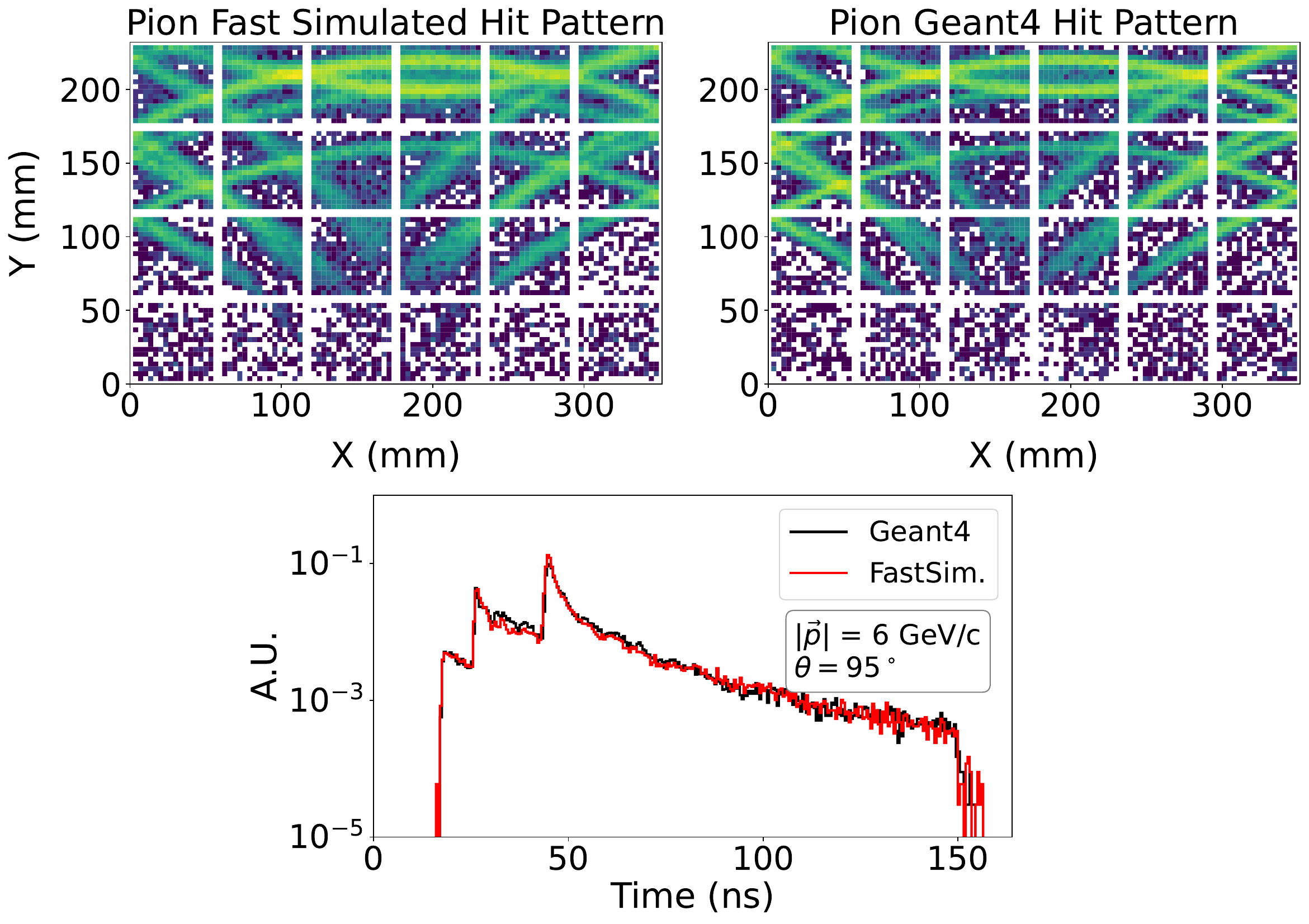} \\
    \includegraphics[width=0.49\textwidth]{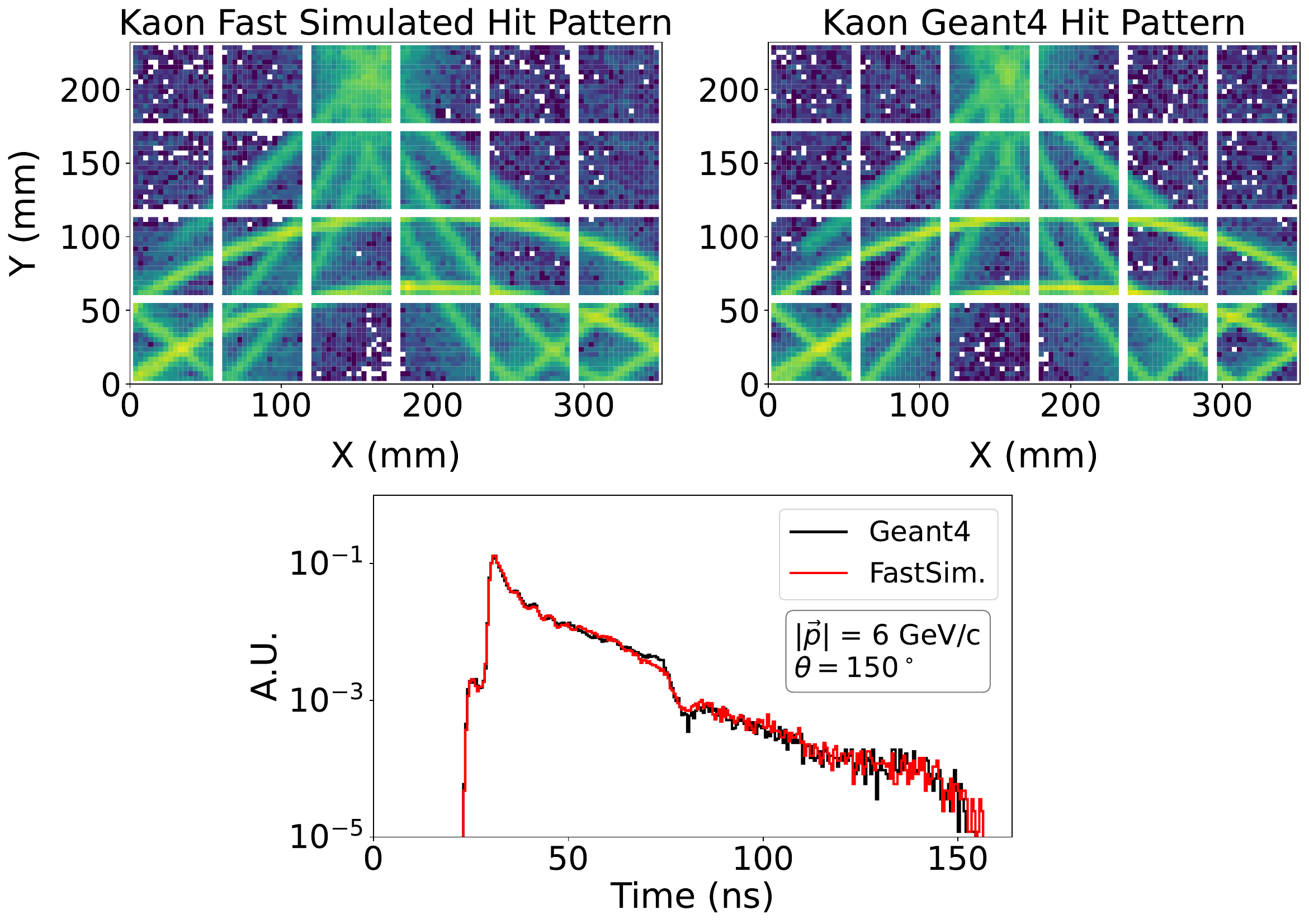} %
    \includegraphics[width=0.49\textwidth]{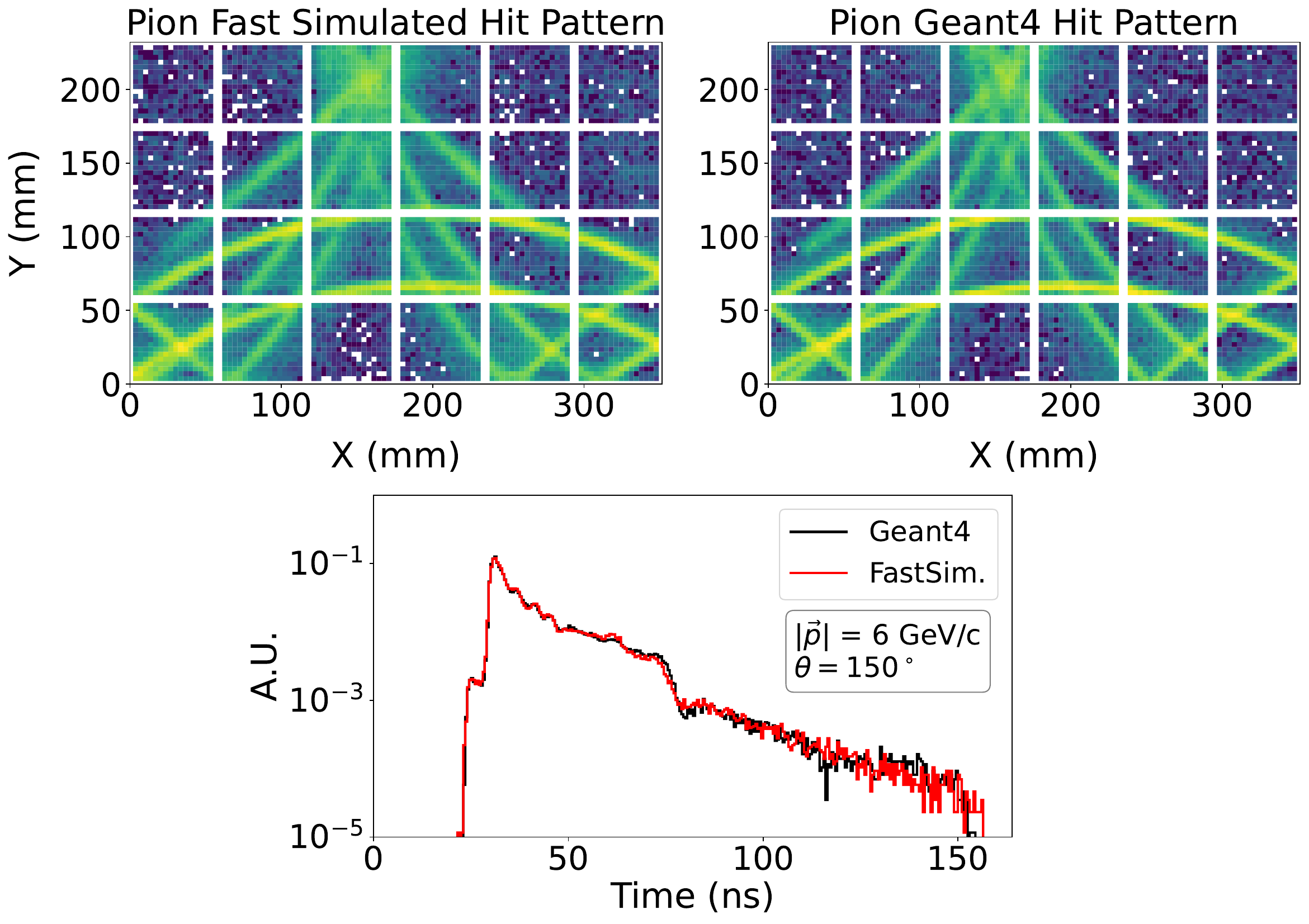} %
    \caption{
    \textbf{Fast Simulation with Continuous Normalizing Flows:} Fast Simulation of Kaons (left column of plots), and Pions (right column of plots) at 6 GeV/c and various polar angles using CNF.}
    \label{fig:CNF_Generations_6GeV}
\end{figure}

%%%%%%%%%%%%%%%%%%% FlowMatching %%%%%%%%%%%%%%%%%%%%%%%%%%%%%%%%%%%%%
\begin{figure}[h]
    \centering
    \includegraphics[width=0.49\textwidth]{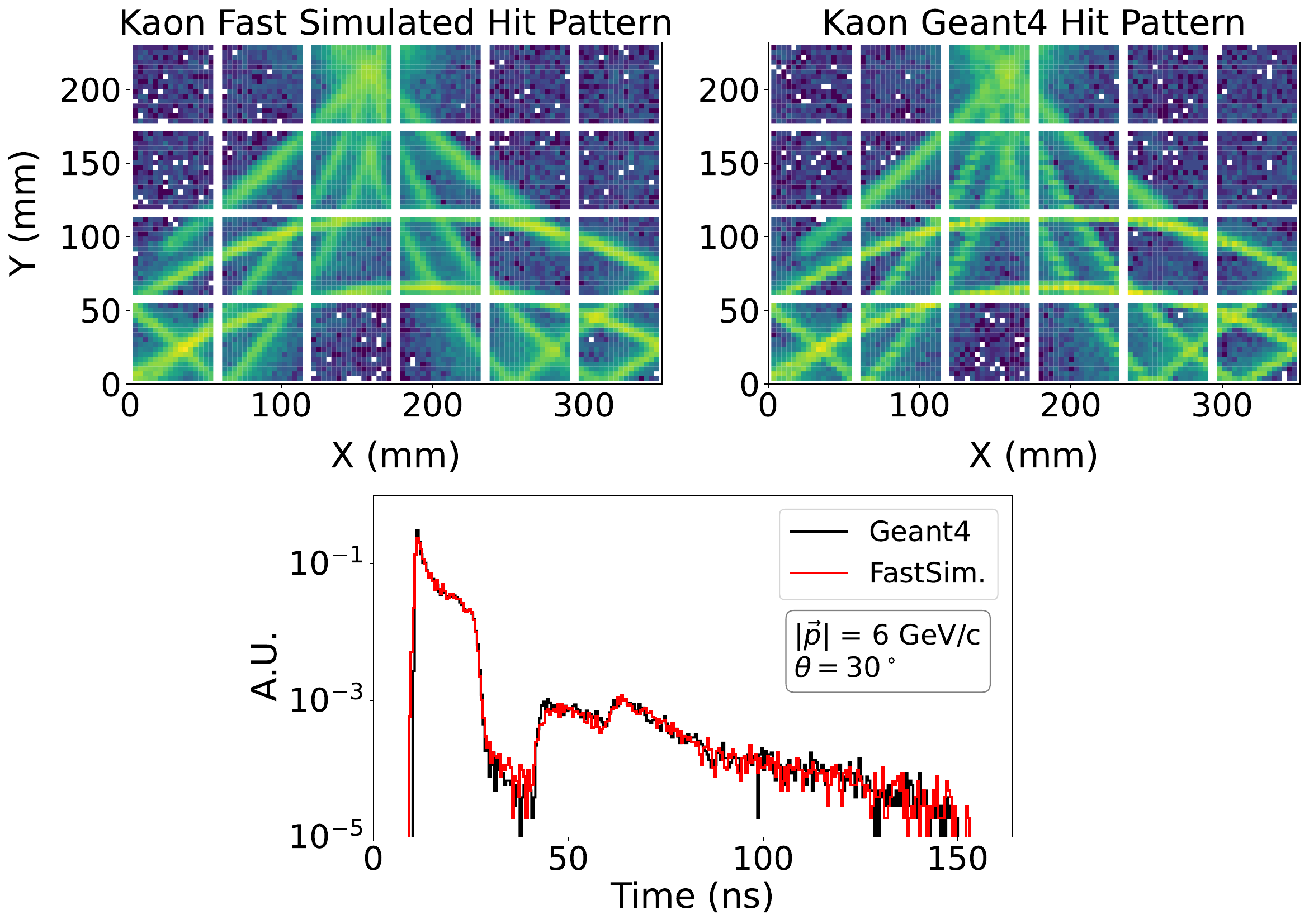}% 
   \includegraphics[width=0.49\textwidth]{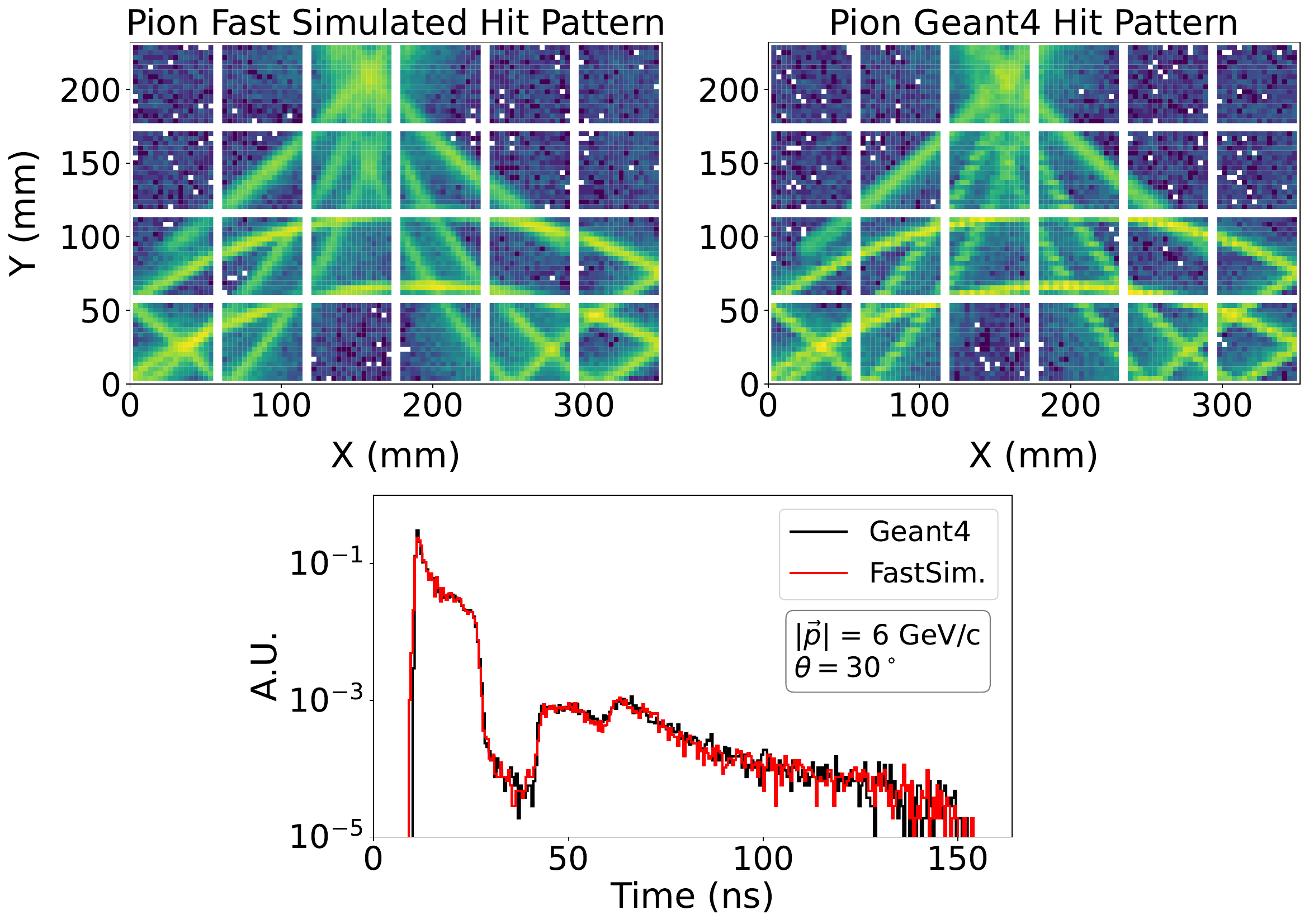} \\
    \includegraphics[width=0.49\textwidth]{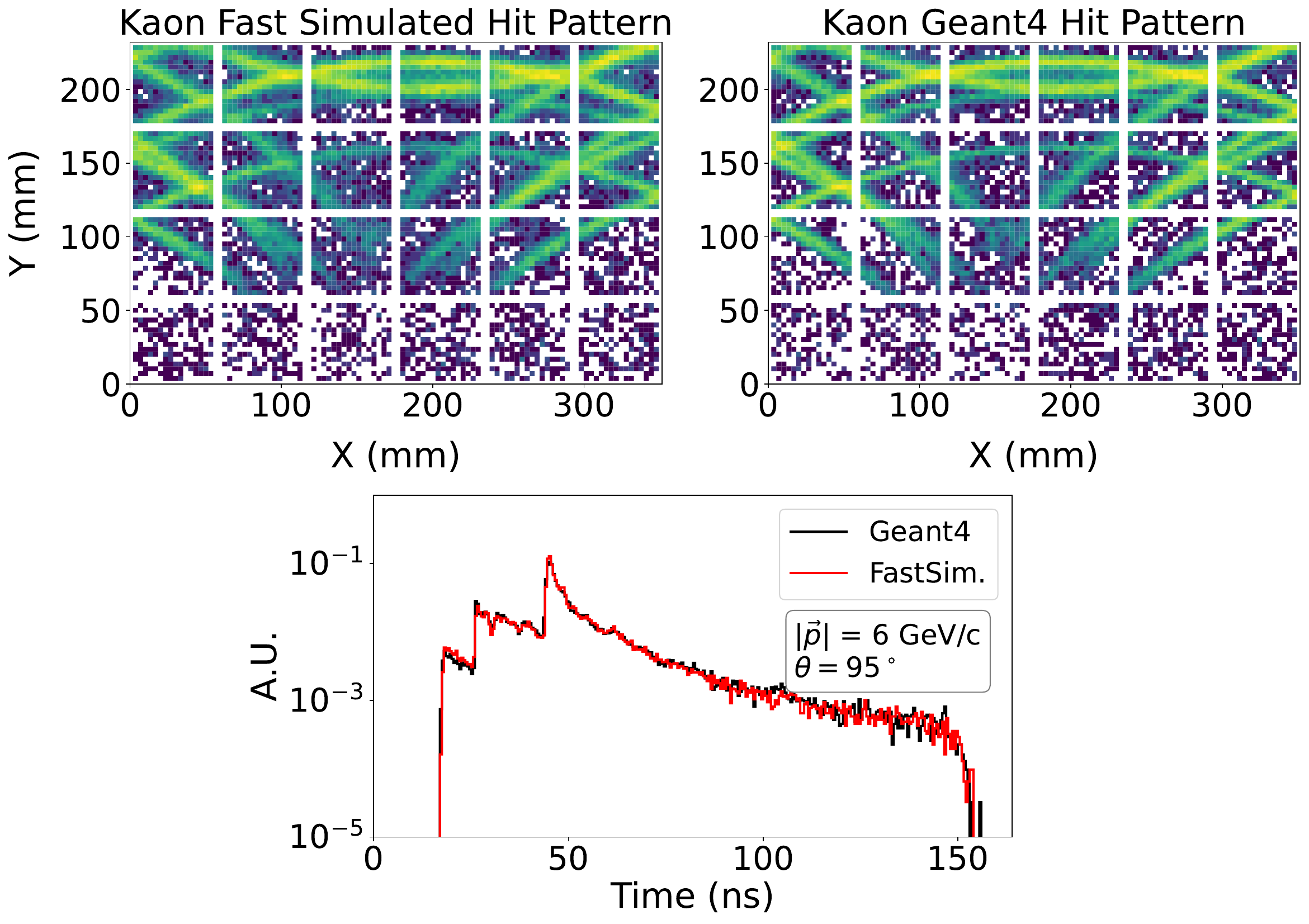} %
    \includegraphics[width=0.49\textwidth]{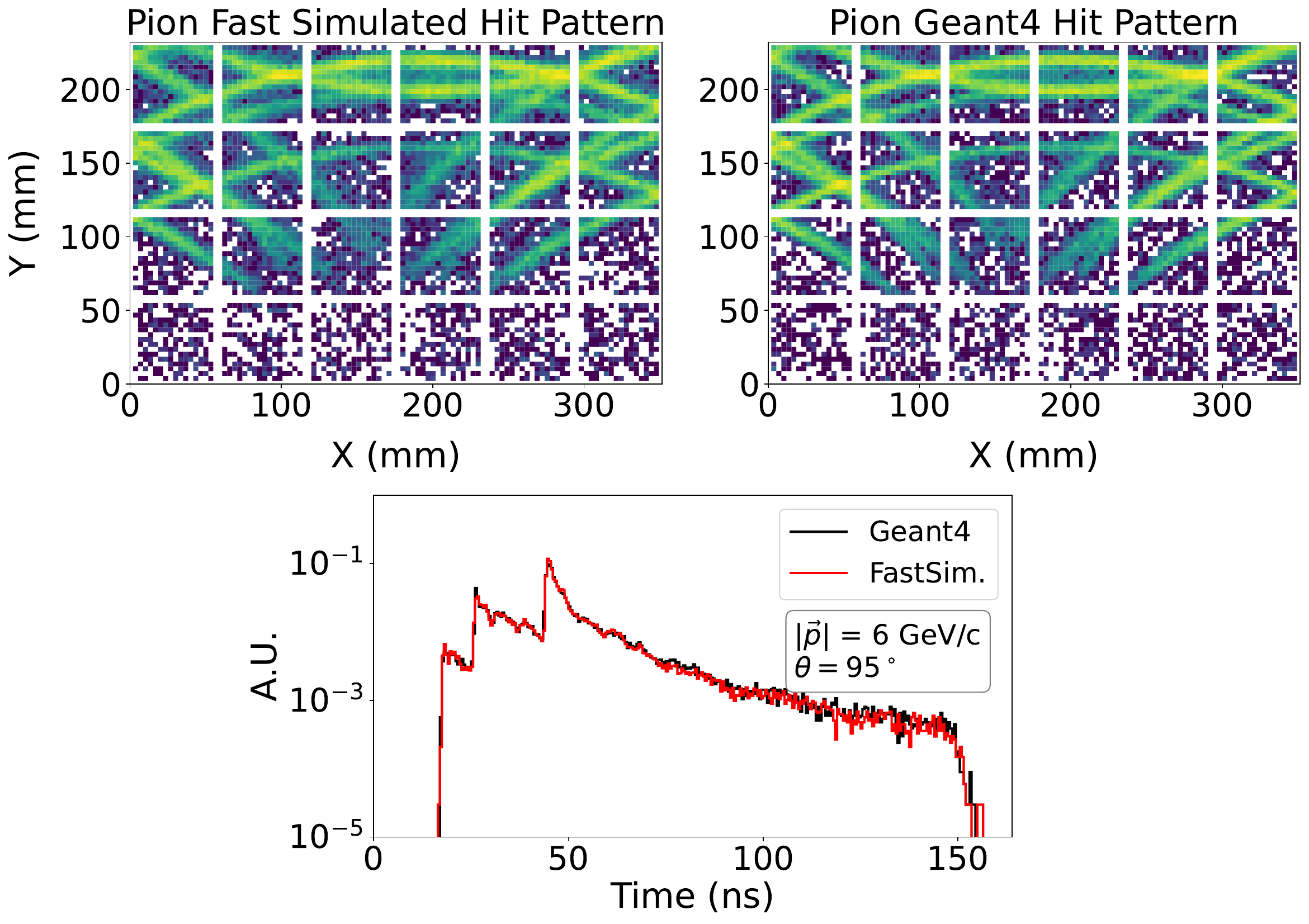} \\
    \includegraphics[width=0.49\textwidth]{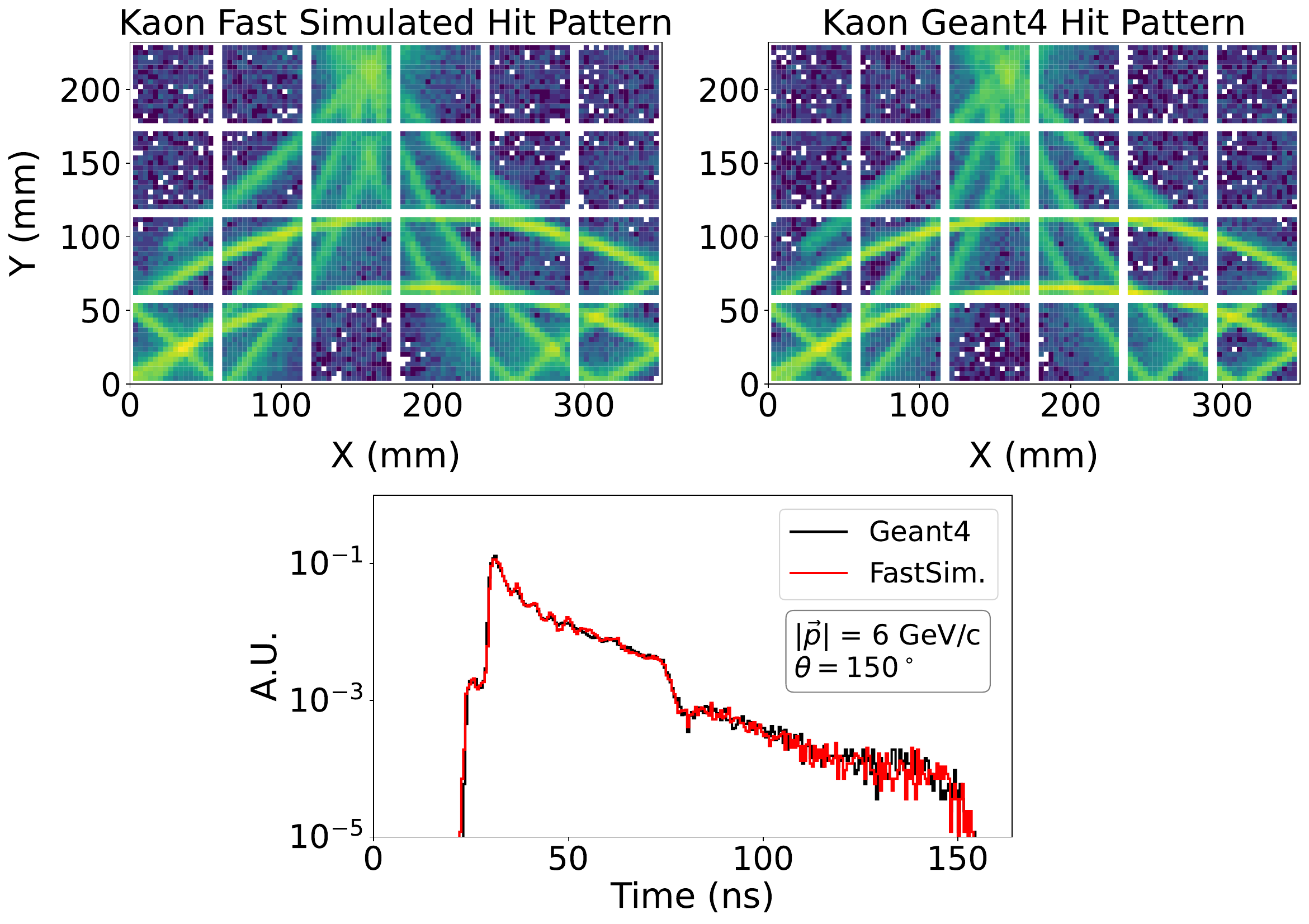} %
    \includegraphics[width=0.49\textwidth]{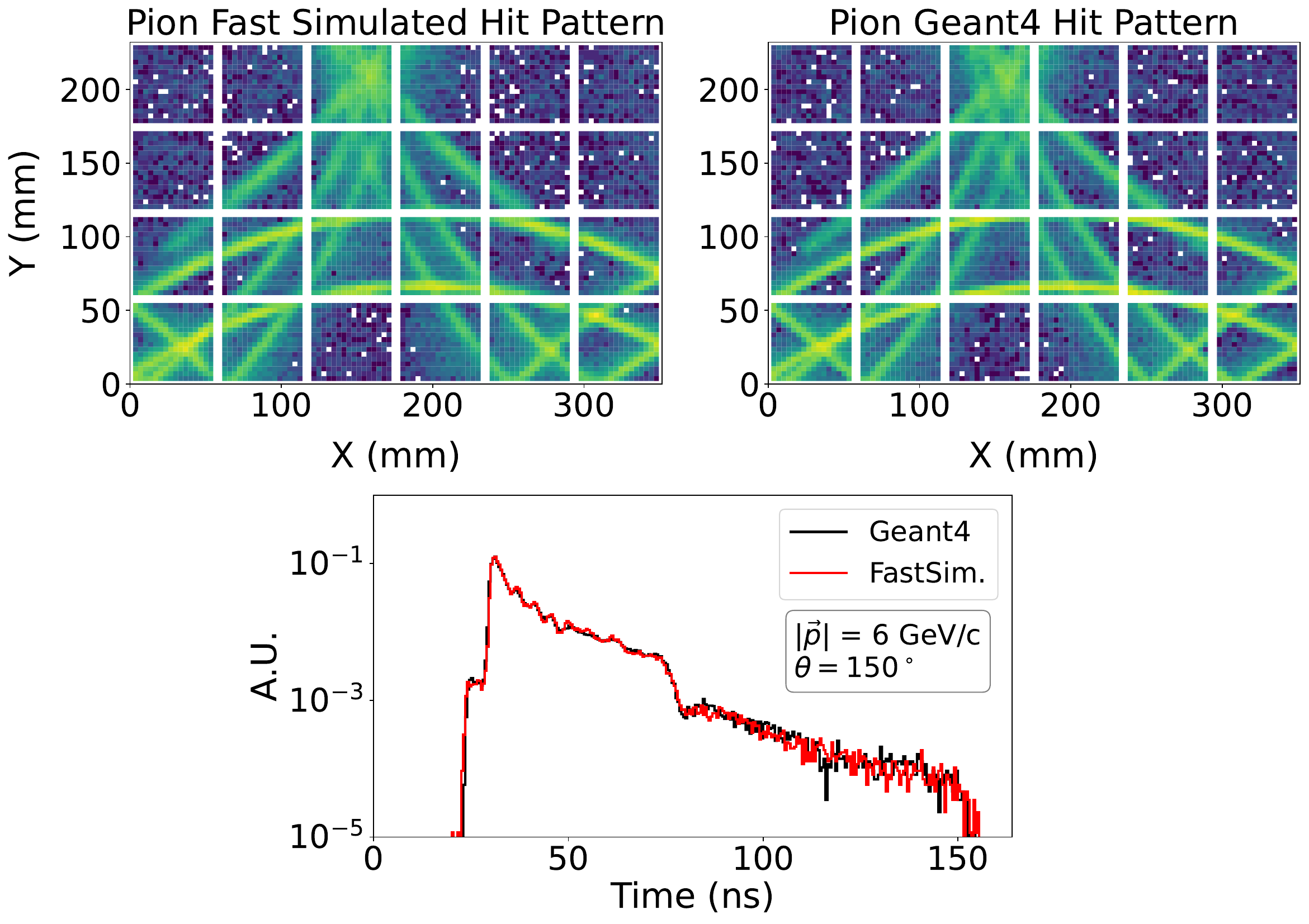} %
    \caption{
    \textbf{Fast Simulation with Flow Matching:} Fast Simulation of Kaons (left column of plots), and Pions (right column of plots) at 6 GeV/c and various polar angles using Flow Matching.}
    \label{fig:FlowMatching_Generations_6GeV}
\end{figure}

%%%%%%%%%%%%%%%%%%% DDPM %%%%%%%%%%%%%%%%%%%%%%%%%%%%%%%%%%%%%
\begin{figure}[h]
    \centering
    \includegraphics[width=0.49\textwidth]{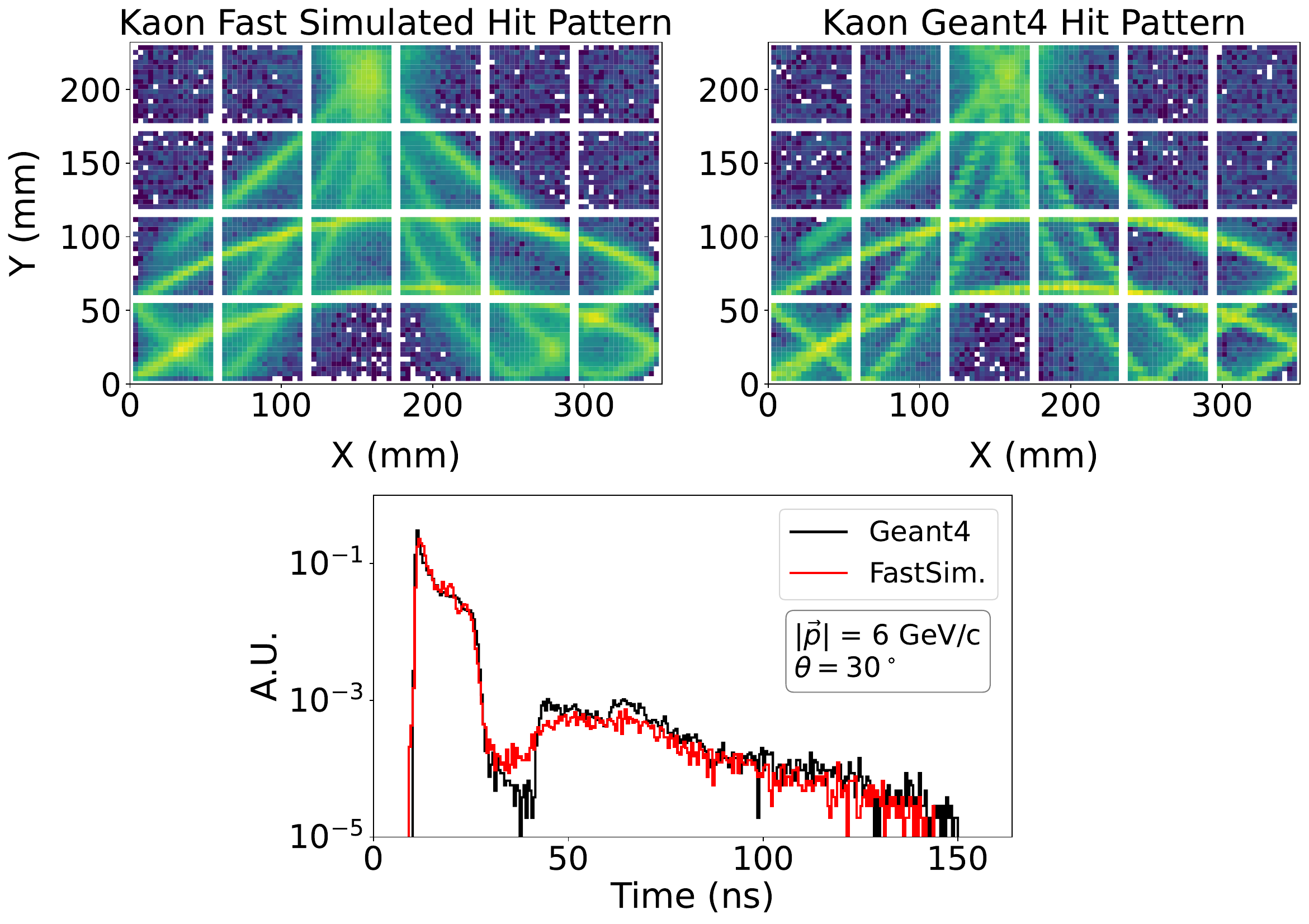}% 
   \includegraphics[width=0.49\textwidth]{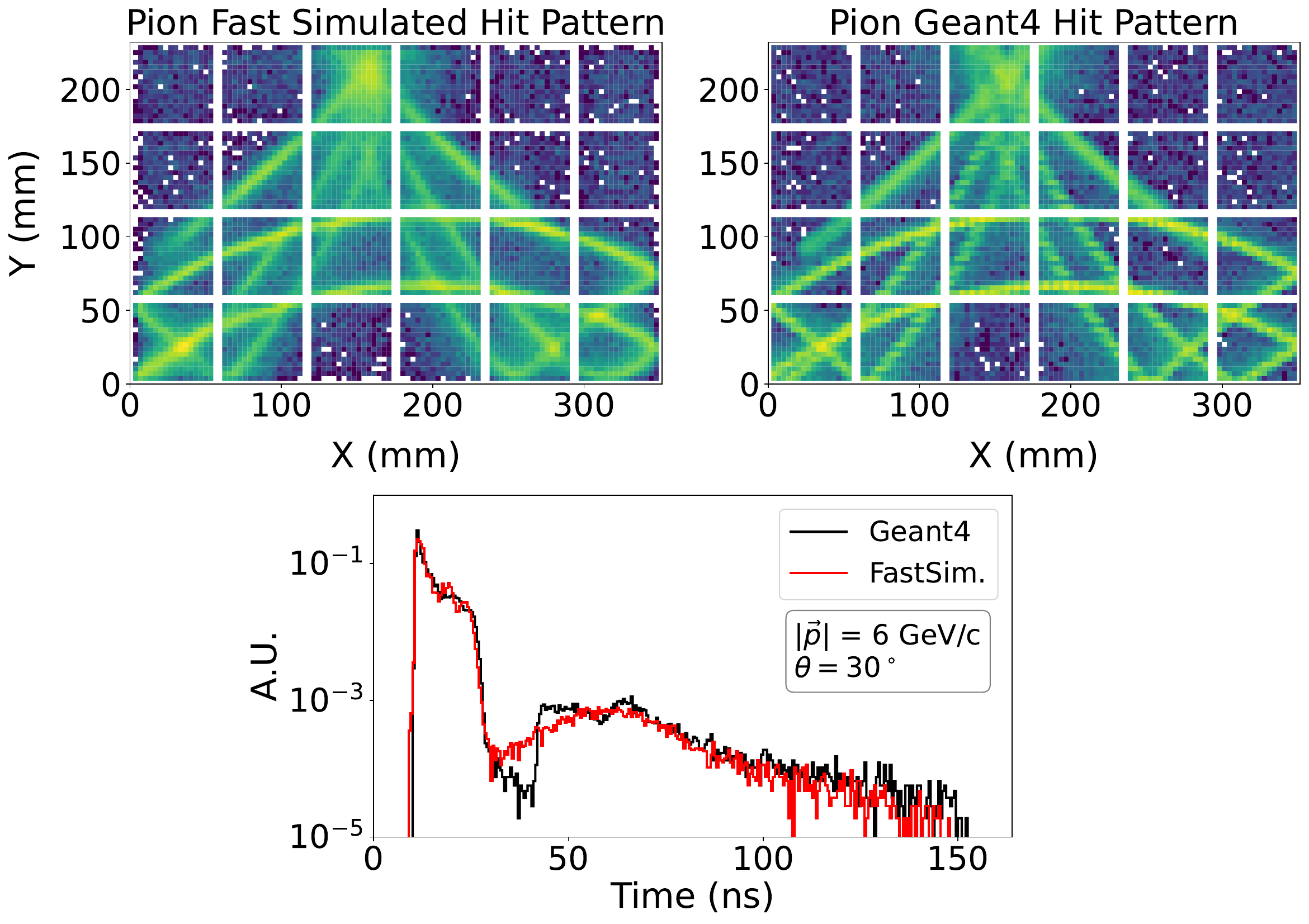} \\
    \includegraphics[width=0.49\textwidth]{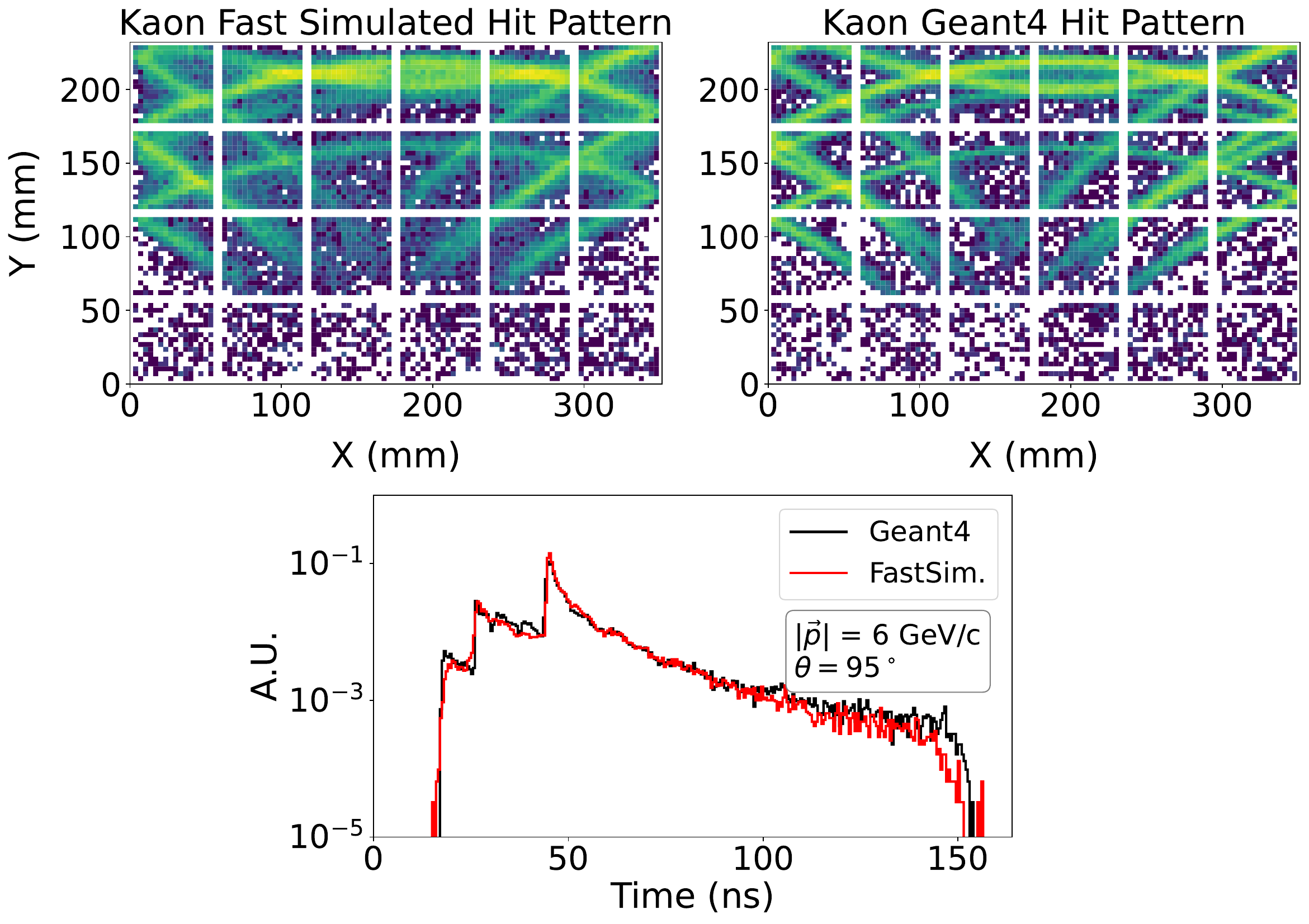} %
    \includegraphics[width=0.49\textwidth]{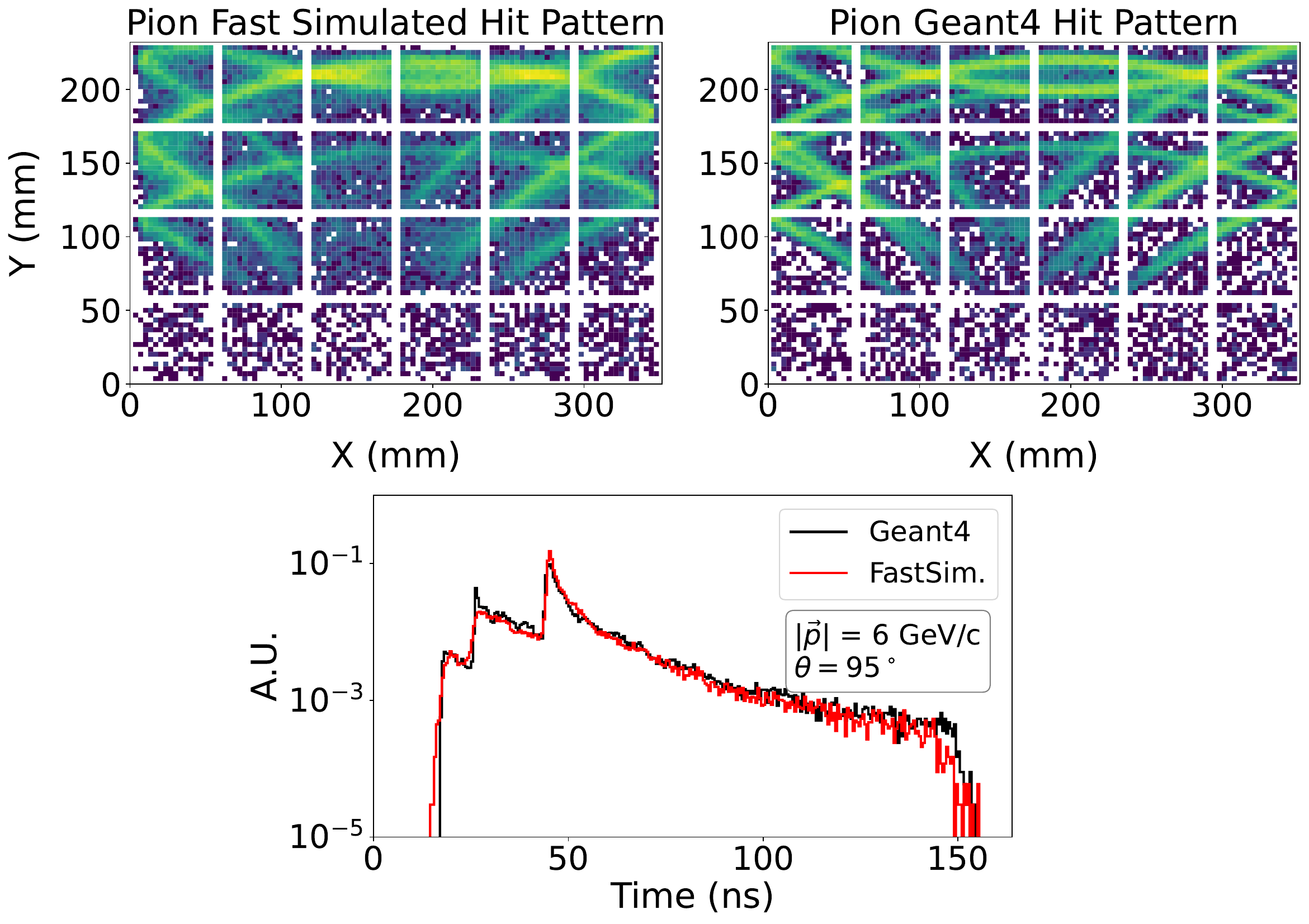} \\
    \includegraphics[width=0.49\textwidth]{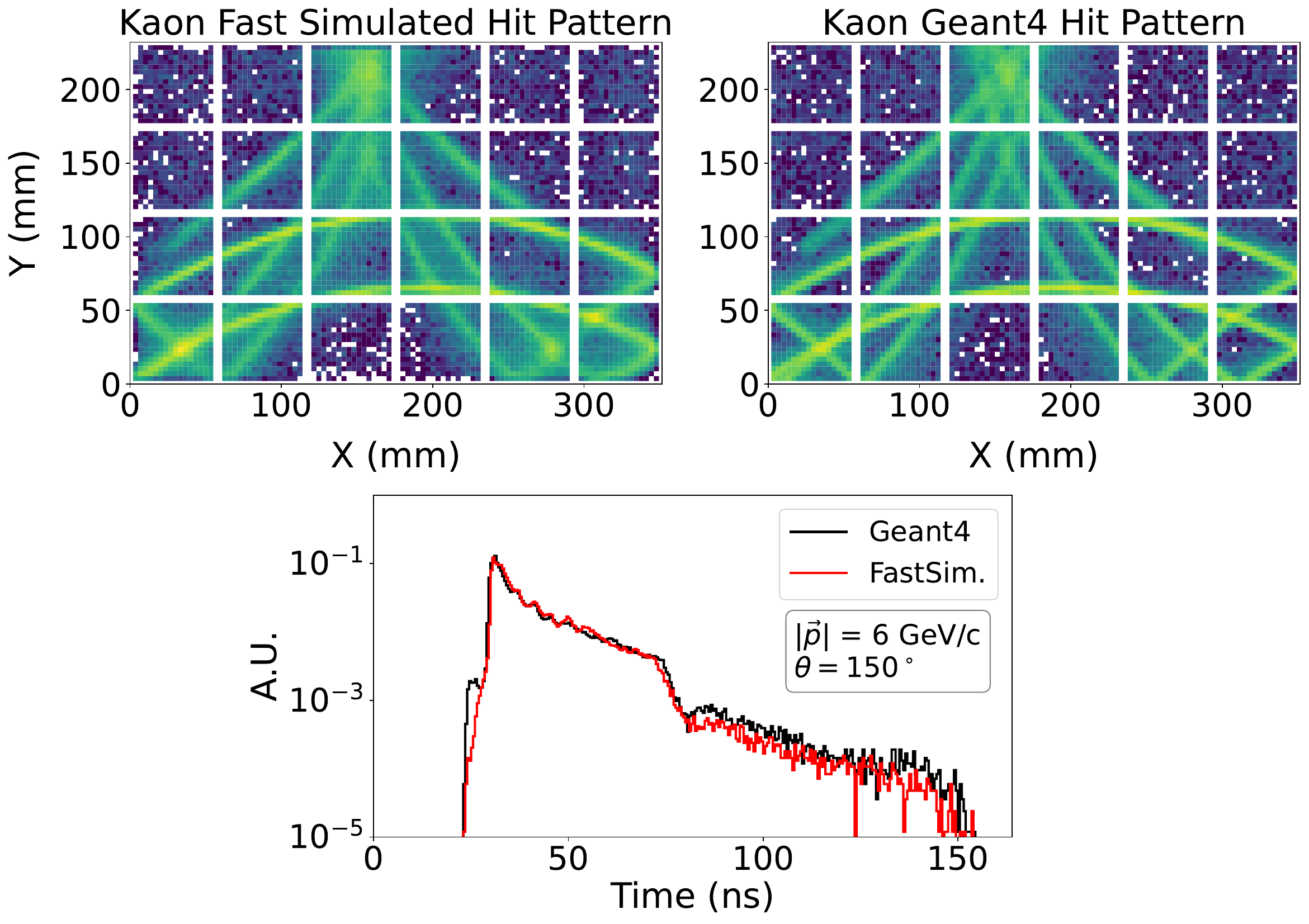} %
    \includegraphics[width=0.49\textwidth]{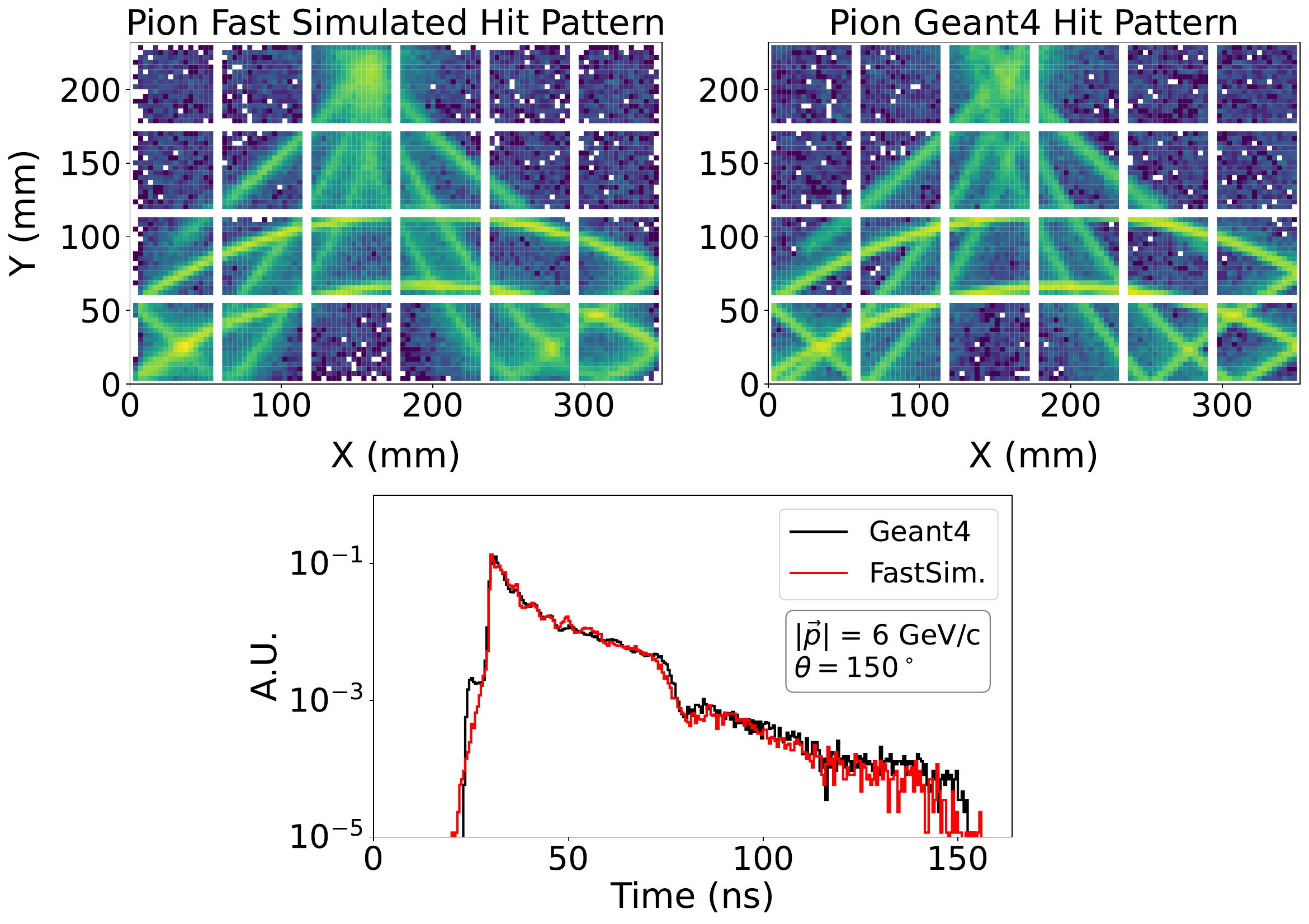} %
    \caption{
    \textbf{Fast Simulation with DDPM:} Fast Simulation of Kaons (left column of plots), and Pions (right column of plots) at 6 GeV/c and various polar angles using DDPM.}
    \label{fig:DDPM_Generations_6GeV}
\end{figure}

%%%%%%%%%%%%%%%%%%% Score Based %%%%%%%%%%%%%%%%%%%%%%%%%%%%%%%%%%%%%
\begin{figure}[h]
    \centering
    \includegraphics[width=0.49\textwidth]{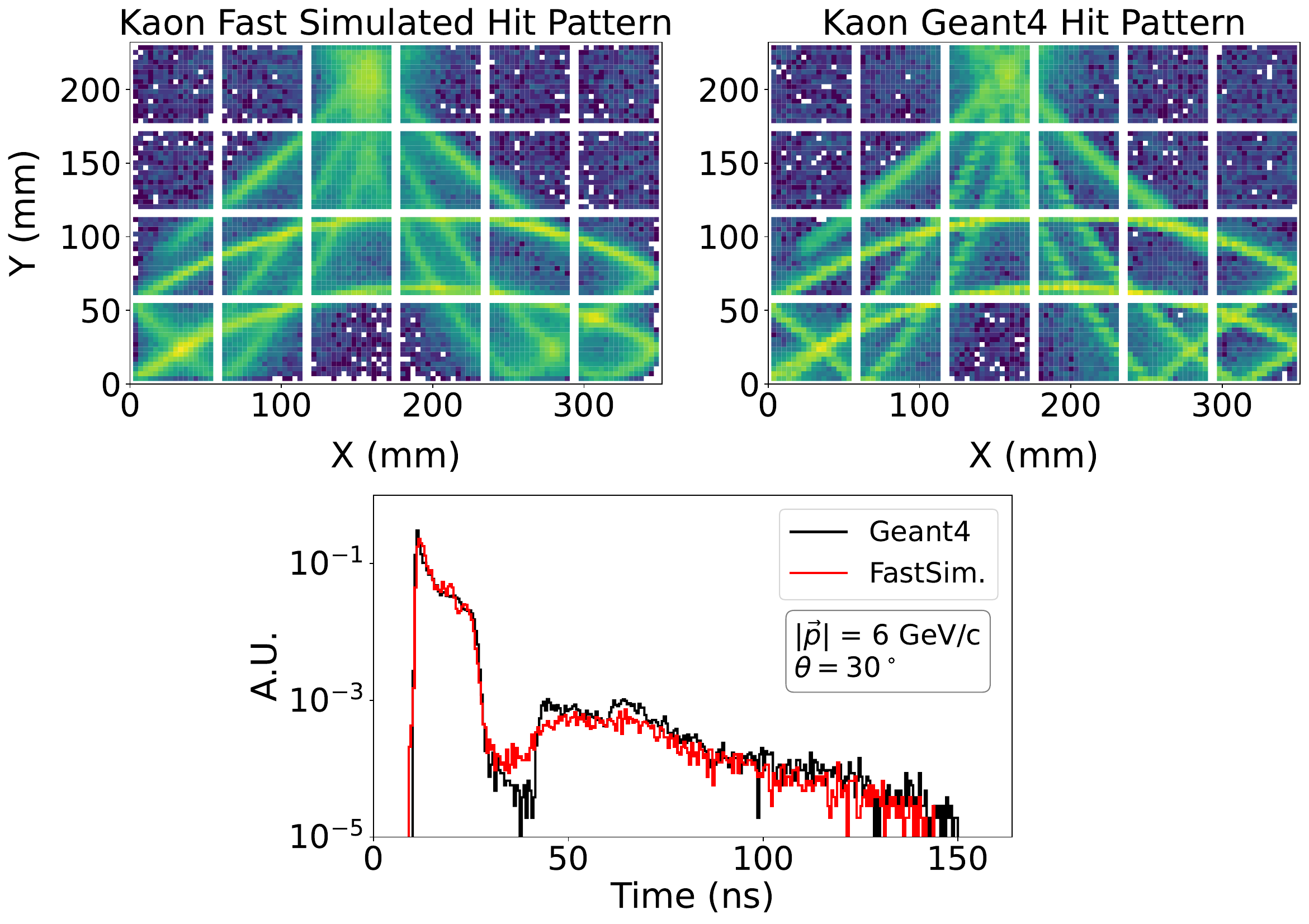}% 
   \includegraphics[width=0.49\textwidth]{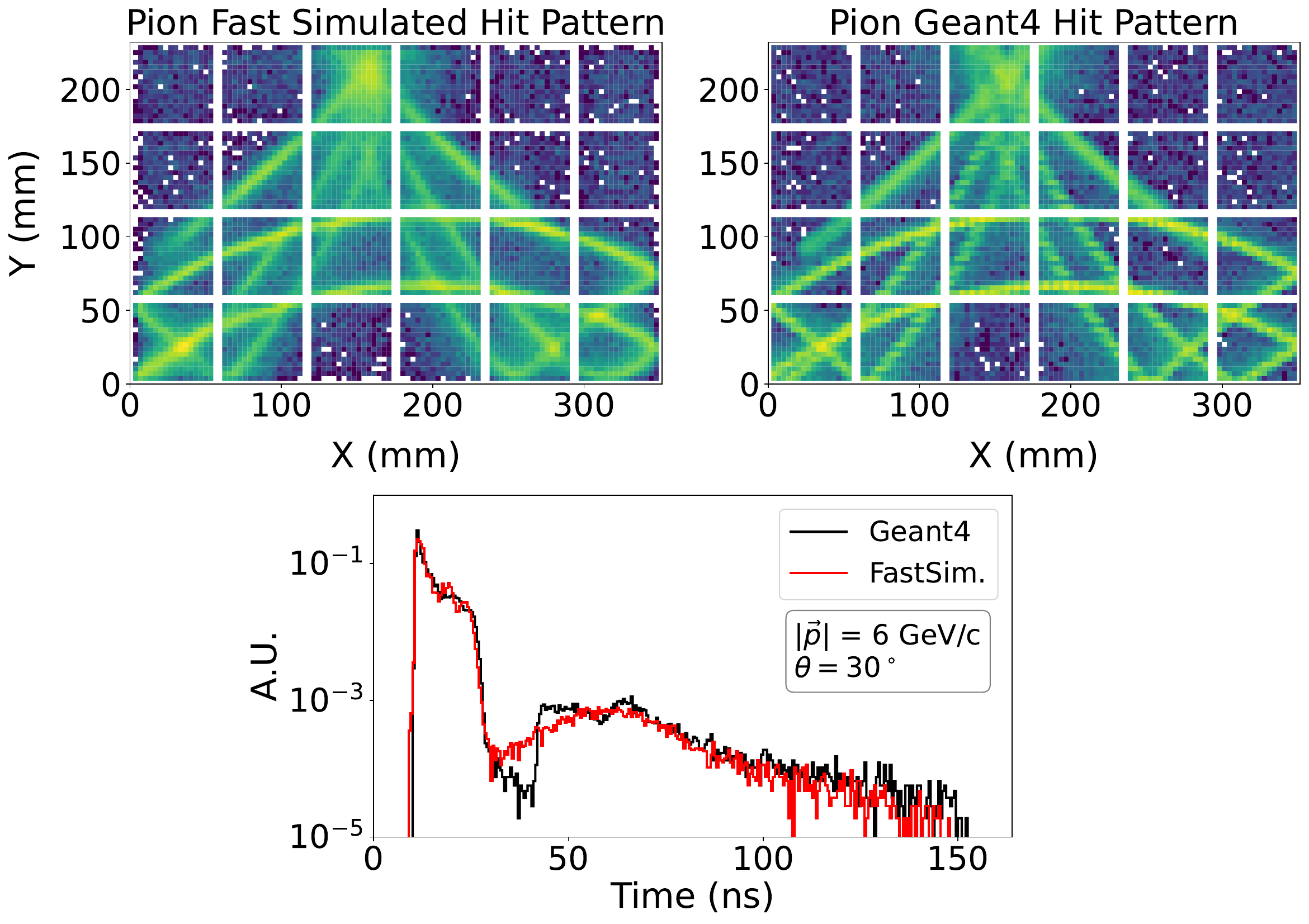} \\
    \includegraphics[width=0.49\textwidth]{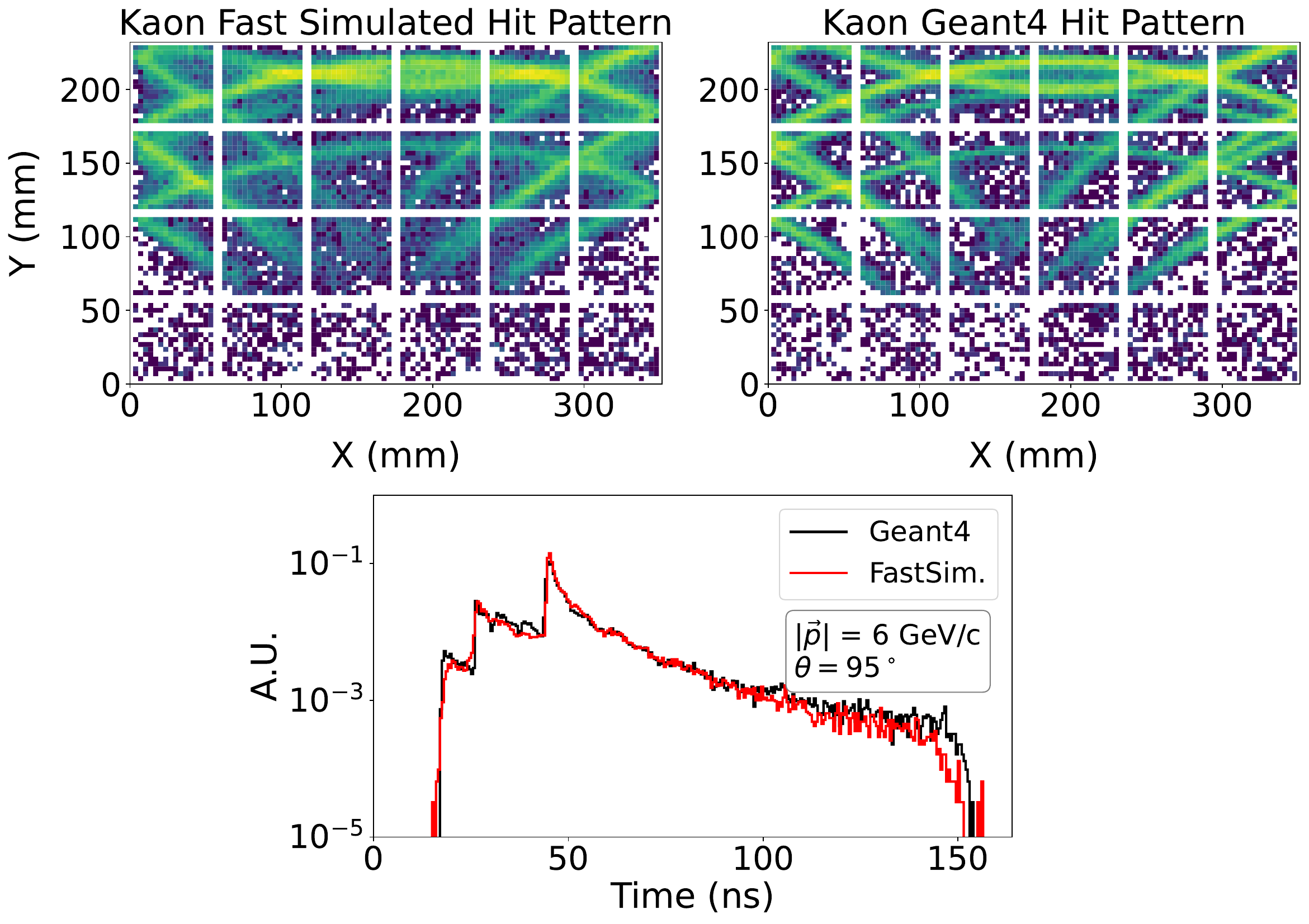} %
    \includegraphics[width=0.49\textwidth]{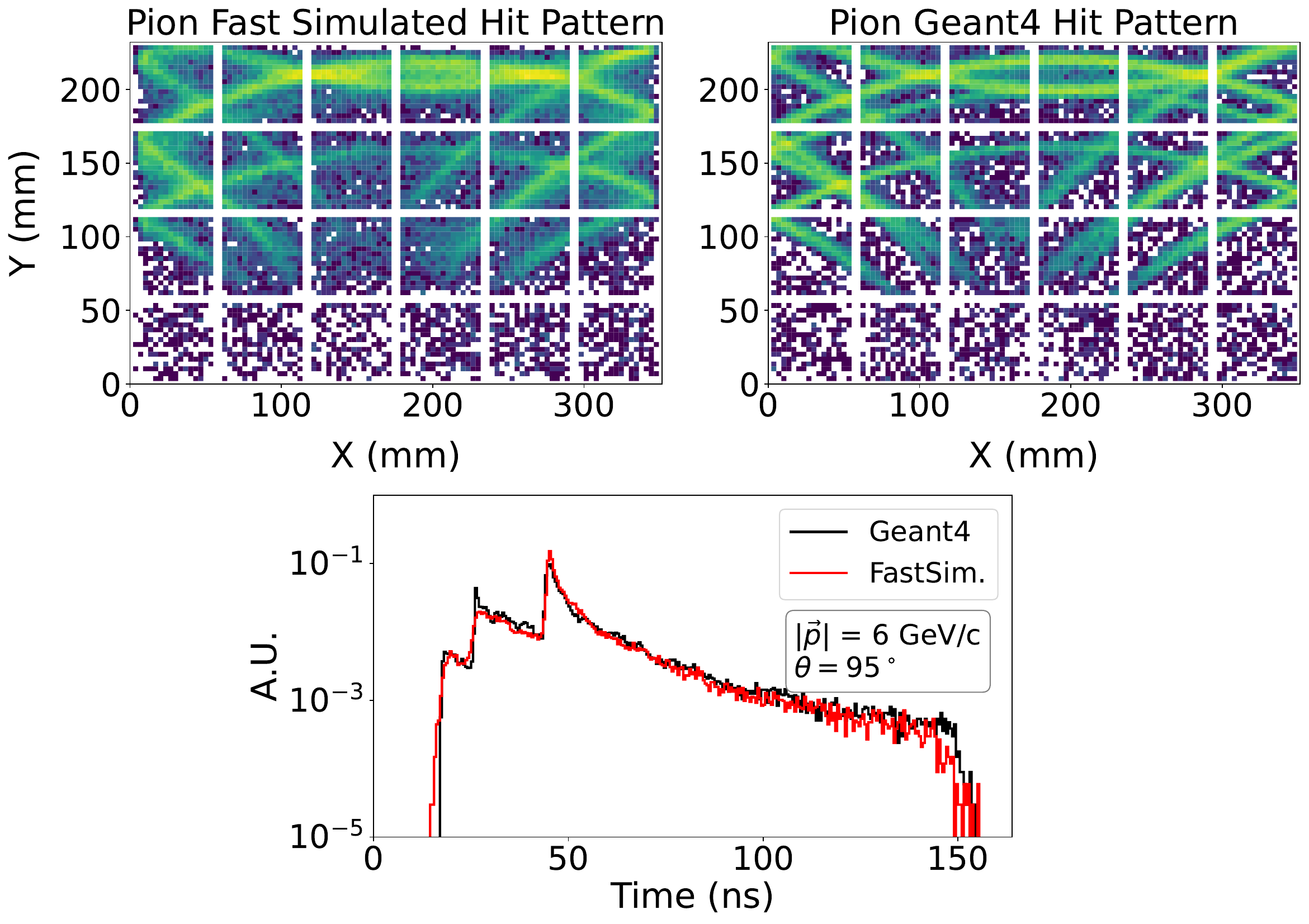} \\
    \includegraphics[width=0.49\textwidth]{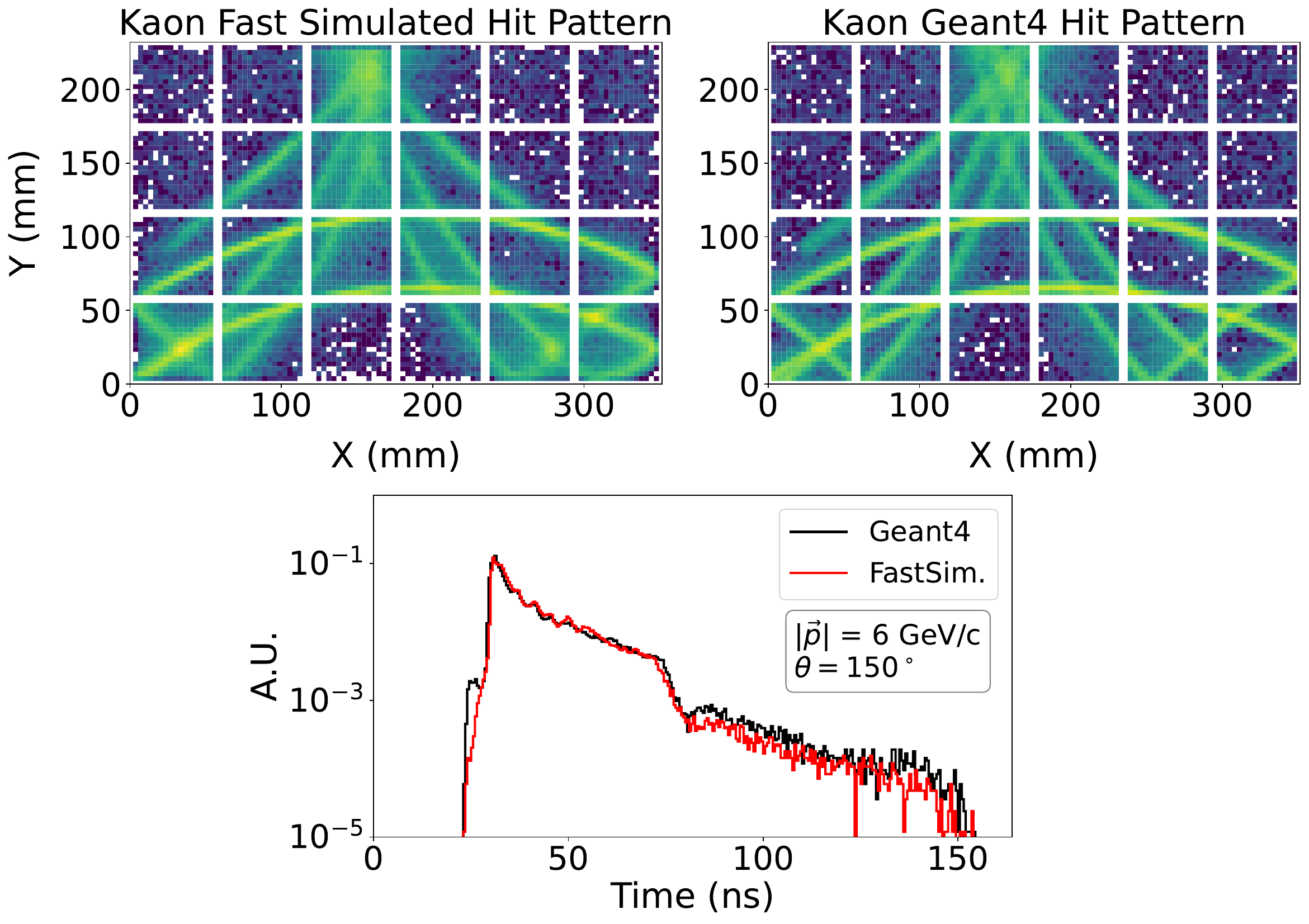} %
    \includegraphics[width=0.49\textwidth]{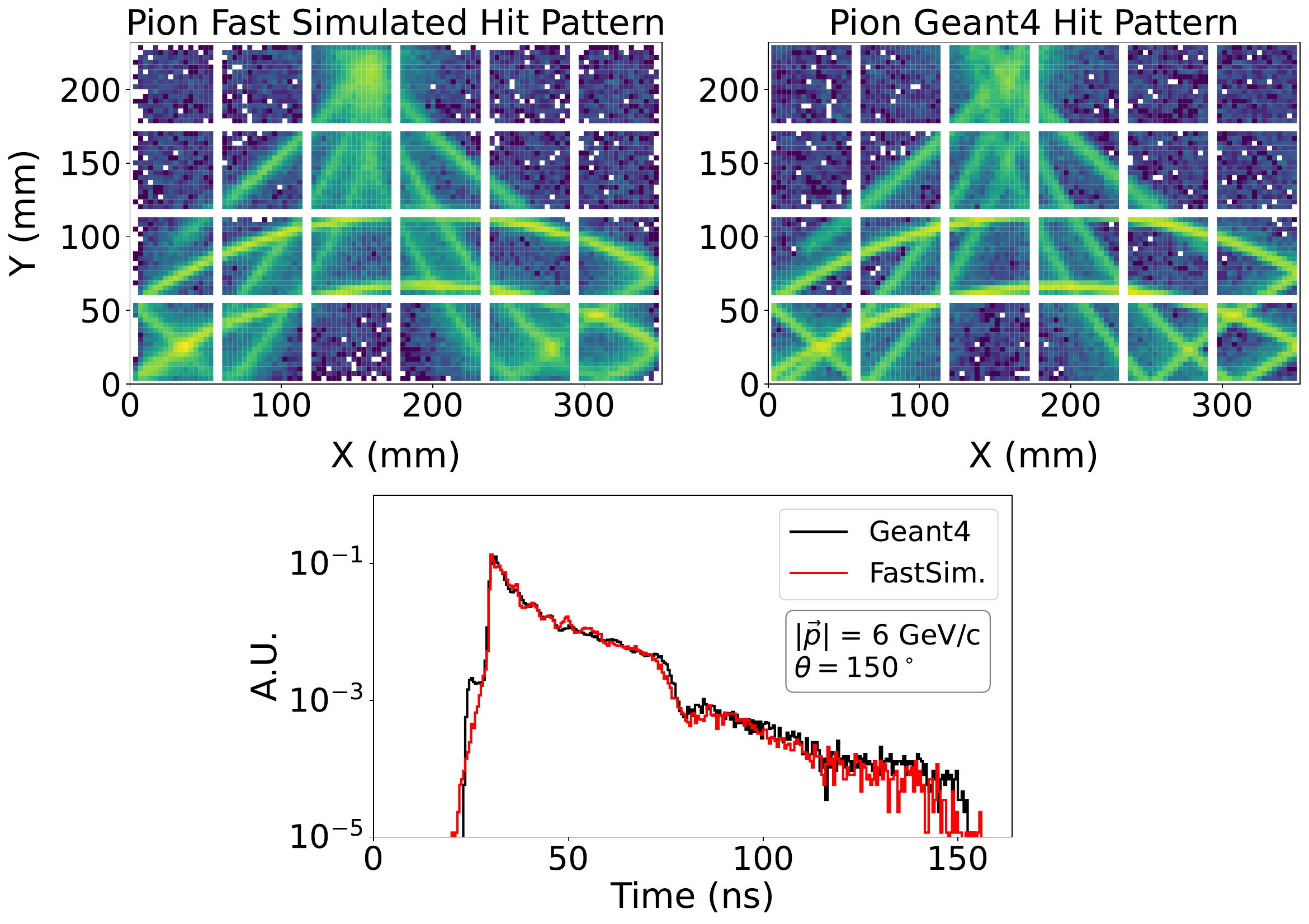} %
    \caption{
    \textbf{Fast Simulation with Score-Based Models:} Fast Simulation of Kaons (left column of plots), and Pions (right column of plots) at 6 GeV/c and various polar angles using Score-Based models.}
    \label{fig:score_Generations_6GeV}
\end{figure}

%%%%%%%%%%%%%% 9 GeV %%%%%%%%%%%%%%%%%%%%%
\clearpage
\section{Evaluation at $\SI[per-mode=symbol]{9}{\giga\eVperc}$} \label{app:9GeV}

%%%%%%%%%%%%%%%%%%%%%%% Cherenkov Ring Plots %%%%%%%%%%%%%%%%%%%%%%%%%%%%%%%
\begin{figure}[h]
    \centering
    \includegraphics[width=0.49\textwidth]{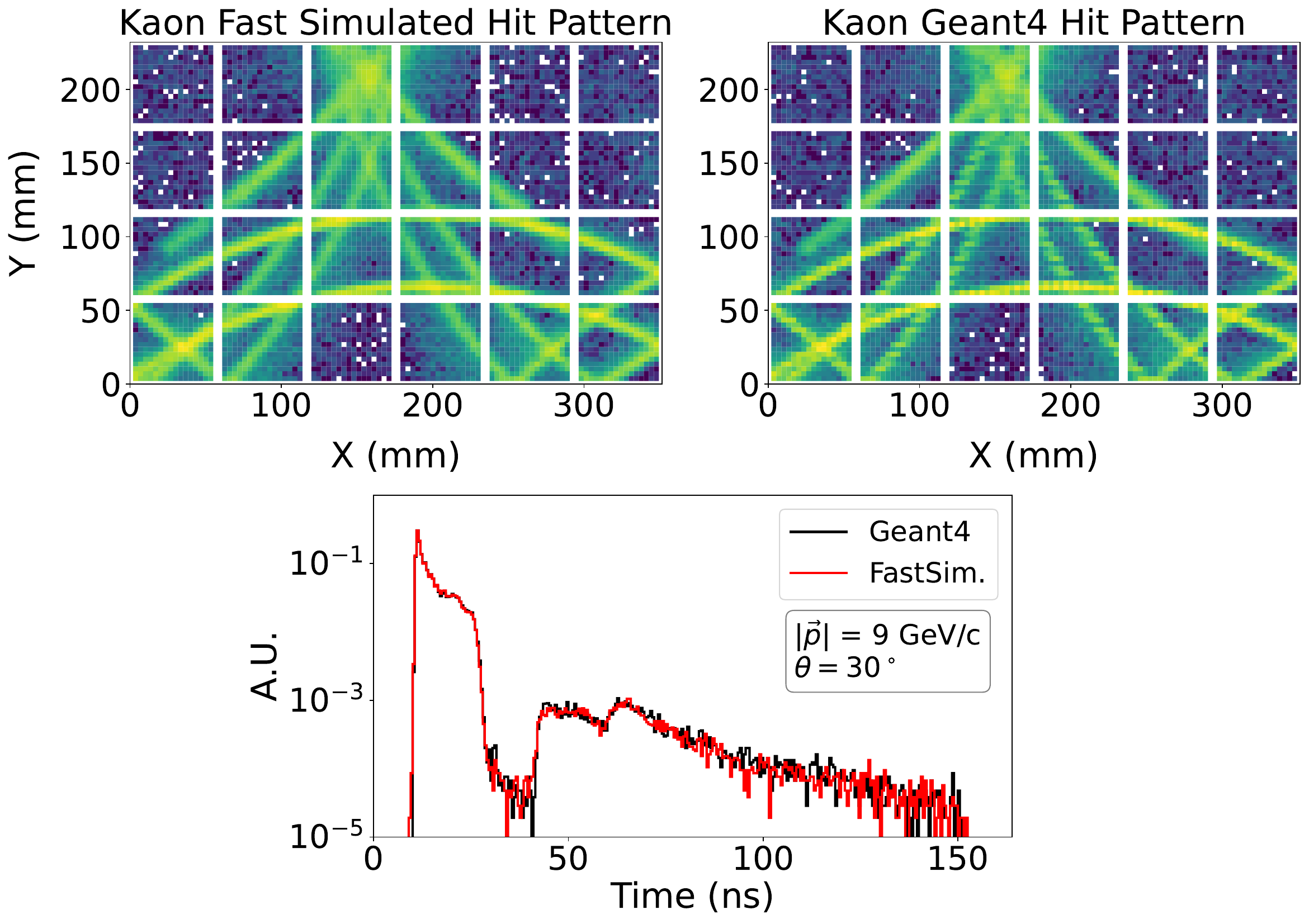}% 
   \includegraphics[width=0.49\textwidth]{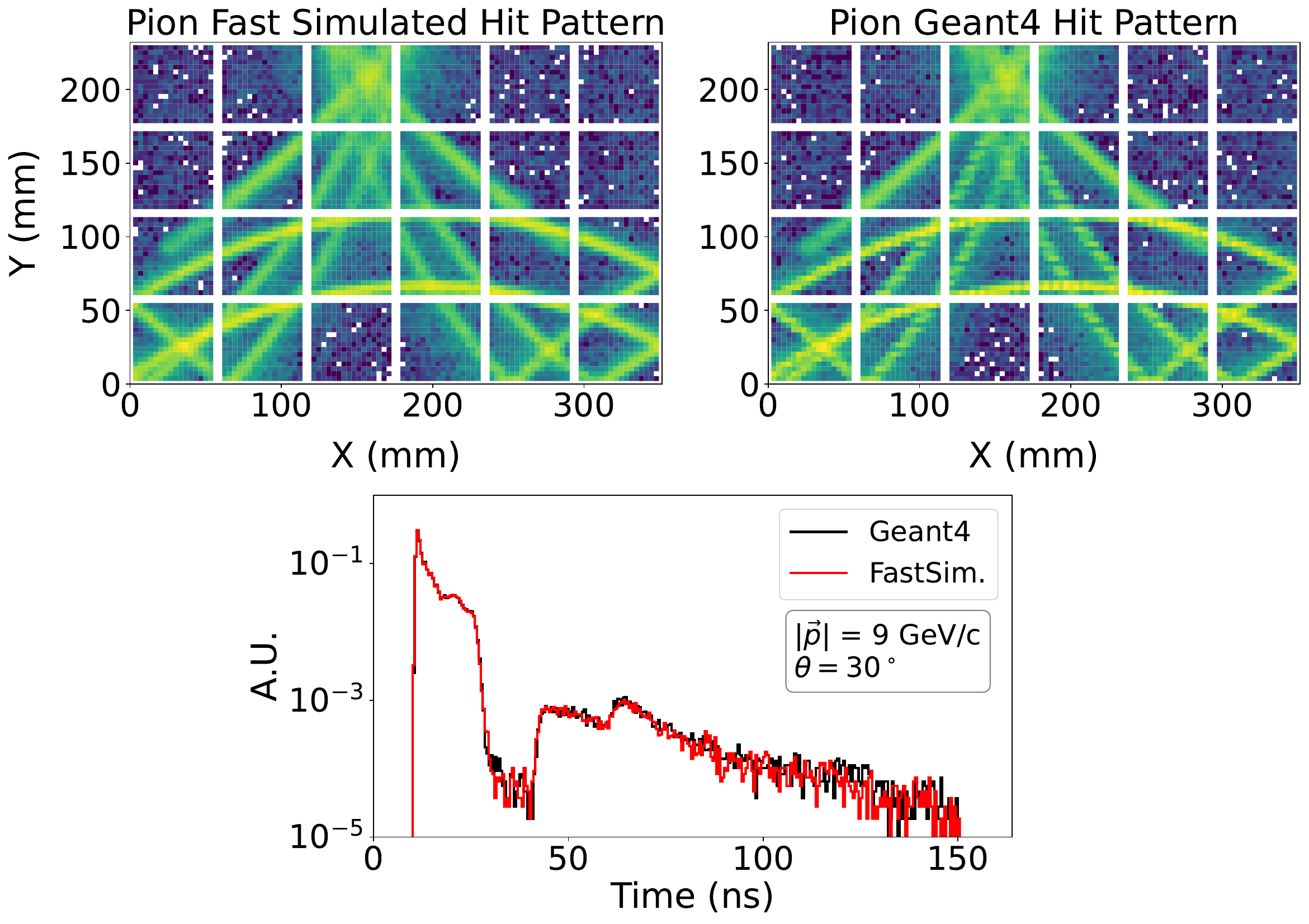} \\
    \includegraphics[width=0.49\textwidth]{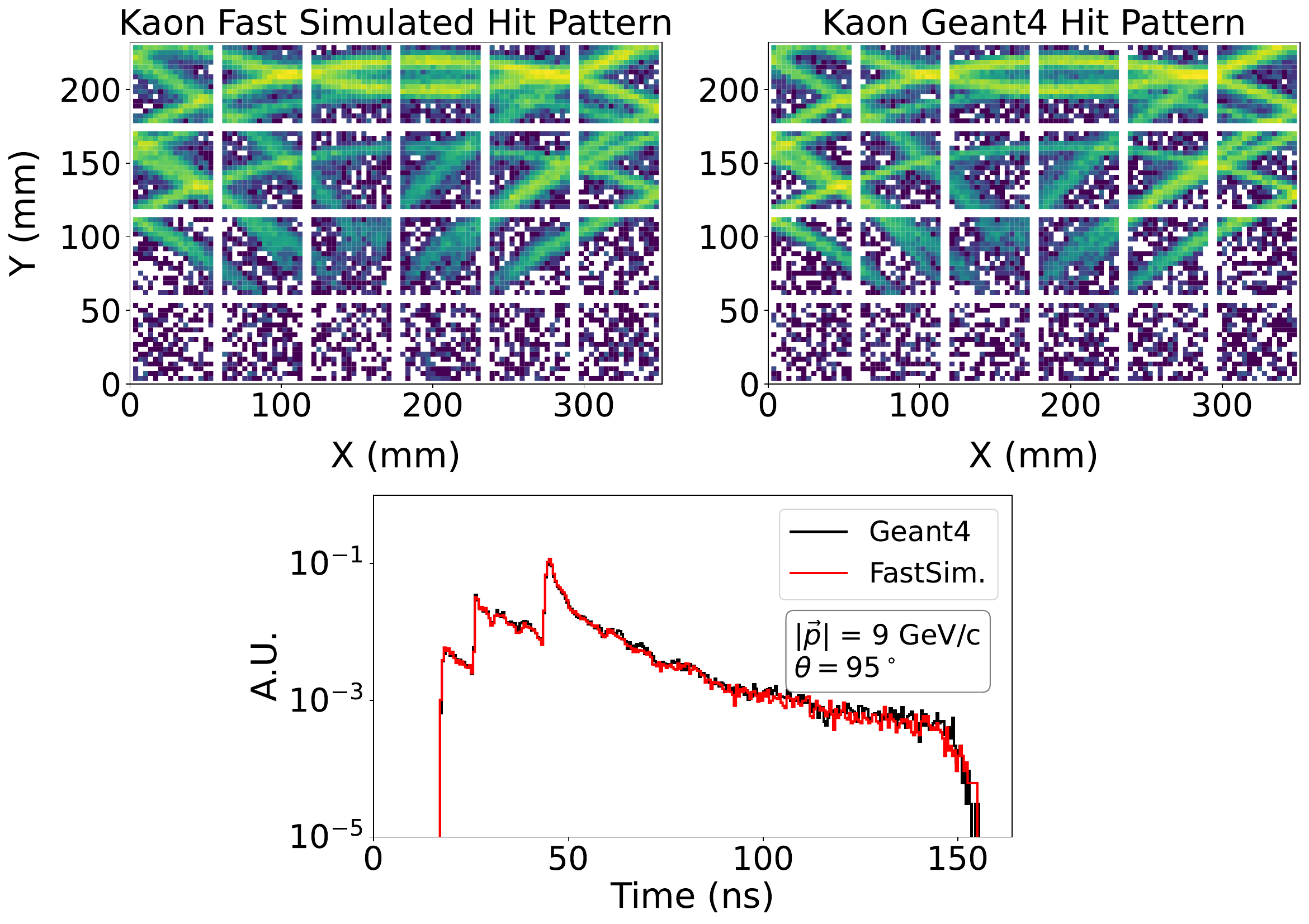} %
    \includegraphics[width=0.49\textwidth]{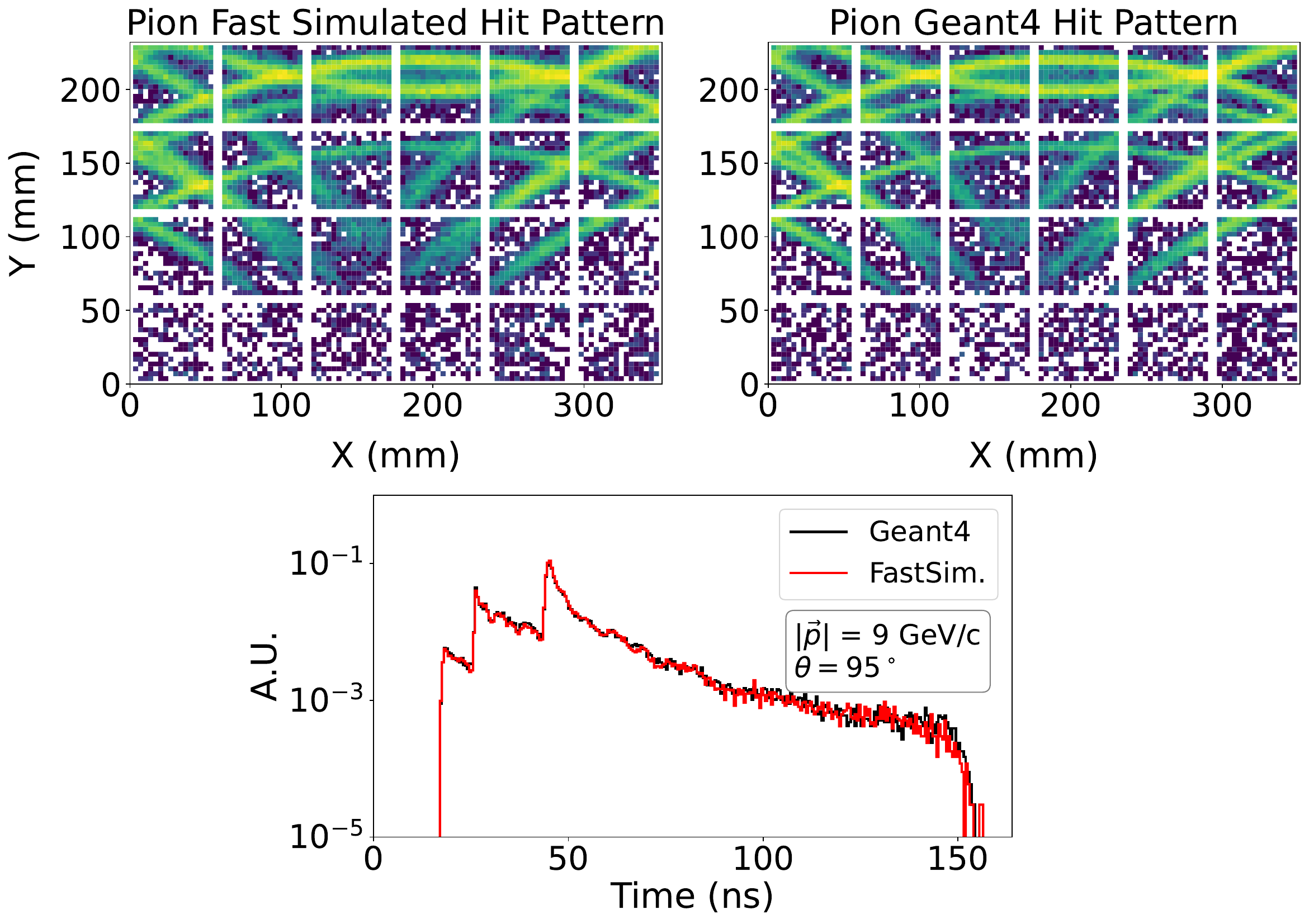} \\
    \includegraphics[width=0.49\textwidth]{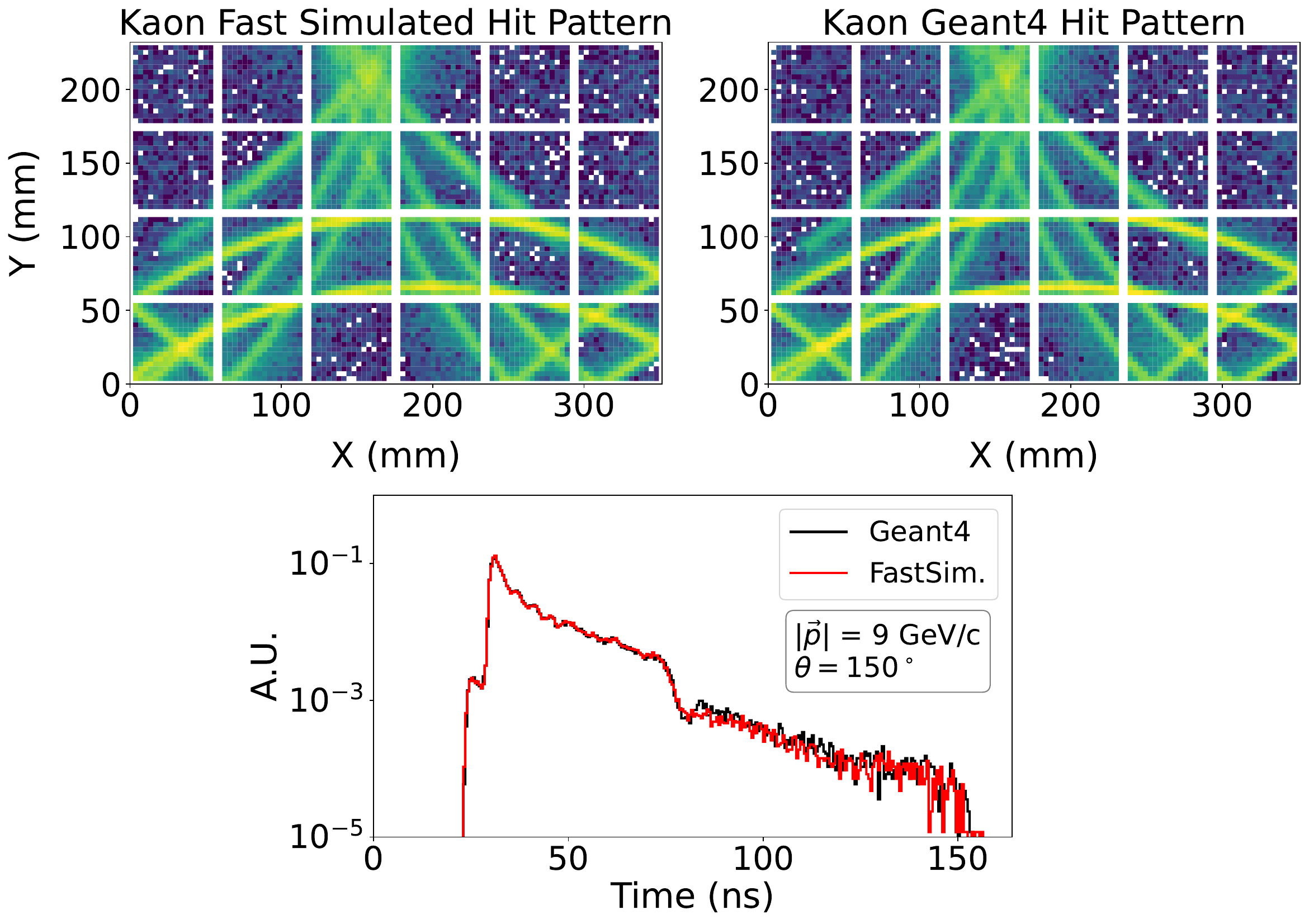} %
    \includegraphics[width=0.49\textwidth]{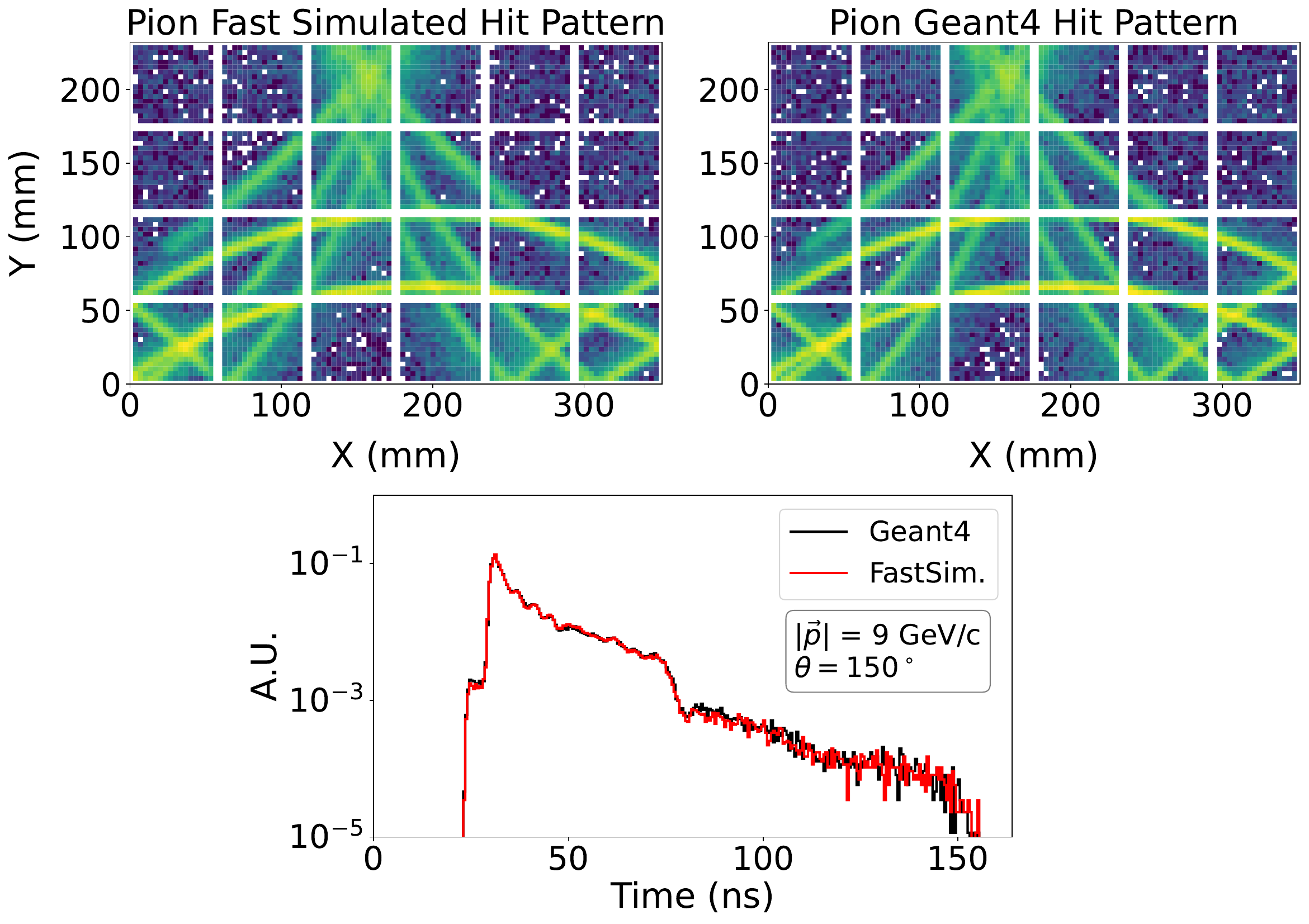} %
    \caption{
    \textbf{Fast Simulation with Discrete Normalizing Flows:} Fast Simulation of Kaons (left column of plots), and Pions (right column of plots) at 9 GeV/c and various polar angles.}
    \label{fig:DNF_Generations_9GeV}
\end{figure}

\begin{figure}[h]
    \centering
    \includegraphics[width=0.49\textwidth]{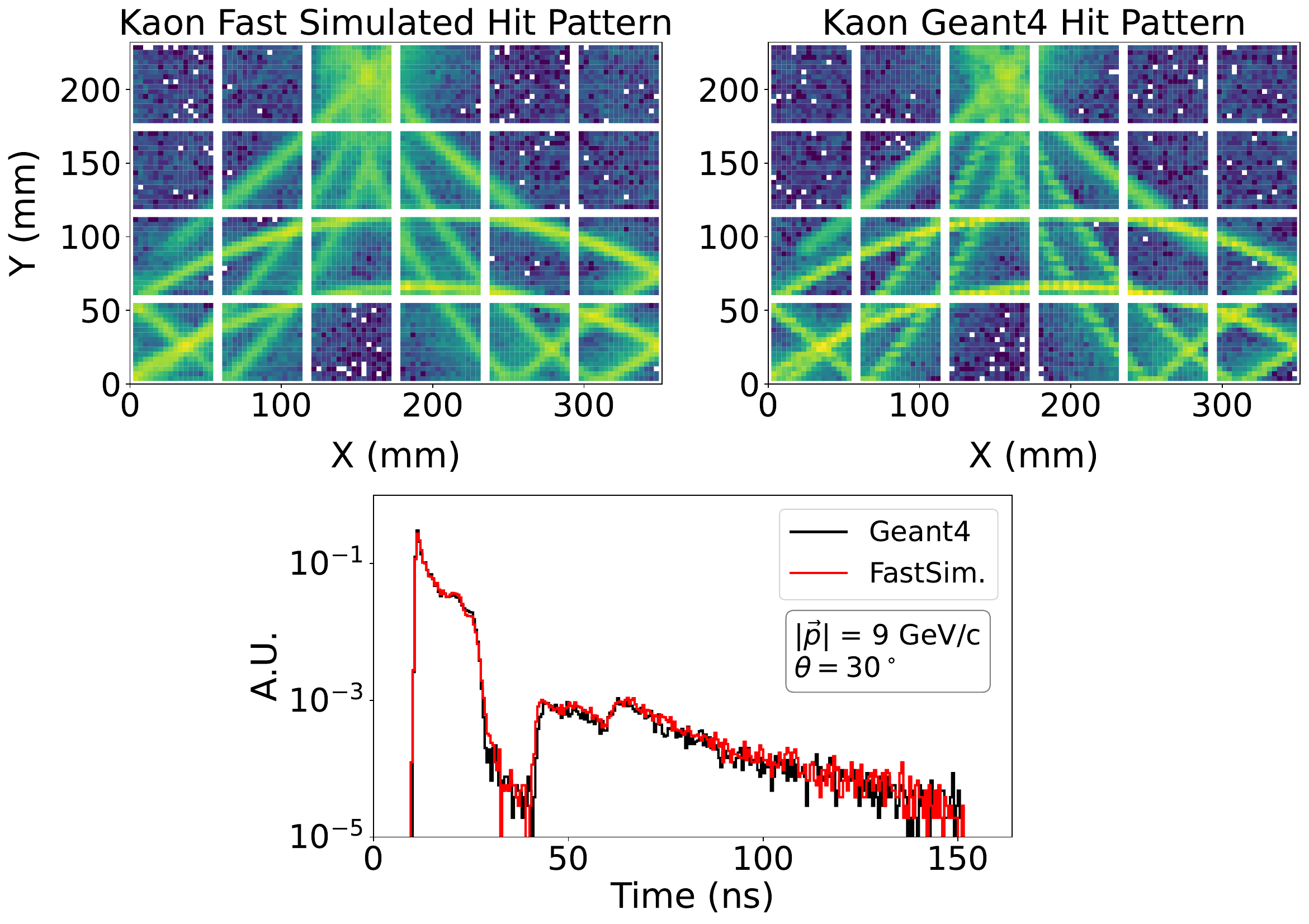}% 
   \includegraphics[width=0.49\textwidth]{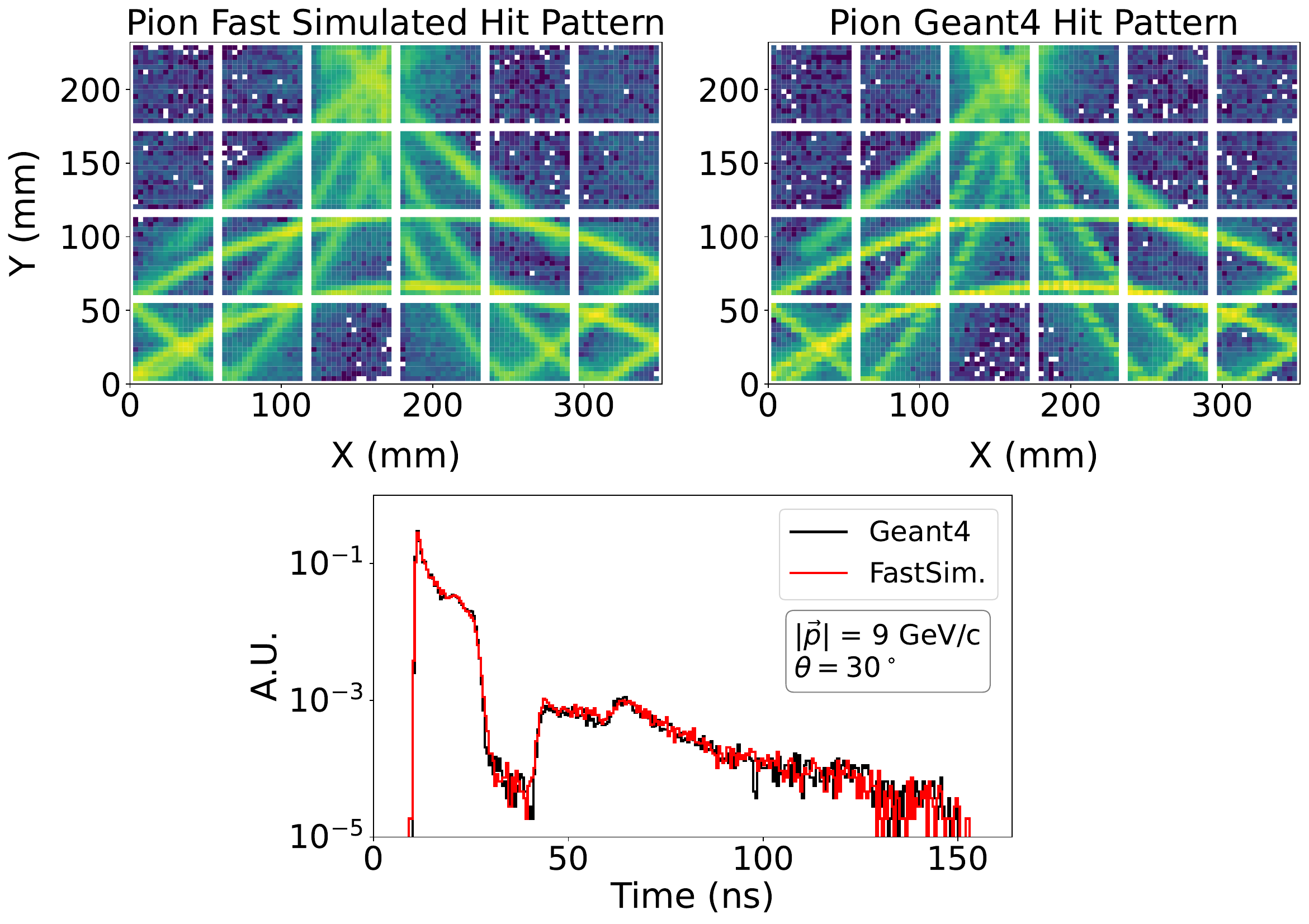} \\
    \includegraphics[width=0.49\textwidth]{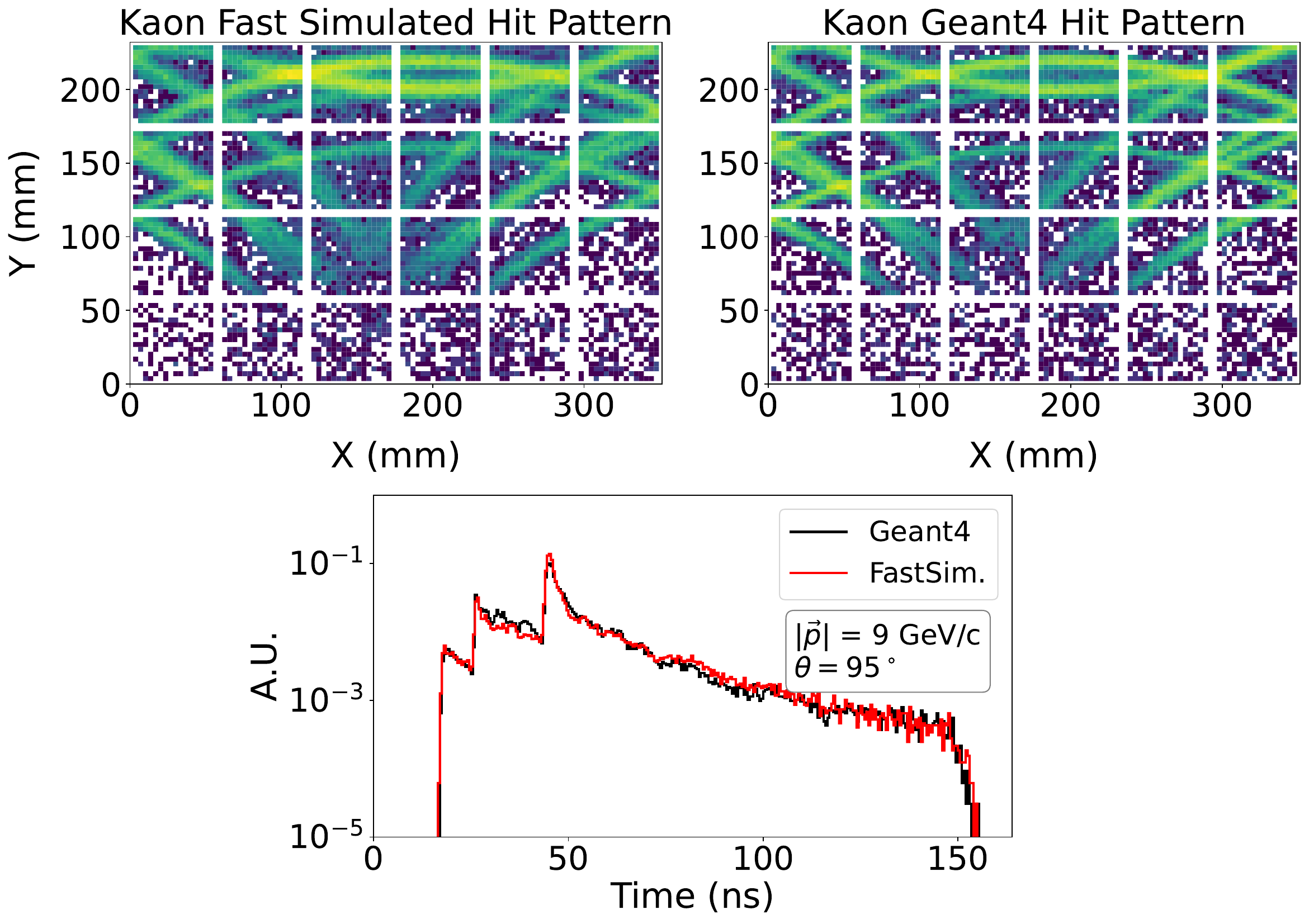} %
    \includegraphics[width=0.49\textwidth]{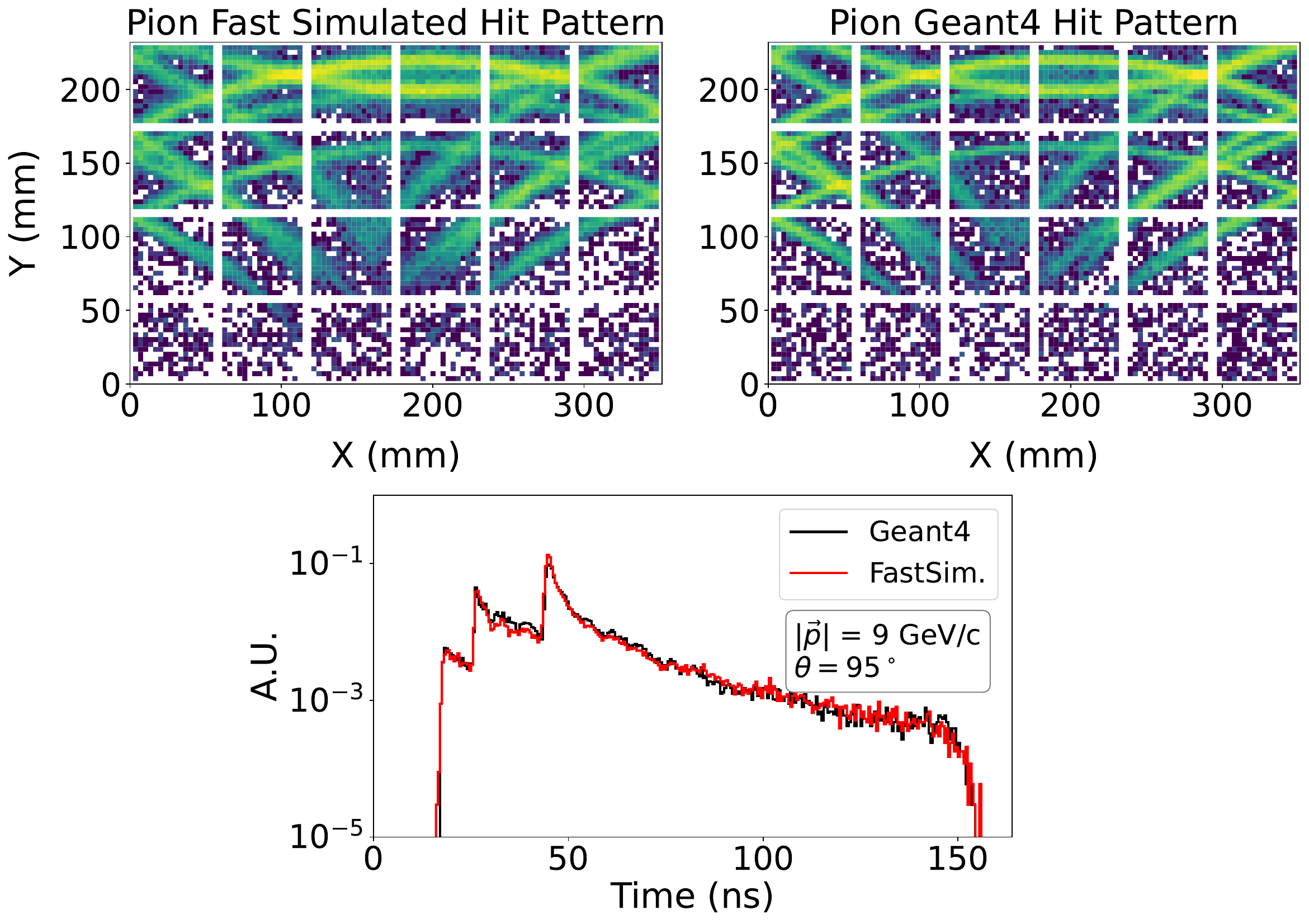} \\
    \includegraphics[width=0.49\textwidth]{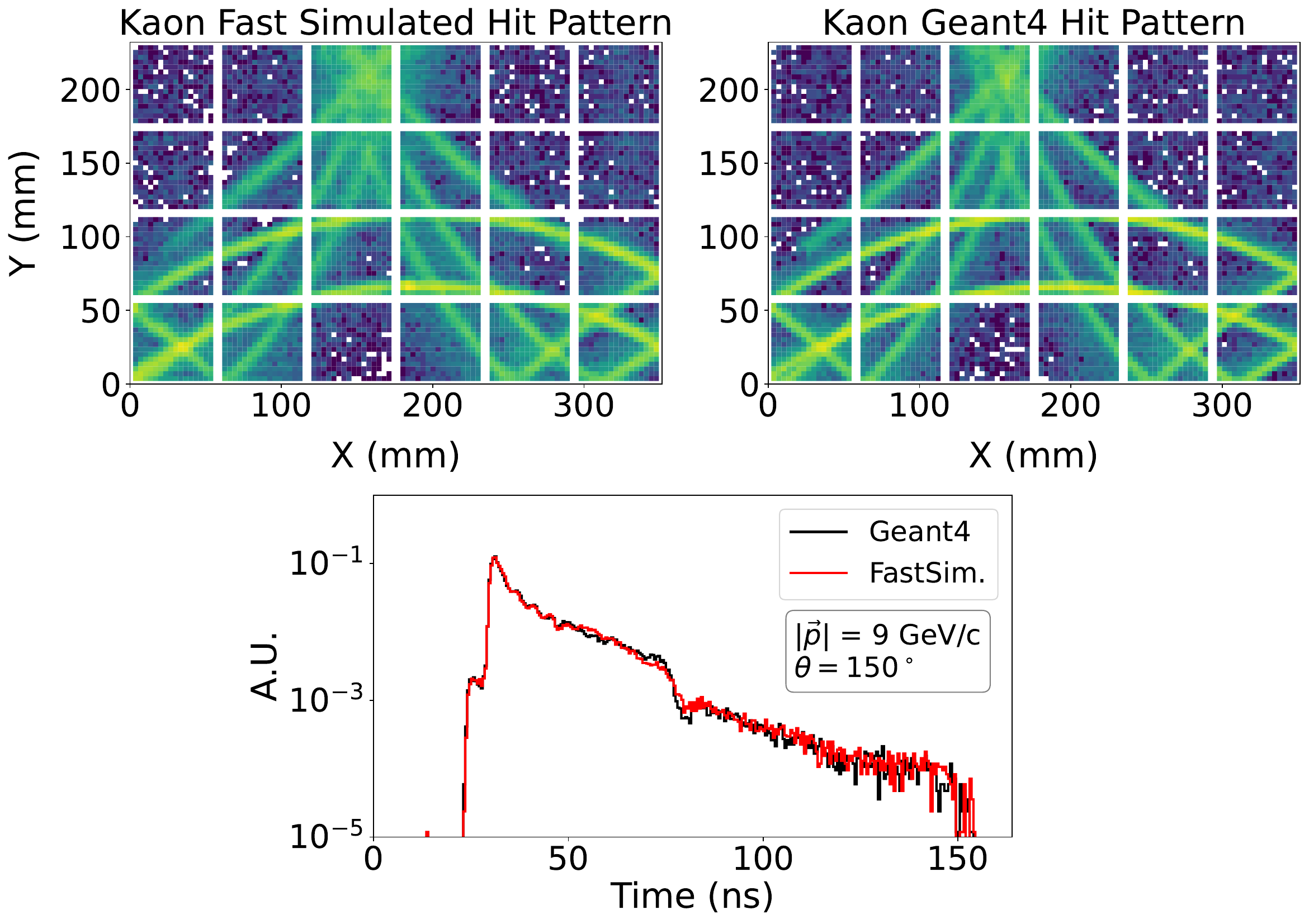} %
    \includegraphics[width=0.49\textwidth]{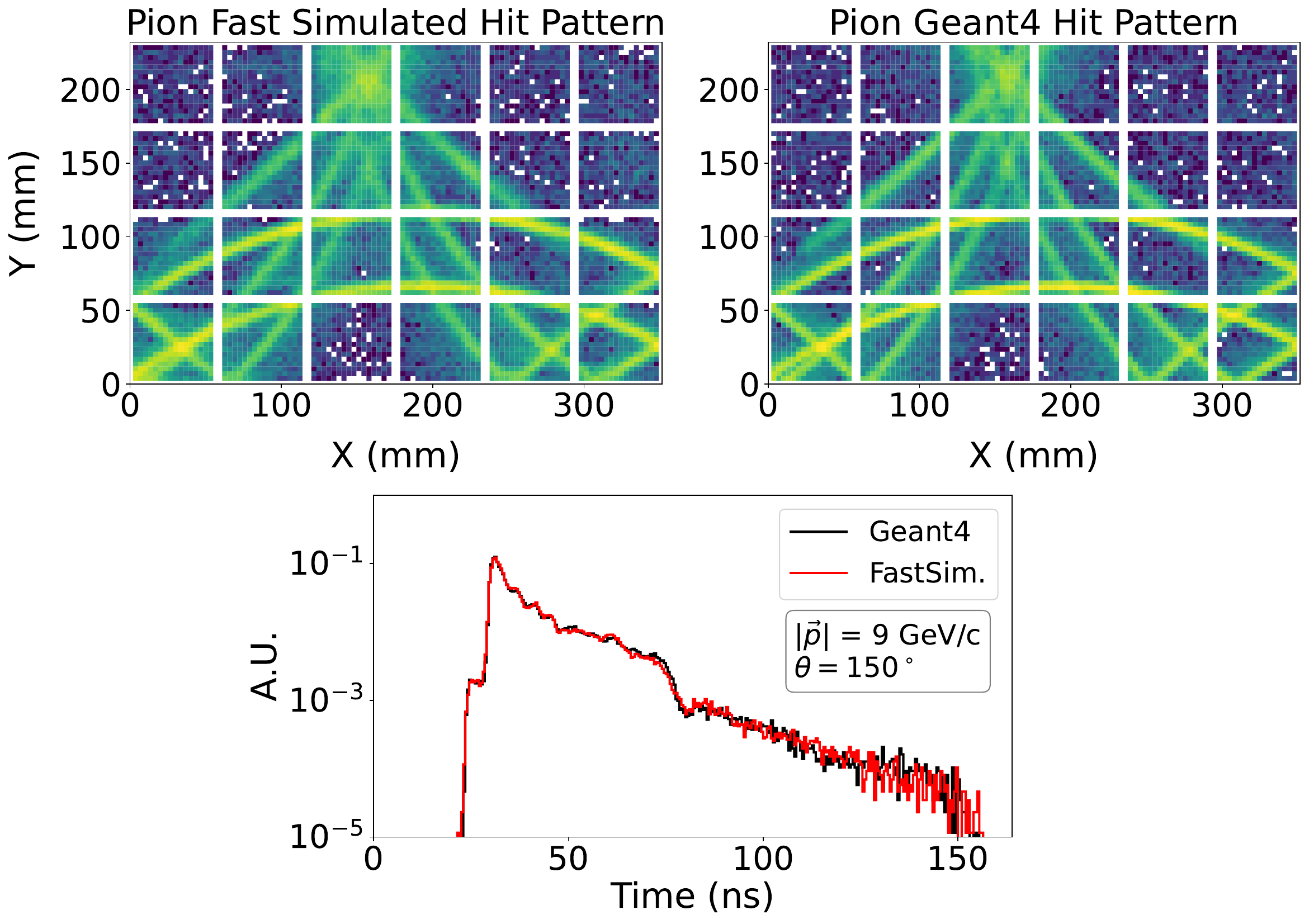} %
    \caption{
    \textbf{Fast Simulation with Continuous Normalizing Flows:} Fast Simulation of Kaons (left column of plots), and Pions (right column of plots) at 9 GeV/c and various polar angles. 20 integration steps with midpoint solver.}
    \label{fig:CNF_Generations_9GeV}
\end{figure}

\begin{figure}[h]
    \centering
    \includegraphics[width=0.49\textwidth]{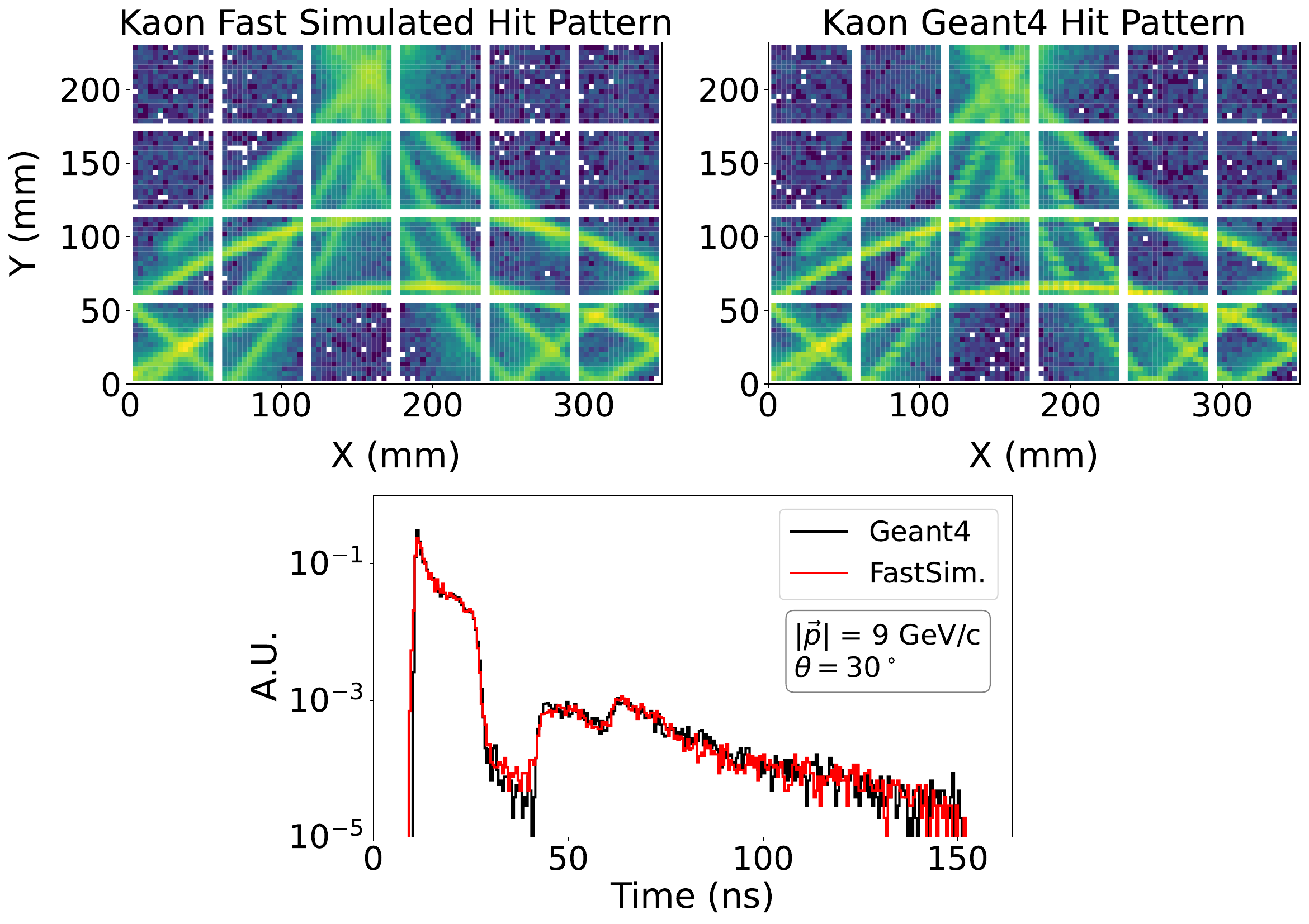}% 
   \includegraphics[width=0.49\textwidth]{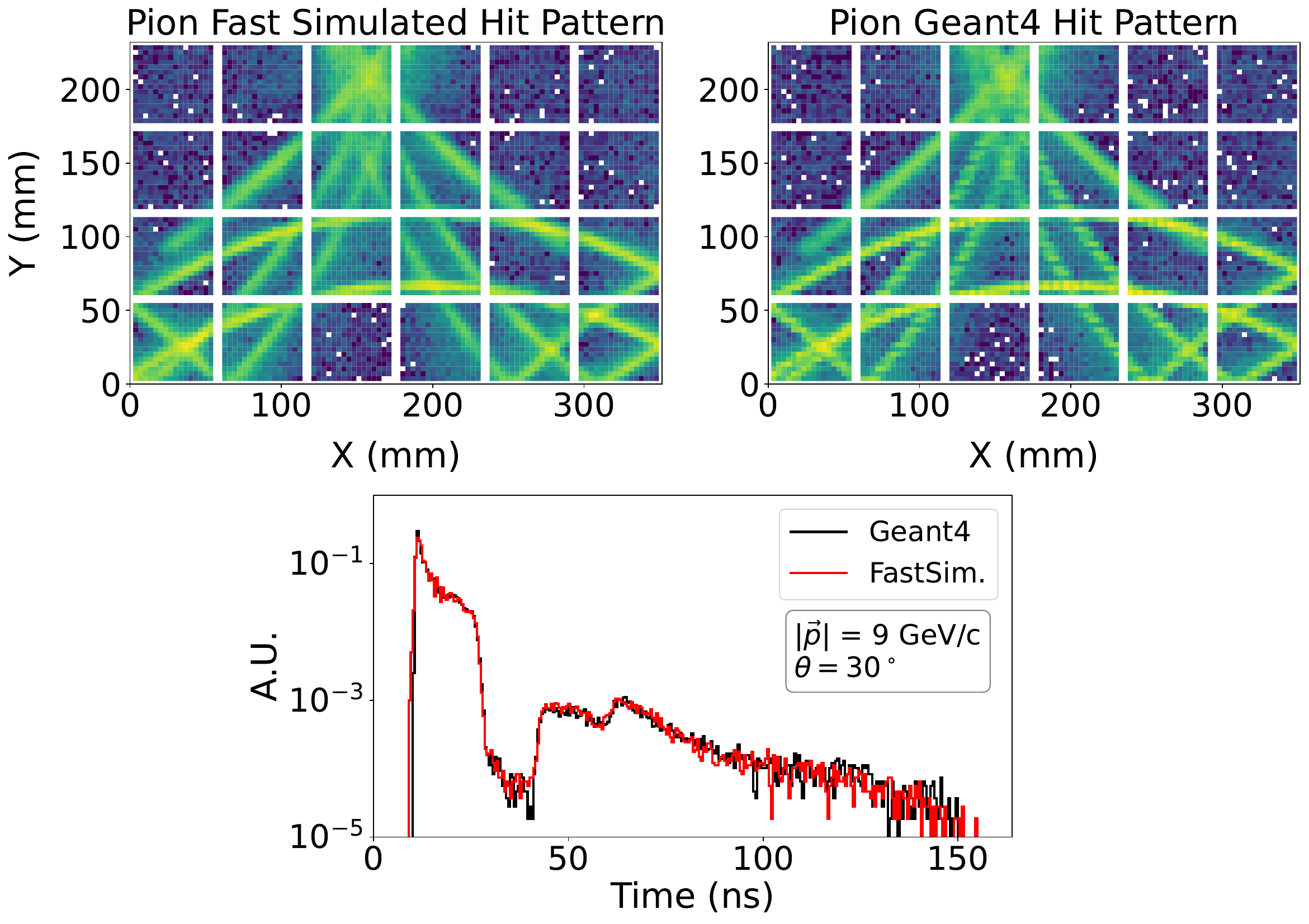} \\
    \includegraphics[width=0.49\textwidth]{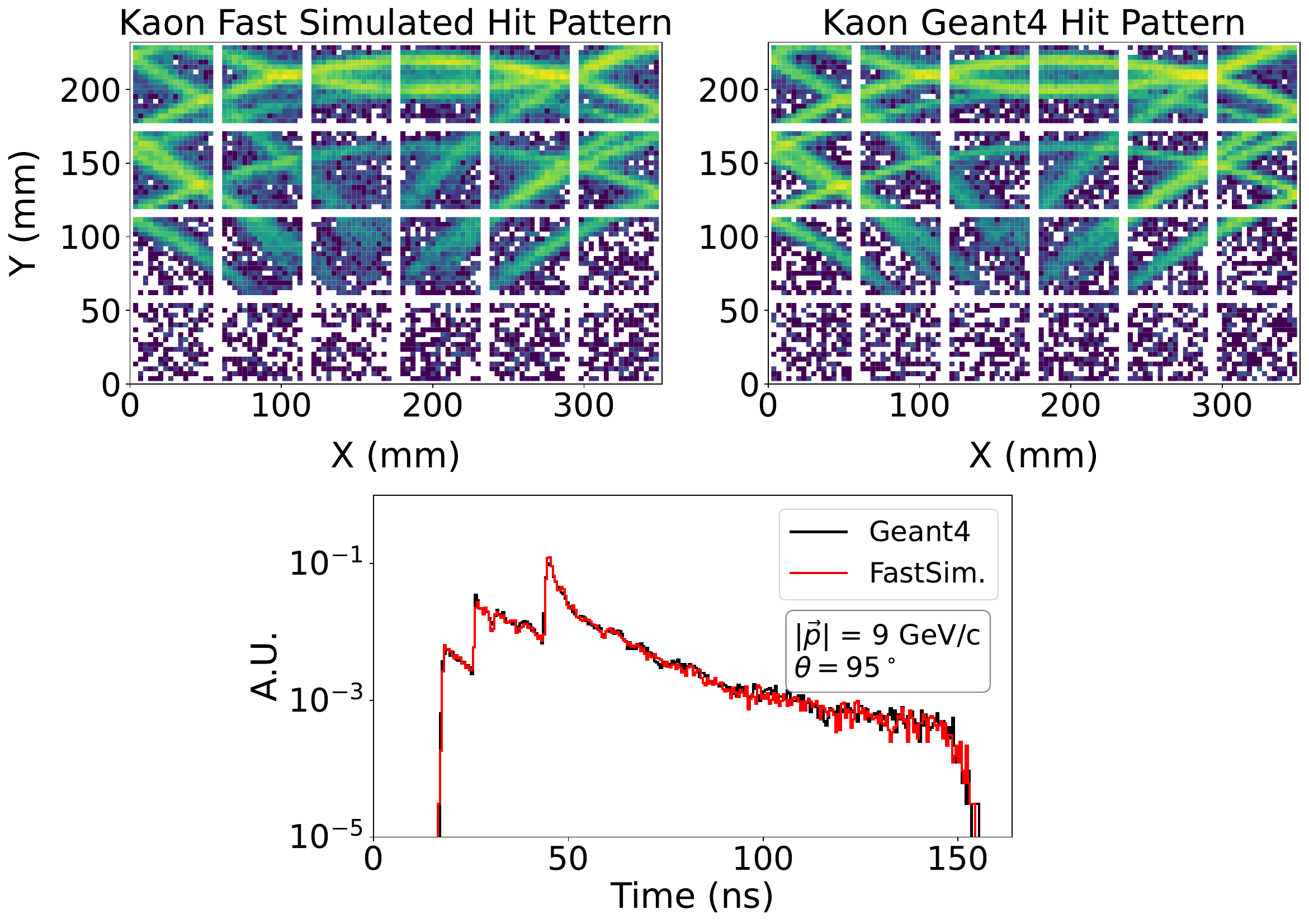} %
    \includegraphics[width=0.49\textwidth]{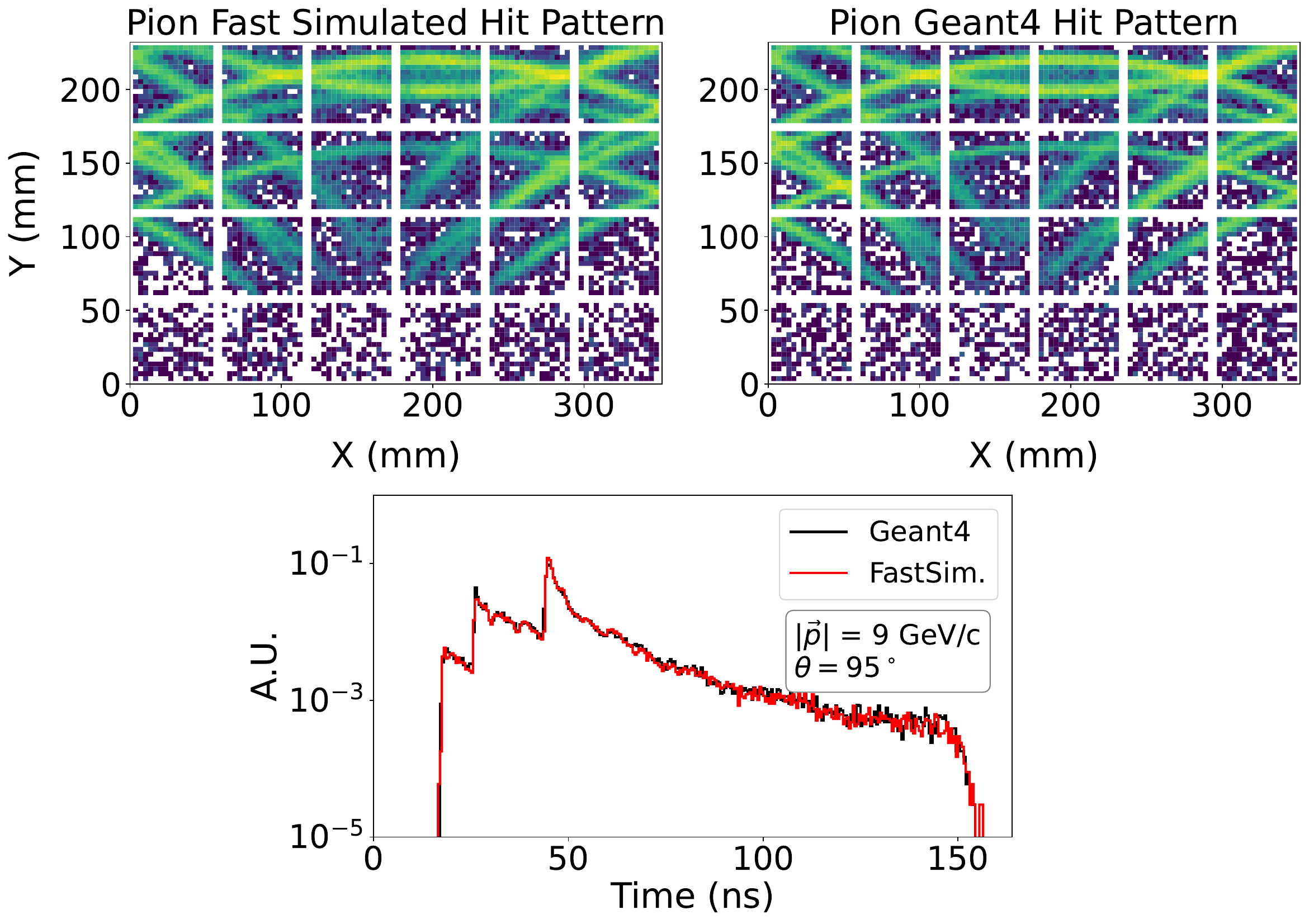} \\
    \includegraphics[width=0.49\textwidth]{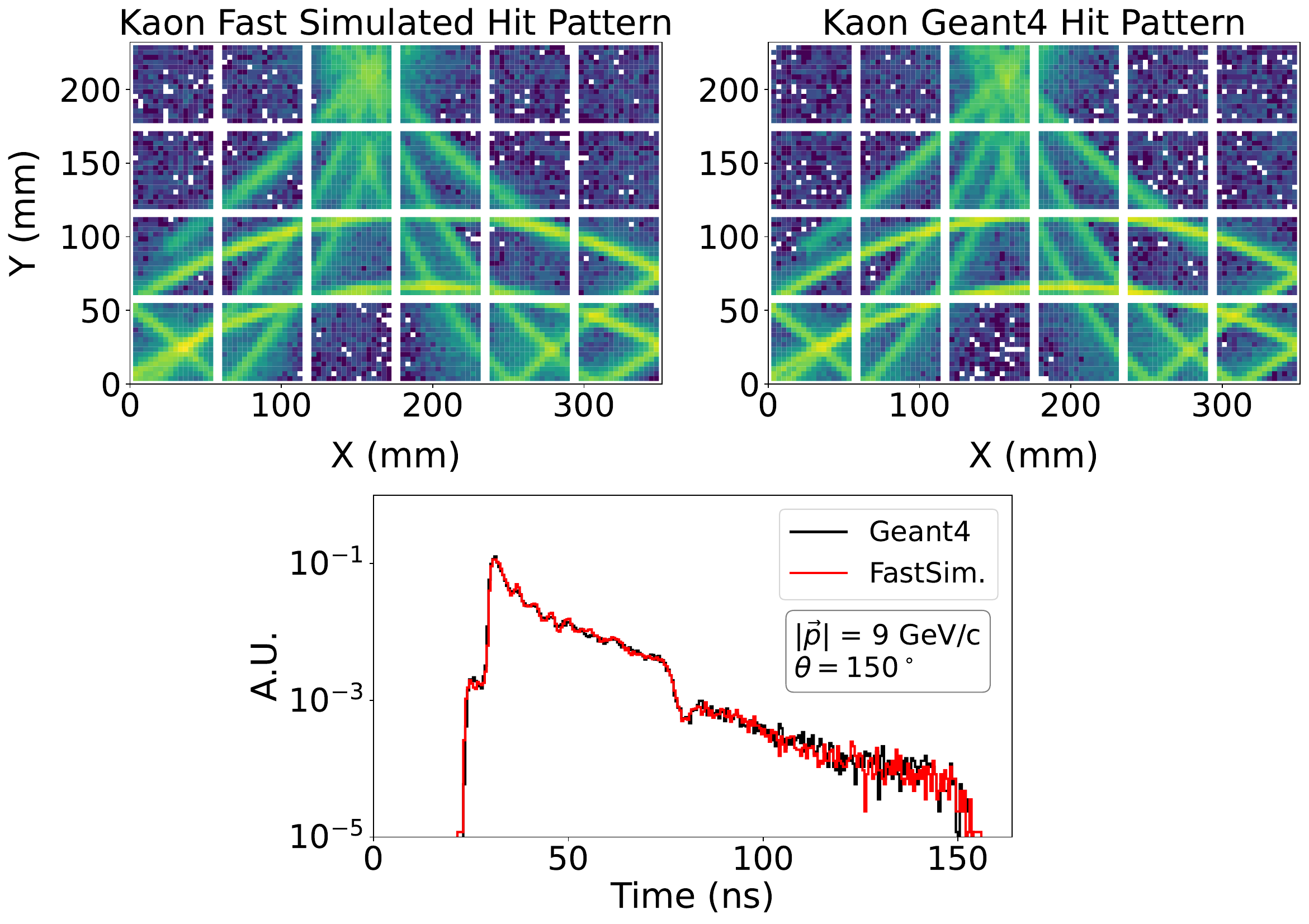} %
    \includegraphics[width=0.49\textwidth]{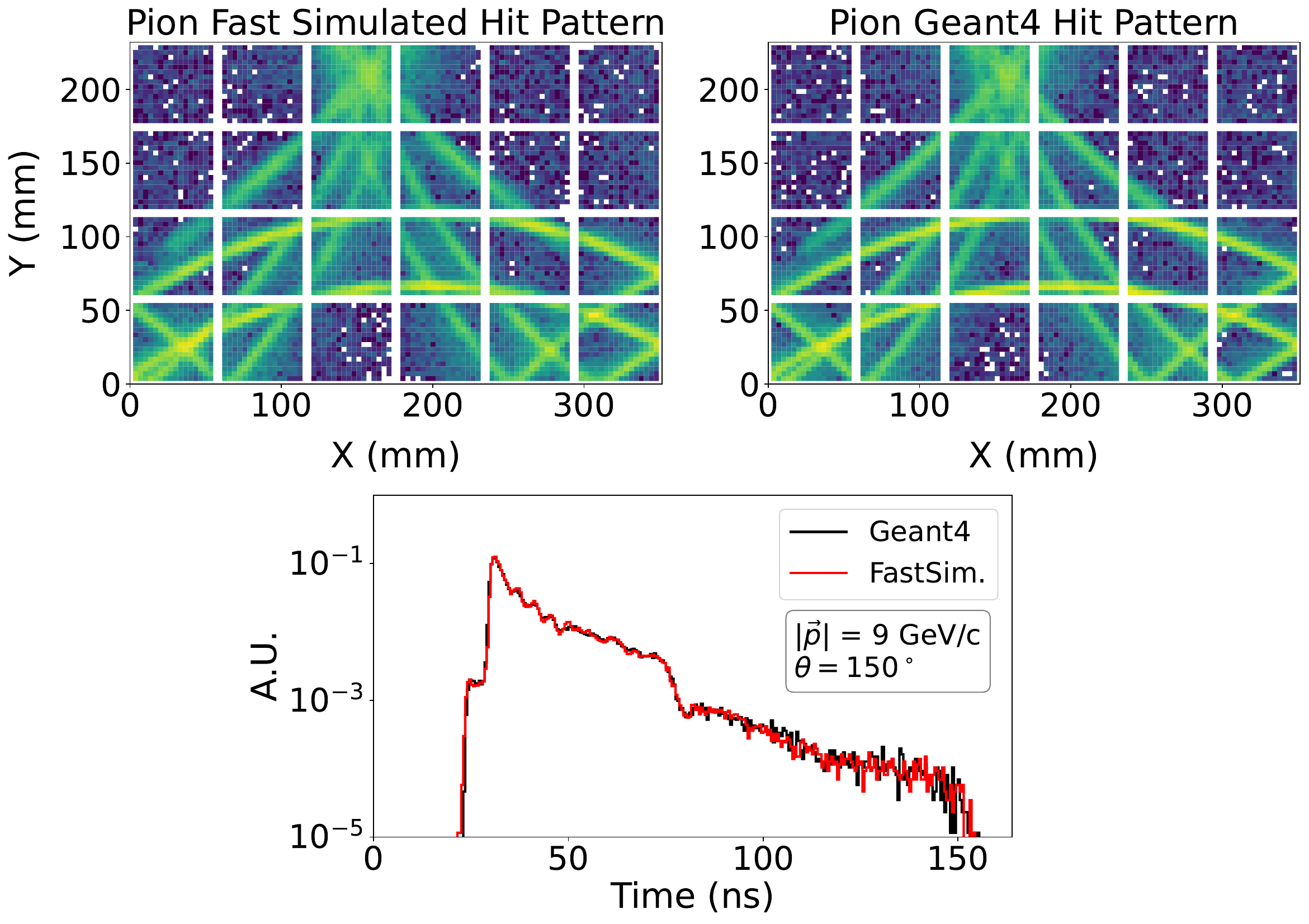} %
    \caption{
    \textbf{Fast Simulation with Flow Matching:} Fast Simulation of Kaons (left column of plots), and Pions (right column of plots) at 9 GeV/c and various polar angles. 20 integration steps with midpoint solver.}
    \label{fig:FlowMatching_Generations_9GeV}
\end{figure}

%%%%%%%%%%%%%%%%%%% DDPM %%%%%%%%%%%%%%%%%%%%%%%%%%%%%%%%%%%%%
\begin{figure}[h]
    \centering
    \includegraphics[width=0.49\textwidth]{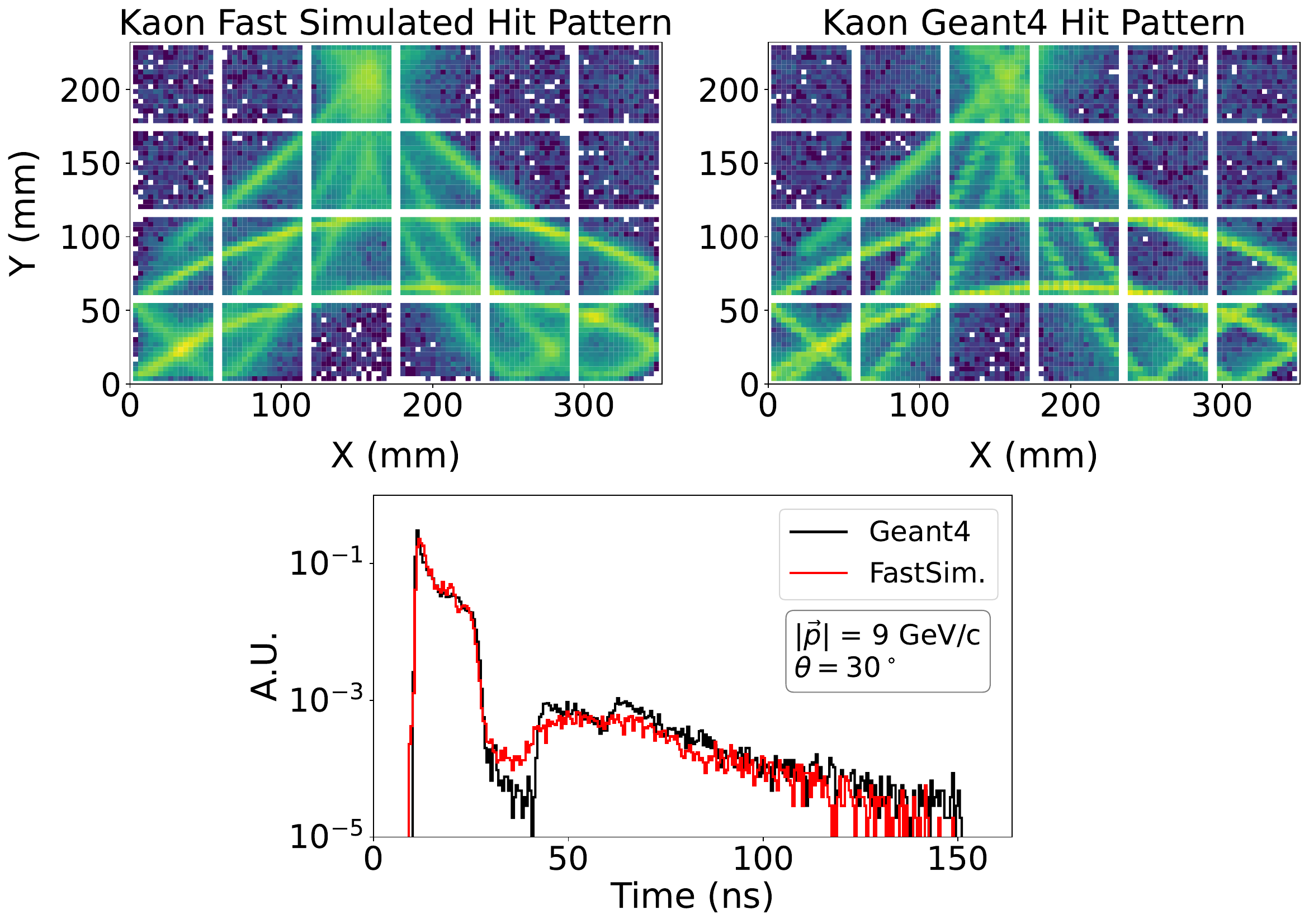}% 
   \includegraphics[width=0.49\textwidth]{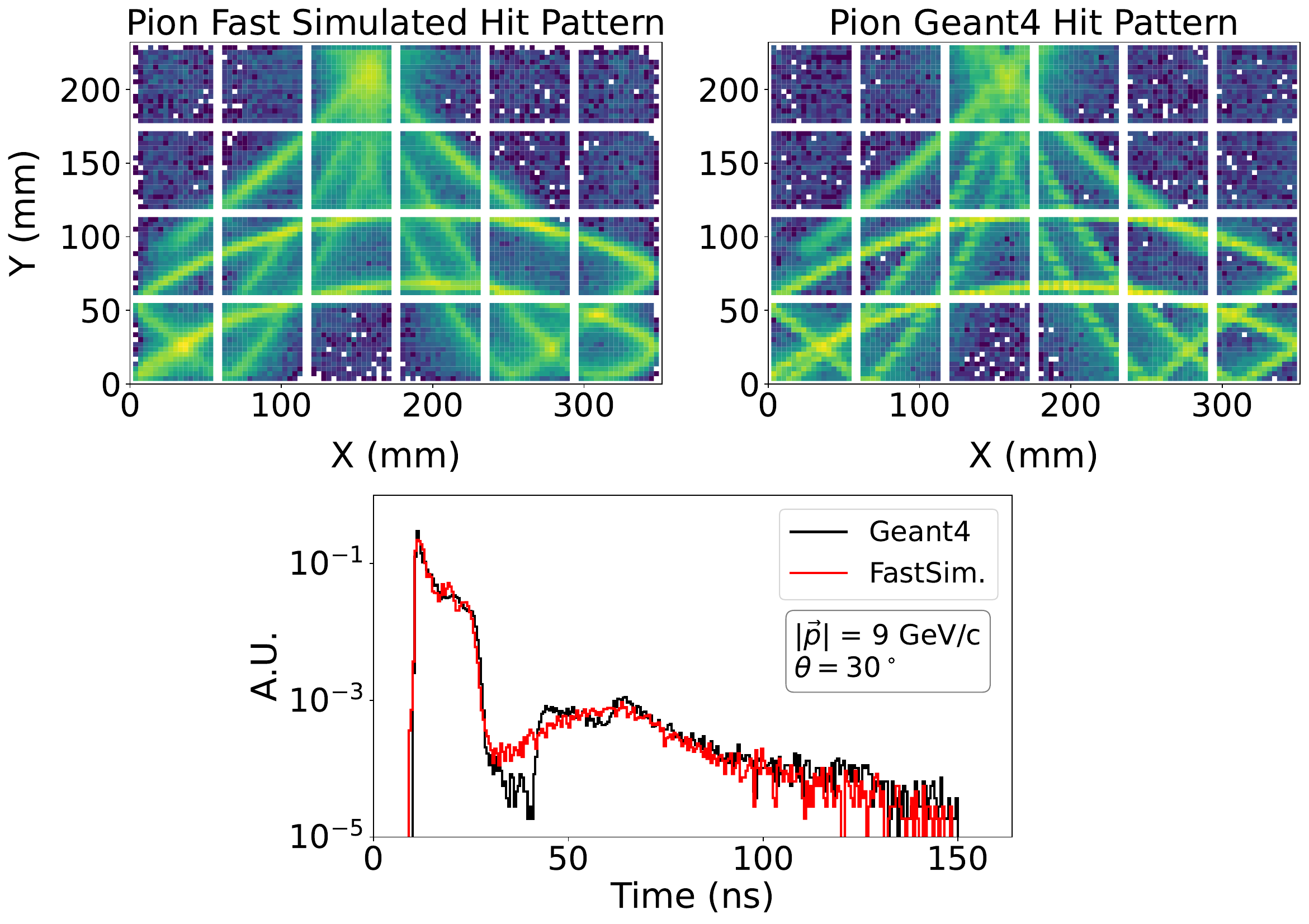} \\
    \includegraphics[width=0.49\textwidth]{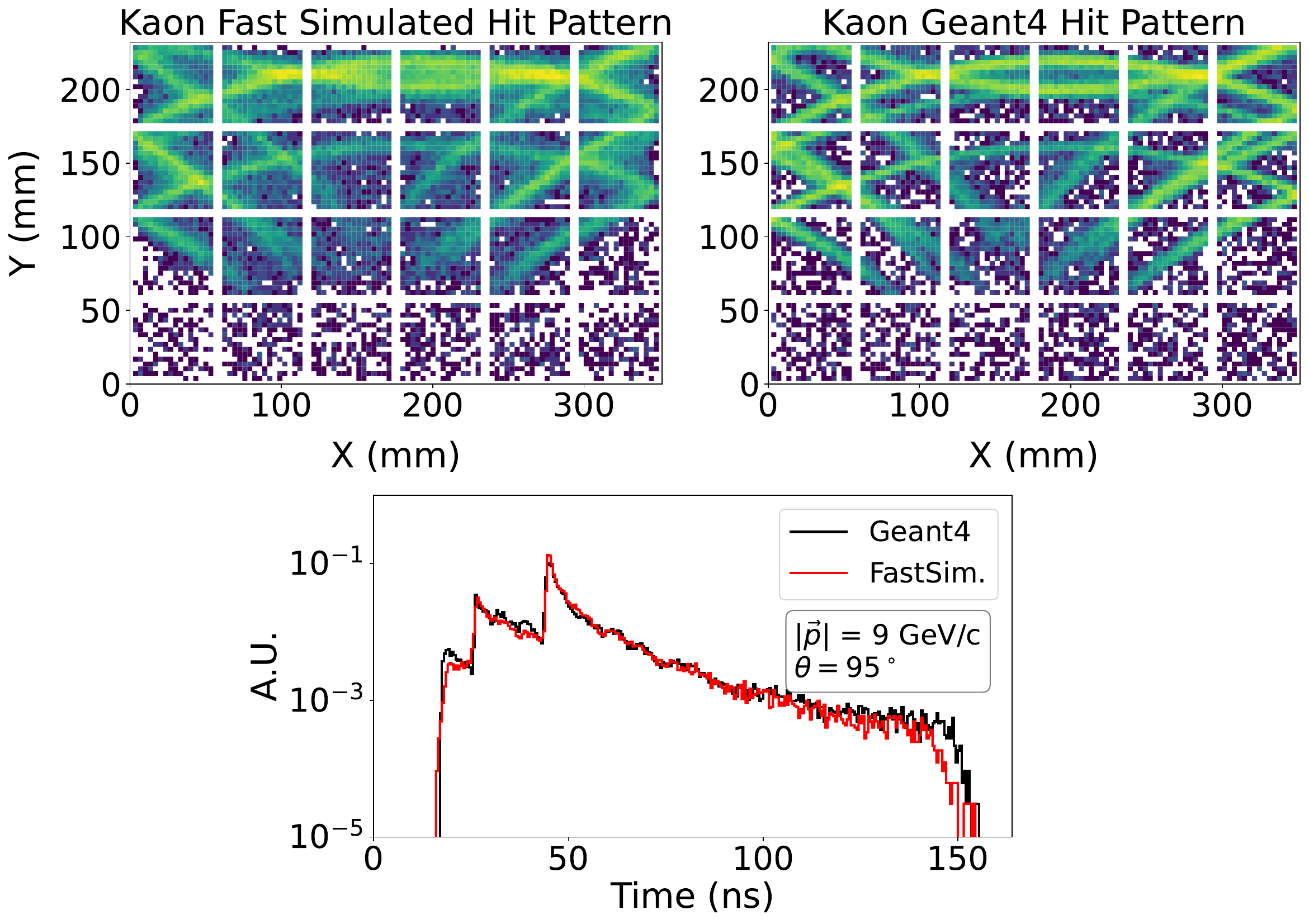} %
    \includegraphics[width=0.49\textwidth]{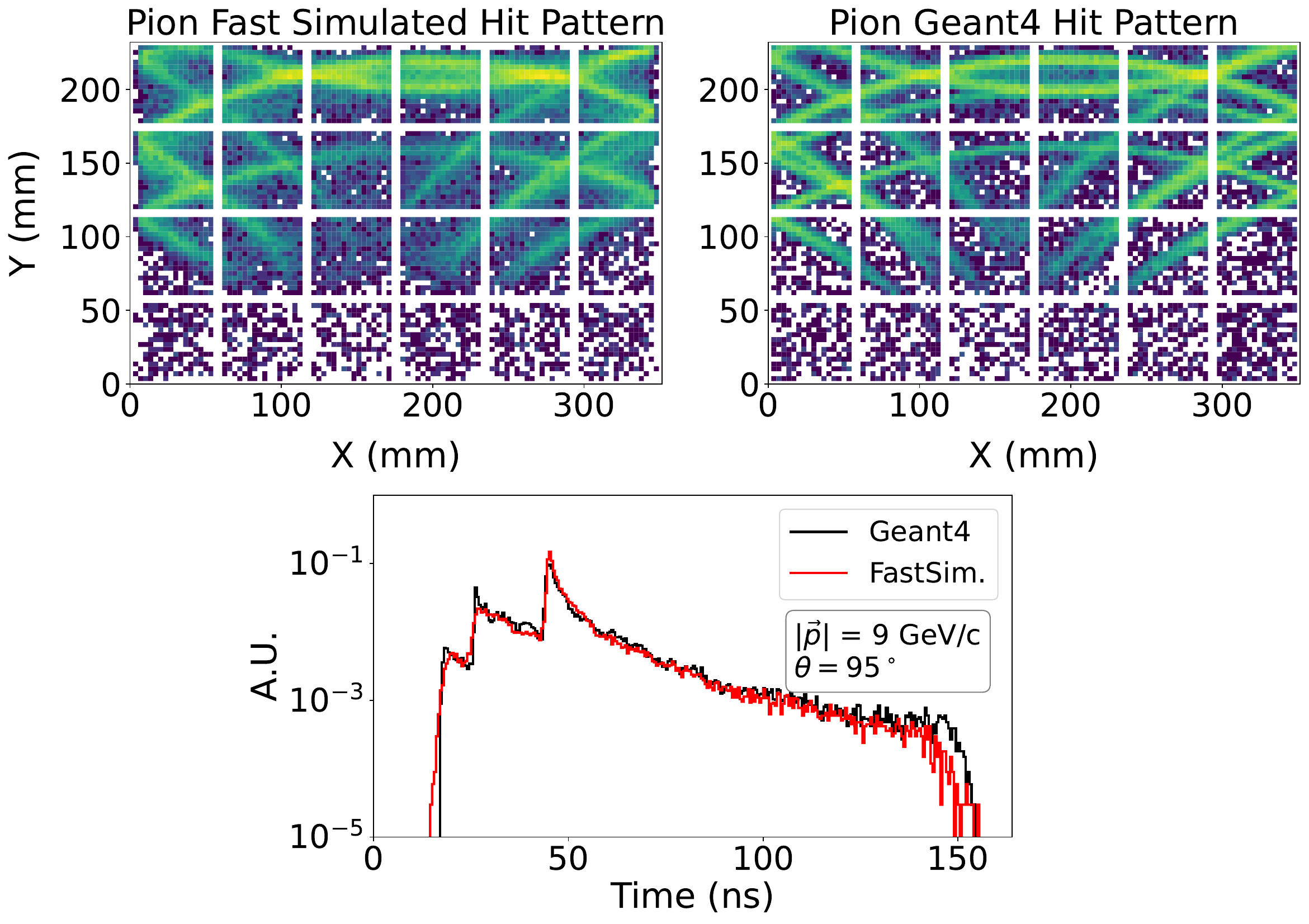} \\
    \includegraphics[width=0.49\textwidth]{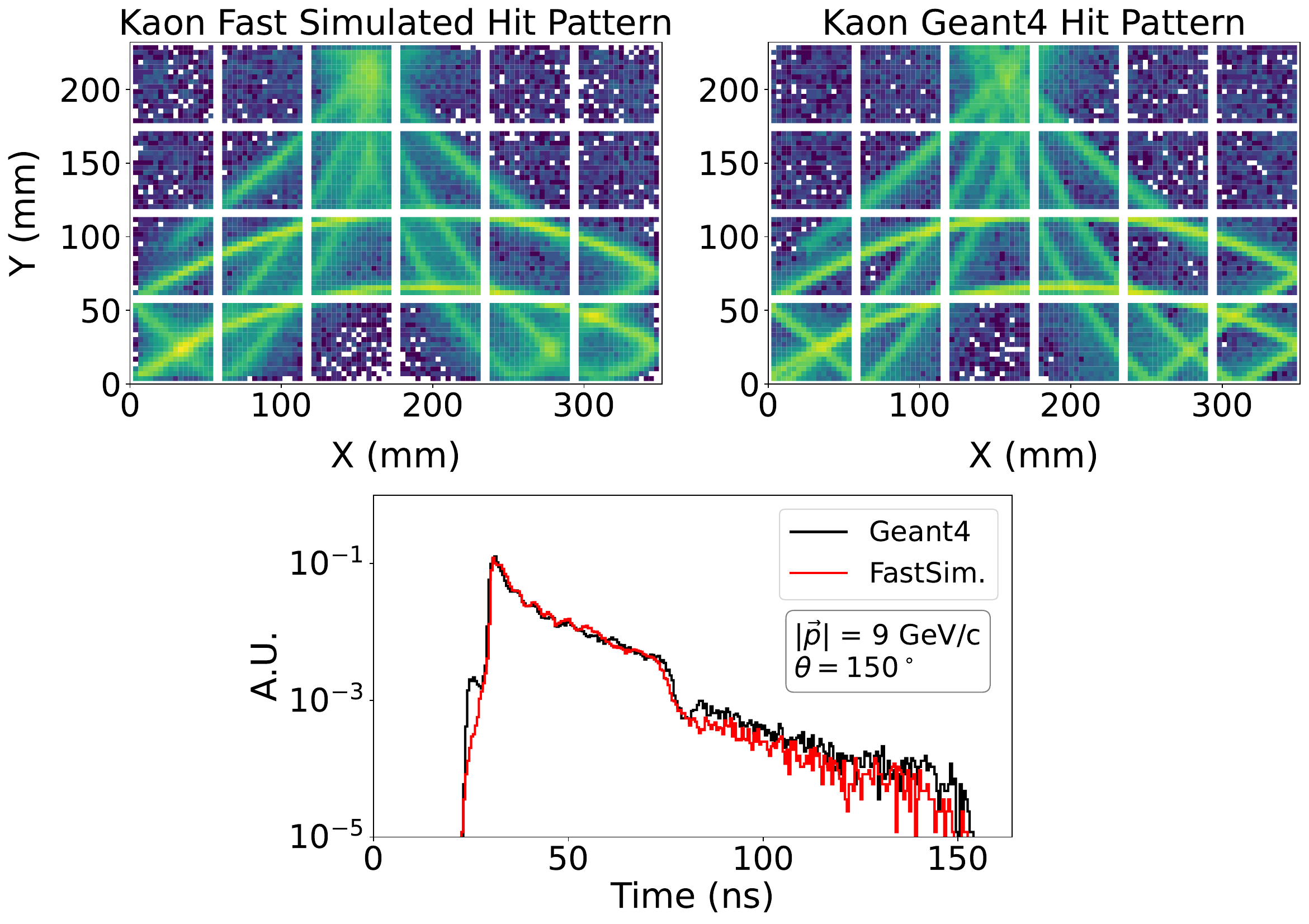} %
    \includegraphics[width=0.49\textwidth]{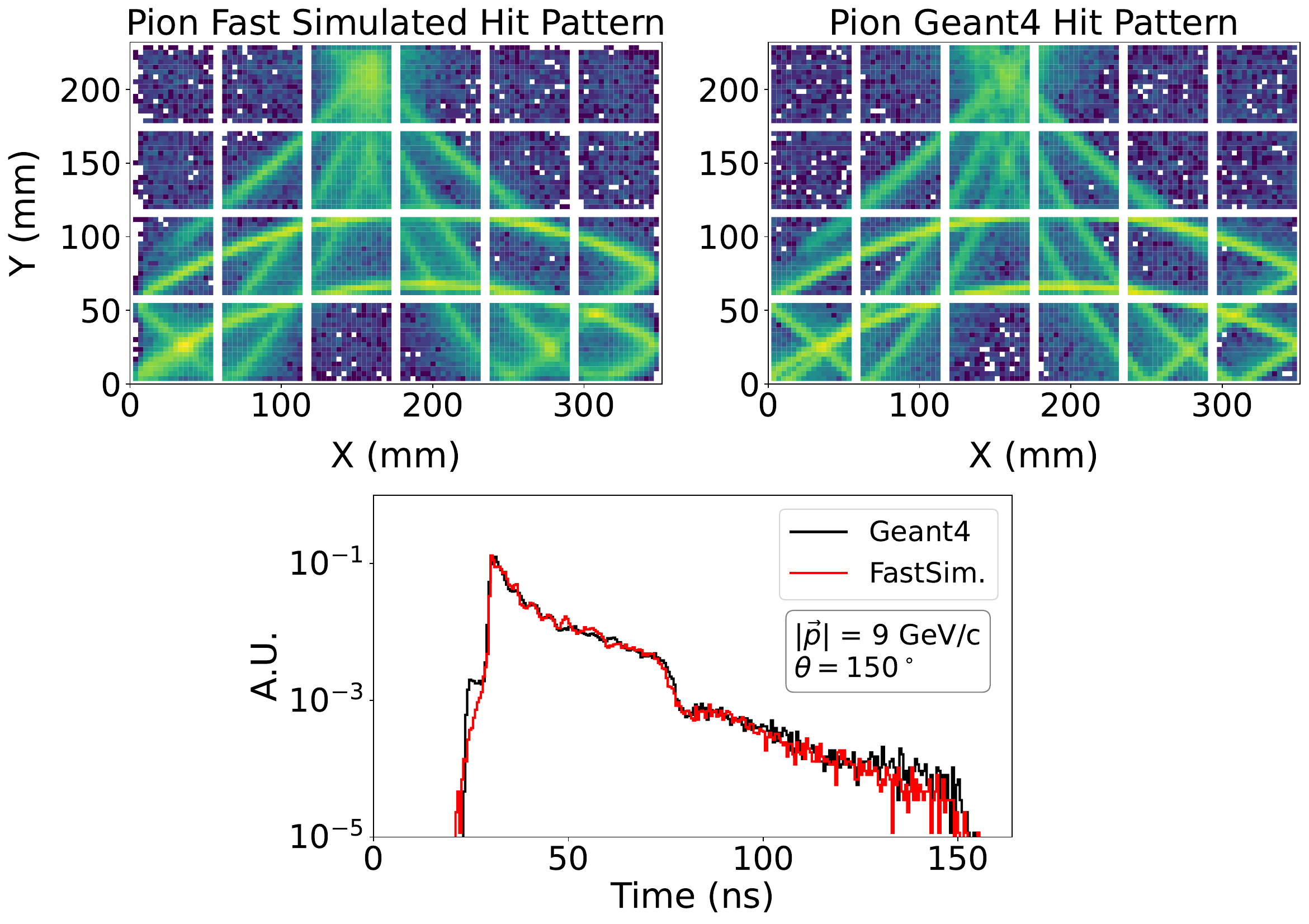} %
    \caption{
    \textbf{Fast Simulation with DDPM:} Fast Simulation of Kaons (left column of plots), and Pions (right column of plots) at 9 GeV/c and various polar angles using DDPM.}
    \label{fig:DDPM_Generations_9GeV}
\end{figure}

%%%%%%%%%%%%%%%%%%% Score Based %%%%%%%%%%%%%%%%%%%%%%%%%%%%%%%%%%%%%
\begin{figure}[h]
    \centering
    \includegraphics[width=0.49\textwidth]{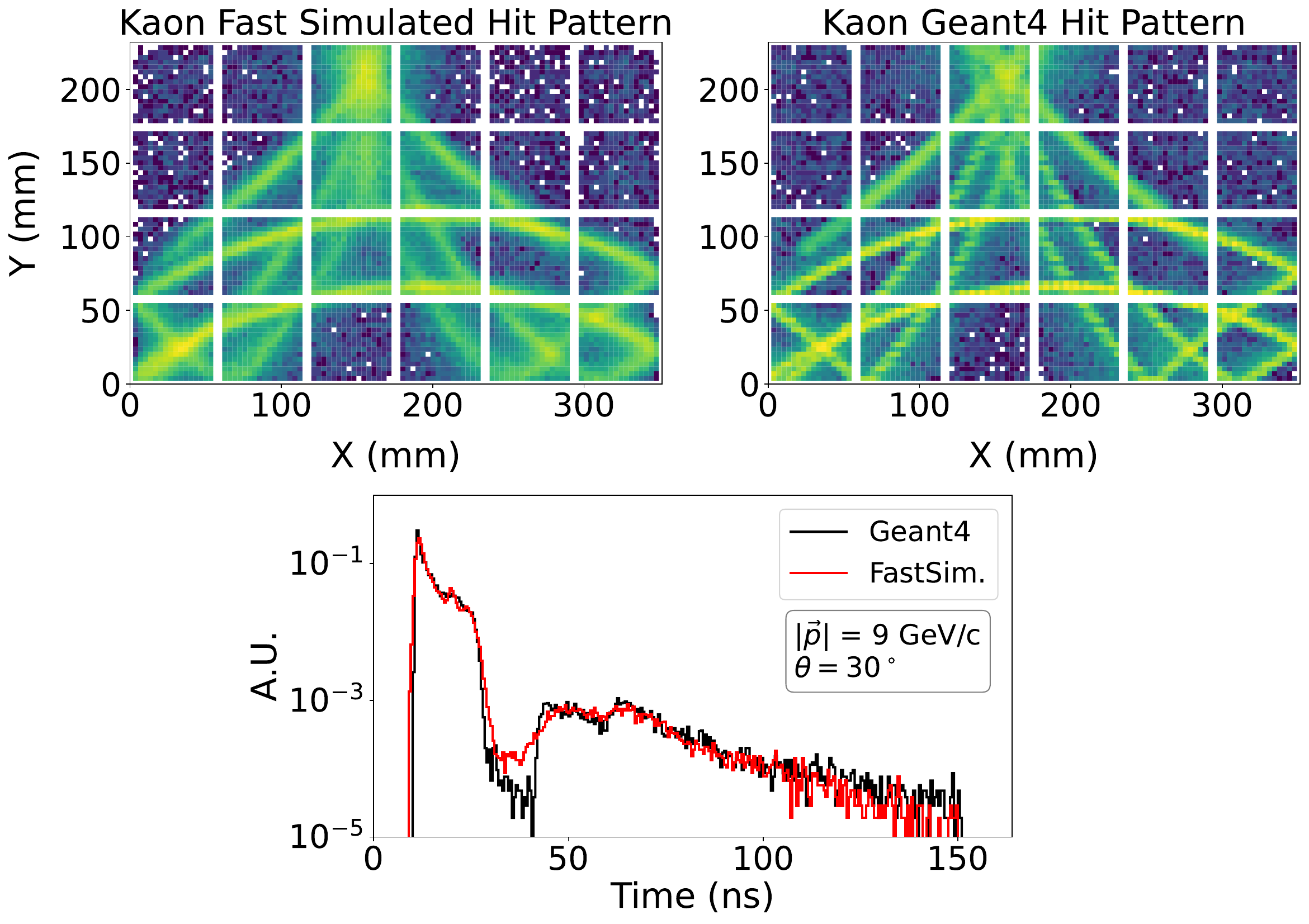}% 
   \includegraphics[width=0.49\textwidth]{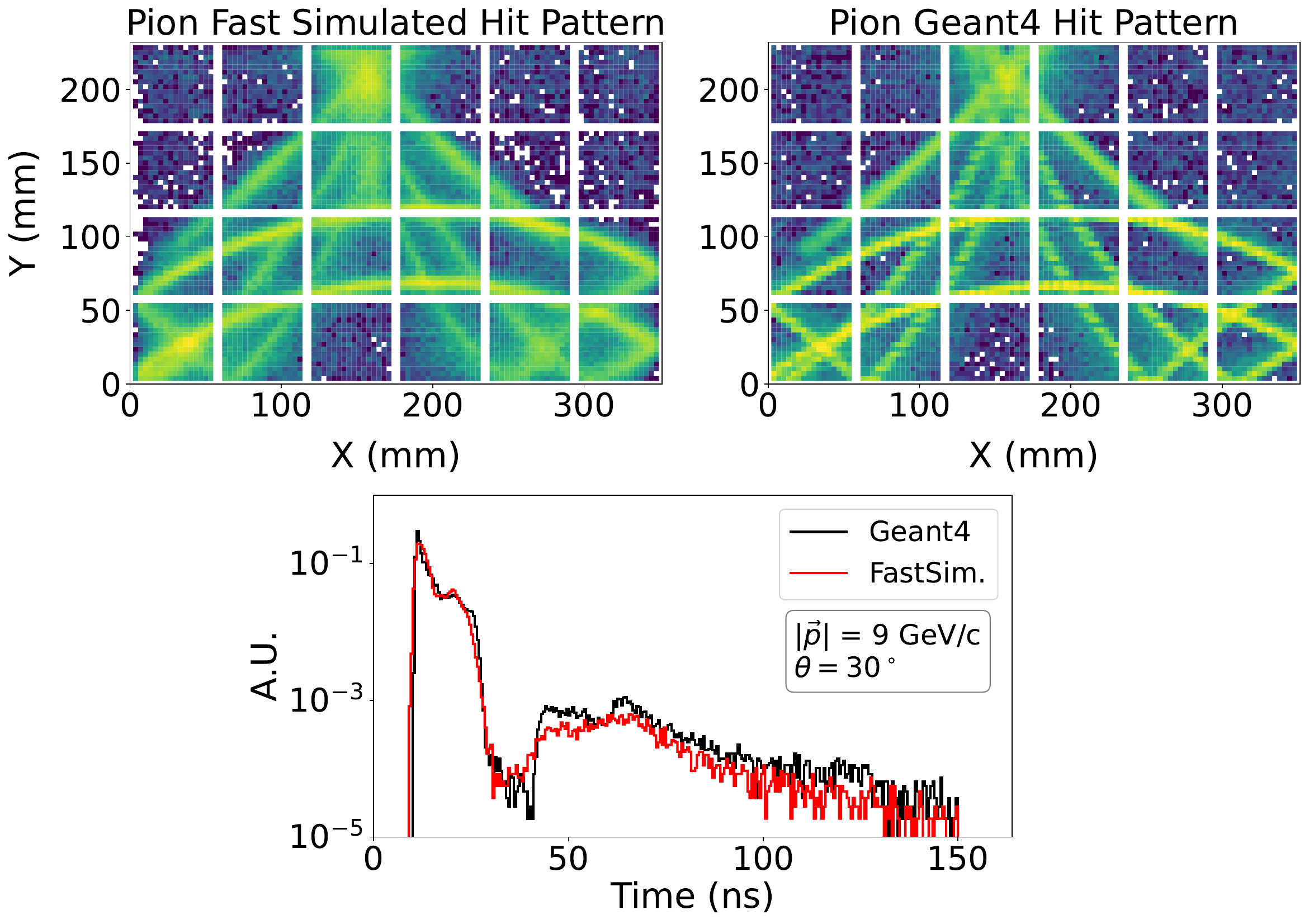} \\
    \includegraphics[width=0.49\textwidth]{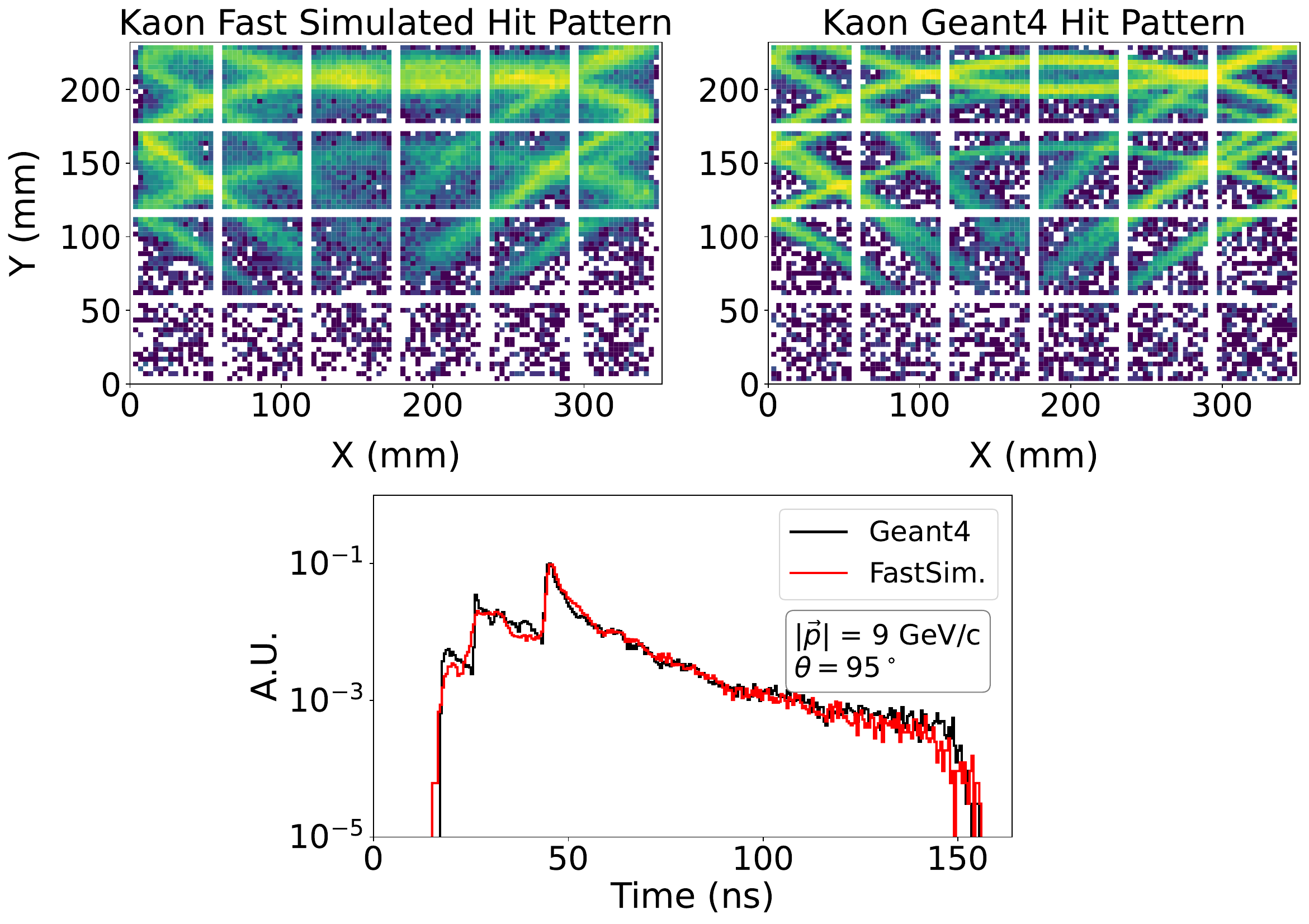} %
    \includegraphics[width=0.49\textwidth]{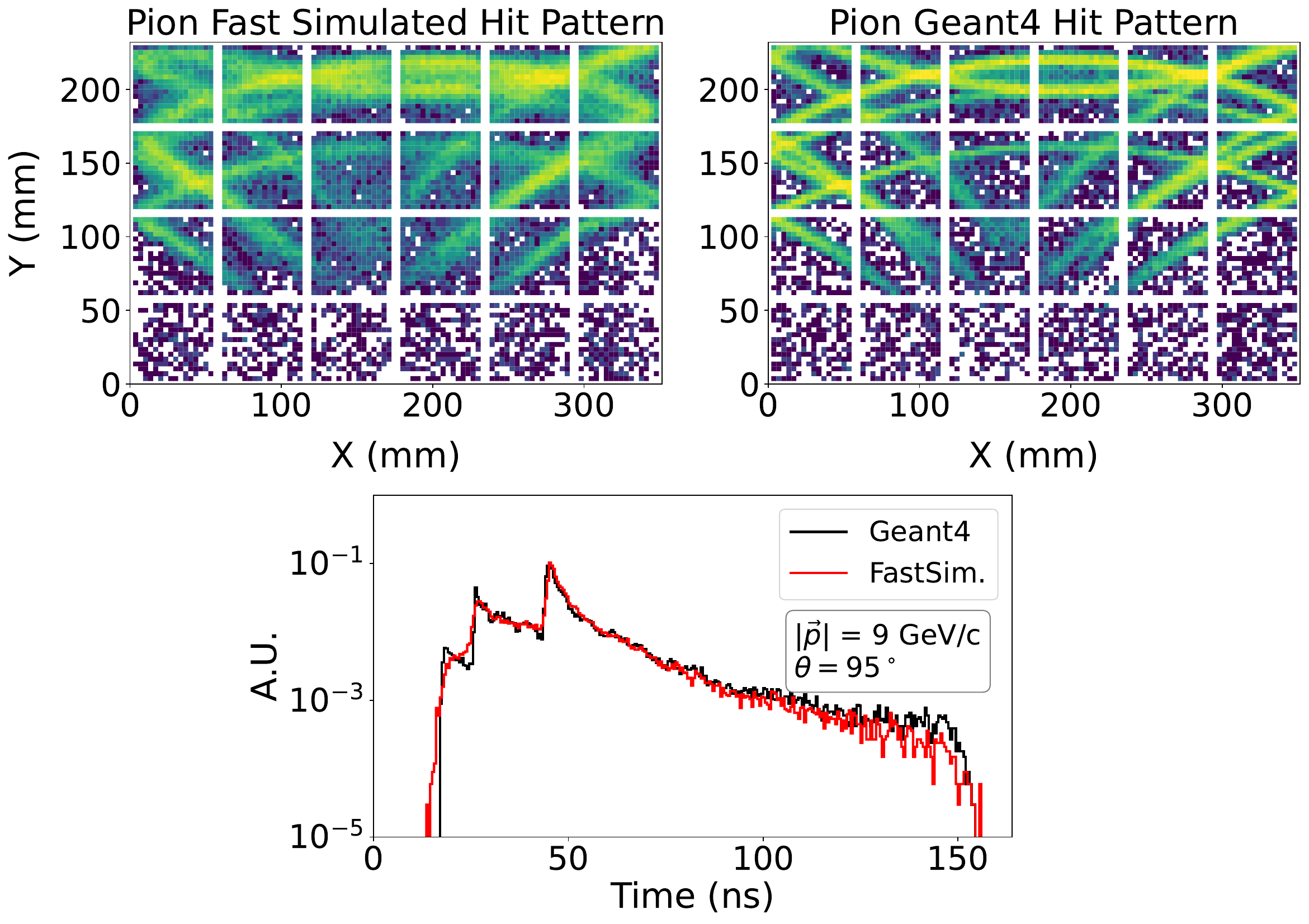} \\
    \includegraphics[width=0.49\textwidth]{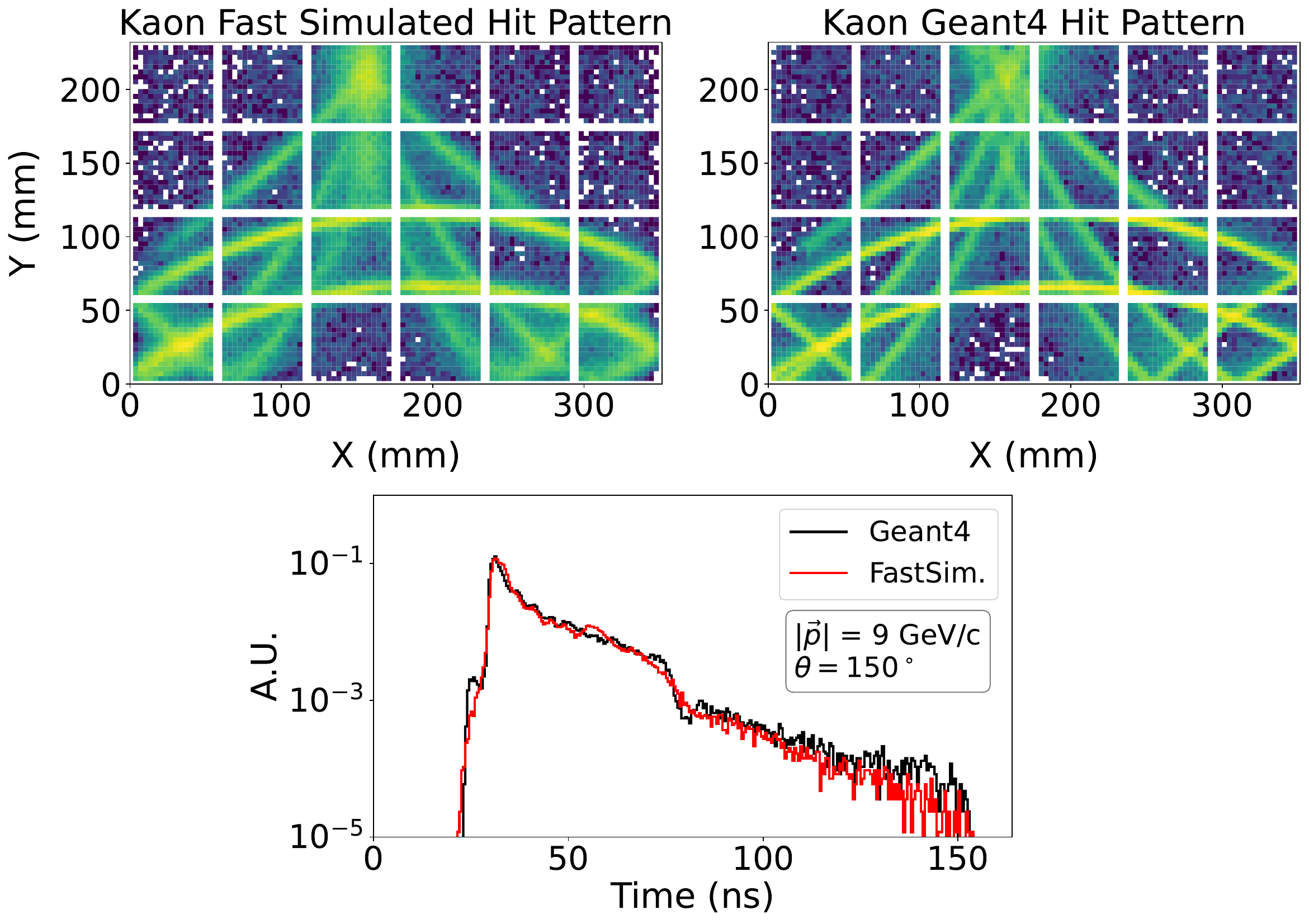} %
    \includegraphics[width=0.49\textwidth]{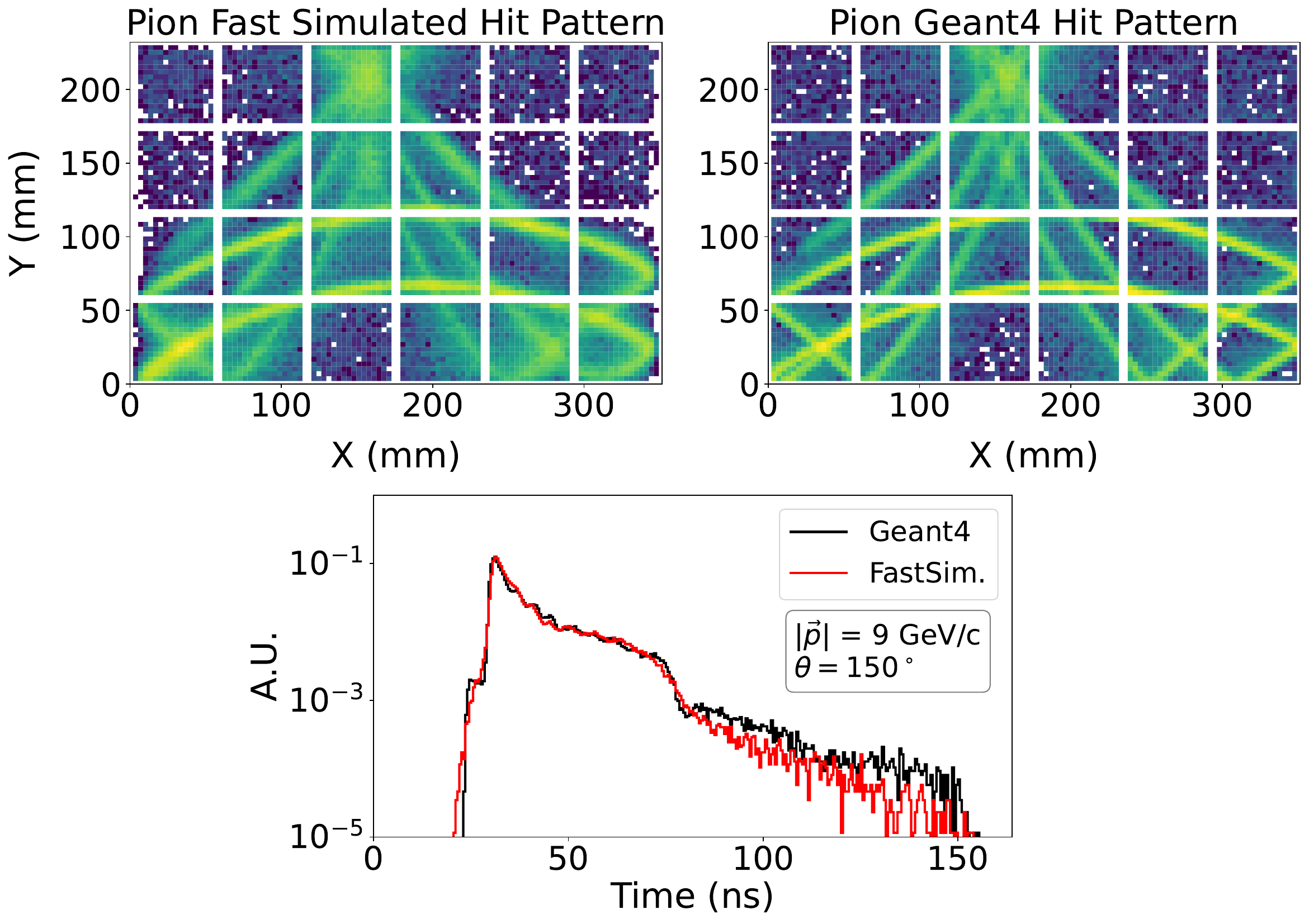} %
    \caption{
    \textbf{Fast Simulation with Score-Based Models:} Fast Simulation of Kaons (left column of plots), and Pions (right column of plots) at 9 GeV/c and various polar angles using Score-Based models.}
    \label{fig:score_Generations_9GeV}
\end{figure}

%%%%%%%%%%%%%%%%%%%%%%% Histogram and Ratios %%%%%%%%%%%%%%%%%%%%%%%%%%%%%%%%%%%

\begin{figure}[h]
    \centering
    \begin{subfigure}[b]{0.49\textwidth}
        \centering
        \includegraphics[width=\textwidth]{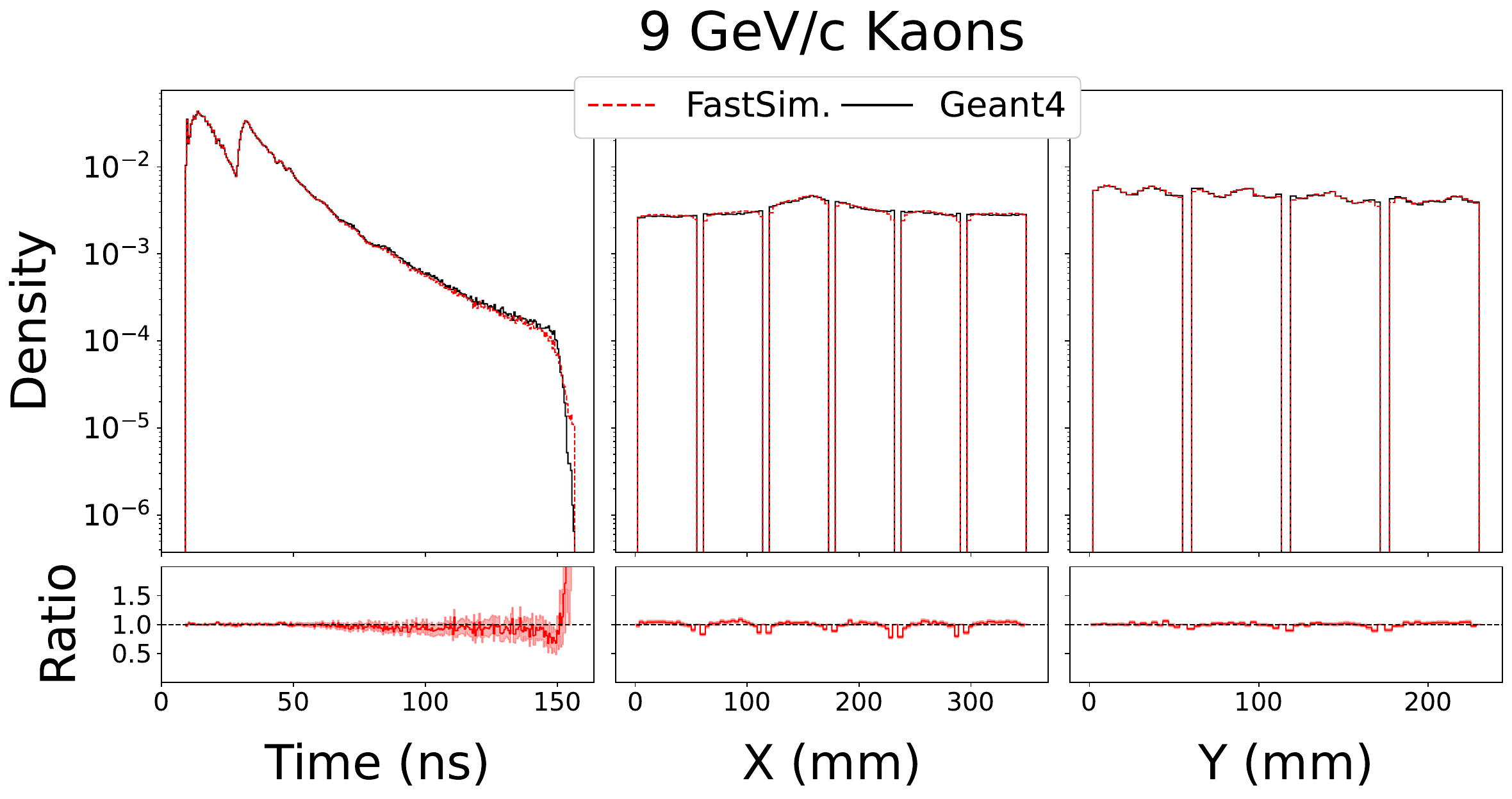}
        \caption{DNF}
    \end{subfigure}
    \begin{subfigure}[b]{0.49\textwidth}
        \centering
        \includegraphics[width=\textwidth]{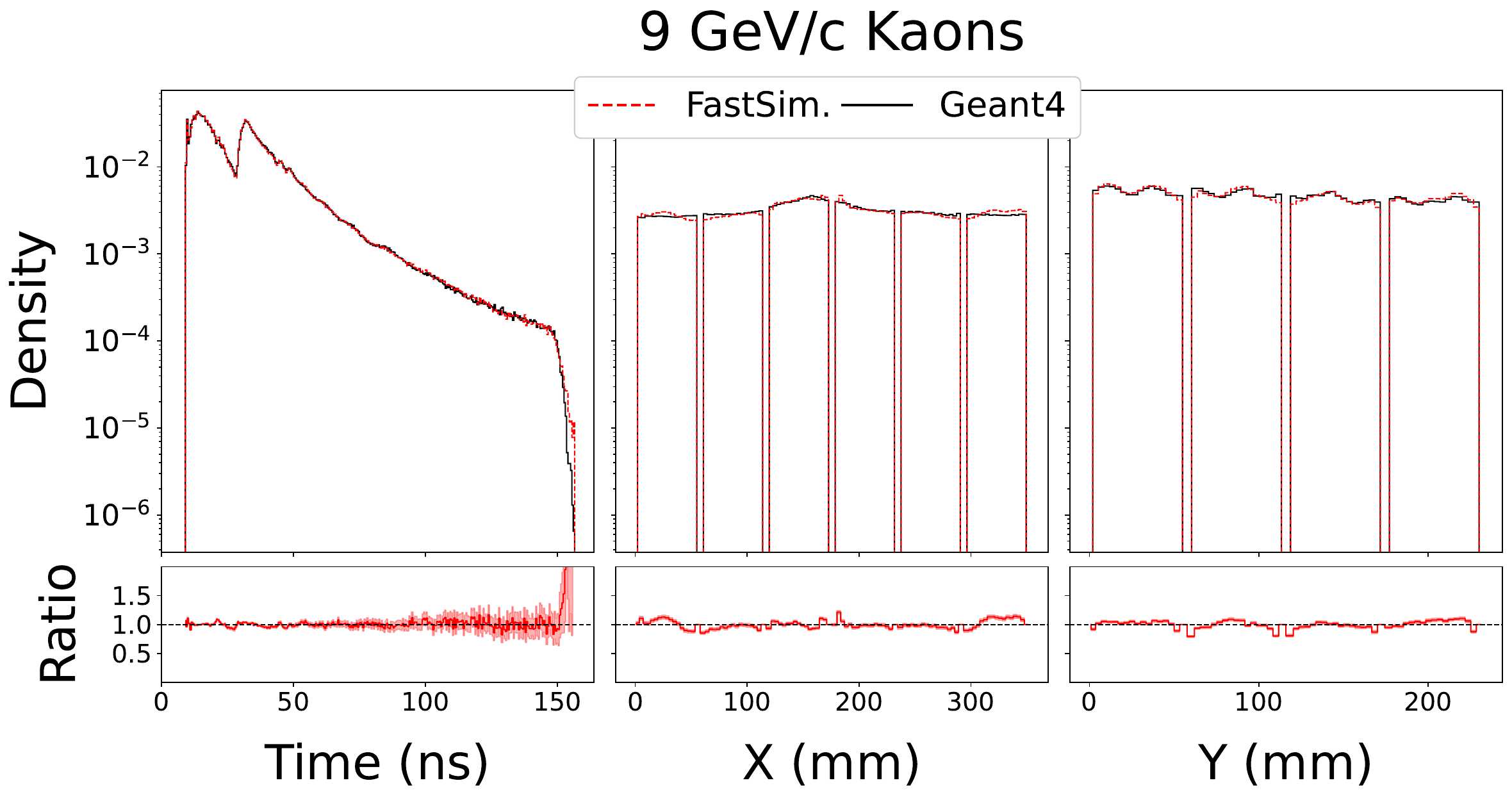}
        \caption{CNF}
    \end{subfigure} \\   
    
    \begin{subfigure}[b]{0.49\textwidth}
        \centering
        \includegraphics[width=\textwidth]{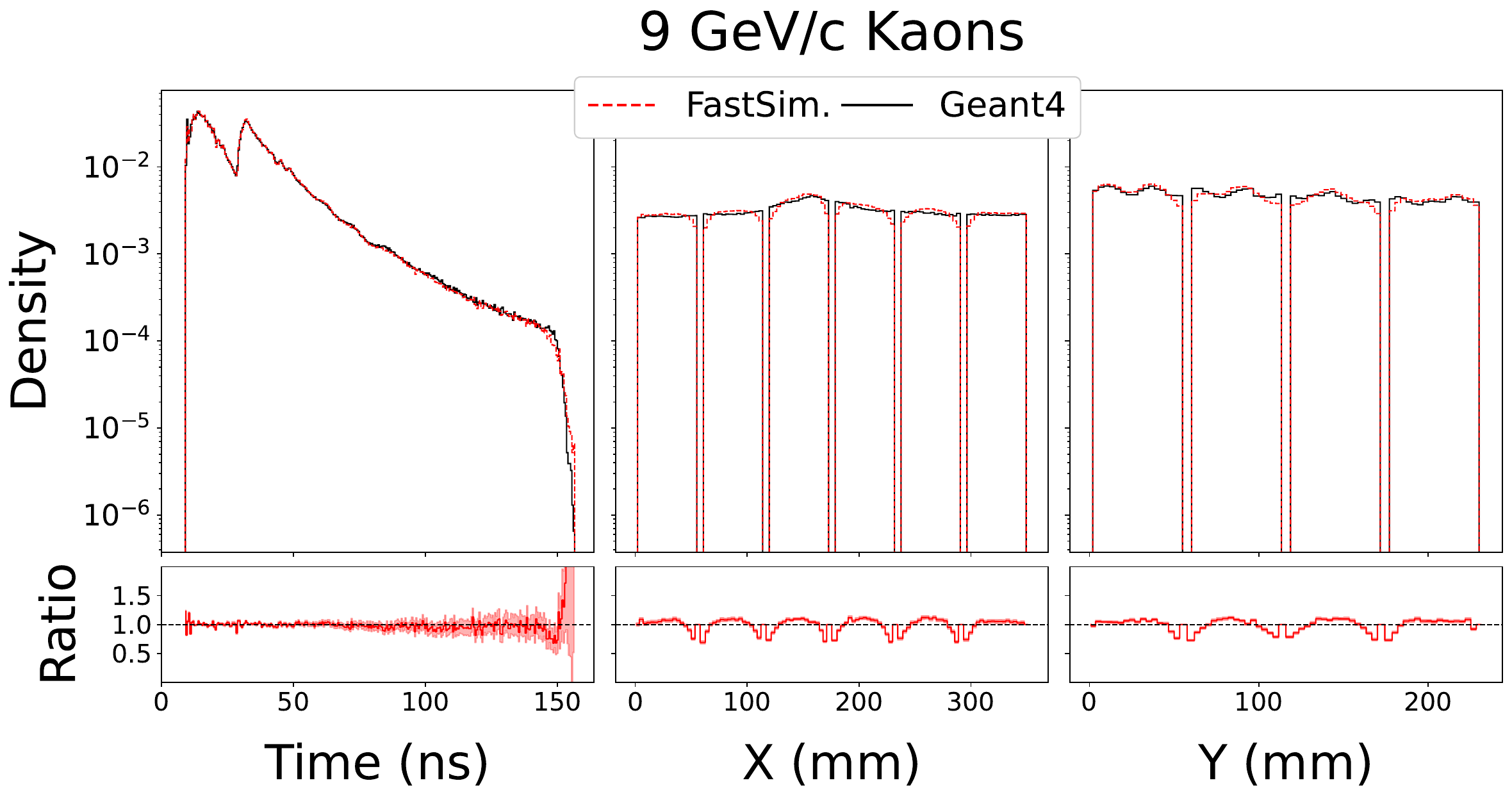}
        \caption{Flow Matching}
    \end{subfigure}
    \begin{subfigure}[b]{0.49\textwidth}
        \centering
        \includegraphics[width=\textwidth]{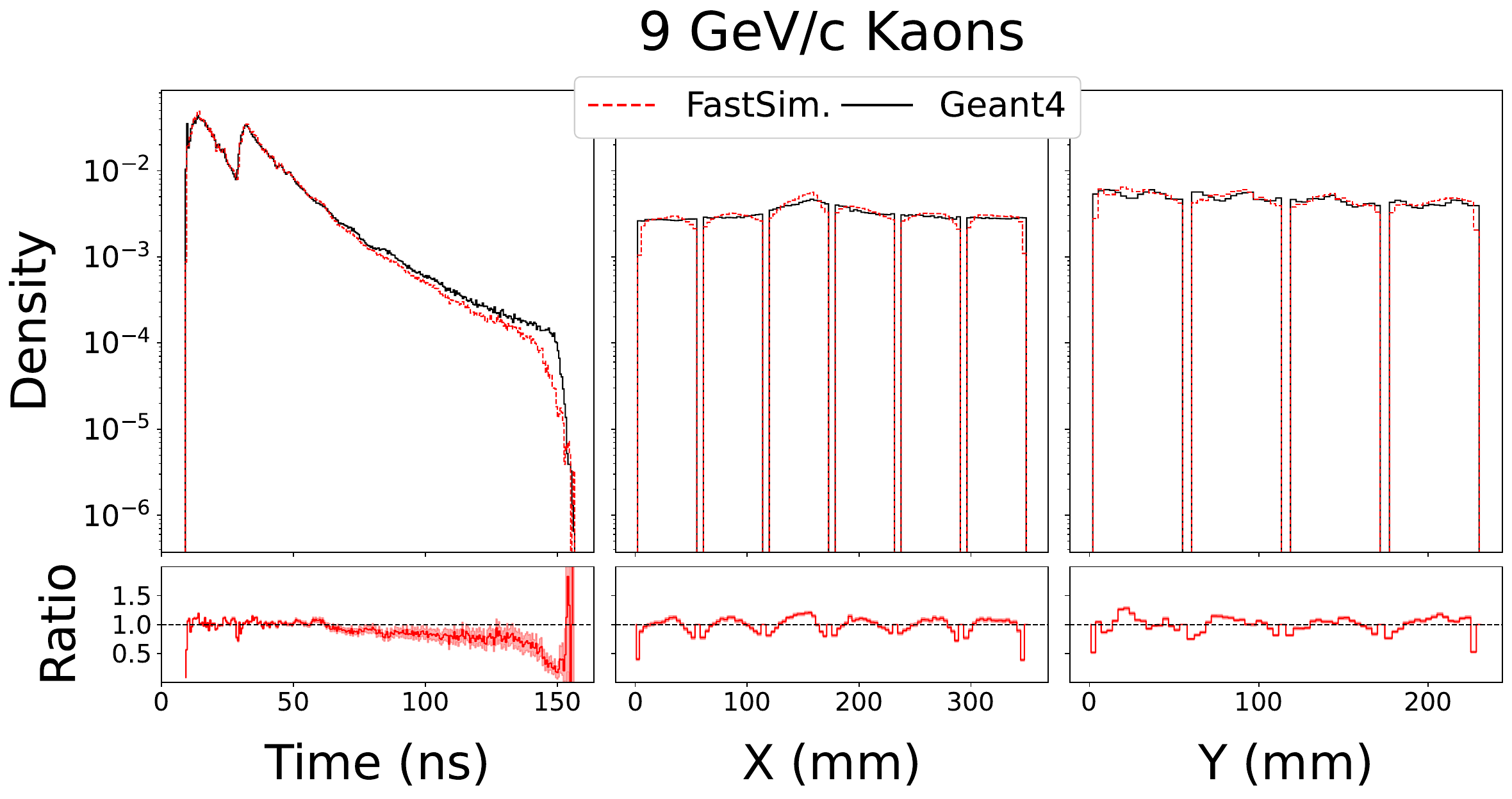}
        \caption{DDPM}
    \end{subfigure} \\  
    
    \begin{subfigure}[b]{0.49\textwidth}
        \centering
        \includegraphics[width=\textwidth]{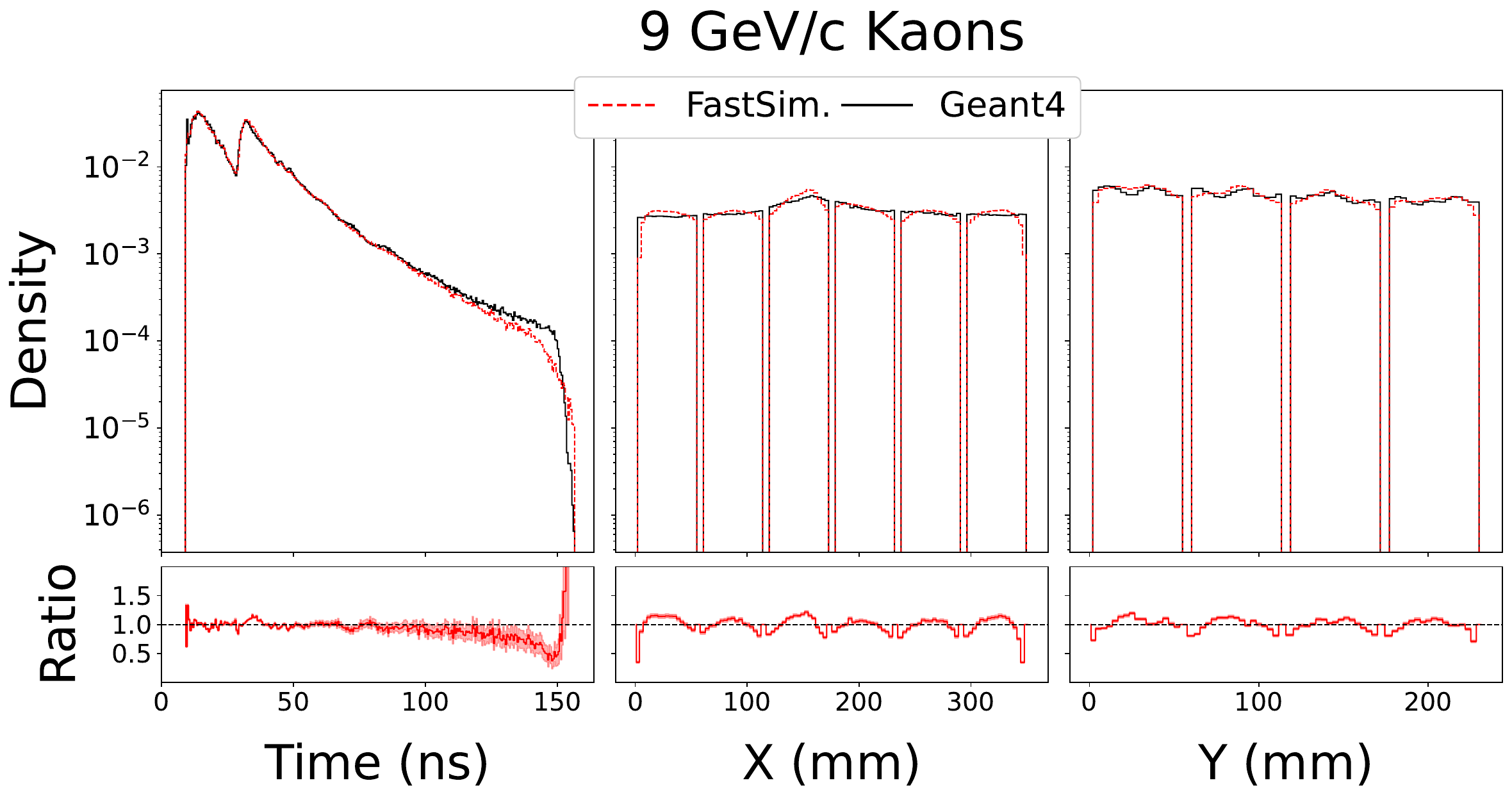}
        \caption{Score-Based}
    \end{subfigure}
    % \begin{subfigure}[b]{0.49\textwidth}
    %     \centering
    %     \includegraphics[width=\textwidth]{Figures/Generations/GSGM/9GeV/Ratios_Kaon.pdf}
    %     \caption{GSGM}
    % \end{subfigure}
    \caption{\textbf{Ratio Plots at 9 GeV/c for Kaons:} Ratio plots for Kaons using the various different models (a) Discrete Normalizing Flows (DNF), (b) Continuous Normalizing Flows (b), (c) Flow Matching, (d) Denoising Diffusion Probabilistic Models (DDPM), and (e) Score-Based Generative Models at 9 GeV/c, integrated over the polar angle.}
    \label{fig:ratio_plots_kaon_9GeV}
\end{figure}

\begin{figure}
    \centering
    \begin{subfigure}[b]{0.49\textwidth}
        \centering
        \includegraphics[width=\textwidth]{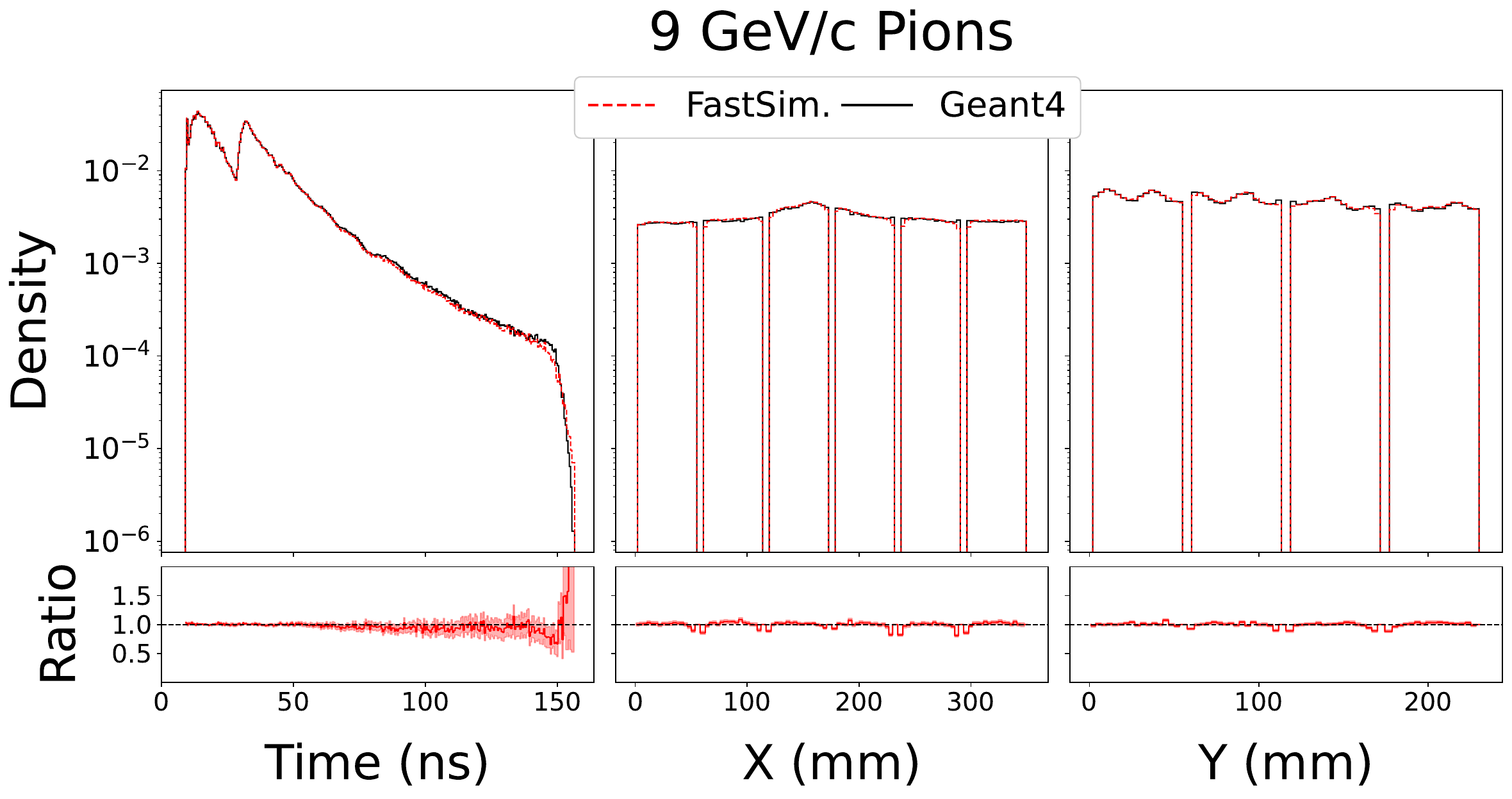}
        \caption{DNF}
    \end{subfigure}
    \begin{subfigure}[b]{0.49\textwidth}
        \centering
        \includegraphics[width=\textwidth]{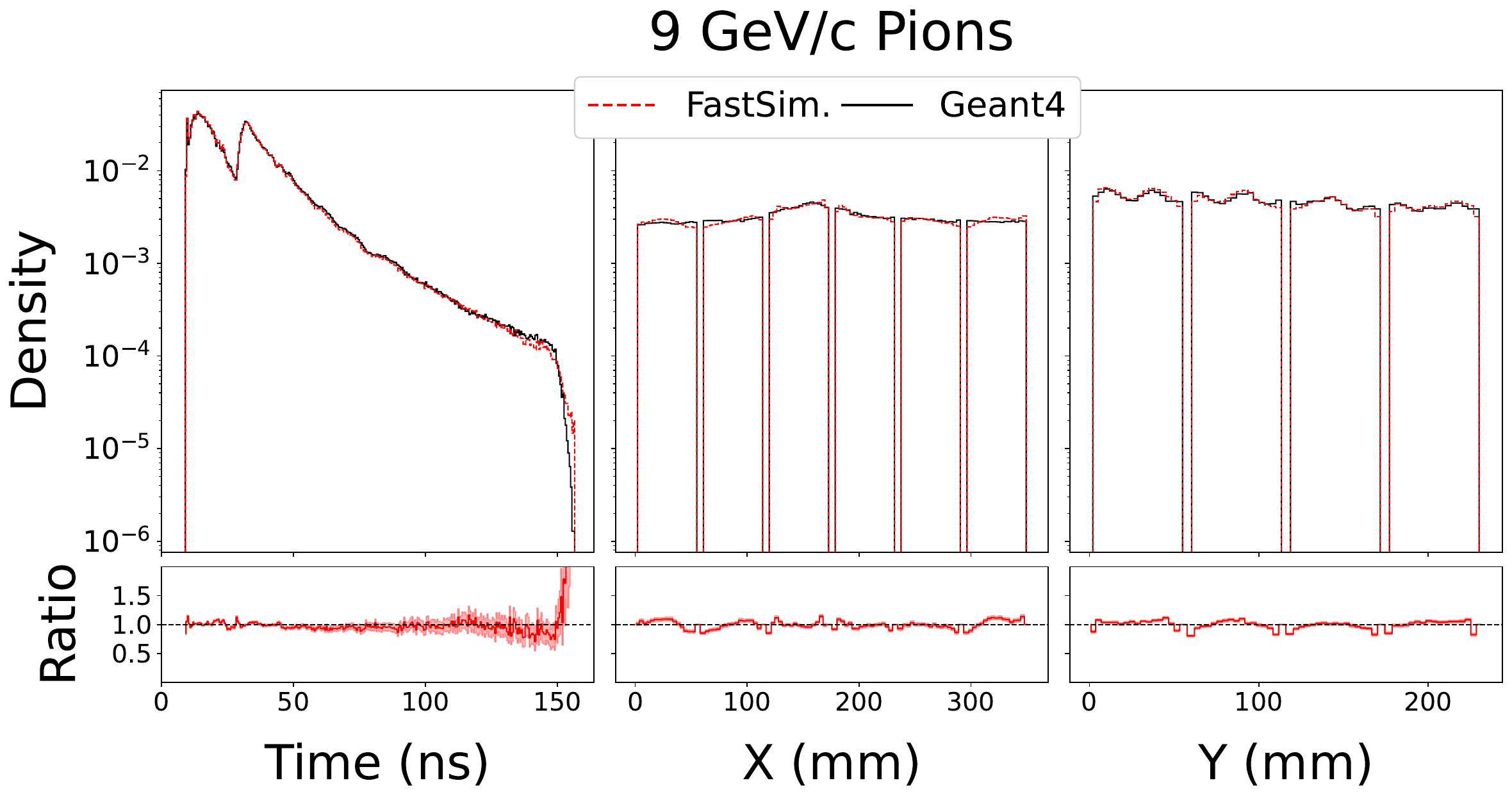}
        \caption{CNF}
    \end{subfigure} \\   
    
    \begin{subfigure}[b]{0.49\textwidth}
        \centering
        \includegraphics[width=\textwidth]{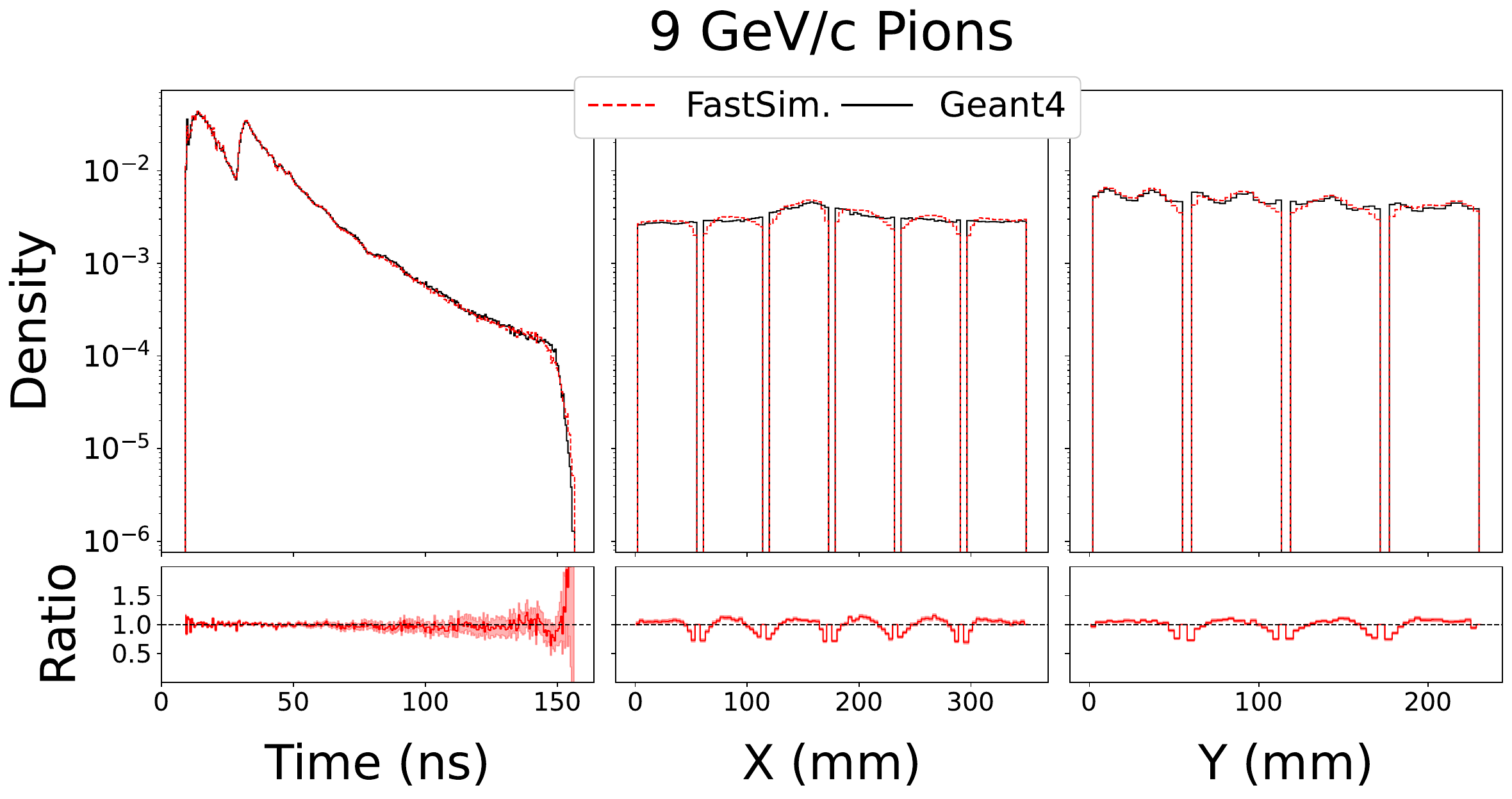}
        \caption{Flow Matching}
    \end{subfigure}
    \begin{subfigure}[b]{0.49\textwidth}
        \centering
        \includegraphics[width=\textwidth]{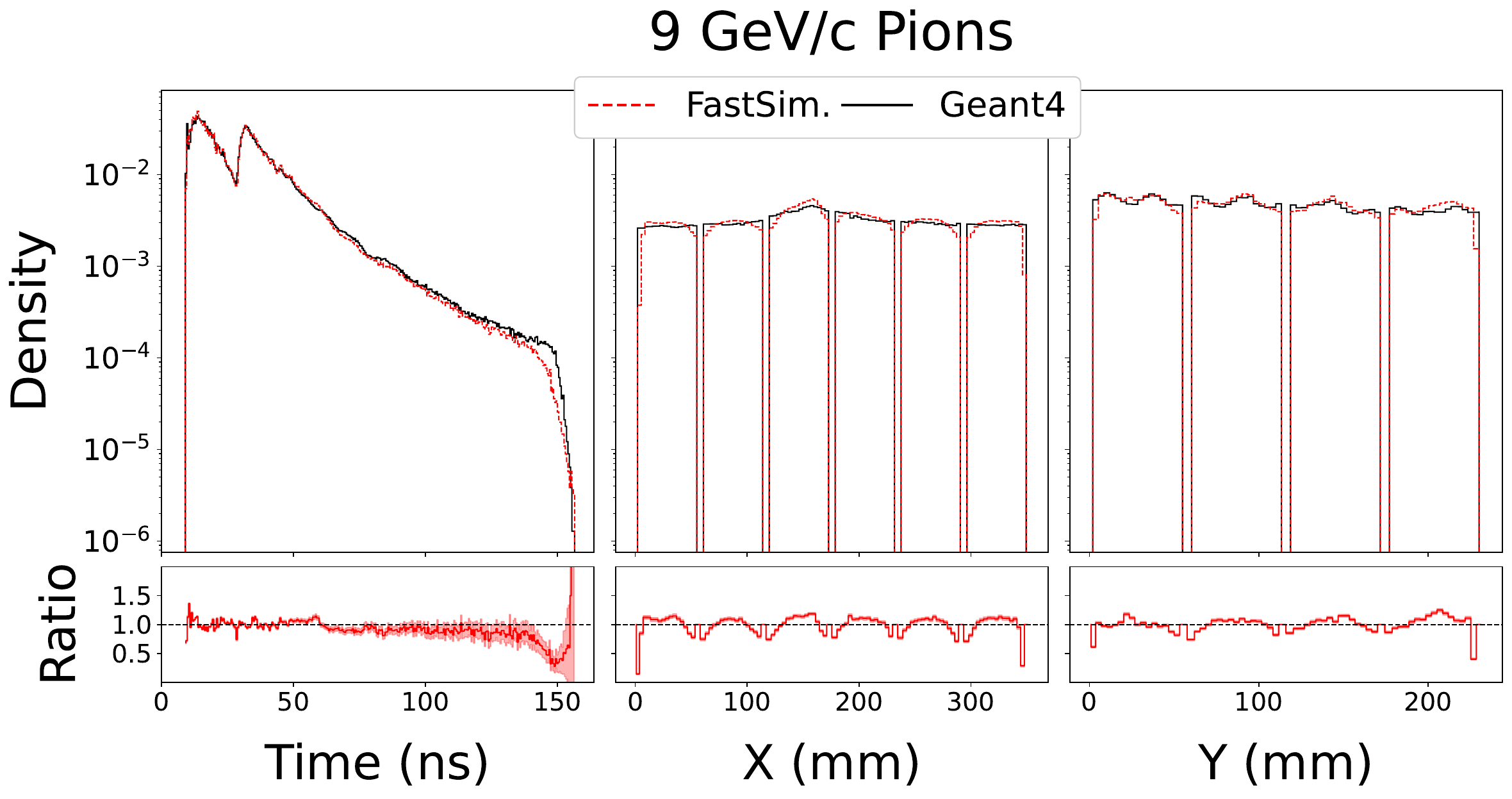}
        \caption{DDPM}
    \end{subfigure} \\  
    
    \begin{subfigure}[b]{0.49\textwidth}
        \centering
        \includegraphics[width=\textwidth]{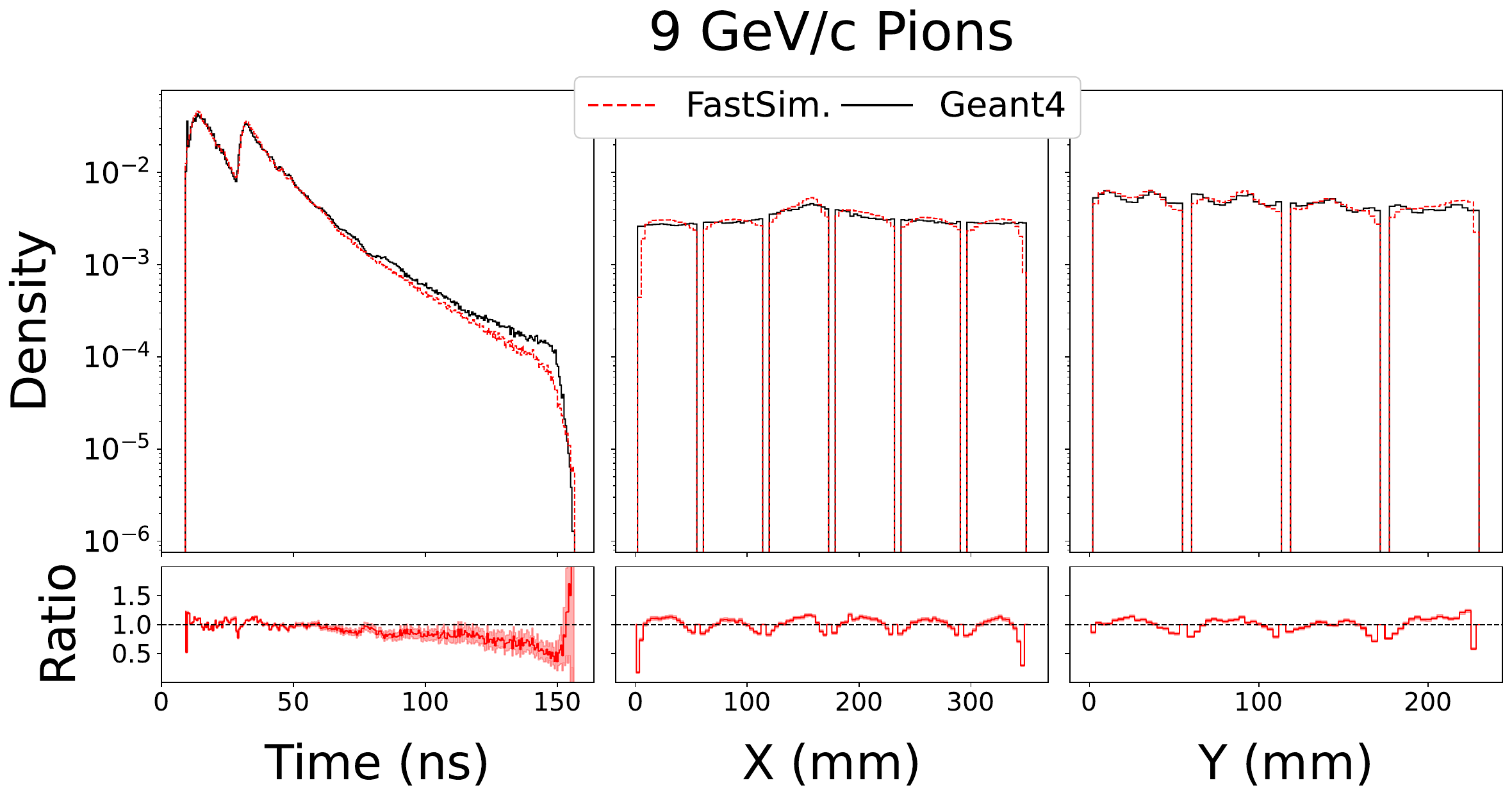}
        \caption{Score-Based}
    \end{subfigure}
    % \begin{subfigure}[b]{0.49\textwidth}
    %     \centering
    %     \includegraphics[width=\textwidth]{Figures/Generations/GSGM/9GeV/Ratios_Pion.pdf}
    %     \caption{GSGM}
    % \end{subfigure}
    \caption{\textbf{Ratio Plots at 9 GeV/c for Pions:} Ratio plots Pions using the various different models (a) Discrete Normalizing Flows (DNF), (b) Continuous Normalizing Flows (b), (c) Flow Matching, (d) Denoising Diffusion Probabilistic Models (DDPM), and (e) Score-Based Generative Models at 9 GeV/c, integrated over the polar angle.}
    \label{fig:ratio_plots_pion_9GeV}
\end{figure}

\clearpage
\newpage
\section{Example Reference Populations for FastDIRC}\label{app:fastDIRC_PDFs}

\begin{figure}[h]
    \centering
    \begin{subfigure}[b]{0.78\textwidth}
        \centering
        \includegraphics[width=0.49\textwidth]{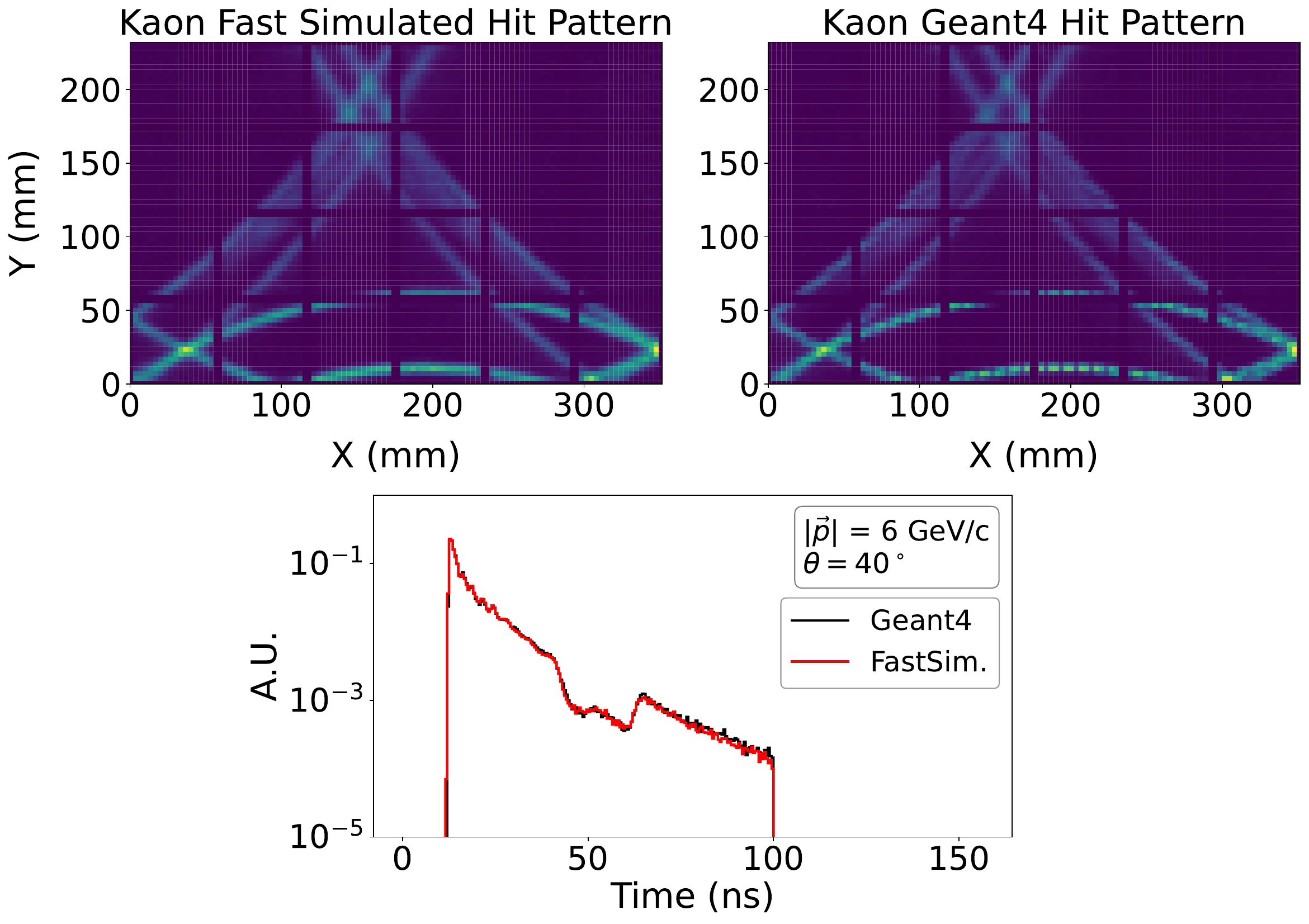} %
        \includegraphics[width=0.49\textwidth]{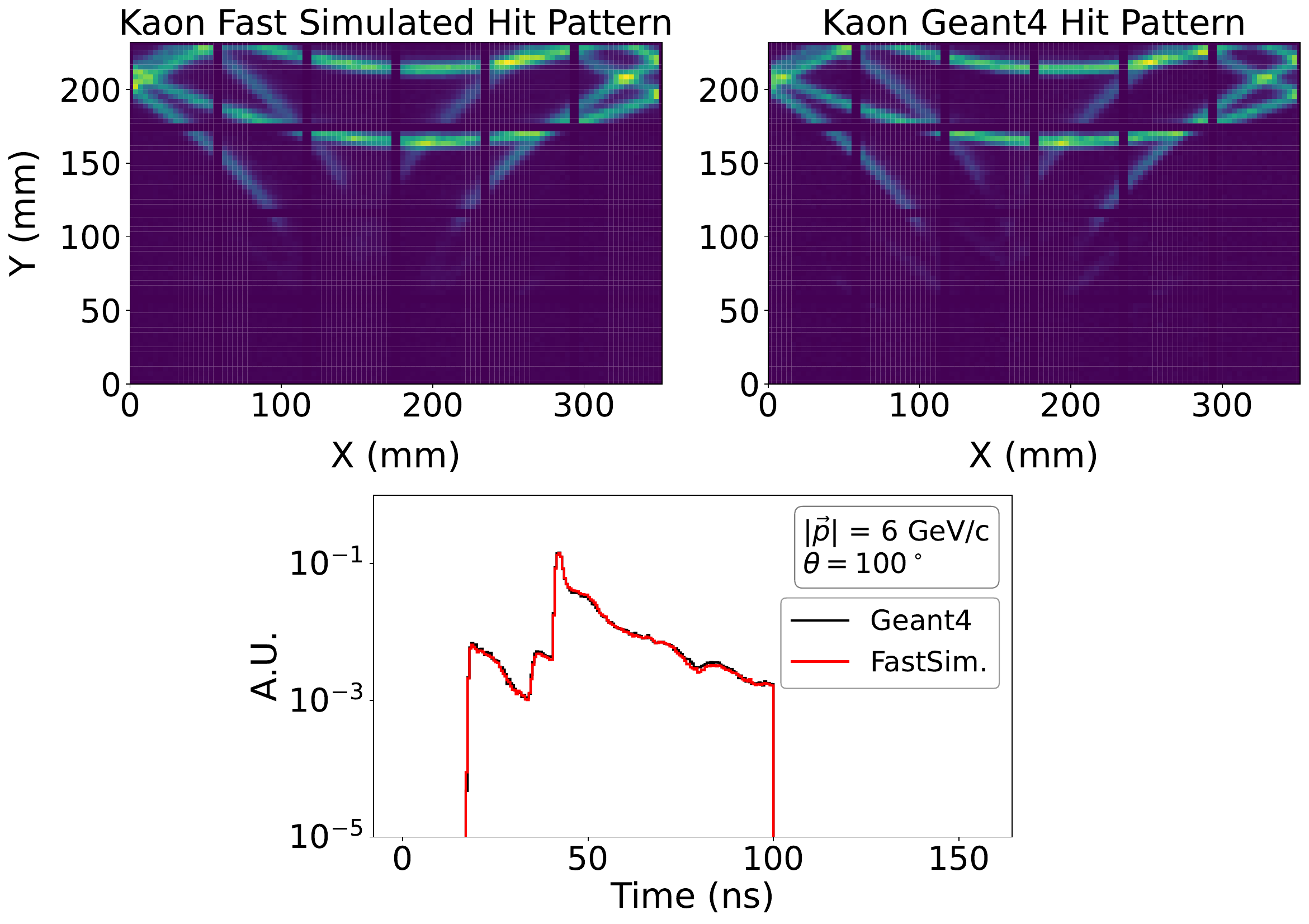} \\
        \includegraphics[width=0.49\textwidth]{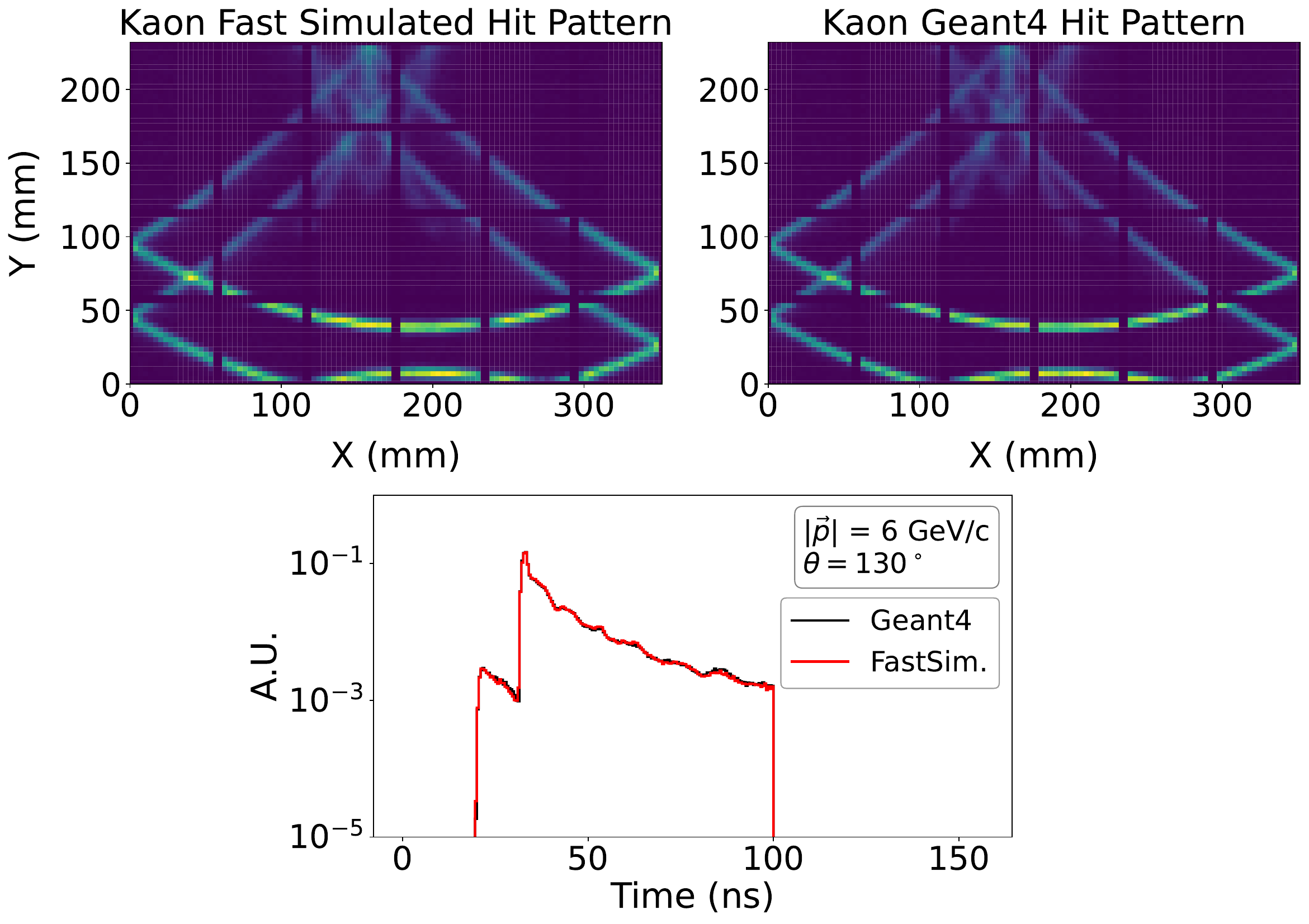}
        \caption{Kaons}
    \end{subfigure}
    \\
    \begin{subfigure}[b]{0.78\textwidth}
    \centering
        \includegraphics[width=0.49\textwidth]{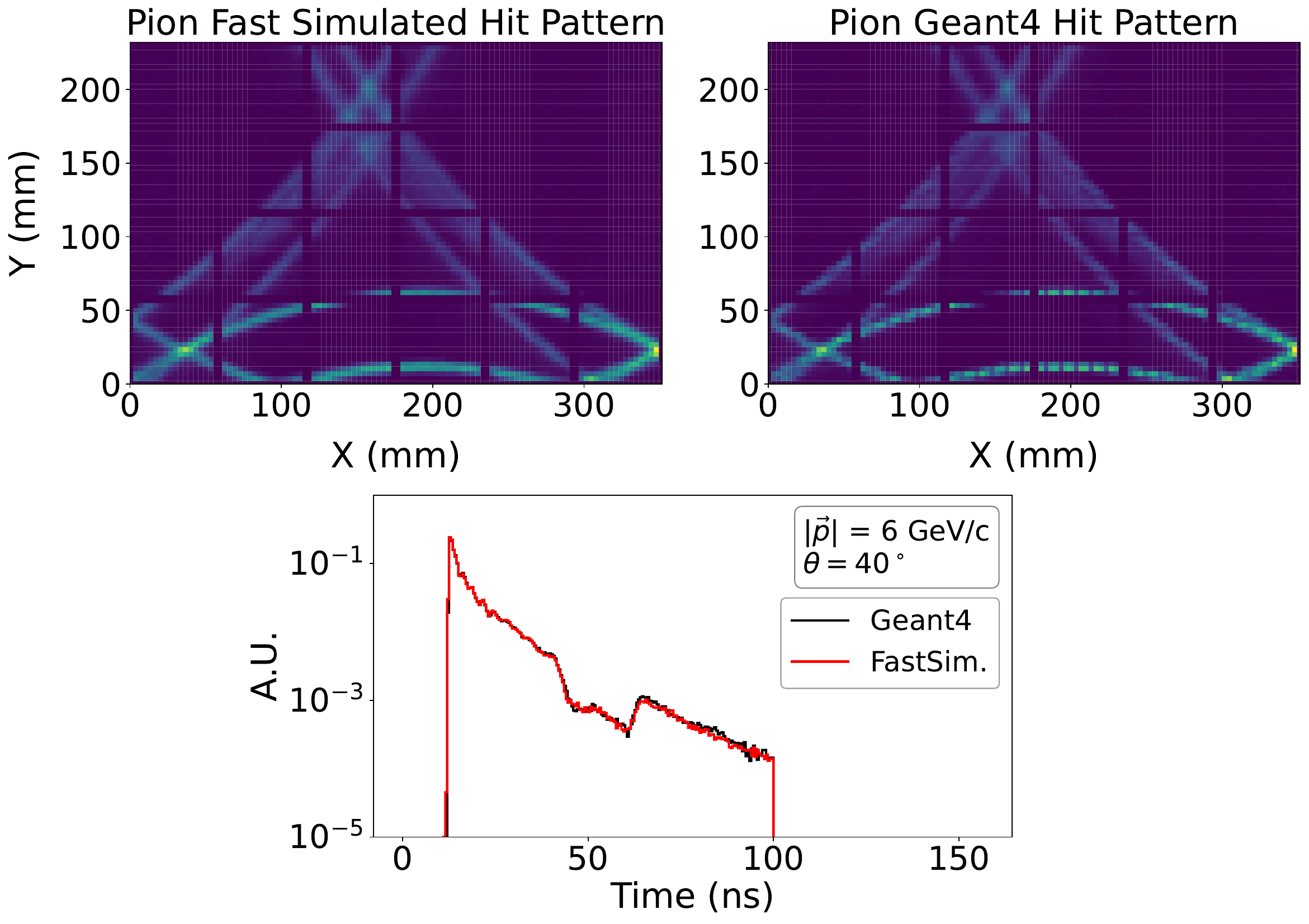} %
        \includegraphics[width=0.49\textwidth]{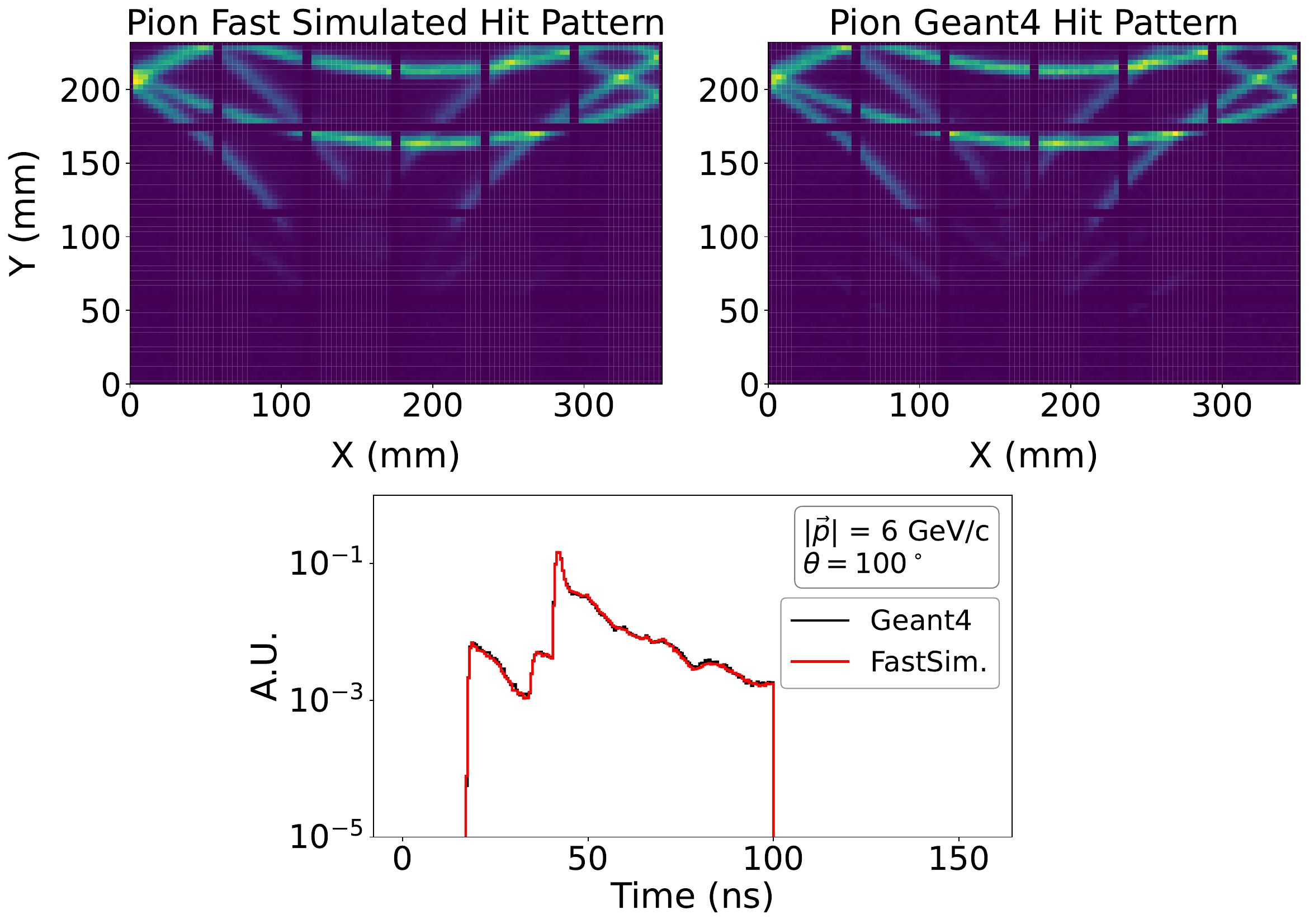} \\
        \includegraphics[width=0.49\textwidth]{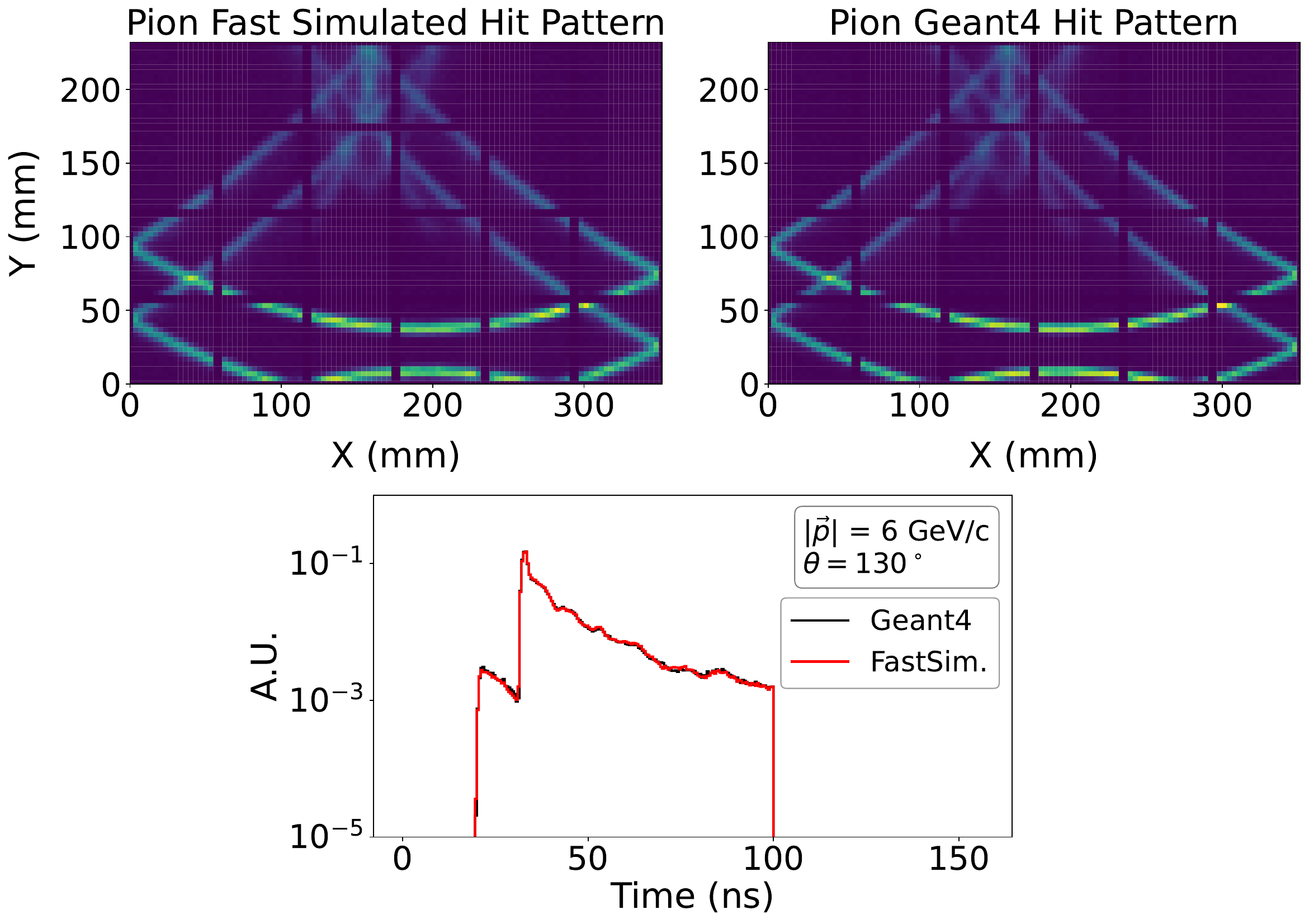}
        \caption{Pions}
    \end{subfigure}
    \caption{\textbf{Example Probability Density Functions for FastDIRC:} Comparison of Probability Density Functions (PDFs) produced by the Discrete Normalizing Flow (DNF) and \geant for (a) Kaons and (b) Pions at 6 GeV/c and various polar angles as indicated within the subplots. }
    \label{fig:example_PDFs}
\end{figure}

%% file: main.bbl
\providecommand{\newblock}{}
\begin{thebibliography}{10}
\expandafter\ifx\csname url\endcsname\relax
  \def\url#1{{\tt #1}}\fi
\expandafter\ifx\csname urlprefix\endcsname\relax\def\urlprefix{URL }\fi
\providecommand{\eprint}[2][]{\url{#2}}

\bibitem{Mikuni_2023}
Mikuni V, Nachman B and Pettee M {Fast point cloud generation with diffusion models in high energy physics} 2023 {\em Phys. Rev. D\/} \href{http://dx.doi.org/10.1103/PhysRevD.108.036025}{{\bf 108}(3) 036025}

\bibitem{araz2024point}
Araz J~Y, Mikuni V, Ringer F, Sato N, Acosta F~T and Whitehill R {Point cloud-based diffusion models for the Electron-Ion Collider} 2024 {\em arXiv preprint arXiv:2410.22421\/}

\bibitem{mikuni2024omnilearn}
Mikuni V and Nachman B {OmniLearn: A method to simultaneously facilitate all jet physics tasks} 2024 {\em arXiv preprint arXiv:2404.16091\/}

\bibitem{birk2024omnijet}
Birk J, Hallin A and Kasieczka G {OmniJet-$\alpha$: the first cross-task foundation model for particle physics} 2024 {\em Machine Learning: Science and Technology\/} {\bf 5} 035031

\bibitem{birk2025flow}
Birk J, Buhmann E, Ewen C, Kasieczka G and Shih D {Flow matching beyond kinematics: Generating jets with particle identification and trajectory displacement information} 2025 {\em Physical Review D\/} {\bf 111} 052008

\bibitem{Buhmann_2023}
Buhmann E, Kasieczka G and Thaler J {EPiC-GAN: Equivariant point cloud generation for particle jets} 2023 {\em SciPost Phys.\/} \href{http://dx.doi.org/10.21468/SciPostPhys.15.4.130}{{\bf 15} 130}

\bibitem{birk2025omnijet}
Birk J, Gaede F, Hallin A, Kasieczka G, Mozzanica M and Rose H {OmniJet-$\{\alpha_C\}$: Learning point cloud calorimeter simulations using generative transformers} 2025 {\em arXiv preprint arXiv:2501.05534\/}

\bibitem{Krause_2023}
Krause C and Shih D {Fast and accurate simulations of calorimeter showers with normalizing flows} 2023 {\em Physical Review D\/} \href{http://dx.doi.org/10.1103/physrevd.107.113003}{{\bf 107}} ISSN 2470-0029

\bibitem{Krause_2023_accel}
Krause C and Shih D {Accelerating accurate simulations of calorimeter showers with normalizing flows and probability density distillation} 2023 {\em Phys. Rev. D\/} \href{http://dx.doi.org/10.1103/PhysRevD.107.113004}{{\bf 107}(11) 113004}

\bibitem{favaro2025calodream}
Favaro L, Ore A, Palacios~Schweitzer S and Plehn T {CaloDREAM--Detector response emulation via attentive flow matching} 2025 {\em SciPost Physics\/} {\bf 18} 088

\bibitem{Paganini_2018}
Paganini M, de~Oliveira L and Nachman B {CaloGAN: Simulating 3D high energy particle showers in multilayer electromagnetic calorimeters with generative adversarial networks} 2018 {\em Physical Review D\/} \href{http://dx.doi.org/10.1103/physrevd.97.014021}{{\bf 97}} ISSN 2470-0029

\bibitem{Mikuni_2022}
Mikuni V and Nachman B {Score-based generative models for calorimeter shower simulation} 2022 {\em Physical Review D\/} \href{http://dx.doi.org/10.1103/physrevd.106.092009}{{\bf 106}} ISSN 2470-0029

\bibitem{Diefenbacher_2023}
Diefenbacher S, Eren E, Gaede F, Kasieczka G, Krause C, Shekhzadeh I and Shih D {L2LFlows: generating high-fidelity 3D calorimeter images} 2023 {\em Journal of Instrumentation\/} \href{http://dx.doi.org/10.1088/1748-0221/18/10/p10017}{{\bf 18} P10017} ISSN 1748-0221

\bibitem{fanelli2022flux+}
Fanelli C, Giroux J and Papandreou Z {‘Flux+ Mutability’: a conditional generative approach to one-class classification and anomaly detection} 2022 {\em Machine Learning: Science and Technology\/} {\bf 3} 045012

\bibitem{devlin2024diffusion}
Devlin P, Qiu J~W, Ringer F and Sato N {Diffusion model approach to simulating electron-proton scattering events} 2024 {\em Physical Review D\/} {\bf 110} 016030

\bibitem{fanelli2024deep}
Fanelli C, Giroux J and Stevens J {Deep er) reconstruction of imaging Cherenkov detectors with swin transformers and normalizing flow models} 2025 {\em Machine Learning: Science and Technology\/} \href{http://dx.doi.org/10.1088/2632-2153/ada8f4}{{\bf 6} 015028}

\bibitem{maevskiy2020fast}
Maevskiy A, Derkach D, Kazeev N, Ustyuzhanin A, Artemev M, Anderlini L, Collaboration L {\em et~al.\/} 2020 {Fast data-driven simulation of cherenkov detectors using generative adversarial networks} Journal of Physics: Conference Series vol 1525 (IOP Publishing) p 012097

\bibitem{kalicy2024high}
Kalicy G {The high-performance DIRC for the ePIC detector at the EIC} 2024 {\em Nuclear Instruments and Methods in Physics Research Section A: Accelerators, Spectrometers, Detectors and Associated Equipment\/} \href{http://dx.doi.org/10.1016/j.nima.2024.169168}{ 169168}

\bibitem{Dzhygadlo_2020}
Dzhygadlo R {\em et~al.\/} {Time imaging reconstruction for the PANDA Barrel DIRC} 2020 {\em Journal of Instrumentation\/} \href{http://dx.doi.org/10.1088/1748-0221/15/09/C09050}{{\bf 15} C09050}

\bibitem{hardin2016fastdirc}
Hardin J and Williams M {FastDIRC: a fast Monte Carlo and reconstruction algorithm for DIRC detectors} 2016 {\em J. Instrum.\/} \href{http://dx.doi.org/10.1088/1748-0221/11/10/P10007}{{\bf 11} P10007} (arXiv:\href{https://arxiv.org/abs/1608.01180}{{\tt 1608.01180}})

\bibitem{fanelli2020deeprich}
Fanelli C and Pomponi J {DeepRICH: learning deeply Cherenkov detectors} 2020 {\em Machine Learning: Science and Technology\/} \href{http://dx.doi.org/10.1088/2632-2153/ab845a}{{\bf 1} 015010}

\bibitem{eicdirc}
Dzhygadlo R 2024 {EICDIRC} accessed: 2025-03-22 \urlprefix\url{https://github.com/rdom/eicdirc}

\bibitem{Kalicy_2020}
Kalicy G {Developing high-performance DIRC detector for the future Electron Ion Collider experiment} 2020 {\em Journal of Instrumentation\/} \href{http://dx.doi.org/10.1088/1748-0221/15/11/C11006}{{\bf 15} C11006}

\bibitem{Affine}
Dinh L, Sohl-Dickstein J and Bengio S 2017 {Density estimation using Real NVP}

\bibitem{he2016deep}
He K, Zhang X, Ren S and Sun J 2016 {Deep residual learning for image recognition} Proceedings of the IEEE conference on computer vision and pattern recognition pp 770--778

\bibitem{NEURIPS2018_69386f6b}
Chen R~T~Q, Rubanova Y, Bettencourt J and Duvenaud D~K 2018 {Neural Ordinary Differential Equations} Advances in Neural Information Processing Systems vol~31 ed Bengio S, Wallach H, Larochelle H, Grauman K, Cesa-Bianchi N and Garnett R (Curran Associates, Inc.)

\bibitem{Skilling1989}
Skilling J 1989 {\em The Eigenvalues of Mega-dimensional Matrices\/} Maximum Entropy and Bayesian Methods: Cambridge, England, 1988 (Dordrecht: Springer Netherlands) \href{http://dx.doi.org/10.1007/978-94-015-7860-8_48}{pp 455--466} ISBN 978-94-015-7860-8

\bibitem{Hutchinson01011990}
Hutchinson M A stochastic estimator of the trace of the influence matrix for laplacian smoothing splines 1990 {\em Communications in Statistics - Simulation and Computation\/} \href{http://dx.doi.org/10.1080/03610919008812866}{{\bf 19} 433--450} (arXiv:\href{https://arxiv.org/abs/https://doi.org/10.1080/03610919008812866}{{\tt https://doi.org/10.1080/03610919008812866}})

\bibitem{onken2021otflowfastaccuratecontinuous}
Onken D, Fung S~W, Li X and Ruthotto L 2021 {OT-Flow: Fast and Accurate Continuous Normalizing Flows via Optimal Transport} (arXiv:\href{https://arxiv.org/abs/2006.00104}{{\tt 2006.00104}}) \urlprefix\url{https://arxiv.org/abs/2006.00104}

\bibitem{evans_2013}
Evans L An introduction to mathematical optimal control theory version 0.2 2013

\bibitem{grathwohl2018ffjordfreeformcontinuousdynamics}
Grathwohl W, Chen R~T~Q, Bettencourt J, Sutskever I and Duvenaud D 2018 {FFJORD: Free-form Continuous Dynamics for Scalable Reversible Generative Models} (arXiv:\href{https://arxiv.org/abs/1810.01367}{{\tt 1810.01367}}) \urlprefix\url{https://arxiv.org/abs/1810.01367}

\bibitem{lipman2023flowmatchinggenerativemodeling}
Lipman Y, Chen R~T~Q, Ben-Hamu H, Nickel M and Le M 2023 {Flow Matching for Generative Modeling} (arXiv:\href{https://arxiv.org/abs/2210.02747}{{\tt 2210.02747}}) \urlprefix\url{https://arxiv.org/abs/2210.02747}

\bibitem{tong2024improvinggeneralizingflowbasedgenerative}
Tong A, Fatras K, Malkin N, Huguet G, Zhang Y, Rector-Brooks J, Wolf G and Bengio Y 2024 {Improving and generalizing flow-based generative models with minibatch optimal transport} (arXiv:\href{https://arxiv.org/abs/2302.00482}{{\tt 2302.00482}}) \urlprefix\url{https://arxiv.org/abs/2302.00482}

\bibitem{sohldickstein2015deepunsupervisedlearningusing}
Sohl-Dickstein J, Weiss E~A, Maheswaranathan N and Ganguli S 2015 {Deep Unsupervised Learning using Nonequilibrium Thermodynamics} (arXiv:\href{https://arxiv.org/abs/1503.03585}{{\tt 1503.03585}}) \urlprefix\url{https://arxiv.org/abs/1503.03585}

\bibitem{ho2020denoisingdiffusionprobabilisticmodels}
Ho J, Jain A and Abbeel P 2020 {Denoising Diffusion Probabilistic Models} (arXiv:\href{https://arxiv.org/abs/2006.11239}{{\tt 2006.11239}}) \urlprefix\url{https://arxiv.org/abs/2006.11239}

\bibitem{song2021scorebasedgenerativemodelingstochastic}
Song Y, Sohl-Dickstein J, Kingma D~P, Kumar A, Ermon S and Poole B 2021 Score-based generative modeling through stochastic differential equations (arXiv:\href{https://arxiv.org/abs/2011.13456}{{\tt 2011.13456}}) \urlprefix\url{https://arxiv.org/abs/2011.13456}

\bibitem{khalek2022science}
Khalek R~A {\em et~al.\/} {Science requirements and detector concepts for the electron-ion collider: EIC yellow report} 2022 {\em Nuclear Physics A\/} \href{http://dx.doi.org/10.1016/j.nuclphysa.2022.122447}{{\bf 1026} 122447}

\bibitem{va2008simulation}
Va'vra J 2008 {Simulation of the focusing DIRC optics with Mathematica} 2008 IEEE Nuclear Science Symposium Conference Record (IEEE) pp 2408--2412

\bibitem{va2014optical}
Va’vra J {Optical properties of RICH detectors} 2014 {\em Nuclear Instruments and Methods in Physics Research Section A: Accelerators, Spectrometers, Detectors and Associated Equipment\/} {\bf 766} 189--198

\end{thebibliography}
